\documentclass[letterpaper,english,reprint, aps, onecolumn, notitlepage]{revtex4-1}
\usepackage[T1]{fontenc}
\usepackage[latin9]{inputenc}
\usepackage{amsmath}
\usepackage{amssymb}
\usepackage{cancel}

\makeatletter

\pdfpageheight\paperheight
\pdfpagewidth\paperwidth

\usepackage[colorlinks]{hyperref}
\usepackage{xcolor}
\definecolor{linkcolor}{RGB}{0,83,166}
\hypersetup{colorlinks=true, allcolors={blue}}

\makeatother

\usepackage{babel}
\begin{document}
\preprint{APS/123-QED}
\title{Essentially exact numerical modelling of flux qubit chains subject
to charge and flux noise}
\author{Matthew R.C.~Fitzpatrick}
\email{mrfitzpa@sfu.ca}

\affiliation{Department of Physics, Simon Fraser University, 8888 University Drive,
Burnaby, British Columbia V5A 1S6, Canada}
\affiliation{D-Wave Systems, 3033 Beta Avenue, Burnaby, BC V5G 4M9, Canada}
\author{Jack Raymond}
\affiliation{D-Wave Systems, 3033 Beta Avenue, Burnaby, BC V5G 4M9, Canada }
\author{Malcolm P. Kennett}
\affiliation{Department of Physics, Simon Fraser University, 8888 University Drive,
Burnaby, British Columbia V5A 1S6, Canada}
\date{\today}
\begin{abstract}
We present an essentially exact numerical method for modelling flux
qubit chains subject to charge and flux noise. We define an essentially
exact method as one that introduces errors that are completely controlled
such that they can be made arbitrarily small by tuning the simulation
parameters. The method adopts the quasi-adiabatic path integral formalism
to express the system's reduced density matrix as a time-discretized
path integral, comprising a series of influence functionals that encode
the non-Markovian dynamics of the system. We present a detailed derivation
of the path integral expression for the system's reduced density matrix
and describe in detail the tensor network algorithm used to evaluate
the path integral expression. We have implemented our method in an
open-sourced Python library called $\texttt{spinbosonchain}$ \citep{sbc1}.
When appropriate, we draw connections between concepts covered in
this manuscript and the library's code.
\end{abstract}

\maketitle

\section{Introduction\label{sec:Introduction}}

Quantum annealing has proven to be an effective approach to a variety
of optimization problems. However, outside the formalism of adiabatic
quantum computing, characterizing and predicting the performance of
quantum annealing has proven to be a great challenge. Accurate modelling
of environmental noise in quantum annealing is therefore of critical
importance to the design of future architectures. One of the primary
sources of noise in D-Wave quantum annealers is flux noise, which
can be decomposed into 1/f-like and ohmic-like components. For fast
anneals on the order of tens of nanoseconds, charge noise also becomes
appreciable. Recently, a method has been developed for simulating
flux qubit chains that are subject to generic flux noise \citep{Strathearn1,Suzuki1,Oshiyama1}.
By combining the quasi-adiabatic path integral (QUAPI) formalism with
tensor network (TN) techniques, the method is computationally efficient,
essentially exact, and captures non-Markovian effects of the flux
noise. We define an essentially exact method as one that introduces
errors that are completely controlled such that they can be made arbitrarily
small by tuning the simulation parameters \footnote{The computational resources required to achieve a particular target
error will depend on the system being studied. In some scenarios,
e.g. systems dominated by strong low-frequency noise, the target error
may scale poorly with computer memory and/or wall time, especially
if one is interested in long-time dynamics.}. While the QUAPI+TN method accurately models flux noise in D-Wave
devices, it does not capture the effects of charge noise.

In this manuscript, we expand on the works of Refs.~\citep{Strathearn1,Suzuki1,Oshiyama1}
by extending the QUAPI approach to include the effects of charge and
flux noise within the same formalism, and to modify the tensor network
algorithm accordingly. We have implemented this generalized approach
in a Python library called $\texttt{spinbosonchain}$ \citep{sbc1},
which makes heavy use of Google's $\texttt{TensorNetwork}$ package
\citep{TensorNetworks1} for its implementation of tensor networks
and related operations. This manuscript describes in detail the generalized
QUAPI formalism as well as the tensor network algorithms on which
$\texttt{spinbosonchain}$ is based. When appropriate, we draw connections
between some of the concepts covered in this manuscript to the library's
code. This manuscript does not describe in detail the application
programming interface of $\texttt{spinbosonchain}$: readers interested
in this can consult the $\texttt{spinbosonchain}$ reference guide
\citep{sbc1}.

This manuscript is structured as follows: In Sec.~\ref{sec:How to read sbc code},
we suggest how those interested in reviewing the code of the $\texttt{spinbosonchain}$
library should do so. In Sec.~\ref{sec:QUAPI formalism} we introduce
our extended QUAPI formalism used to describe the dynamics of a generalized
one-dimensional spin-boson chain model, where both the $z$- and $y$-components
of the quantum Ising spins are coupled to bosonic baths, rather than
only the $z$-components. The main result of this section is our time-discretized
path integral expression of the system's reduced state operator. In
Sec.~\ref{sec:Tensor network algorithm}, we develop the tensor network
algorithm used to evaluate the path integral, and discuss what quantities
can be calculated using our approach. Finally, in Sec.~\ref{sec:Conclusion},
we make some remarks about our method and present our conclusions.

\newpage{}

\section{How to read $\texttt{spinbosonchain}$ code\label{sec:How to read sbc code}}

For those that are interested in reviewing the code of the $\texttt{spinbosonchain}$
library in considerable detail, below is one suggested order for reading
the code.

\noindent $\texttt{spinbosonchain}$ installation instructions:
\begin{enumerate}
\item $\texttt{README.rst}$
\item $\texttt{INSTALL.rst}$
\end{enumerate}
\noindent $\texttt{spinbosonchain}$ source code files:
\begin{enumerate}
\item $\texttt{spinbosonchain/\_\_init\_\_.py}$
\item $\texttt{spinbosonchain/scalar.py}$
\begin{enumerate}
\item[Note:] Read first Sec.~\ref{subsec:Model}.
\end{enumerate}
\item $\texttt{spinbosonchain/system.py}$
\begin{enumerate}
\item[Note:] Read first suggested reading for $\texttt{spinbosonchain/scalar.py}$.
\end{enumerate}
\item $\texttt{spinbosonchain/bath.py}$
\begin{enumerate}
\item[Note:] Read first Sec.~\ref{sec:QUAPI formalism} and Appendix \ref{sec:Evaluating numerically the spectral densities of noise}.
\end{enumerate}
\item $\texttt{spinbosonchain/\_influence/\_\_init\_\_.py}$
\item $\texttt{spinbosonchain/\_influence/eta.py}$
\begin{enumerate}
\item[Note:] Read first suggested reading for $\texttt{spinbosonchain/bath.py}$;
read also Appendix~\ref{sec:How to evaluate eta -- 1}.
\end{enumerate}
\item $\texttt{spinbosonchain/\_base4.py}$
\begin{enumerate}
\item[Note:] Read first suggested reading for $\texttt{spinbosonchain/\_influence/eta.py}$;
read also Secs.~\ref{subsec:Base 4 variables -- 1} and \ref{subsec:tilde sigma variables -- 1}.
\end{enumerate}
\item $\texttt{spinbosonchain/\_influence/twopt.py}$
\begin{enumerate}
\item[Note:] Read first suggested reading for $\texttt{spinbosonchain/\_influence/\_base4.py}$;
read also Sec.~\ref{subsec:Rearranging terms in the path integral expression of rho^(A) -- 1}.
\end{enumerate}
\item $\texttt{spinbosonchain/\_influence/tensorfactory.py}$
\begin{enumerate}
\item[Note:] Read first suggested reading for $\texttt{spinbosonchain/\_influence/twopt.py}$;
read also Sec.~\ref{subsec:Tensor network representation of influence functional -- 1}.
\end{enumerate}
\item $\texttt{spinbosonchain/compress.py}$
\begin{enumerate}
\item[Note:] Read first suggested reading for $\texttt{spinbosonchain/\_influence/tensorfactory.py}$;
read also Appendix~\ref{sec:MPS compression techniques used in sbc}
and the references therein.
\end{enumerate}
\item $\texttt{spinbosonchain/\_backend.py}$
\item $\texttt{spinbosonchain/\_svd.py}$
\begin{enumerate}
\item[Note:] Read first suggested reading for $\texttt{spinbosonchain/\_influence/compress.py}$.
\end{enumerate}
\item $\texttt{spinbosonchain/\_qr.py}$
\begin{enumerate}
\item[Note:] Read first suggested reading for $\texttt{spinbosonchain/\_svd.py}$.
\end{enumerate}
\item $\texttt{spinbosonchain/\_arnoldi.py}$
\begin{enumerate}
\item[Note:] Read first suggested reading for $\texttt{spinbosonchain/\_qr.py}$;
read also Ref.~\citep{Lehoucq1}.
\end{enumerate}
\item $\texttt{spinbosonchain/\_mpomps.py}$
\begin{enumerate}
\item[Note:] Read first suggested reading for $\texttt{spinbosonchain/\_arnoldi.py}$.
\end{enumerate}
\item $\texttt{spinbosonchain/\_influence/path.py}$
\begin{enumerate}
\item[Note:] Read first suggested reading for $\texttt{spinbosonchain/\_mpomps.py}$.
\end{enumerate}
\item $\texttt{spinbosonchain/phasefactor/\_\_init\_\_.py}$
\item $\texttt{spinbosonchain/\_phasefactor/tensorfactory.py}$
\begin{enumerate}
\item[Note:] Read first suggested reading for $\texttt{spinbosonchain/\_influence/path.py}$;
read also Secs.~\ref{subsec:Rearranging terms in the path integral expression of rho^(A) further -- 1}-\ref{subsec:MPS representation of e^(i phi) -- 1}.
\end{enumerate}
\item $\texttt{spinbosonchain/alg.py}$
\begin{enumerate}
\item[Note:] Read first suggested reading for $\texttt{spinbosonchain/\_phasefactor/tensorfactory.py}$;
read also Secs.~\ref{subsec:MPO representations of U quantities -- 1}
and \ref{subsec:Remaining elements of tensor network algorithm -- 1}.
\end{enumerate}
\item $\texttt{spinbosonchain/state.py}$
\begin{enumerate}
\item[Note:] Read first suggested reading for $\texttt{spinbosonchain/alg.py}$;
read also Secs.~\ref{subsec:Computable quantities -- 1}.
\end{enumerate}
\item $\texttt{spinbosonchain/ev.py}$
\begin{enumerate}
\item[Note:] Read first suggested reading for $\texttt{spinbosonchain/state.py}$.
\end{enumerate}
\item $\texttt{spinbosonchain/report.py}$
\begin{enumerate}
\item[Note:] Read first suggested reading for $\texttt{spinbosonchain/ev.py}$.
\end{enumerate}
\item $\texttt{spinbosonchain/version.py}$
\end{enumerate}
\newpage{}

\section{QUAPI formalism\label{sec:QUAPI formalism}}

\subsection{Model\label{subsec:Model}}

$\texttt{spinbosonchain}$ is a library for simulating the dynamics
of a generalized one-dimensional spin-boson chain model, where both
the $z$- and $y$-components of the quantum Ising spins are coupled
to bosonic baths, rather than only the $z$-components. A convenient
way to discuss both finite and infinite chains is to express the Hamiltonian
of the aforementioned spin-boson model as a sum of 2$N$+1 \textquoteleft unit
cell\textquoteright{} Hamiltonians:

\begin{equation}
\hat{H}\left(t\right)\equiv\sum_{u=-N}^{N}\hat{H}_{u}\left(t\right),\label{eq:total Hamiltonian -- 1}
\end{equation}

\noindent where $N$ is a non-negative integer, and $\hat{H}_{u}\left(t\right)$
is the Hamiltonian of the $u^{\text{th}}$ unit cell of the model:

\begin{equation}
\hat{H}_{u}\left(t\right)=\hat{H}_{u}^{\left(A\right)}\left(t\right)+\hat{H}_{u}^{\left(B\right)}+\hat{H}_{u}^{\left(AB\right)}\left(t\right),\label{eq:total Hamiltonian -- 2}
\end{equation}

\noindent with $\hat{H}_{u}^{\left(A\right)}\left(t\right)$ being
the system part of $\hat{H}_{u}\left(t\right)$, which encodes all
information regarding energies associated exclusively with the spins;
$\hat{H}_{u}^{\left(B\right)}$ being the bath part of $\hat{H}_{u}\left(t\right)$,
which encodes all information regarding energies associated with the
components of the bosonic environment; and $\hat{H}_{u}^{\left(AB\right)}\left(t\right)$
is the system-bath coupling part of $\hat{H}_{u}\left(t\right)$,
which describes all energies associated with the coupling between
the system and the environment.

$\hat{H}_{u}^{\left(A\right)}\left(t\right)$ is simply an instance
of the one-dimensional transverse-field Ising model:

\begin{equation}
\hat{H}_{u}^{\left(A\right)}\left(t\right)=\hat{H}_{u;x}^{\left(A\right)}\left(t\right)+\hat{H}_{u;z}^{\left(A\right)}\left(t\right),\label{eq:H_u^(A) -- 1}
\end{equation}

\begin{equation}
\hat{H}_{u;x}^{\left(A\right)}\left(t\right)\equiv\sum_{r=0}^{L-1}h_{x;r}\left(t\right)\hat{\sigma}_{x;r+uL},\label{eq:H_=00007Bu; x=00007D^(A) -- 1}
\end{equation}

\begin{equation}
\hat{H}_{u;z}^{\left(A\right)}\left(t\right)\equiv\sum_{r=0}^{L-1}h_{z;r}\left(t\right)\hat{\sigma}_{z;r+uL}+\sum_{r=0}^{L-1}J_{z,z;r,r+1}\left(t\right)\hat{\sigma}_{z;r+uL}\hat{\sigma}_{z;r+1+uL},\label{eq:H_=00007Bu; z=00007D^(A) -- 1}
\end{equation}

\noindent with $\hat{\sigma}_{\nu;r}$ being the $\nu^{\text{th}}$
Pauli operator acting on site $r$, $h_{x;r}\left(t\right)$ being
the transverse field energy scale for site $r$ at time $t$, $h_{z;r}\left(t\right)$
being the longitudinal field energy scale for site $r$ at time $t$,
$J_{z,z;r,r+1}\left(t\right)$ being the longitudinal coupling energy
scale for sites $r$ and $r+1$ at time $t$, and $L$ being the number
of sites in every unit cell. 

In $\texttt{spinbosonchain}$, the system's model parameters $h_{x;r}\left(t\right)$,
$h_{z;r}\left(t\right)$, and $J_{z,z;r,r+1}\left(t\right)$ are represented
by \\
$\texttt{spinbosonchain.scalar.Scalar}$ objects. Generally speaking,
the $\texttt{spinbosonchain.scalar.Scalar}$ class represents time-dependent
scalar model parameters. The system's model parameter set is represented
by the \\
$\texttt{spinbosonchain.system.Model}$ class.

\noindent $\hat{H}_{u}^{\left(B\right)}$ describes a collection of
decoupled harmonic oscillators:

\begin{equation}
\hat{H}_{u}^{\left(B\right)}=\sum_{\nu\in\left\{ y,z\right\} }\hat{H}_{u;\nu}^{\left(B\right)},\label{eq:H_u^(B) -- 1}
\end{equation}

\begin{equation}
\hat{H}_{u;\nu}^{\left(B\right)}=\sum_{r=0}^{L-1}\sum_{\epsilon}\omega_{\nu;\epsilon}\hat{b}_{\nu;r+uL;\epsilon}^{\dagger}\hat{b}_{\nu;r+uL;\epsilon}^{\vphantom{\dagger}},\label{eq:H_=00007Bu; v=00007D^(B) -- 1}
\end{equation}

\noindent with $\epsilon$ being the oscillator mode index, $\hat{b}_{\nu;r;\epsilon}^{\dagger}$
and $\hat{b}_{\nu;r;\epsilon}^{\vphantom{\dagger}}$ being the bosonic
creation and annihilation operators respectively for the harmonic
oscillator at site $r$ in mode $\epsilon$ with angular frequency
$\omega_{\nu;\epsilon}>0$ coupled to the $\nu^{\text{th}}$ component
of the spin at the same site.

\noindent $\hat{H}_{u}^{\left(AB\right)}\left(t\right)$ is given
by

\begin{equation}
\hat{H}_{u}^{\left(AB\right)}\left(t\right)=\hat{H}_{u;y}^{\left(AB\right)}\left(t\right)+\hat{H}_{u;z}^{\left(AB\right)}\left(t\right),\label{eq:H_u^(AB) -- 1}
\end{equation}

\begin{equation}
\hat{H}_{u;\nu}^{\left(AB\right)}\left(t\right)=-\sum_{r=0}^{L-1}\hat{\sigma}_{\nu;r+uL}\hat{\mathcal{Q}}_{\nu;r+uL}\left(t\right),\label{eq:H_=00007Bu; v=00007D^(AB) -- 1}
\end{equation}

\noindent where $\hat{\mathcal{Q}}_{\nu;r}\left(t\right)$ is the
generalized reservoir force operator at site $r$ that acts on the
$\nu^{\text{th}}$ component of the spin at the same site:

\begin{equation}
\hat{\mathcal{Q}}_{\nu;r+uL}\left(t\right)=\mathcal{E}_{v;r}^{\left(\lambda\right)}\left(t\right)\hat{Q}_{\nu;r+uL},\label{eq:mathcal Q_=00007Bv; r=00007D(t) -- 1}
\end{equation}

\noindent with $\mathcal{E}_{v;r}^{\left(\lambda\right)}\left(t\right)$
being a time-dependent energy scale, $\hat{Q}_{\nu;r}$ being a rescaled
generalized reservoir force:

\begin{equation}
\hat{Q}_{\nu;r+uL}=-\sum_{\epsilon}\lambda_{\nu;r;\epsilon}\left\{ \hat{b}_{\nu;r+uL;\epsilon}^{\dagger}+\hat{b}_{\nu;r+uL;\epsilon}^{\vphantom{\dagger}}\right\} ,\label{eq:Q_=00007Bv; r=00007D -- 1}
\end{equation}

\noindent with $\lambda_{\nu;\epsilon}$ being the coupling strength
between the $\nu^{\text{th}}$ component of the spin at site $r$
and the harmonic oscillator at the same site in mode $\epsilon$. 

Note that the bosonic creation and annihilations operators satisfy
the following commutation relation:

\begin{equation}
\left[\hat{b}_{\nu_{1};r_{1};\epsilon_{1}}^{\vphantom{\dagger}},\hat{b}_{\nu_{2};r_{2};\epsilon_{2}}^{\dagger}\right]=\delta_{\nu_{1},\nu_{2}}\delta_{r_{1},r_{2}}\delta_{\epsilon_{1},\epsilon_{2}}\hat{1}_{\nu_{1};r_{1};\epsilon_{1}}^{\left(\mathcal{B}\right)}\hat{1}_{\nu_{2};r_{2};\epsilon_{2}}^{\left(\mathcal{B}\right)},\label{eq:bosonic commutation relation -- 1}
\end{equation}

\noindent where $\hat{1}_{\nu;r;\epsilon}^{\left(\mathcal{B}\right)}$
is the identity operator of the Hilbert space associated with the
harmonic oscillator at site $r$ in mode $\epsilon$, coupled to the
$\nu^{\text{th}}$ component of the spin at the same site, and $\delta_{i,j}$
is the Kronecker delta function.

For finite chains, we set $N=0$ {[}see Eq.~\eqref{eq:total Hamiltonian -- 1}{]}
and $J_{z,z;L-1,L}\left(t\right)=0$ {[}see Eq.~\eqref{eq:H_=00007Bu; z=00007D^(A) -- 1}{]},
whereas for infinite chains, we take the limit of $N\to\infty$.

\subsection{Initial state of system\label{subsec:Initial state of system}}

We assume that the spin system and bath together are initially prepared
in a state at time $t=0$ corresponding to the state operator:

\begin{equation}
\hat{\rho}^{\left(i\right)}\equiv\hat{\rho}^{\left(i,A\right)}\otimes\hat{\rho}^{\left(i,B\right)},\label{eq:rho^(i) -- 1}
\end{equation}

\noindent where $\hat{\rho}^{\left(i,A\right)}$ is the system's reduced
state operator at $t=0$:

\begin{equation}
\hat{\rho}^{\left(i,A\right)}\equiv\left|\Psi^{\left(i,A\right)}\right\rangle \left\langle \Psi^{\left(i,A\right)}\right|,\label{eq:rho^(i, A) -- 1}
\end{equation}

\noindent with

\begin{equation}
\left|\Psi^{\left(i,A\right)}\right\rangle \equiv\bigotimes_{u=-N}^{N}\left|\Psi_{u}^{\left(i,A\right)}\right\rangle ,\label{eq:|Psi^(i, A)> -- 1}
\end{equation}

\noindent and $\left|\Psi_{u}^{\left(i,A\right)}\right\rangle $ being
a state vector describing the reduced state of the $u^{\text{th}}$
unit cell; $\hat{\rho}^{\left(i,B\right)}$ is the bath's reduced
state operator at $t=0$:

\begin{equation}
\hat{\rho}^{\left(i,B\right)}\equiv\bigotimes_{u=-N}^{N}\hat{\rho}_{u}^{\left(i,B\right)},\label{eq:rho^(i, B) -- 1}
\end{equation}

\begin{equation}
\hat{\rho}_{u}^{\left(i,B\right)}\equiv\frac{e^{-\beta\hat{H}_{u}^{\left(B\right)}}}{\mathcal{Z}_{u}^{\left(B\right)}},\label{eq:rho_u^(i, B) -- 1}
\end{equation}

\noindent where $\beta=1/\left(k_{B}T\right)$, $k_{B}$ is the Boltzmann
constant, $T$ is the temperature, $\mathcal{Z}_{u}^{\left(B\right)}$
is the partition function of the bath:

\begin{equation}
\mathcal{Z}_{u}^{\left(B\right)}=\text{Tr}^{\left(B\right)}\left\{ e^{-\beta\hat{H}_{u}^{\left(B\right)}}\right\} ,\label{eq:bosonic partition function -- 1}
\end{equation}
and $\text{Tr}^{\left(B\right)}\left\{ \cdots\right\} $ is the partial
trace with respect to the bath degrees of freedom. Using Eq.~\eqref{eq:bosonic commutation relation -- 1},
we can write:

\begin{equation}
\hat{\rho}_{u}^{\left(i,B\right)}=\hat{\rho}_{u;y}^{\left(i,B\right)}\hat{\rho}_{u;z}^{\left(i,B\right)},\label{eq:rho_u^(i, B) -- 2}
\end{equation}

\begin{equation}
\hat{\rho}_{u;\nu}^{\left(i,B\right)}=\frac{e^{-\beta\hat{H}_{u;\nu}^{\left(B\right)}}}{\mathcal{Z}_{u;\nu}^{\left(B\right)}},\label{eq:rho_=00007Bu; v=00007D^(i, B) -- 1}
\end{equation}

\noindent where

\begin{equation}
\mathcal{Z}_{u;\nu}^{\left(B\right)}=\text{Tr}_{\nu}^{\left(B\right)}\left\{ e^{-\beta\hat{H}_{u;\nu}^{\left(B\right)}}\right\} ,\label{eq:Z_=00007Bu; v=00007D^(B) -- 1}
\end{equation}

\noindent and $\text{Tr}_{\nu}^{\left(B\right)}\left\{ \cdots\right\} $
is the partial trace with respect to the bath degrees of freedom coupled
to the $\nu^{\text{th}}$ components of the spins. The trace $\text{Tr}_{\nu}^{\left(B\right)}\left\{ \cdots\right\} $
can be evaluated using the bosonic occupation number basis states
$\left|n_{\nu;r;\epsilon}\right\rangle $, where 

\begin{equation}
\hat{b}_{\nu;r;\epsilon}^{\dagger}\hat{b}_{\nu;r;\epsilon}^{\vphantom{\dagger}}\left|n_{\nu^{\prime};r^{\prime};\epsilon^{\prime}}\right\rangle =\begin{cases}
n_{\nu^{\prime};r^{\prime};\epsilon^{\prime}}\left|n_{\nu^{\prime};r^{\prime};\epsilon^{\prime}}\right\rangle , & \text{if }\nu=\nu^{\prime}\text{ \& }r=r^{\prime}\text{ \& }\epsilon=\epsilon^{\prime},\\
0, & \text{otherwise}.
\end{cases}\label{eq:introducing occupation basis states -- 1}
\end{equation}

\noindent Using Eqs.~\eqref{eq:H_=00007Bu; v=00007D^(B) -- 1}, and
\eqref{eq:introducing occupation basis states -- 1}, we can evaluate
Eq.~\eqref{eq:Z_=00007Bu; v=00007D^(B) -- 1} as follows:

\begin{align}
\mathcal{Z}_{u;\nu}^{\left(B\right)} & =\text{Tr}_{\nu}^{\left(B\right)}\left\{ e^{-\beta\hat{H}_{u;\nu}^{\left(B\right)}}\right\} \nonumber \\
 & =\text{Tr}_{\nu}^{\left(B\right)}\left\{ \prod_{r=0}^{L-1}\prod_{\epsilon}\left[e^{-\beta\omega_{\nu;\epsilon}\hat{b}_{\nu;r+uL;\epsilon}^{\dagger}\hat{b}_{\nu;r+uL;\epsilon}^{\vphantom{\dagger}}}\right]\right\} \nonumber \\
 & =\prod_{r=0}^{L-1}\prod_{\epsilon}\left[\sum_{n_{\nu;r+uL;\epsilon}=0}^{\infty}\right]\prod_{r=0}^{L-1}\prod_{\epsilon}\left[\left\langle n_{\nu;r+uL;\epsilon}\right|e^{-\beta\omega_{\nu;\epsilon}\hat{b}_{\nu;r+uL;\epsilon}^{\dagger}\hat{b}_{\nu;r+uL;\epsilon}^{\vphantom{\dagger}}}\left|n_{\nu;r+uL;\epsilon}\right\rangle \right]\nonumber \\
 & =\prod_{r=0}^{L-1}\prod_{\epsilon}\left[\sum_{n_{\nu;r+uL;\epsilon}=0}^{\infty}\left\langle n_{\nu;r+uL;\epsilon}\right|e^{-\beta\omega_{\nu;\epsilon}\hat{b}_{\nu;r+uL;\epsilon}^{\dagger}\hat{b}_{\nu;r+uL;\epsilon}^{\vphantom{\dagger}}}\left|n_{\nu;r+uL;\epsilon}\right\rangle \right]\nonumber \\
 & =\prod_{r=0}^{L-1}\prod_{\epsilon}\left[\sum_{n_{\nu;r+uL;\epsilon}=0}^{\infty}e^{-\beta\omega_{\nu;\epsilon}n_{\nu;r+uL;\epsilon}}\right]\nonumber \\
 & =\prod_{r=0}^{L-1}\prod_{\epsilon}\left[\sum_{n_{\nu;r+uL;\epsilon}=0}^{\infty}\left(e^{-\beta\omega_{\nu;\epsilon}}\right)^{n_{\nu;r+uL;\epsilon}}\right]\nonumber \\
 & =\prod_{r=0}^{L-1}\prod_{\epsilon}\left[\frac{1}{\left\{ 1-e^{-\beta\omega_{\nu;\epsilon}}\right\} }\right],\label{eq:Z_=00007Bu; v=00007D^(B) -- 2}
\end{align}

\noindent where in the last step we used the geometric series identity

\begin{equation}
\sum_{n=0}^{\infty}a^{n}=\frac{1}{1-a},\quad\text{for }0<\left|a\right|<1\label{eq:geometric series identity -- 1}
\end{equation}

\noindent and we introduced the $\Pi$ product notation. When limits
are present explicitly in the $\Pi$ product notation, the following
convention is adopted:

\begin{equation}
\prod_{l=l_{i}}^{l_{f}}\left\{ x_{l_{\vphantom{i}}}\right\} =\begin{cases}
x_{l_{i}}x_{l_{i}+1}\cdots x_{l_{f}-1}x_{l_{f}}, & \text{if }l_{f}\ge l_{i},\\
1, & \text{if }l_{f}<l_{i}.
\end{cases}\label{eq:capital Pi notation -- 1}
\end{equation}

\subsection{Evolution operator\label{subsec:Time evolution operator -- 1}}

The evolution operator $\hat{U}\left(t,t^{\prime}\right)$ can be
expressed as \citep{Stefanucci1}:

\begin{align}
\hat{U}\left(t,t^{\prime}\right) & \equiv\begin{cases}
T\left\{ e^{-i\int_{t^{\prime}}^{t}dt^{\prime\prime}\hat{H}\left(t^{\prime\prime}\right)}\right\} , & \text{if }t\ge t^{\prime},\\
\tilde{T}\left\{ e^{i\int_{t}^{t^{\prime}}dt^{\prime\prime}\hat{H}\left(t^{\prime\prime}\right)}\right\} , & \text{if }t<t^{\prime},
\end{cases}\nonumber \\
 & \equiv\begin{cases}
\sum_{n=0}^{\infty}\frac{\left(-i\right)^{n}}{n!}\prod_{m=1}^{n}\left\{ \int_{t^{\prime}}^{t}dt^{\left(m\right)}\right\} T\left\{ \prod_{m=1}^{n}\left[\hat{H}\left(t^{\left(m\right)}\right)\right]\right\} , & \text{if }t\ge t^{\prime},\\
\sum_{n=0}^{\infty}\frac{\left(i\right)^{n}}{n!}\prod_{m=1}^{n}\left\{ \int_{t}^{t^{\prime}}dt^{\left(m\right)}\right\} \tilde{T}\left\{ \prod_{m=1}^{n}\left[\hat{H}\left(t^{\left(m\right)}\right)\right]\right\} , & \text{if }t<t^{\prime},
\end{cases}\label{eq:U -- 1}
\end{align}

\noindent where $\hat{H}\left(t\right)$ is the total Hamiltonian
{[}see Eq.~\eqref{eq:total Hamiltonian -- 1}{]} $T\left\{ \cdots\right\} $
is the time ordering symbol, which specifies that the string of time-dependent
operators contained within $T\left\{ \cdots\right\} $ be rearranged
as a time-descending sequence, and $\tilde{T}\left\{ \cdots\right\} $
is the anti-time ordering symbol, which orders strings of time-dependent
operators in the reverse order with respect to $T\left\{ \cdots\right\} $.
For example, in the case of three operators, we have for $T\left\{ \cdots\right\} $

\begin{equation}
T\left\{ \hat{A}\left(t^{\left(1\right)}\right)\hat{B}\left(t^{\left(2\right)}\right)\hat{C}\left(t^{\left(3\right)}\right)\right\} =\begin{cases}
\hat{A}\left(t^{\left(1\right)}\right)\hat{B}\left(t^{\left(2\right)}\right)\hat{C}\left(t^{\left(3\right)}\right), & \text{if }t^{\left(1\right)}>t^{\left(2\right)}>t^{\left(3\right)},\\
\hat{A}\left(t^{\left(1\right)}\right)\hat{C}\left(t^{\left(3\right)}\right)\hat{B}\left(t^{\left(2\right)}\right), & \text{if }t^{\left(1\right)}>t^{\left(3\right)}>t^{\left(2\right)},\\
\hat{B}\left(t^{\left(2\right)}\right)\hat{A}\left(t^{\left(1\right)}\right)\hat{C}\left(t^{\left(3\right)}\right), & \text{if }t^{\left(2\right)}>t^{\left(1\right)}>t^{\left(3\right)},\\
\hat{B}\left(t^{\left(2\right)}\right)\hat{C}\left(t^{\left(3\right)}\right)\hat{A}\left(t^{\left(1\right)}\right), & \text{if }t^{\left(2\right)}>t^{\left(3\right)}>t^{\left(1\right)},\\
\hat{C}\left(t^{\left(3\right)}\right)\hat{A}\left(t^{\left(1\right)}\right)\hat{B}\left(t^{\left(2\right)}\right), & \text{if }t^{\left(3\right)}>t^{\left(1\right)}>t^{\left(2\right)},\\
\hat{C}\left(t^{\left(3\right)}\right)\hat{B}\left(t^{\left(2\right)}\right)\hat{A}\left(t^{\left(1\right)}\right), & \text{if }t^{\left(3\right)}>t^{\left(2\right)}>t^{\left(1\right)}.
\end{cases}\label{eq:time-ordering for three operators - 1}
\end{equation}

\noindent One can show by differentiating both sides of Eq.~\eqref{eq:U -- 1}
with respect to $t$ that $\hat{U}\left(t,t^{\prime}\right)$ satisfies
the following equations of motion \citep{Stefanucci1}:

\begin{equation}
\frac{\partial\hat{U}\left(t,t^{\prime}\right)}{\partial t}=-i\hat{H}\left(t\right)\hat{U}\left(t,t^{\prime}\right),\label{eq:U equation of motion -- 1}
\end{equation}

\begin{equation}
\frac{\partial\hat{U}\left(t^{\prime},t\right)}{\partial t}=i\hat{U}\left(t^{\prime},t\right)\hat{H}\left(t\right).\label{eq:U conjugate equation of motion -- 1}
\end{equation}

\subsection{Spectral densities of noise\label{sec:Spectral densities of noise}}

Rather than specify the bath model parameters $\omega_{\nu;\epsilon}$
and $\lambda_{\nu;r;\epsilon}$ {[}introduced in Eqs.~\eqref{eq:H_=00007Bu; v=00007D^(B) -- 1}
and \eqref{eq:Q_=00007Bv; r=00007D -- 1}{]}, one can alternatively
specify the spectral densities of noise coupled to the $y$- and $z$-components
of the system\textquoteright s spins at temperature $T$, which contain
the same information as the aforementioned model parameters. This
alternative approach is easier to incorporate into the QUAPI formalism,
upon which $\texttt{spinbosonchain}$ is based on.

We define the spectral density of the noise coupled to the $\nu$-component
of the spin at site $r$ and temperature $T$ as:

\begin{equation}
A_{\nu;r;T}\left(\omega\right)=\int_{-\infty}^{\infty}dt\,e^{i\omega t}C_{\nu;r;T}\left(t\right),\label{eq:A_=00007Bv; r; T=00007D(omega) in main manuscript -- 1}
\end{equation}

\noindent where $C_{\nu;r;T}\left(t\right)$ is the bath correlation
function of the noise coupled to the $\nu$-component of the spin
at site $r$ and temperature $T$:

\begin{equation}
C_{\nu;r;T}\left(t\right)=\text{Tr}^{\left(B\right)}\left\{ \hat{\rho}^{\left(i,B\right)}\hat{\mathcal{Q}}_{\nu;r}^{\left(B\right)}\left(t\right)\hat{\mathcal{Q}}_{\nu;r}^{\left(B\right)}\left(0\right)\right\} ,\label{eq:C_=00007Bv; r; T=00007D(t) in main manuscript -- 1}
\end{equation}

\noindent with $\hat{\mathcal{Q}}_{\nu;r}^{\left(B\right)}\left(t\right)$
being the rescaled reservoir force operator in the Heisenberg picture:

\begin{equation}
\hat{\mathcal{Q}}_{\nu;r}^{\left(B\right)}\left(t\right)=e^{i\hat{H}^{\left(B\right)}t}\hat{Q}_{\nu;r}e^{-i\hat{H}^{\left(B\right)}t},\label{eq:Q_=00007Bv; r=00007D^=00007B(B)=00007D(t) in main manuscript -- 1}
\end{equation}

\noindent $\hat{Q}_{\nu;r}$ being given by Eq.~\eqref{eq:Q_=00007Bv; r=00007D -- 1}, 

\begin{equation}
\hat{H}^{\left(B\right)}=\sum_{u=-N}^{N}\hat{H}_{u}^{\left(B\right)},\label{eq:H^(B) in main manuscript -- 1}
\end{equation}

\noindent $\hat{H}_{u}^{\left(B\right)}$ being given by Eq.~\eqref{eq:H_u^(B) -- 1},
and $\hat{\rho}^{\left(i,B\right)}$ being given by Eq.~\eqref{eq:rho^(i, B) -- 1}.
In other words, $A_{\nu;r;T}\left(\omega\right)$ is the Fourier transform
of $C_{\nu;r;T}\left(t\right)$.

Let us calculate $A_{\nu;r;T}\left(\omega\right)$ for the case of
the generalized spin-boson model defined in Sec.~\eqref{subsec:Model}.
First, we calculate $C_{\nu;r;T}\left(t\right)$ using Eqs.~\eqref{eq:H_u^(B) -- 1},
\eqref{eq:H_=00007Bu; v=00007D^(B) -- 1}, \eqref{eq:mathcal Q_=00007Bv; r=00007D(t) -- 1}-\eqref{eq:bosonic commutation relation -- 1},
\eqref{eq:rho^(i, B) -- 1}, \eqref{eq:rho_u^(i, B) -- 2}, \eqref{eq:rho_=00007Bu; v=00007D^(i, B) -- 1},
\eqref{eq:introducing occupation basis states -- 1}, \eqref{eq:Z_=00007Bu; v=00007D^(B) -- 2},
\eqref{eq:C_=00007Bv; r; T=00007D(t) in main manuscript -- 1}, \eqref{eq:Q_=00007Bv; r=00007D^=00007B(B)=00007D(t) in main manuscript -- 1}:

\begin{equation}
C_{\nu;r;T}\left(t\right)=\sum_{\epsilon}\lambda_{\nu;r;\epsilon}^{2}\left\{ 1-e^{-\beta\omega_{\nu;\epsilon}}\right\} \sum_{n=0}^{\infty}e^{-\beta\omega_{\nu;\epsilon}n}\left[ne^{i\omega_{\nu;\epsilon}t}+\left\{ n+1\right\} e^{-i\omega_{\nu;\epsilon}t}\right].\label{eq:C_=00007Bv; r; T=00007D(t) in main manuscript -- 2}
\end{equation}

\noindent Next, using Eqs.~\eqref{eq:geometric series identity -- 1},
\eqref{eq:A_=00007Bv; r; T=00007D(omega) in main manuscript -- 1},
and \eqref{eq:C_=00007Bv; r; T=00007D(t) in main manuscript -- 2},
we can evaluate $A_{\nu;r;T}\left(\omega\right)$ as follows:

\begin{align}
A_{\nu;r;T}\left(\omega\right) & =2\pi\sum_{\epsilon}\lambda_{\nu;r;\epsilon}^{2}\left\{ 1-e^{-\beta\omega_{\nu;\epsilon}}\right\} \sum_{n=0}^{\infty}e^{-\beta\omega_{\nu;\epsilon}n}\left[n\delta\left(\omega+\omega_{\nu;\epsilon}\right)\right.\nonumber \\
 & \phantom{\mathrel{=}2\pi\sum_{\epsilon}\lambda_{\nu;r;\epsilon}^{2}\left\{ 1-e^{-\beta\omega_{\nu;\epsilon}}\right\} \sum_{n=0}^{\infty}e^{-\beta\omega_{\nu;\epsilon}n}}\left.\quad+\left\{ n+1\right\} \delta\left(\omega-\omega_{\nu;\epsilon}\right)\right]\nonumber \\
 & =2\pi\sum_{\epsilon}\delta\left(\omega+\omega_{\nu;\epsilon}\right)\lambda_{\nu;r;\epsilon}^{2}\left\{ 1-e^{-\beta\omega_{\nu;\epsilon}}\right\} \sum_{n=0}^{\infty}e^{-\beta\omega_{\nu;\epsilon}n}n\nonumber \\
 & \mathrel{\phantom{=}}\mathop{+}2\pi\sum_{\epsilon}\delta\left(\omega-\omega_{\nu;\epsilon}\right)\lambda_{\nu;r;\epsilon}^{2}\left\{ 1-e^{-\beta\omega_{\nu;\epsilon}}\right\} \sum_{n=0}^{\infty}e^{-\beta\omega_{\nu;\epsilon}n}n\nonumber \\
 & \mathrel{\phantom{=}}\mathop{+}2\pi\sum_{\epsilon}\delta\left(\omega-\omega_{\nu;\epsilon}\right)\lambda_{\nu;r;\epsilon}^{2}\left\{ 1-e^{-\beta\omega_{\nu;\epsilon}}\right\} \sum_{n=0}^{\infty}e^{-\beta\omega_{\nu;\epsilon}n}\nonumber \\
 & =2\pi\sum_{\epsilon}\delta\left(\omega+\omega_{\nu;\epsilon}\right)\lambda_{\nu;r;\epsilon}^{2}\left\{ 1-e^{-\beta\omega_{\nu;\epsilon}}\right\} \left\{ -\frac{1}{\beta}\frac{\partial}{\partial\omega_{\nu;\epsilon}}\left[\sum_{n=0}^{\infty}e^{-\beta\omega_{\nu;\epsilon}n}\right]\right\} \nonumber \\
 & \mathrel{\phantom{=}}\mathop{+}2\pi\sum_{\epsilon}\delta\left(\omega-\omega_{\nu;\epsilon}\right)\lambda_{\nu;r;\epsilon}^{2}\left\{ 1-e^{-\beta\omega_{\nu;\epsilon}}\right\} \left\{ -\frac{1}{\beta}\frac{\partial}{\partial\omega_{\nu;\epsilon}}\left[\sum_{n=0}^{\infty}e^{-\beta\omega_{\nu;\epsilon}n}\right]\right\} \nonumber \\
 & \mathrel{\phantom{=}}\mathop{+}2\pi\sum_{\epsilon}\delta\left(\omega-\omega_{\nu;\epsilon}\right)\lambda_{\nu;r;\epsilon}^{2}\left\{ 1-e^{-\beta\omega_{\nu;\epsilon}}\right\} \left\{ \sum_{n=0}^{\infty}e^{-\beta\omega_{\nu;\epsilon}n}\right\} \nonumber \\
 & =2\pi\sum_{\epsilon}\delta\left(\omega+\omega_{\nu;\epsilon}\right)\lambda_{\nu;r;\epsilon}^{2}\left\{ 1-e^{-\beta\omega_{\nu;\epsilon}}\right\} \left\{ -\frac{1}{\beta}\frac{\partial}{\partial\omega_{\nu;\epsilon}}\left[\frac{1}{1-e^{-\beta\omega_{\nu;\epsilon}}}\right]\right\} \nonumber \\
 & \mathrel{\phantom{=}}\mathop{+}2\pi\sum_{\epsilon}\delta\left(\omega-\omega_{\nu;\epsilon}\right)\lambda_{\nu;r;\epsilon}^{2}\left\{ 1-e^{-\beta\omega_{\nu;\epsilon}}\right\} \left\{ -\frac{1}{\beta}\frac{\partial}{\partial\omega_{\nu;\epsilon}}\left[\frac{1}{1-e^{-\beta\omega_{\nu;\epsilon}}}\right]\right\} \nonumber \\
 & \mathrel{\phantom{=}}\mathop{+}2\pi\sum_{\epsilon}\delta\left(\omega-\omega_{\nu;\epsilon}\right)\lambda_{\nu;r;\epsilon}^{2}\left\{ 1-e^{-\beta\omega_{\nu;\epsilon}}\right\} \left\{ \frac{1}{1-e^{-\beta\omega_{\nu;\epsilon}}}\right\} \nonumber \\
 & =2\pi\sum_{\epsilon}\delta\left(\omega+\omega_{\nu;\epsilon}\right)\lambda_{\nu;r;\epsilon}^{2}\cancel{\left\{ 1-e^{-\beta\omega_{\nu;\epsilon}}\right\} }\left\{ -\frac{1}{\cancel{\beta}}\left[-\frac{\left(\cancel{\beta}e^{-\beta\omega_{\nu;\epsilon}}\right)}{\left(1-e^{-\beta\omega_{\nu;\epsilon}}\right)^{\cancel{2}}}\right]\right\} \nonumber \\
 & \mathrel{\phantom{=}}\mathop{+}2\pi\sum_{\epsilon}\delta\left(\omega-\omega_{\nu;\epsilon}\right)\lambda_{\nu;r;\epsilon}^{2}\cancel{\left\{ 1-e^{-\beta\omega_{\nu;\epsilon}}\right\} }\left\{ -\frac{1}{\cancel{\beta}}\left[-\frac{\left(\cancel{\beta}e^{-\beta\omega_{\nu;\epsilon}}\right)}{\left(1-e^{-\beta\omega_{\nu;\epsilon}}\right)^{\cancel{2}}}\right]\right\} \nonumber \\
 & \mathrel{\phantom{=}}\mathop{+}2\pi\sum_{\epsilon}\delta\left(\omega-\omega_{\nu;\epsilon}\right)\lambda_{\nu;r;\epsilon}^{2}\cancel{\left\{ 1-e^{-\beta\omega_{\nu;\epsilon}}\right\} }\left\{ \frac{1}{\cancel{1-e^{-\beta\omega_{\nu;\epsilon}}}}\right\} \nonumber \\
 & \text{*Use: }\delta\left(\omega\pm\omega_{\nu;\epsilon}\right)f\left(\omega_{\nu;\epsilon}\right)=\delta\left(\omega\pm\omega_{\nu;\epsilon}\right)f\left(\mp\omega\right),\nonumber \\
 & =2\pi\sum_{\epsilon}\delta\left(\omega+\omega_{\nu;\epsilon}\right)\lambda_{\nu;r;\epsilon}^{2}\frac{e^{\beta\omega}}{1-e^{\beta\omega}}\nonumber \\
 & \mathrel{\phantom{=}}\mathop{+}2\pi\sum_{\epsilon}\delta\left(\omega-\omega_{\nu;\epsilon}\right)\lambda_{\nu;r;\epsilon}^{2}\frac{e^{-\beta\omega}}{1-e^{-\beta\omega}}\nonumber \\
 & \mathrel{\phantom{=}}\mathop{+}2\pi\sum_{\epsilon}\delta\left(\omega-\omega_{\nu;\epsilon}\right)\lambda_{\nu;r;\epsilon}^{2}\nonumber \\
 & =2\pi\sum_{\epsilon}\frac{\lambda_{\nu;r;\epsilon}^{2}}{1-e^{-\beta\omega}}\left[\delta\left(\omega-\omega_{\nu;\epsilon}\right)-\delta\left(\omega+\omega_{\nu;\epsilon}\right)\right].\label{eq:A_=00007Bv; r; T=00007D(omega) in main manuscript -- 2}
\end{align}

\noindent Note that the zero-temperature limit of $A_{\nu;r;T}\left(\omega\right)$
yields:

\begin{equation}
A_{\nu;r;T=0}\left(\omega\right)=\lim_{T\to0}A_{\nu;r;T}\left(\omega\right)=2\pi\sum_{\epsilon}\lambda_{\nu;r;\epsilon}^{2}\delta\left(\omega-\omega_{\nu;\epsilon}\right).\label{eq:A_=00007Bv; r; T=00003D0=00007D(omega) in main manuscript -- 1}
\end{equation}

\noindent Lastly, using Eqs.~\eqref{eq:A_=00007Bv; r; T=00007D(omega) in main manuscript -- 2}
and \eqref{eq:A_=00007Bv; r; T=00003D0=00007D(omega) in main manuscript -- 1},
we get

\begin{equation}
A_{\nu;r;T}\left(\omega\right)=\text{sign}\left(\omega\right)\frac{A_{\nu;r;T=0}\left(\left|\omega\right|\right)}{1-e^{-\beta\omega}},\label{eq:A_=00007Bv; r; T=00007D(omega) in main manuscript -- 3}
\end{equation}

\noindent where

\begin{equation}
\text{sign}\left(\omega\right)=\begin{cases}
1 & \text{if }\omega>0,\\
-1 & \text{if }\omega<0.
\end{cases}\label{eq:sign function in main manuscript -- 1}
\end{equation}

In many applications of interest, $A_{\nu;r;T=0}\left(\omega\right)$
will comprise of multiple components:

\begin{equation}
A_{\nu;r;T=0}\left(\omega\right)=\sum_{\varsigma}A_{\nu;r;T=0;\varsigma}\left(\omega\right),\label{eq:decomposing A_=00007Bv; r; T=00003D0=00007D(omega) in main manuscript -- 1}
\end{equation}

\noindent where $A_{\nu;r;T=0;\varsigma}\left(\omega\right)$ is the
$\varsigma^{\text{th}}$ component of $A_{\nu;r;T=0}\left(\omega\right)$.
As an example, $A_{\nu;r;T=0}\left(\omega\right)$ may be strongly
peaked at multiple frequencies, in which case it naturally decomposes
into multiple components. If $A_{\nu;r;T=0}\left(\omega\right)$ naturally
decomposes into multiple components, it is best to treat each component
separately in our QUAPI approach rather than treat $A_{\nu;r;T=0}\left(\omega\right)$
as a single entity. 

\noindent From Eqs.~\eqref{eq:A_=00007Bv; r; T=00007D(omega) in main manuscript -- 3}
and \eqref{eq:decomposing A_=00007Bv; r; T=00003D0=00007D(omega) in main manuscript -- 1},
we can write

\begin{equation}
A_{\nu;r;T}\left(\omega\right)=\sum_{\varsigma}A_{\nu;r;T;\varsigma}\left(\omega\right),\label{eq:decomposing A_=00007Bv; r; T=00007D(omega) in main manuscript -- 1}
\end{equation}

\noindent where

\begin{equation}
A_{\nu;r;T;\varsigma}\left(\omega\right)=\text{sign}\left(\omega\right)\frac{A_{\nu;r;T=0;\varsigma}\left(\left|\omega\right|\right)}{1-e^{-\beta\omega}}.\label{eq:decomposing A_=00007Bv; r; T; varsigma=00007D(omega) in main manuscript -- 1}
\end{equation}

\noindent In the continuum limit, $A_{\nu;r;T=0;\varsigma}\left(\omega\right)$
becomes a continuous function of $\omega$ satisfying

\begin{equation}
A_{\nu;r;T=0;\varsigma}\left(\omega\right)\ge0,\label{eq:continuous A_=00007Bv; r; T=00003D0; varsigma=00007D(omega) properties -- 1}
\end{equation}

\noindent and

\begin{equation}
A_{\nu;r;T=0;\varsigma}\left(\omega\le0\right)=0.\label{eq:continuous A_=00007Bv; r; T=00003D0; varsigma=00007D(omega) properties -- 2}
\end{equation}

\noindent In order for the noisy dynamics to be ``well-behaved'',
$A_{\nu;r;T=0;\varsigma}\left(\omega\right)$ must also satisfy

\begin{equation}
\left|\lim_{\omega\to0^{+}}\frac{A_{\nu;r;T=0;\varsigma}\left(\omega\right)}{\omega}\right|<\infty.\label{eq:continuous A_=00007Bv; r; T=00003D0; varsigma=00007D(omega) requirement -- 1}
\end{equation}

For a given source of noise, described by the spectral density of
noise $A_{\nu;r;T}\left(\omega\right)$ at temperature $T$, we can
characterize the strength of this noise by calculating:

\begin{equation}
W_{\nu;r;T}=\int_{-\infty}^{\infty}\frac{d\omega}{2\pi}A_{\nu;r;T}\left(\omega\right).\label{eq:strength of noise -- 1}
\end{equation}

In $\texttt{spinbosonchain}$, the quantities $A_{\nu;r;T}\left(\omega\right)$,
$A_{\nu;r;T=0}\left(\omega\right)$, $A_{\nu;r;T;\varsigma}\left(\omega\right)$,
and $A_{\nu;r;T=0;\varsigma}\left(\omega\right)$ are represented
by the classes $\texttt{spinbosonchain.bath.SpectralDensity}$, $\texttt{spinbosonchain.bath.SpectralDensity0T}$,
\\
$\texttt{spinbosonchain.bath.SpectralDensityCmpnt}$, and $\texttt{spinbosonchain.bath.SpectralDensityCmpnt0T}$
respectively. For a discussion on how these quantities are evaluated
numerically in $\texttt{spinbosonchain}$, see Appendix~\ref{sec:Evaluating numerically the spectral densities of noise}.
The quantities $C_{\nu;r;T}\left(t\right)$ and $W_{\nu;r;T}$ can
be calculated using the functions $\texttt{spinbosonchain.bath.correlation}$
and $\texttt{spinbosonchain.bath.noise\_strength}$ respectively.

\subsection{Reduced state operator of the device\label{subsec:Reduced state operator of the device -- 1}}

\noindent The full state operator at time $t$, $\hat{\rho}\left(t\right)$,
can be expressed as

\begin{equation}
\hat{\rho}\left(t\right)=\hat{U}\left(t,0\right)\hat{\rho}^{\left(i\right)}\hat{U}\left(0,t\right),\label{eq:rho -- 1}
\end{equation}

\noindent where $\hat{\rho}^{\left(i\right)}$ and $\hat{U}\left(t,0\right)$
are given by Eqs.~\eqref{eq:rho^(i) -- 1} and \eqref{eq:U -- 1}
respectively. We are ultimately interested in calculating the reduced
state operator of the device:

\begin{equation}
\hat{\rho}^{\left(A\right)}\left(t\right)=\text{Tr}^{\left(B\right)}\left\{ \hat{\rho}\left(t\right)\right\} .\label{eq:reduced state operator of device -- 1}
\end{equation}

To calculate $\hat{\rho}^{\left(A\right)}\left(t\right)$ efficiently
using tensor network methods, we discretize time over a uniform grid
of step size $\Delta t$ and express the reduced state operator in
path integral form. There are various ways to express $\hat{\rho}^{\left(A\right)}\left(t\right)$
as a path integral. Here we loosely draw from Refs.~\citep{Makri1,Makri2,Strathearn1,Suzuki1,Oshiyama1},
wherein a single particle or spin is considered. In Appendix~\ref{sec:Path integral representation derivation -- 1},
we derive the following path integral representation of $\hat{\rho}^{\left(A\right)}\left(t\right)$
that is accurate to second order in the time step size $\Delta t$:

\begin{align}
\rho^{\left(A\right)}\left(\boldsymbol{\sigma}_{z;1;n+1},\boldsymbol{\sigma}_{z;-1;n+1}\right) & \equiv\left\langle \boldsymbol{\sigma}_{z;1;n+1}z\right|\hat{\rho}^{\left(A\right)}\left(t_{n}\right)\left|\boldsymbol{\sigma}_{z;-1;n+1}z\right\rangle \nonumber \\
 & =\prod_{\alpha=\pm1}\prod_{u=-N}^{N}\prod_{r=0}^{L-1}\left[\prod_{q=0}^{q_{y;n}}\left\{ \sum_{\sigma_{y;r+uL;\alpha;q}=\pm1}\right\} \prod_{q=0}^{q_{z;n}-1}\left\{ \sum_{\sigma_{z;r+uL;\alpha;q}=\pm1}\right\} \right]\nonumber \\
 & \mathrel{\phantom{=}}\quad\mathop{\times}e^{i\sum_{\alpha=\pm1}\phi_{\alpha;n}^{\left(\text{lcafc}\right)}\left(\boldsymbol{\sigma}_{z;\alpha;q\in\left[0,q_{z;n}\right]}\right)}I_{n}\left(\boldsymbol{\sigma}_{\nu\in\left\{ y,z\right\} ;\alpha=\pm1;q\in\left[0,q_{\nu;n}\right]}\right)\nonumber \\
 & \mathrel{\phantom{=}}\quad\mathop{\times}\rho^{\left(i,A\right)}\left(\boldsymbol{\sigma}_{z;1;0},\boldsymbol{\sigma}_{z;-1;0}\right)+\mathcal{O}\left[\Delta t^{2}\right],\label{eq:rho^(A) in main manuscript -- 1}
\end{align}

\noindent where $2N+1$ is the number of unit cells; ``lcafc'' is
an initialism for ``longitudinal coupler and field components'';
$n\ge1$;

\begin{equation}
q_{y;n}=\begin{cases}
-1, & \text{if }A_{y;r;T}\left(\omega\right)=0,\\
2n+1, & \text{otherwise},
\end{cases}\label{eq:q_=00007By; n=00007D in main manuscript -- 1}
\end{equation}

\noindent with $A_{\nu;r;T}\left(\omega\right)$ being the spectral
density of the noise coupled to the $\nu$-component of the spin at
site $r$ and temperature $T$ {[}given by Eq.~\eqref{eq:A_=00007Bv; r; T=00007D(omega) in main manuscript -- 1}{]},

\begin{equation}
q_{z;n}=n+1;\label{eq:q_=00007Bz; n=00007D in main manuscript -- 1}
\end{equation}

\begin{equation}
\hat{\sigma}_{\nu;r}\left|\boldsymbol{\sigma}_{\nu;\alpha;q}\nu\right\rangle =\sigma_{\nu;r;\alpha;q}\left|\boldsymbol{\sigma}_{\nu;\alpha;q}\nu\right\rangle ,\label{eq:sigma_=00007Bv; r=00007D eigenvalue equation in main manuscript -- 1}
\end{equation}

\noindent with

\begin{equation}
\boldsymbol{\sigma}_{\nu;\alpha;q}=\left(\sigma_{\nu;r=-NL;\alpha;q},\ldots,\sigma_{\nu;r=NL+L-1;\alpha;q}\right),\label{eq:sigma vector in main manuscript -- 1}
\end{equation}

\noindent $\nu\in\left\{ y,z\right\} $, $\hat{\sigma}_{\nu;r}$ being
the $\nu^{\text{th}}$ Pauli operator acting on site $r$, $\alpha=\pm1$,
and $q\in\left[0,q_{\nu;n}\right]$;

\begin{equation}
\rho^{\left(i,A\right)}\left(\boldsymbol{\sigma}_{z;1;0},\boldsymbol{\sigma}_{z;-1;0}\right)=\left\langle \boldsymbol{\sigma}_{z;1;0}z\vphantom{\Psi^{\left(i,A\right)}}\right.\left|\Psi^{\left(i,A\right)}\right\rangle \left\langle \Psi^{\left(i,A\right)}\right|\left.\boldsymbol{\sigma}_{z;-1;0}z\vphantom{\Psi^{\left(i,A\right)}}\right\rangle ,\label{eq:matrix elements of rho^(i, A) in main manuscript -- 1}
\end{equation}

\noindent with $\hat{\rho}^{\left(i,A\right)}=\left|\Psi^{\left(i,A\right)}\right\rangle \left\langle \Psi^{\left(i,A\right)}\right|$
being the system's reduced state operator at the initial time, $t=0$;

\begin{equation}
\phi_{\alpha;n}^{\left(\text{lcafc}\right)}\left(\boldsymbol{\sigma}_{z;\alpha;q\in\left[0,q_{z;n}\right]}\right)=\sum_{q=0}^{q_{z;n}}\phi_{\alpha;n;k=q}^{\left(\text{lcafc}\right)}\left(\boldsymbol{\sigma}_{z;\alpha;q}\right),\label{eq:phi_=00007Balpha; n=00007D^(lcafc) in main manuscript -- 1}
\end{equation}

\noindent with

\begin{equation}
\phi_{\alpha;n;k}^{\left(\text{lcafc}\right)}\left(\boldsymbol{\sigma}_{z;\alpha;q}\right)=-\frac{\alpha\Delta t}{2}\sum_{k^{\prime}=k-1}^{k}w_{n;k^{\prime}}H_{z}^{\left(A\right)}\left(t_{k^{\prime}};\boldsymbol{\sigma}_{z;\alpha;q}\right),\label{eq:phi_=00007Balpha; n; k=00007D^(lcafc) in main manuscript -- 1}
\end{equation}

\begin{equation}
H_{z}^{\left(A\right)}\left(t_{k^{\prime}};\boldsymbol{\sigma}_{z;\alpha;q}\right)=\sum_{u=-N}^{N}H_{u;z}^{\left(A\right)}\left(t_{k^{\prime}};\boldsymbol{\sigma}_{z;\alpha;q}\right),\label{eq:H_z^(A)(t; sigma) in main manuscript -- 1}
\end{equation}

\begin{equation}
H_{u;z}^{\left(A\right)}\left(t;\boldsymbol{\sigma}_{z;\alpha;q}\right)=\sum_{r=0}^{L-1}h_{z;r}\left(t\right)\sigma_{z;r+uL;\alpha;q}+\sum_{r=1}^{L-1}J_{z,z;r,r+1}\left(t\right)\sigma_{z;r+uL;\alpha;q}\sigma_{z;r+uL+1;\alpha;q},\label{eq:H_=00007Bu; z=00007D^(A)(t; sigma) in main manuscript -- 1}
\end{equation}

\begin{equation}
w_{n;k}=\begin{cases}
0, & \text{if }k=-1,n+1,\\
\frac{1}{2}, & \text{if }k=0,n,\\
1, & \text{if }0<k<n,
\end{cases}\label{eq:k composite quadrature rule in main manuscript -- 1}
\end{equation}

\noindent $h_{z;r}\left(t\right)$ being the longitudinal field energy
scale for site $r$ at time $t$, $J_{z,z;r,r+1}\left(t\right)$ being
the longitudinal coupling energy scale for sites $r$ and $r+1$ at
time $t$, and $L$ being the size of the unit cell; 

\begin{equation}
I_{n}\left(\boldsymbol{\sigma}_{\nu\in\left\{ y,z\right\} ;\alpha=\pm1;q\in\left[0,q_{\nu;n}\right]}\right)=\prod_{u=-N}^{N}\prod_{r=0}^{L-1}\left\{ I_{r+uL;n}\left(\sigma_{\nu\in\left\{ y,z\right\} ;r+uL;\alpha=\pm1;q\in\left[0,q_{\nu;n}\right]}\right)\right\} ,\label{eq:I_n in main manuscript -- 1}
\end{equation}

\noindent with $I_{r;n}\left(\sigma_{\nu\in\left\{ y,z\right\} ;r;\alpha=\pm1;q\in\left[0,q_{\nu;n}\right]}\right)$
being the influence functional that couples the Ising spin variables
\\
$\sigma_{\nu\in\left\{ y,z\right\} ;r;\alpha=\pm1;q\in\left[0,q_{\nu;n}\right]}$
with one another {[}including Ising spin variables associated with
different times{]}:

\begin{equation}
I_{r;n}\left(\sigma_{\nu\in\left\{ y,z\right\} ;r;\alpha=\pm1;q\in\left[0,q_{\nu;n}\right]}\right)=\begin{cases}
I_{r;n}^{\left(\text{z-noise}\right)}\left(\sigma_{z;r;\alpha=\pm1;q\in\left[0,q_{z;n}\right]}\right), & \text{if }A_{y;r;T}\left(\omega\right)=0,\\
I_{r;n}^{\left(\text{yz-noise}\right)}\left(\sigma_{\nu\in\left\{ y,z\right\} ;r;\alpha=\pm1;q\in\left[0,q_{\nu;n}\right]}\right), & \text{otherwise},
\end{cases}\label{eq:I_=00007Br; n=00007D in main manuscript -- 1}
\end{equation}

\begin{align}
I_{r;n}^{\left(\text{yz-noise}\right)}\left(\sigma_{\nu\in\left\{ y,z\right\} ;r;\alpha=\pm1;q\in\left[0,q_{\nu;n}\right]}\right) & =I_{n}^{\left(y\leftrightarrow z\right)}\left(\sigma_{\nu\in\left\{ y,z\right\} ;r;\alpha=\pm1;q\in\left[0,q_{\nu;n}\right]}\right)\nonumber \\
 & \mathrel{\phantom{=}}\mathop{\times}I_{y;r;n}^{\left(\text{tfc}\right)}\left(\sigma_{y;r;\alpha=\pm1;q\in\left[0,q_{y;n}\right]}\right)\nonumber \\
 & \mathrel{\phantom{=}}\mathop{\times}\prod_{\nu\in\left\{ y,z\right\} }\left\{ I_{\nu;r;n}^{\left(\text{bath}\right)}\left(\sigma_{\nu;r;\alpha=\pm1;q\in\left[0,q_{\nu;n}\right]}\right)\right\} ,\label{eq:I_=00007Br; n=00007D^(yz-noise) in main manuscript -- 1}
\end{align}

\begin{equation}
I_{r;n}^{\left(\text{z-noise}\right)}\left(\sigma_{z;r;\alpha=\pm1;q\in\left[0,q_{z;n}\right]}\right)=I_{z;r;n}^{\left(\text{tfc}\right)}\left(\sigma_{z;r;\alpha=\pm1;q\in\left[0,q_{z;n}\right]}\right)I_{z;r;n}^{\left(\text{bath}\right)}\left(\sigma_{z;r;\alpha=\pm1;q\in\left[0,q_{z;n}\right]}\right),\label{eq:I_=00007Br; n=00007D^(z-noise) in main manuscript -- 1}
\end{equation}

\begin{align}
I_{n}^{\left(y\leftrightarrow z\right)}\left(\sigma_{\nu\in\left\{ y,z\right\} ;r;\alpha=\pm1;q\in\left[0,q_{\nu;n}\right]}\right) & =\prod_{l=0}^{n}\left\{ I^{\left(y\leftrightarrow z,1\right)}\left(\sigma_{y;r;1,2l+1},\sigma_{y;r;-1,2l+1},\sigma_{z;r;1,l+1},\sigma_{z;r;-1,l+1}\right)\right.\nonumber \\
 & \phantom{=\prod_{l=0}^{n}}\left.\quad\mathop{\times}I^{\left(y\leftrightarrow z,2\right)}\left(\sigma_{y;r;1,2l},\sigma_{y;r;-1,2l},\sigma_{z;r;1,l},\sigma_{z;r;-1,l}\right)\right\} ,\label{eq:I_n^(y<->z) in main manuscript -- 1}
\end{align}

\begin{align}
 & I^{\left(y\leftrightarrow z,1\right)}\left(\sigma_{y;r;1,2l+1},\sigma_{y;r;-1,2l+1},\sigma_{z;r;1,l+1},\sigma_{z;r;-1,l+1}\right)\nonumber \\
 & \quad=I^{\left(y\to z\right)}\left(\sigma_{z;r;1,l+1},\sigma_{y;r;1,2l+1}\right)I^{\left(z\to y\right)}\left(\sigma_{y;r;-1,2l+1},\sigma_{z;r;-1,l+1}\right),\label{eq:I^(y<->z, 1) in main manuscript -- 1}
\end{align}

\begin{equation}
I^{\left(y\leftrightarrow z,2\right)}\left(\sigma_{y;r;1,2l},\sigma_{y;r;-1,2l},\sigma_{z;r;1,l},\sigma_{z;r;-1,l}\right)=I^{\left(z\to y\right)}\left(\sigma_{y;r;1,2l},\sigma_{z;r;1,l}\right)I^{\left(y\to z\right)}\left(\sigma_{z;r;-1,l},\sigma_{y;r;-1,2l}\right),\label{eq:I^(y<->z, 2) in main manuscript -- 1}
\end{equation}

\begin{align}
I^{\left(z\to y\right)}\left(\sigma_{y;r;\alpha;q^{\vphantom{\prime}}},\sigma_{z;r;\alpha;q^{\prime}}\right) & =\frac{1}{2\sqrt{2}}\left\{ 1-i\sigma_{y;r;\alpha;q^{\vphantom{\prime}}}+\sigma_{z;r;\alpha;q^{\prime}}+i\sigma_{y;r;\alpha;q^{\vphantom{\prime}}}\sigma_{z;r;\alpha;q^{\prime}}\right\} ,\label{eq:I^(z->y) in main manuscript -- 1}\\
I^{\left(y\to z\right)}\left(\sigma_{z;r;\alpha;q^{\prime}},\sigma_{y;r;\alpha;q^{\vphantom{\prime}}}\right) & =\left\{ I^{\left(z\to y\right)}\left(\sigma_{y;r;\alpha;q^{\vphantom{\prime}}},\sigma_{z;r;\alpha;q^{\prime}}\right)\right\} ^{*},\label{eq:I^(y->z) in main manuscript -- 1}
\end{align}

\begin{equation}
I_{y;r;n}^{\left(\text{tfc}\right)}\left(\sigma_{y;r;\alpha=\pm1;q\in\left[0,q_{y;n}\right]}\right)=\prod_{l=0}^{n}\left\{ I_{y;r;n;k=l}^{\left(\text{tfc}\right)}\left(\sigma_{y;r;1;2l},\sigma_{y;r;-1;2l},\sigma_{y;r;1;2l+1},\sigma_{y;r;-1;2l+1}\right)\right\} ,\label{eq:I_=00007By; r; n=00007D^(tfc) in main manuscript -- 1}
\end{equation}

\begin{equation}
I_{z;r;n}^{\left(\text{tfc}\right)}\left(\sigma_{z;r;\alpha=\pm1;q\in\left[0,q_{z;n}\right]}\right)=\prod_{l=0}^{n}\left\{ I_{z;r;n;k=l}^{\left(\text{tfc}\right)}\left(\sigma_{z;r;1;l},\sigma_{z;r;-1;l},\sigma_{z;r;1;l+1},\sigma_{z;r;-1;l+1}\right)\right\} ,\label{eq:I_=00007Bz; r; n=00007D^(tfc) in main manuscript -- 1}
\end{equation}

\begin{align}
 & I_{\nu;r;n;k}^{\left(\text{tfc}\right)}\left(\sigma_{\nu;r;1;q_{1}},\sigma_{\nu;r;-1;q_{1}},\sigma_{\nu;r;1;q_{2}},\sigma_{\nu;r;-1;q_{2}}\right)\nonumber \\
 & \quad=I_{\nu;r;1;n;k}^{\left(\text{tfc}\right)}\left(\sigma_{\nu;r;1;q_{2}},\sigma_{\nu;r;1;q_{1}}\right)I_{\nu;r;-1;n;k}^{\left(\text{tfc}\right)}\left(\sigma_{\nu;r;-1;q_{1}},\sigma_{\nu;r;-1;q_{2}}\right),\label{eq:I_=00007Bv; r; n; k=00007D^(tfc) in main manuscript -- 1}
\end{align}

\begin{align}
I_{\nu;r;\alpha;n;k}^{\left(\text{tfc}\right)}\left(\sigma_{y;r;\alpha;q_{1}},\sigma_{y;r;\alpha;q_{2}}\right) & =\frac{1}{4}\left(\sigma_{\nu;r;\alpha;q_{1}}+\sigma_{\nu;r;\alpha;q_{2}}\right)^{2}\cos\left(\frac{\theta_{r;n;k}}{2}\right)\nonumber \\
 & \phantom{=}\mathop{+}i^{1+c_{\nu}}\alpha\left\{ \frac{1}{2}\left(\sigma_{\nu;r;\alpha;q_{1}}-\sigma_{\nu;r;\alpha;q_{2}}\right)\right\} ^{c_{\nu}}\sin\left(\frac{\theta_{r;n;k}}{2}\right),\label{eq:I_=00007Bv; r; alpha; n; k=00007D^(tfc) in main manuscript -- 1}
\end{align}

\begin{equation}
c_{\nu}=\begin{cases}
1, & \text{if }\nu=y,\\
2, & \text{if }\nu=z,
\end{cases}\label{eq:c_v -- 1}
\end{equation}

\begin{equation}
\theta_{r;n;k}=2\Delta tw_{n;k}h_{x;r}\left(t_{k}\right),\label{eq:theta_=00007Br; n; k=00007D in main manuscript -- 1}
\end{equation}

\noindent ``tfc'' being an initialism for ``transverse field components'',
$h_{x;r}\left(t\right)$ being the transverse field energy scale for
site $r$ at time $t$,

\begin{equation}
I_{\nu;r;n}^{\left(\text{bath}\right)}\left(\sigma_{\nu;r;\alpha=\pm1;q\in\left[0,q_{\nu;n}\right]}\right)=\prod_{q_{2}=0}^{q_{\nu;n}}\prod_{q_{1}=0}^{q_{2}}\left\{ I_{\nu;r;n;q_{1},q_{2}}^{\left(\text{bath}\right)}\left(\sigma_{\nu;r;1;q_{1}},\sigma_{\nu;r;-1;q_{1}},\sigma_{\nu;r;1;q_{2}},\sigma_{\nu;r;-1;q_{2}}\right)\right\} ,\label{eq:I_=00007Bv; r; n=00007D^(bath) in main manuscript -- 1}
\end{equation}

\begin{align}
 & I_{\nu;r;n;q_{1},q_{2}}^{\left(\text{bath}\right)}\left(\sigma_{\nu;r;1;q_{1}},\sigma_{\nu;r;-1;q_{1}},\sigma_{\nu;r;1;q_{2}},\sigma_{\nu;r;-1;q_{2}}\right)\nonumber \\
 & \quad=\prod_{\left(l_{1},l_{2}\right)\in\Upsilon_{\nu;n;q_{1},q_{2}}}\left\{ e^{-\gamma_{\nu;r;n;l_{1},l_{2}}\left(\sigma_{\nu;r;1;q_{1}},\sigma_{\nu;r;-1;q_{1}},\sigma_{\nu;r;1;q_{2}},\sigma_{\nu;r;-1;q_{2}}\right)}\right\} ,\label{eq:I_=00007Bv; r; n; q_1, q_2=00007D^(bath) in main manuscript -- 1}
\end{align}

\begin{equation}
\Upsilon_{y;n;q_{1},q_{2}}=\left\{ \left(q_{1},q_{2}\right)\right\} ,\label{eq:Upsilon_=00007By; n; q_1; q_2=00007D in main manuscript -- 1}
\end{equation}

\begin{align}
 & \Upsilon_{z;n;q_{1},q_{2}}\nonumber \\
 & \quad=\begin{cases}
\left\{ \left(0,0\right)\right\} , & \text{if }q_{1}=q_{2}=0,\\
\left\{ \left(0,2q_{2}-1\right),\left(0,2q_{2}\right)\right\} , & \text{if }q_{1}=0\text{ and }1\le q_{2}\le n,\\
\left\{ \left(0,2n+1\right)\right\} , & \text{if }q_{1}=0\text{ and }q_{2}=n+1,\\
\left\{ \left(2q_{1}-1,2q_{2}-1\right),\left(2q_{1},2q_{2}-1\right),\left(2q_{1}-1,2q_{2}\right),\left(2q_{1},2q_{2}\right)\right\} , & \text{if }1\le q_{1}<q_{2}\le n,\\
\left\{ \left(2q_{1}-1,2q_{2}-1\right),\left(2q_{1}-1,2q_{2}\right),\left(2q_{1},2q_{2}\right)\right\} , & \text{if }1\le q_{1}=q_{2}\le n,\\
\left\{ \left(2q_{1}-1,2n+1\right),\left(2q_{1},2n+1\right)\right\} , & \text{if }1\le q_{1}\le n\text{ and }q_{2}=n+1,\\
\left\{ \left(2n+1,2n+1\right)\right\} , & \text{if }q_{1}=q_{2}=n+1,
\end{cases}\label{eq:Upsilon_=00007Bz; n; q_1; q_2=00007D in main manuscript -- 1}
\end{align}

\begin{align}
 & \gamma_{\nu;r;n;l_{1},l_{2}}\left(\sigma_{\nu;r;1;q_{1}},\sigma_{\nu;r;-1;q_{1}},\sigma_{\nu;r;1;q_{2}},\sigma_{\nu;r;-1;q_{2}}\right)\nonumber \\
 & \quad=\mathcal{E}_{\nu;r;l_{1}}^{\left(\lambda\right)}\mathcal{E}_{\nu;r;l_{2}}^{\left(\lambda\right)}\left(\sigma_{\nu;r;1;q_{2}}-\sigma_{\nu;r;-1;q_{2}}\right)\nonumber \\
 & \quad\mathrel{\phantom{=}}\mathop{\times}\left\{ \left(\sigma_{\nu;r;1;q_{1}}-\sigma_{\nu;r;-1;q_{1}}\right)\text{Re}\left[\eta_{\nu;r;n;l_{2},l_{1}}\right]+i\left(\sigma_{\nu;r;1;q_{1}}+\sigma_{\nu;r;-1;q_{1}}\right)\text{Im}\left[\eta_{\nu;r;n;l_{2},l_{1}}\right]\right\} ,\label{eq:gamma_=00007Bv; r; n; l_1; l_2=00007D in main manuscript -- 1}
\end{align}

\begin{equation}
\mathcal{E}_{\nu;r;l}^{\left(\lambda\right)}=\mathcal{E}_{\nu;r}^{\left(\lambda\right)}\left(t=\left\lfloor l/2\right\rfloor \Delta t\right),\label{eq:mathcal E_=00007Bv; r; l=00007D^(lambda) in main manuscript -- 1}
\end{equation}

\noindent $\mathcal{E}_{\nu;r}^{\left(\lambda\right)}\left(t\right)$
being the energy scale associated with the coupling between the environment
and the $\nu$-component of the spin at site $r$ {[}introduced in
Eq.~\eqref{eq:mathcal Q_=00007Bv; r=00007D(t) -- 1}{]},

\begin{equation}
\eta_{\nu;r;n;l_{1},l_{2}}=\int_{-\infty}^{\infty}\frac{d\omega}{2\pi}\,A_{\nu;r;T}\left(\omega\right)\frac{f_{n;l_{1},l_{2}}^{\left(\eta\right)}\left(\omega\right)}{\omega^{2}},\label{eq:eta_=00007Bv; r; n; l_1, l_2=00007D in main manuscript -- 1}
\end{equation}

\begin{equation}
f_{n;l_{1},l_{2}}^{\left(\eta\right)}\left(\omega\right)=\begin{cases}
f_{n;l_{1}}^{\left(\eta,\Delta l=0\right)}\left(\omega\right), & \text{if }l_{1}-l_{2}=0,\\
f_{n;l_{1},l_{2}}^{\left(\eta,\Delta l\ge1\right)}\left(\omega\right), & \text{if }l_{1}-l_{2}\ge1,
\end{cases}\label{eq:f_=00007Bn; l_1, l_2=00007D^=00007B(eta)=00007D(omega) in main manuscript -- 1}
\end{equation}

\begin{equation}
f_{n;l_{1}}^{\left(\eta,\Delta l=0\right)}\left(\omega\right)=2\sin^{2}\left(\frac{\Delta t\omega\tilde{w}_{n;l_{1}}}{2}\right),\label{eq:f_=00007Bn; l_1, l_2=00007D^=00007B(eta, dl=00003D0)=00007D(omega) in main manuscript -- 1}
\end{equation}

\begin{equation}
f_{n;l_{1},l_{2}}^{\left(\eta,\Delta l\ge1\right)}\left(\omega\right)=4e^{-i\Delta t\omega\left\{ \frac{1}{2}\tilde{w}_{n;l_{1}}+\sum_{l=l_{2}+1}^{l_{1}-1}\tilde{w}_{n;l}+\frac{1}{2}\tilde{w}_{n;l_{2}}\right\} }\sin\left(\frac{\Delta t\omega\tilde{w}_{n;l_{1}}}{2}\right)\sin\left(\frac{\Delta t\omega\tilde{w}_{n;l_{2}}}{2}\right),\label{eq:f_=00007Bn; l_1, l_2=00007D^=00007B(eta, dl=00003D>1)=00007D(omega) in main manuscript -- 1}
\end{equation}

\begin{equation}
\tilde{w}_{n;l}=\begin{cases}
\frac{1}{4}, & \text{if }l=0,1,2n,2n+1,\\
\frac{1}{2}, & \text{if }1<l<2n,
\end{cases}\label{eq:l composite quadrature rule in main manuscript -- 1}
\end{equation}

\begin{equation}
\sum_{l=a}^{b}F_{l}=\begin{cases}
0, & \text{if }a>b,\\
F_{a}+F_{a+1}+\ldots+F_{b} & \text{otherwise},
\end{cases}\label{eq:summing convention in main manuscript -- 1}
\end{equation}

\noindent and $F_{l}$ being an arbitrary sequence. For a detailed
discussion on how to evaluate numerically $\eta_{\nu;r;n;l_{1},l_{2}}$,
see Appendix~\ref{sec:How to evaluate eta -- 1}.

\subsection{Memory of the system\label{subsec:Memory of the system -- 1}}

Because the system is coupled to its environment, the dynamics of
the system not only depend on its current state, but also its history.
This should be clear from the path integral representation of $\hat{\rho}^{\left(A\right)}\left(t\right)$
that we derived in Appendix~\ref{sec:Path integral representation derivation -- 1}.
The degree to which the dynamics depend on the history of the system
can be characterized by a quantity known as the bath correlation time,
or ``memory''. Informally, one may define the system's memory $\tau$
as the time beyond which both the $y$- and the $z$-components of
the bath correlation function {[}see Eqs.~\eqref{eq:C_=00007Bv; r; T=00007D(t) in main manuscript -- 1}-\eqref{eq:C_=00007Bv; r; T=00007D(t) in main manuscript -- 2}{]}
are negligibly small:

\begin{equation}
\left|C_{\nu;T}\left(t>\tau\right)\right|\ll1.\label{eq:informal definition of bath correlation time -- 1}
\end{equation}

\noindent All information regarding the non-Markovian dynamics of
the system is encoded in the $\eta_{\nu;n;k_{1},k_{2}}$, which appear
in the path integral representation of $\hat{\rho}^{\left(A\right)}\left(t\right)$
{[}see Eqs.~\eqref{eq:rho^(A) in main manuscript -- 1}-\eqref{eq:summing convention in main manuscript -- 1}
\textendash{} in particular Eq.~\eqref{eq:eta_=00007Bv; r; n; l_1, l_2=00007D in main manuscript -- 1}{]}.
A system with finite memory implies that there exists an integer $K_{\tau}>0$
such that the $\eta_{\nu;n;l_{1},l_{2}}$ for $2n+1\ge l_{1}\ge l_{2}+2K_{\tau}$
contribute negligibly to the dynamics. In Appendix~\ref{sec:How to evaluate eta -- 1}
we determine lower bounds for the $K_{\tau}$ given the system's memory
$\tau$, namely that

\begin{equation}
K_{\tau}=\max\left(0,\left\lceil \frac{\tau-\frac{7}{4}\Delta t}{\Delta t}\right\rceil \right)+3,\label{eq:K_=00007Btau=00007D in main manuscript -- 1}
\end{equation}

\noindent and $\Delta t$ being the time step size. We return to the
topic of memory in Sec.~\ref{sec:Tensor network algorithm}, where
we develop the tensor network algorithm used to calculate the path
integral representation of $\hat{\rho}^{\left(A\right)}\left(t\right)$.

At this point, we have discussed all the model components that define
the bath/environment. These components are the unit cell size $L$,
the inverse temperature $\beta$, the system's memory $\tau$, the
energy scales $\mathcal{E}_{\nu;r}^{\left(\lambda\right)}\left(t\right)$
associated with the couplings between the environment and the spins
{[}introduced in Eq.~\eqref{eq:mathcal Q_=00007Bv; r=00007D(t) -- 1}{]},
and the zero-temperature limits $A_{\nu;r;T=0}\left(\omega\right)$
of the spectral densities of noise {[}introduced in Sec.~\ref{sec:Spectral densities of noise}{]}.
In $\texttt{spinbosonchain}$, this bath model component set is represented
by the $\texttt{spinbosonchain.bath.Model}$ class.

\newpage{}

\section{Tensor network algorithm\label{sec:Tensor network algorithm}}

In this section, we describe in detail the tensor network algorithm
used to calculate the path integral representation of $\hat{\rho}^{\left(A\right)}\left(t\right)$
{[}Eqs.~\eqref{eq:rho^(A) in main manuscript -- 1}-\eqref{eq:summing convention in main manuscript -- 1}{]}.
The method we develop here draws from Refs.~\citep{Strathearn1,Suzuki1}.

\subsection{$\tilde{\sigma}$ variables\label{subsec:tilde sigma variables -- 1}}

To reduce notational complexity, we map the Ising spin variable $\sigma_{\nu\in\left\{ y,z\right\} ;r\in\left[0,L-1\right];\alpha=\pm1;q\in\left[0,q_{\nu;n}\right]}$
to a variable $\tilde{\sigma}_{r\in\left[0,L-1\right];\alpha=\pm1;m\in\left[0,\left(n+1\right)\Delta m\right]}$:

\begin{equation}
\tilde{\sigma}_{r;\alpha;m}\equiv\sigma_{\tilde{\nu}_{m};r;\alpha;\tilde{q}_{m}},\label{eq:tilde sigma -- 1}
\end{equation}

\noindent where

\begin{equation}
\Delta m=\begin{cases}
1, & \text{if }A_{y;r;T}\left(\omega\right)=0,\\
3, & \text{otherwise};
\end{cases}\label{eq:Delta m -- 1}
\end{equation}

\begin{equation}
\tilde{\nu}_{m}=\begin{cases}
z, & \text{if }m\text{ mod }3=0\text{ or }A_{y;r;T}\left(\omega\right)=0,\\
y, & \text{otherwise},
\end{cases}\label{eq:tilde nu_m -- 1}
\end{equation}

\noindent and

\begin{equation}
\tilde{q}_{m}=\begin{cases}
m, & \text{if }A_{y;r;T}\left(\omega\right)=0,\\
\left(2-\left\lfloor \frac{\left\{ m+2\right\} \text{ mod }3}{2}\right\rfloor \right)\left\lfloor \frac{m}{3}\right\rfloor +\left\lfloor \frac{\left(m\text{ mod }3\right)+1}{3}\right\rfloor , & \text{otherwise},
\end{cases}\label{eq:tilde q_m -- 1}
\end{equation}

\noindent We also introduce the vector notation:

\begin{equation}
\tilde{\boldsymbol{\sigma}}_{\alpha;m}=\left(\tilde{\sigma}_{r=-NL;\alpha;m},\ldots,\tilde{\sigma}_{r=NL+L-1;\alpha;m}\right),\label{eq:tilde sigma vector -- 1}
\end{equation}

\noindent where as a reminder, $2N+1$ is the number of unit cells.

\subsection{Base-4 variables\label{subsec:Base 4 variables -- 1}}

To reduce notational complexity even further, we map the ordered pair
$\left(\tilde{\sigma}_{r;1;m},\tilde{\sigma}_{r;-1;m}\right)$ to
a variable $j_{r;m}$ with four possible values:

\begin{equation}
\left(\tilde{\sigma}_{r;1;m},\tilde{\sigma}_{r;-1;m}\right)\to\begin{cases}
\left(1,1\right), & \text{if }j_{r;m}=0,\\
\left(1,-1\right), & \text{if }j_{r;m}=1,\\
\left(-1,1\right), & \text{if }j_{r;m}=2,\\
\left(-1,-1\right), & \text{if }j_{r;m}=3.
\end{cases}\label{eq:mapping spin pairs to j -- 1}
\end{equation}

\noindent By introducing the function $g_{\alpha}$:

\noindent 
\begin{equation}
g_{\alpha}\left(j\right)=\begin{cases}
1-2\left\lfloor j/2\right\rfloor , & \text{if }\alpha=1,\\
1-2\left(j\mod2\right), & \text{if }\alpha=-1.
\end{cases}\label{eq:g_=00007Balpha=00007D -- 1}
\end{equation}

\noindent we may write

\begin{equation}
\tilde{\sigma}_{r;\alpha;m}=g_{\alpha}\left(j_{r;m}\right).\label{eq:tilde sigma related to j by g_=00007Balpha=00007D -- 1}
\end{equation}

\noindent We also introduce the vector notation:

\begin{equation}
\mathbf{j}_{m}=\left(j_{r=-NL;m},\ldots,j_{r=NL+L-1;m}\right).\label{eq:j vector -- 1}
\end{equation}

\noindent and extend the definition of $g_{\alpha}$ to accept $\mathbf{j}_{m}$
vectors:

\begin{equation}
g_{\alpha}\left(\mathbf{j}_{m}\right)=\left(g_{\alpha}\left(j_{r=-NL;m}\right),\ldots,g_{\alpha}\left(j_{r=NL+L-1;m}\right)\right).\label{eq:applying g_=00007Balpha=00007D to j vector -- 1}
\end{equation}

In $\texttt{spinbosonchain}$, the module $\texttt{spinbosonchain.\_base4}$
contains functions for mapping between Ising variable pairs $\left(\tilde{\sigma}_{r;1;m},\tilde{\sigma}_{r;-1;m}\right)$
and base-4 variables $j_{r;m}$.

\subsection{Rearranging terms in the path integral expression of $\hat{\rho}^{\left(A\right)}\left(t\right)$\label{subsec:Rearranging terms in the path integral expression of rho^(A) -- 1}}

In this section, we rearrange terms in the path integral expression
of $\hat{\rho}^{\left(A\right)}\left(t\right)$ {[}Eq.~\eqref{eq:rho^(A) in main manuscript -- 1}{]}
to make the subsequent discussion regarding the tensor network algorithm
easier to follow by the reader. By rearranging terms in Eq.~\eqref{eq:rho^(A) in main manuscript -- 1},
and taking into account the system's memory {[}see Sec.~\ref{subsec:Memory of the system -- 1}{]},
we may rewrite the path integral expression of $\hat{\rho}^{\left(A\right)}\left(t\right)$
as:

\begin{align}
\mathring{\rho}_{n}^{\left(A\right)}\left(\mathbf{j}_{\left(n+1\right)\Delta m}\right) & =\left\langle g_{1}\left(\mathbf{j}_{\left(n+1\right)\Delta m}\right)z\right|\hat{\rho}^{\left(A\right)}\left(t_{n}\right)\left|g_{-1}\left(\mathbf{j}_{\left(n+1\right)\Delta m}\right)z\right\rangle \nonumber \\
 & =\prod_{u=-N}^{N}\prod_{r=0}^{L-1}\left[\sum_{b_{r+uL;0}=0}^{3}\prod_{m=0}^{\left(n+1\right)\Delta m-1}\left\{ \sum_{j_{r+uL;m}=0}^{3}\right\} \right]\nonumber \\
 & \mathrel{\phantom{=}}\quad\mathop{\times}\mathring{\rho}^{\left(i,A\right)}\left(\mathbf{b}_{0}\right)\mathring{I}_{n}\left(\mathbf{b}_{0};\mathbf{j}_{m\in\left[0,\left(n+1\right)\Delta m\right]}\right)e^{i\mathring{\phi}_{n}^{\left(\text{lcafc}\right)}\left(\mathbf{j}_{m\in\left\{ k\Delta m\right\} _{k=0}^{q_{z;n}}}\right)}+\mathcal{O}\left[\Delta t^{2}\right],\label{eq:rearranging terms in rho^(A) -- 1}
\end{align}

\noindent where $\Delta m$ is given by Eq.~\eqref{eq:Delta m -- 1},
we have used the compact $j$-notation introduced in Secs.~\ref{subsec:tilde sigma variables -- 1}
and \ref{subsec:Base 4 variables -- 1};

\begin{equation}
\mathbf{b}_{0}=\left(b_{r=-NL;0},\ldots,b_{r=NL+L-1;0}\right);\label{eq:b_0 vector -- 1}
\end{equation}

\begin{equation}
\mathring{\rho}^{\left(i,A\right)}\left(\mathbf{b}_{0}\right)=\rho^{\left(i,A\right)}\left(g_{1}\left(\mathbf{b}_{0}\right),g_{-1}\left(\mathbf{b}_{0}\right)\right),\label{eq:ring rho^(i, A) -- 1}
\end{equation}

\noindent with $g_{\alpha}\left(\mathbf{j}\right)$ given by Eqs\@.~\eqref{eq:g_=00007Balpha=00007D -- 1}
and \eqref{eq:applying g_=00007Balpha=00007D to j vector -- 1};

\begin{equation}
\mathring{\phi}_{n}^{\left(\text{lcafc}\right)}\left(\mathbf{j}_{m\in\left\{ k\Delta m\right\} _{k=0}^{q_{z;n}}}\right)=\sum_{k=0}^{q_{z;n}}\sum_{\alpha=\pm1}\phi_{\alpha;n;k}^{\left(\text{lcafc}\right)}\left(g_{\alpha}\left(\mathbf{j}_{k\Delta m}\right)\right),\label{eq:ring phi_n^(lcafc) -- 1}
\end{equation}

\noindent with $\phi_{\alpha;n;k}^{\left(\text{lcafc}\right)}\left(g_{\alpha}\left(\mathbf{j}_{3k}\right)\right)$
given by Eq.~\eqref{eq:phi_=00007Balpha; n; k=00007D^(lcafc) in main manuscript -- 1};

\begin{equation}
\mathring{I}_{n}\left(\mathbf{b}_{0};\mathbf{j}_{m\in\left[0,\left(n+1\right)\Delta m\right]}\right)=\prod_{u=-N}^{N}\prod_{r=0}^{L-1}\left\{ \mathring{I}_{r;n}\left(b_{r;0};j_{r;m\in\left[0,\left(n+1\right)\Delta m\right]}\right)\right\} ,\label{eq:ring I_n -- 1}
\end{equation}

\noindent with $\mathring{I}_{r;n}\left(b_{r;0};j_{r;m\in\left[0,\left(n+1\right)\Delta m\right]}\right)$
being the influence functional that couples the base-$4$ variables
$j_{r;m\in\left[0,\left(n+1\right)\Delta m\right]}$ with one another
{[}including base-$4$ variables associated with different times{]}:

\begin{equation}
\mathring{I}_{r;n}\left(b_{r;0};j_{r;m\in\left[0,\left(n+1\right)\Delta m\right]}\right)=\mathring{I}_{r;n\Delta m-1}^{\left(1\right)}\left(b_{r;0};j_{r;m\in\left[0,n\Delta m-1\right]}\right)\mathring{I}_{r;n}^{\left(2\right)}\left(j_{r;m\in\left[\mu_{n\Delta m;\tau},\left(n+1\right)\Delta m\right]}\right),\label{eq:ring I_=00007Br; n=00007D -- 1}
\end{equation}

\begin{equation}
\mathring{I}_{r;0}^{\left(1\right)}\left(b_{r;0};j_{r;0}\right)=\delta_{b_{r;0},j_{r;0}}\mathring{I}_{r;n;0}\left(j_{r;0}\right),\label{eq:ring I_=00007Br; 0=00007D^(1) -- 1}
\end{equation}

\begin{equation}
\mathring{I}_{r;m_{2}+1}^{\left(1\right)}\left(b_{r;0};j_{r;m\in\left[0,m_{2}+1\right]}\right)=\mathring{I}_{r;n=\infty;m_{2}+1}\left(j_{r;m\in\left[\mu_{m_{2}+1;\tau},m_{2}+1\right]}\right)\mathring{I}_{r;m_{2}}^{\left(1\right)}\left(b_{r;0};j_{r;m\in\left[0,m_{2}\right]}\right),\label{eq:ring I_=00007Br; m_2+1=00007D^(1) -- 1}
\end{equation}

\begin{equation}
\mathring{I}_{r;n}^{\left(2\right)}\left(j_{r;m\in\left[\mu_{n\Delta m;\tau},\left(n+1\right)\Delta m\right]}\right)=\prod_{m_{2}=n\Delta m}^{\left(n+1\right)\Delta m}\left\{ \mathring{I}_{r;n;m_{2}}\left(j_{r;m\in\left[\mu_{m_{2};\tau},m_{2}\right]}\right)\right\} ,\label{eq:ring I_=00007Br; n=00007D^(2) -- 1}
\end{equation}

\begin{equation}
\mathring{I}_{r;n;m_{2}}\left(j_{r;m\in\left[\mu_{m_{2};\tau},m_{2}\right]}\right)=\prod_{m_{1}=\mu_{m_{2};\tau}}^{m_{2}}\left\{ \mathring{I}_{r;n;m_{1},m_{2}}\left(j_{r;m_{1}},j_{r;m_{2}}\right)\right\} ,\label{eq:ring I_=00007Br; n; m_2=00007D -- 1}
\end{equation}

\begin{equation}
\mu_{m;\tau}=\max\left(0,m-K_{\tau}\Delta m+1\right),\label{eq:mu_=00007Bm; tau=00007D -- 1}
\end{equation}

\noindent $K_{\tau}$ given by Eq.~\eqref{eq:K_=00007Btau=00007D -- 1},
$\tau$ being the system's ``memory'',

\begin{equation}
\mathring{I}_{r;n;m_{1},m_{2}}\left(j_{r;m_{1}},j_{r;m_{2}}\right)=\begin{cases}
\mathring{I}_{r;n;m_{1},m_{2}}^{\left(\text{z-noise+tfc}\right)}\left(j_{r;m_{1}},j_{r;m_{2}}\right), & \text{if }A_{y;r;T}\left(\omega\right)=0,\\
\mathring{I}_{r;n;m_{1},m_{2}}^{\left(\text{yz-noise+tfc}\right)}\left(j_{r;m_{1}},j_{r;m_{2}}\right), & \text{otherwise},
\end{cases}\label{eq:ring I_=00007Br; n; m_1, m_2=00007D -- 1}
\end{equation}

\begin{align}
 & \mathring{I}_{r;n;m_{1},m_{2}}^{\left(\text{yz-noise+tfc}\right)}\left(j_{r;m_{1}},j_{r;m_{2}}\right)\nonumber \\
 & \quad=\begin{cases}
\mathring{I}_{z;r;n;m_{1},m_{2}}^{\left(\text{bath}\right)}\left(j_{r;m_{1}},j_{r;m_{2}}\right), & \text{if }m_{1}\text{ mod }3=0\text{ \& }m_{2}\text{ mod }3=0,\\
\mathring{I}^{\left(y\leftrightarrow z,1\right)}\left(j_{r;m_{1}},j_{r;m_{2}}\right), & \text{if }m_{1}=m_{2}-1\text{ \& }m_{2}\text{ mod }3=0,\\
1, & \text{if }m_{1}\neq m_{2}-1\text{ \& }m_{1}\text{ mod }3\neq0\text{ \& }m_{2}\text{ mod }3=0,\\
1, & \text{if }m_{1}\text{ mod }3=0\text{ \& }m_{1}\neq m_{2}-1\text{ \& }m_{2}\text{ mod }3=1,\\
\mathring{I}^{\left(y\leftrightarrow z,2\right)}\left(j_{r;m_{1}},j_{r;m_{2}}\right), & \text{if }m_{1}=m_{2}-1\text{ \& }m_{2}\text{ mod }3=1,\\
\mathring{I}_{y;r;n;m_{1},m_{2}}^{\left(\text{bath}\right)}\left(j_{r;m_{1}},j_{r;m_{2}}\right), & \text{if }m_{1}\text{ mod }3\neq0\text{ \& }m_{2}\text{ mod }3=1,\\
1, & \text{if }m_{1}\text{ mod }3=0\text{ \& }m_{2}\text{ mod }3=2,\\
\mathring{I}_{r;n;m_{1},m_{2}}^{\left(\text{tfc+y-bath}\right)}\left(j_{r;m_{1}},j_{r;m_{2}}\right), & \text{if }m_{1}=m_{2}-1\text{ \& }m_{2}\text{ mod }3=2,\\
\mathring{I}_{y;r;n;m_{1},m_{2}}^{\left(\text{bath}\right)}\left(j_{r;m_{1}},j_{r;m_{2}}\right), & \text{if }m_{1}\neq m_{2}-1\text{ \& }m_{1}\text{ mod }3\neq0\text{ \& }m_{2}\text{ mod }3=2,
\end{cases}\label{eq:ring I_=00007Br; n; m_1, m_2=00007D^=00007Byz-noise+tfc=00007D -- 1}
\end{align}

\begin{equation}
\mathring{I}_{r;n;m_{1},m_{2}}^{\left(\text{z-noise+tfc}\right)}\left(j_{r;m_{1}},j_{r;m_{2}}\right)=\begin{cases}
\mathring{I}_{z;r;n;m_{1},m_{2}}^{\left(\text{bath}\right)}\left(j_{r;m_{1}},j_{r;m_{2}}\right), & \text{if }m_{1}\neq m_{2}-1,\\
\mathring{I}_{r;n;m_{1},m_{2}}^{\left(\text{tfc+z-bath}\right)}\left(j_{r;m_{1}},j_{r;m_{2}}\right), & \text{if }m_{1}=m_{2}-1,
\end{cases}\label{eq:ring I_=00007Br; n; m_1, m_2=00007D^=00007Bz-noise+tfc=00007D -- 1}
\end{equation}

\begin{equation}
\mathring{I}_{\nu;r;n;m_{1},m_{2}}^{\left(\text{bath}\right)}\left(j_{r;m_{1}},j_{r;m_{2}}\right)=I_{\nu;r;n;\tilde{q}_{m_{1}},\tilde{q}_{m_{2}}}^{\left(\text{bath}\right)}\left(g_{1}\left(j_{r;m_{1}}\right),g_{-1}\left(j_{r;m_{1}}\right),g_{1}\left(j_{r;m_{2}}\right),g_{-1}\left(j_{r;m_{2}}\right)\right),\label{eq:ring I_=00007Bv; r; n; m_1; m_2=00007D^(bath) -- 1}
\end{equation}

\noindent $I_{\nu;r;n;q_{1},q_{2}}^{\left(\text{bath}\right)}\left(\cdots\right)$
given by Eq.~\eqref{eq:I_=00007Bv; r; n; q_1, q_2=00007D^(bath) in main manuscript -- 1},
$\tilde{q}_{m}$ given by Eq.~\eqref{eq:tilde q_m -- 1},

\begin{equation}
\mathring{I}_{r;n;m_{1},m_{2}}^{\left(\text{tfc+y-bath}\right)}\left(j_{r;m_{1}},j_{r;m_{2}}\right)=\mathring{I}_{y;r;n;k=\frac{1}{2}\tilde{q}_{m_{1}}}^{\left(\text{tfc}\right)}\left(j_{r;m_{1}},j_{r;m_{2}}\right)\mathring{I}_{y;r;n;m_{1},m_{2}}^{\left(\text{bath}\right)}\left(j_{r;m_{1}},j_{r;m_{2}}\right),\label{eq:ring I_=00007Bv; r; n; m_1; m_2=00007D^(tfc+y-bath) -- 1}
\end{equation}

\begin{equation}
\mathring{I}_{r;n;m_{1},m_{2}}^{\left(\text{tfc+z-bath}\right)}\left(j_{r;m_{1}},j_{r;m_{2}}\right)=\mathring{I}_{z;r;n;k=\tilde{q}_{m_{1}}}^{\left(\text{tfc}\right)}\left(j_{r;m_{1}},j_{r;m_{2}}\right)\mathring{I}_{z;r;n;m_{1},m_{2}}^{\left(\text{bath}\right)}\left(j_{r;m_{1}},j_{r;m_{2}}\right),\label{eq:ring I_=00007Bv; r; n; m_1; m_2=00007D^(tfc+z-bath) -- 1}
\end{equation}

\begin{equation}
\mathring{I}_{\nu;r;n;k}^{\left(\text{tfc}\right)}\left(j_{r;m_{1}},j_{r;m_{2}}\right)=I_{\nu;r;n;k}^{\left(\text{tfc}\right)}\left(g_{1}\left(j_{r;m_{1}}\right),g_{-1}\left(j_{r;m_{1}}\right),g_{1}\left(j_{r;m_{2}}\right),g_{-1}\left(j_{r;m_{2}}\right)\right),\label{eq:ring I_=00007Bv; r; n; k=00007D^(tfc) -- 1}
\end{equation}

\noindent $I_{\nu;r;n;k}^{\left(\text{tfc}\right)}\left(\cdots\right)$
given by Eq.~\eqref{eq:I_=00007Bv; r; n; k=00007D^(tfc) in main manuscript -- 1},

\begin{equation}
\mathring{I}^{\left(y\leftrightarrow z,1\right)}\left(j_{r;m_{1}},j_{r;m_{2}}\right)=I^{\left(y\leftrightarrow z,1\right)}\left(g_{1}\left(j_{r;m_{1}}\right),g_{-1}\left(j_{r;m_{1}}\right),g_{1}\left(j_{r;m_{2}}\right),g_{-1}\left(j_{r;m_{2}}\right)\right),\label{eq:ring I^(y<->z, 1) -- 1}
\end{equation}

\begin{equation}
\mathring{I}^{\left(y\leftrightarrow z,2\right)}\left(j_{r;m_{1}},j_{r;m_{2}}\right)=I^{\left(y\leftrightarrow z,2\right)}\left(g_{1}\left(j_{r;m_{2}}\right),g_{-1}\left(j_{r;m_{2}}\right),g_{1}\left(j_{r;m_{1}}\right),g_{-1}\left(j_{r;m_{1}}\right)\right),\label{eq:ring I^(y<->z, 2) -- 1}
\end{equation}

\noindent and $I^{\left(y\leftrightarrow z,1\right)}\left(\cdots\right)$
and $I^{\left(y\leftrightarrow z,2\right)}\left(\cdots\right)$ given
by Eqs.~\eqref{eq:I^(y<->z, 1) in main manuscript -- 1} and \eqref{eq:I^(y<->z, 2) in main manuscript -- 1}
respectively.

In $\texttt{spinbosonchain}$, the quantities $\mathring{I}_{\nu;r;n;m_{1},m_{2}}^{\left(\text{bath}\right)}\left(j_{r;m_{1}},j_{r;m_{2}}\right)$,
$\mathring{I}_{\nu;r;n;k}^{\left(\text{tfc}\right)}\left(j_{r;m_{1}},j_{r;m_{2}}\right)$,
$\mathring{I}^{\left(y\leftrightarrow z,1\right)}\left(j_{r;m_{1}},j_{r;m_{2}}\right)$,
\\
$\mathring{I}^{\left(y\leftrightarrow z,2\right)}\left(j_{r;m_{1}},j_{r;m_{2}}\right)$,
and $\mathring{I}_{r;n;m_{1},m_{2}}\left(j_{r;m_{1}},j_{r;m_{2}}\right)$
are represented by instances of the classes \\
$\texttt{spinbosonchain.\_influence.twopt.Bath}$, $\texttt{spinbosonchain.\_influence.twopt.TF}$,
\\
$\texttt{spinbosonchain.\_influence.twopt.YZ}$, $\texttt{spinbosonchain.\_influence.twopt.ZY}$,
and \\
$\texttt{spinbosonchain.\_influence.twopt.Total}$ respectively.

\subsection{Tensor network representation of the influence functionals\label{subsec:Tensor network representation of influence functional -- 1}}

In this section, we show how the influence functional $\mathring{I}_{r;n}\left(b_{r;0};j_{r;m\in\left[0,\left(n+1\right)\Delta m\right]}\right)$
given by Eq.~\eqref{eq:ring I_=00007Br; n=00007D -- 1} can be calculated
efficiently using tensor networks.

First, we introduce the two rank-$3$ tensors:

\begin{equation}
\left[M_{r;n;m_{2}}^{\left(\mathring{I}\right)}\right]_{b_{r;m_{2}};0}^{j_{r;m_{2}}}=\delta_{b_{r;m_{2}},j_{r;m_{2}}}\mathring{I}_{r;n;m_{2},m_{2}}\left(j_{r;m_{2}},j_{r;m_{2}}\right),\label{eq:M_=00007Br; n; m_2=00007D^(I) -- 1}
\end{equation}

\begin{equation}
\left[M^{\left(\mathring{I}\right)}\right]_{0;0}^{j_{r;m_{2}}}=1,\label{eq:M^(I) -- 1}
\end{equation}

\noindent where $\mathring{I}_{r;n;m_{1},m_{2}}\left(j_{r;m_{1}},j_{r;m_{2}}\right)$
is given by Eq.~\eqref{eq:ring I_=00007Br; n; m_1, m_2=00007D^=00007Byz-noise+tfc=00007D -- 1}.
We also introduce four rank-$4$ tensors:

\begin{equation}
\left[W_{r;n;m_{1}=\mu_{m_{2};\tau},m_{2}}^{\left(\mathring{I}\right)}\right]_{0;b_{r;\mu_{m_{2};\tau}+1}}^{j_{r;\mu_{m_{2};\tau}};j_{r;\mu_{m_{2};\tau}}^{\prime}}=\delta_{j_{r;\mu_{m_{2};\tau}},j_{r;\mu_{m_{2};\tau}}^{\prime}}\mathring{I}_{r;n;\mu_{m_{2};\tau},m_{2}}\left(j_{r;\mu_{m_{2};\tau}},b_{r;\mu_{m_{2};\tau}+1}\right),\label{eq:W_=00007Br; n; m_1, m_2=00007D^(I) -- 1}
\end{equation}

\begin{equation}
\left[W_{r;n;m_{1},m_{2}}^{\left(\mathring{I}\right)}\right]_{b_{r;m_{1}};b_{r;m_{1}+1}}^{j_{r;m_{1}};j_{r;m_{1}}^{\prime}}=\delta_{b_{r;m_{1}},b_{r;m_{1}+1}}\delta_{j_{r;m_{1}},j_{r;m_{1}}^{\prime}}\mathring{I}_{r;n;m_{1},m_{2}}\left(j_{r;m_{1}},b_{r;m_{1}+1}\right),\quad\text{for }\mu_{m_{2};\tau}+1\le m_{1}\le m_{2}-1,\label{eq:W_=00007Br; n; m_1, m_2=00007D^(I) -- 2}
\end{equation}

\begin{equation}
\left[W_{r;n;m_{1}=m_{2},m_{2}<\left(n+1\right)\Delta m}^{\left(\mathring{I}\right)}\right]_{b_{r;m_{1}};0}^{j_{r;m_{1}};j_{r;m_{1}}^{\prime}}=\delta_{b_{r;m_{1}},j_{r;m_{1}}}\delta_{j_{r;m_{1}},j_{r;m_{1}}^{\prime}}\mathring{I}_{r;n;m_{2},m_{2}}\left(j_{r;m_{2}},j_{r;m_{2}}\right),\label{eq:W_=00007Br; n; m_1, m_2=00007D^(I) -- 3}
\end{equation}

\noindent and

\begin{equation}
\left[W_{r;n;m_{1}=m_{2},m_{2}=\left(n+1\right)\Delta m}^{\left(\mathring{I}\right)}\right]_{b_{r;m_{1}};b_{r;m_{1}+1}}^{j_{r;m_{1}};j_{r;m_{1}}^{\prime}}=\delta_{b_{r;m_{1}},j_{r;m_{1}}}\delta_{j_{r;m_{1}},j_{r;m_{1}}^{\prime}}\delta_{j_{r;m_{1}}^{\prime},b_{r;m_{1}+1}}\mathring{I}_{r;n;m_{2},m_{2}}\left(j_{r;m_{2}},j_{r;m_{2}}\right),\label{eq:W_=00007Br; n; m_1, m_2=00007D^(I) -- 4}
\end{equation}

\noindent where $\mu_{m;\tau}$ is given by Eq.~\eqref{eq:mu_=00007Bm; tau=00007D -- 1}.
Note that any tensor indices labelled ``$0$'' indicate trivial
indices that are of dimension $1$. In $\texttt{spinbosonchain}$,
the tensors $M_{r;n;m_{2}}^{\left(\mathring{I}\right)}$ and $W_{r;n;m_{1},m_{2}}^{\left(\mathring{I}\right)}$
are built by instances of the ``factory'' classes $\texttt{spinbosonchain.\_influence.tensorfactory.InfluenceNodeRank3}$
and \\
$\texttt{spinbosonchain.\_influence.tensorfactory.InfluenceNodeRank4}$
respectively.

The rank-$4$ tensors introduced above are used to construct MPO's
of the form:

\begin{equation}
\left[\Omega_{r;n;m_{2}}^{\left(\mathring{I}\right)}\right]_{b_{r;\mu_{m_{2};\tau}}=0;b_{r;m_{2}+1}}^{j_{r;\mu_{m_{2};\tau}},\ldots,j_{r;m_{2}};j_{r;\mu_{m_{2};\tau}}^{\prime},\ldots,j_{r;m_{2}}^{\prime}}=\prod_{m_{1}=\mu_{m_{2};\tau}+1}^{m_{2}}\left\{ \sum_{b_{r;m_{1}}}\right\} \prod_{m_{1}=\mu_{m_{2};\tau}}^{m_{2}}\left\{ \left[W_{r;n;m_{1},m_{2}}^{\left(\mathring{I}\right)}\right]_{b_{r;m_{1}};b_{r;m_{1}+1}}^{j_{r;m_{1}};j_{r;m_{1}}^{\prime}}\right\} .\label{eq:Ohm_=00007Br; n; m_2=00007D^(I) -- 1}
\end{equation}

\noindent Note that from Eqs.~\eqref{eq:ring I_=00007Br; n; m_2=00007D -- 1}
and \eqref{eq:M_=00007Br; n; m_2=00007D^(I) -- 1}-\eqref{eq:Ohm_=00007Br; n; m_2=00007D^(I) -- 1},
we have:

\begin{align}
 & \mathring{I}_{r;n=\infty;m_{2}+1}\left(j_{r;m\in\left[\mu_{m_{2}+1;\tau},m_{2}+1\right]}\right)F\left(j_{r;0},\ldots,j_{r;m_{2}}\right)\nonumber \\
 & \quad=\prod_{m=\mu_{m_{2}+1;\tau}}^{m_{2}+1}\left\{ \sum_{j_{r;m}^{\prime}}\right\} \left[\Omega_{r;n=\infty;m_{2}+1}^{\left(\mathring{I}\right)}\right]_{b_{r;\mu_{m_{2}+1;\tau}}=0;b_{r;m_{2}+2}}^{j_{r;\mu_{m_{2}+1;\tau}},\ldots,j_{r;m_{2}+1};j_{r;\mu_{m_{2}+1;\tau}}^{\prime},\ldots,j_{r;m_{2}+1}^{\prime}}\nonumber \\
 & \quad\mathrel{\phantom{=}}\mathop{\times}F\left(j_{r;0},\ldots,j_{r;\mu_{m_{2}+1;\tau}-1},j_{r;\mu_{m_{2}+1;\tau}}^{\prime},\ldots,j_{r;m_{2}}^{\prime}\right)\left[M^{\left(\mathring{I}\right)}\right]_{0;0}^{j_{r;m_{2}+1}^{\prime}},\label{eq:expressing ring I_=00007Br; infty; m_2+1=00007D * F as a MPO+MPS -- 1}
\end{align}

\noindent where $F\left(j_{r;0},\ldots,j_{r;m_{2}}\right)$ is an
arbitrary function. In $\texttt{spinbosonchain}$, the MPO's $\Omega_{r;n;m_{2}}^{\left(\mathring{I}\right)}$
are built by instances of the ``factory'' classes $\texttt{spinbosonchain.\_influence.tensorfactory.InfluenceMPO}$.

Next, we show how the influence functional $\mathring{I}_{r;m_{2}}^{\left(1\right)}\left(b_{r;0};j_{r;m\in\left[0,m_{2}\right]}\right)$
{[}Eq.~\eqref{eq:ring I_=00007Br; m_2+1=00007D^(1) -- 1}{]} can
be calculated using MPS's and MPO's in an interative fashion. The
$\mathring{I}_{r;m_{2}}^{\left(1\right)}\left(b_{r;0};j_{r;m\in\left[0,m_{2}\right]}\right)$
can be expressed generally as two MPS's that are joined together: 

\begin{equation}
\left[\Xi_{r;m_{2}}^{\left(\mathring{I},1\right)}\right]_{b_{r;0};b_{r;m_{2}+1}=0}^{j_{r;0},\ldots,j_{r;m_{2}}}=\sum_{b_{r;\mu_{m_{2}+1;\tau}}^{\prime}}\left[\Xi_{r;0,\mu_{m_{2}+1;\tau}-1}^{\left(\mathring{I},1,1\right)}\right]_{b_{r;0};b_{r;\mu_{m_{2}+1;\tau}}^{\prime}}^{j_{r;0},\ldots,j_{r;\mu_{m_{2}+1;\tau}-1}}\left[\Xi_{r;m_{2}}^{\left(\mathring{I},1,2\right)}\right]_{b_{r;\mu_{m_{2}+1;\tau}}^{\prime};b_{r;m_{2}+1}=0}^{j_{r;\mu_{m_{2}+1;\tau}},\ldots,j_{r;m_{2}}},\label{eq:Xi_=00007Br; m_2=00007D^(I, 1) -- 1}
\end{equation}

\noindent where $\Xi_{r;m_{1},m_{2}}^{\left(\mathring{I},1,1\right)}$
is the first MPS:

\begin{equation}
\left[\Xi_{r;m_{1},m_{2}}^{\left(\mathring{I},1,1\right)}\right]_{b_{r;m_{1}};b_{r;m_{2}+1}}^{j_{r;m_{1}},\ldots,j_{r;m_{2}}}=\begin{cases}
\prod_{m=m_{1}+1}^{m_{2}}\left\{ \sum_{b_{r;m}}\right\} \prod_{m=m_{1}}^{m_{2}}\left\{ \left[M_{r;m}^{\left(\mathring{I},1,1\right)}\right]_{b_{r;m};b_{r;m+1}}^{j_{r;m}}\right\} , & \text{if }m_{1}\le m_{2},\\
\delta_{b_{r;m_{1}},b_{r;m_{2}+1}}, & \text{otherwise},
\end{cases}\label{eq:Xi_=00007Br; m_1, m_2=00007D^(I, 1, 1) -- 1}
\end{equation}

\noindent with the rank-$3$ tensors $M_{r;m}^{\left(\mathring{I},1,1\right)}$
determined from previous steps in the iteration procedure; and $\Xi_{r;m_{2}}^{\left(\mathring{I},1,2\right)}$
is the second MPS:

\begin{equation}
\left[\Xi_{r;m_{2}}^{\left(\mathring{I},1,2\right)}\right]_{b_{r;\mu_{m_{2}+1;\tau}};b_{r;m_{2}+1}=0}^{j_{r;\mu_{m_{2}+1;\tau}},\ldots,j_{r;m_{2}}}=\prod_{m_{1}=\mu_{m_{2}+1;\tau}+1}^{m_{2}}\left\{ \sum_{b_{r;m_{1}}}\right\} \prod_{m_{1}=\mu_{m_{2}+1;\tau}}^{m_{2}}\left\{ \left[M_{r;m_{1},m_{2}}^{\left(\mathring{I},1,2\right)}\right]_{b_{r;m_{1}};b_{r;m_{1}+1}}^{j_{r;m_{1}}}\right\} ,\label{eq:Xi_=00007Br; m_1, m_2=00007D^(I, 1, 2) -- 1}
\end{equation}

\noindent with the rank-$3$ tensors $M_{r;m_{1},m_{2}}^{\left(\mathring{I},1,2\right)}$
determined from previous steps in the iteration procedure. It is important
to note that we use the following product notation convention:

\begin{equation}
\prod_{l=l_{1}}^{l_{2}}\left\{ F_{l}\right\} =\begin{cases}
F_{l_{1}}F_{l_{1}+1}\cdots F_{l_{2}}, & \text{if }l_{1}\le l_{2},\\
1, & \text{if }l_{1}>l_{2},
\end{cases}\label{eq:product convention -- 1}
\end{equation}

\noindent where $F_{l}$ is an arbitrary sequence. The initial step
to our iterative procedure is to set:

\begin{equation}
\left[M_{r;0,0}^{\left(\mathring{I},1,2\right)}\right]_{b_{r;0};b_{r;1}}^{j_{r;0}}=\left[M_{r;n=\infty;0}^{\left(\mathring{I}\right)}\right]_{b_{r;0};0}^{j_{r;0}}.\label{eq:M_=00007Br; 0, 0=00007D^(I, 1, 2) -- 1}
\end{equation}

\noindent The iterative step consists of calculating first:

\begin{align}
\left[\Xi_{r;m_{2}+1}^{\left(\mathring{I},1,2,1\right)}\right]_{b_{r;\mu_{m_{2}+1;\tau}};b_{r;m_{2}+2}=0}^{j_{r;\mu_{m_{2}+1;\tau}},\ldots,j_{r;m_{2}+1}} & =\prod_{m_{1}=\mu_{m_{2}+1;\tau}}^{m_{2}+1}\left\{ \sum_{j_{r;m}^{\prime}}\right\} \left[\Omega_{r;n=\infty;m_{2}+1}^{\left(\mathring{I}\right)}\right]_{0;0}^{j_{r;\mu_{m_{2}+1;\tau}},\ldots,j_{r;m_{2}+1};j_{r;\mu_{m_{2}+1;\tau}}^{\prime},\ldots,j_{r;m_{2}+1}^{\prime}}\nonumber \\
 & \mathrel{\phantom{=}}\mathop{\times}\left[\Xi_{r;m_{2}}^{\left(\mathring{I},1,2\right)}\right]_{b_{r;\mu_{m_{2}+1;\tau}};0}^{j_{r;\mu_{m_{2}+1;\tau}}^{\prime},\ldots,j_{r;m_{2}}^{\prime}}\left[M^{\left(\mathring{I}\right)}\right]_{0;0}^{j_{r;m_{2}+1}^{\prime}}\nonumber \\
 & \quad=\prod_{m_{1}=\mu_{m_{2}+1;\tau}+1}^{m_{2}+1}\left\{ \sum_{b_{r;m_{1}}}\right\} \prod_{m_{1}=\mu_{m_{2}+1;\tau}}^{m_{2}+1}\left\{ \left[M_{r;m_{1},m_{2}+1}^{\left(\mathring{I},1,2,1\right)}\right]_{b_{r;m_{1}};b_{r;m_{1}+1}}^{j_{r;m_{1}}}\right\} ,\label{eq:Xi_=00007Br; m_2+1=00007D^(I, 1, 2, 1) -- 1}
\end{align}

\noindent where

\begin{align}
 & \left[M_{r;m_{1}=\mu_{m_{2}+1;\tau},m_{2}+1}^{\left(\mathring{I},1,2,1\right)}\right]_{b_{r;\mu_{m_{2}+1;\tau}}=b_{r;\mu_{m_{2}+1;\tau}}^{\prime};b_{r;\mu_{m_{2}+1;\tau}+1}=\left(b_{r;\mu_{m_{2}+1;\tau}+1}^{\prime\prime},b_{r;\mu_{m_{2}+1;\tau}+1}^{\prime}\right)}^{j_{r;\mu_{m_{2}+1;\tau}}}\nonumber \\
 & \quad=\sum_{j_{r;\mu_{m_{2}+1;\tau}}^{\prime}}\left[W_{r;n=\infty;\mu_{m_{2}+1;\tau},m_{2}+1}^{\left(\mathring{I}\right)}\right]_{0;b_{r;\mu_{m_{2}+1;\tau}+1}^{\prime\prime}}^{j_{r;\mu_{m_{2}+1;\tau}};j_{r;\mu_{m_{2}+1;\tau}}^{\prime}}\left[M_{r;\mu_{m_{2}+1;\tau},m_{2}}^{\left(\mathring{I},1,2\right)}\right]_{b_{r;\mu_{m_{2}+1;\tau}}^{\prime};b_{r;\mu_{m_{2}+1;\tau}+1}^{\prime}}^{j_{r;\mu_{m_{2}+1;\tau}}^{\prime}},\label{eq:M_=00007Br; m_1, m_2+1=00007D^(I, 1, 2, 1) -- 1}
\end{align}

\begin{align}
 & \left[M_{r;m_{1},m_{2}+1}^{\left(\mathring{I},1,2,1\right)}\right]_{b_{r;m_{1}}=\left(b_{r;m_{1}}^{\prime\prime},b_{r;m_{1}}^{\prime}\right);b_{r;m_{1}+1}=\left(b_{r;m_{1}+1}^{\prime\prime},b_{r;m_{1}+1}^{\prime}\right)}^{j_{r;m_{1}}}\nonumber \\
 & \quad\mathrel{\phantom{=}}=\sum_{j_{r;m_{1}}^{\prime}}\left[W_{r;n=\infty;m_{1},m_{2}+1}^{\left(\mathring{I}\right)}\right]_{b_{r;m_{1}}^{\prime\prime};b_{r;m_{1}+1}^{\prime\prime}}^{j_{r;m_{1}};j_{r;m_{1}}^{\prime}}\left[M_{r;m_{1},m_{2}}^{\left(\mathring{I},1,2\right)}\right]_{b_{r;m_{1}}^{\prime};b_{r;m_{1}+1}^{\prime}}^{j_{r;m_{1}}^{\prime}},\quad\text{for }\mu_{m_{2}+1;\tau}+1\le m_{1}\le m_{2}-1,\label{eq:M_=00007Br; m_1, m_2+1=00007D^(I, 1, 2, 1) -- 2}
\end{align}

\begin{equation}
\left[M_{r;m_{1}=m_{2},m_{2}+1}^{\left(\mathring{I},1,2,1\right)}\right]_{b_{r;m_{2}}=\left(b_{r;m_{2}}^{\prime\prime},b_{r;m_{2}}^{\prime}\right);b_{r;m_{2}+1}=b_{r;m_{2}+1}^{\prime\prime}}^{j_{r;m_{2}}}=\sum_{j_{r;m_{2}}^{\prime}}\left[W_{r;n=\infty;m_{2},m_{2}+1}^{\left(\mathring{I}\right)}\right]_{b_{r;m_{2}}^{\prime\prime};b_{r;m_{2}+1}^{\prime\prime}}^{j_{r;m_{2}};j_{r;m_{2}}^{\prime}}\left[M_{r;m_{2},m_{2}}^{\left(\mathring{I},1,2\right)}\right]_{b_{r;m_{2}}^{\prime};0}^{j_{r;m_{2}}^{\prime}},\label{eq:M_=00007Br; m_1, m_2+1=00007D^(I, 1, 2, 1) -- 3}
\end{equation}

\noindent and

\begin{equation}
\left[M_{r;m_{1}=m_{2}+1,m_{2}+1}^{\left(\mathring{I},1,2,1\right)}\right]_{b_{r;m_{2}+1}=b_{r;m_{2}+1}^{\prime\prime};b_{r;m_{2}+2}=0}^{j_{r;m_{2}+1}}=\sum_{j_{r;m_{2}+1}^{\prime}}\left[W_{r;n=\infty;m_{2}+1,m_{2}+1}^{\left(\mathring{I}\right)}\right]_{b_{r;m_{2}+1}^{\prime\prime};0}^{j_{r;m_{2}+1};j_{r;m_{2}+1}^{\prime}}\left[M^{\left(\mathring{I}\right)}\right]_{0;0}^{j_{r;m_{2}+1}^{\prime}}.\label{eq:M_=00007Br; m_1, m_2+1=00007D^(I, 1, 2, 1) -- 4}
\end{equation}

\noindent The internal bond dimensions associated with the indices
$\left\{ b_{r;m}\right\} _{m=\mu_{m_{2}+1;\tau}+1}^{m_{2}}$ in $\Xi_{r;m_{2}+1}^{\left(\mathring{I},1,2,1\right)}$
{[}Eq.~\eqref{eq:Xi_=00007Br; m_2+1=00007D^(I, 1, 2, 1) -- 1}{]}
are four times larger than those in $\Xi_{r;m_{2}}^{\left(\mathring{I},1,2\right)}$
{[}Eq.~\eqref{eq:Xi_=00007Br; m_1, m_2=00007D^(I, 1, 2) -- 1}{]},
therefore in order to avoid an exponential growth in the bond dimensions
upon repeated applications of the iterative procedure, one must compress
$\Xi_{r;m_{2}+1}^{\left(\mathring{I},1,2,1\right)}$. For a brief
discussion on the compression techniques available in $\texttt{spinbosonchain}$,
see Appendix~\ref{sec:MPS compression techniques used in sbc}. After
performing the MPS compression, we set

\begin{equation}
\left[M_{r;m_{1},m_{2}+1}^{\left(\mathring{I},1,2\right)}\right]_{b_{r;m_{1}};b_{r;m_{1}+1}}^{j_{r;m_{1}}}=\left[M_{r;m_{1},m_{2}+1}^{\left(\mathring{I},1,2,1\right)}\right]_{b_{r;m_{1}};b_{r;m_{1}+1}}^{j_{r;m_{1}}},\quad\text{for }\mu_{m_{2}+2;\tau}\le m_{1}\le m_{2}+1,\label{eq:M_=00007Br; m_1, m_2+1=00007D^(I, 1, 2) -- 1}
\end{equation}

\noindent and

\begin{equation}
\left[M_{r;\mu_{m_{2}+1;\tau}}^{\left(\mathring{I},1,1\right)}\right]_{b_{r;\mu_{m_{2}+1;\tau}};b_{r;\mu_{m_{2}+1;\tau}+1}}^{j_{r;\mu_{m_{2}+1;\tau}}}=\left[M_{r;\mu_{m_{2}+1;\tau},m_{2}+1}^{\left(\mathring{I},1,2,1\right)}\right]_{b_{r;\mu_{m_{2}+1;\tau}};b_{r;\mu_{m_{2}+1;\tau}+1}}^{j_{r;\mu_{m_{2}+1;\tau}}},\quad\text{if }\mu_{m_{2}+2;\tau}\ge1,\label{eq:M_=00007Br; mu=00007D^(I, 1, 1) -- 1}
\end{equation}

\noindent which completes the iterative step.

Note that Eqs.~\eqref{eq:Xi_=00007Br; m_2=00007D^(I, 1) -- 1}-\eqref{eq:M_=00007Br; mu=00007D^(I, 1, 1) -- 1}
imply that

\begin{equation}
\left[\Xi_{r;0}^{\left(\mathring{I},1\right)}\right]_{b_{r;0};b_{r;1}=0}^{j_{r;0}}=\left[M_{r;n=\infty;0}^{\left(\mathring{I}\right)}\right]_{b_{r;0};0}^{j_{r;0}},\label{eq:consise iterative identity for Xi_=00007Br; m_2+1=00007D^(I, 1) -- 1}
\end{equation}

\noindent and

\begin{align}
\left[\Xi_{r;m_{2}+1}^{\left(\mathring{I},1\right)}\right]_{b_{r;0};b_{r;m_{2}+2}=0}^{j_{r;0},\ldots,j_{r;m_{2}+1}} & \quad=\sum_{b_{r;m_{2}+1}}\prod_{m=\mu_{m_{2}+1;\tau}}^{m_{2}+1}\left\{ \sum_{j_{r;m}^{\prime}}\right\} \left[\Omega_{r;n=\infty;m_{2}+1}^{\left(\mathring{I}\right)}\right]_{0;0}^{j_{r;\mu_{m_{2}+1;\tau}},\ldots,j_{r;m_{2}+1};j_{r;\mu_{m_{2}+1;\tau}}^{\prime},\ldots,j_{r;m_{2}+1}^{\prime}}\nonumber \\
 & \quad\mathrel{\phantom{=}}\mathop{\times}\left[\Xi_{r;m_{2}}^{\left(\mathring{I},1\right)}\right]_{b_{r;0};0}^{j_{r;0},\ldots,j_{r;\mu_{m_{2}+1;\tau}-1},j_{r;\mu_{m_{2}+1;\tau}}^{\prime},\ldots,j_{r;m_{2}}^{\prime}}\left[M^{\left(\mathring{I}\right)}\right]_{0;0}^{j_{r;m_{2}+1}^{\prime}}.\label{eq:consise iterative identity for Xi_=00007Br; m_2+1=00007D^(I, 1) -- 2}
\end{align}

\noindent Furthermore, Eqs.~\eqref{eq:mu_=00007Bm; tau=00007D -- 1},
\eqref{eq:M_=00007Br; n; m_2=00007D^(I) -- 1}, \eqref{eq:M^(I) -- 1},
\eqref{eq:expressing ring I_=00007Br; infty; m_2+1=00007D * F as a MPO+MPS -- 1},
\eqref{eq:consise iterative identity for Xi_=00007Br; m_2+1=00007D^(I, 1) -- 1}
and \eqref{eq:consise iterative identity for Xi_=00007Br; m_2+1=00007D^(I, 1) -- 2}
imply that

\begin{equation}
\left[\Xi_{r;0}^{\left(\mathring{I},1\right)}\right]_{b_{r;0};b_{r;1}=0}^{j_{r;0}}=\delta_{b_{r;0},j_{r;0}}\mathring{I}_{r;n;0,0}\left(j_{r;0},j_{r;0}\right),\label{eq:consise iterative identity for Xi_=00007Br; m_2+1=00007D^(I, 1) -- 3}
\end{equation}

\noindent and

\begin{equation}
\left[\Xi_{r;m_{2}+1}^{\left(\mathring{I},1\right)}\right]_{b_{r;0};b_{r;m_{2}+2}=0}^{j_{r;0},\ldots,j_{r;m_{2}+1}}=\mathring{I}_{r;n=\infty;m_{2}+1}\left(j_{r;m\in\left[\mu_{m_{2}+1;\tau},m_{2}+1\right]}\right)\left[\Xi_{r;m_{2}}^{\left(\mathring{I},1\right)}\right]_{b_{r;0};0}^{j_{r;0},\ldots,j_{r;m_{2}}}\left[M^{\left(\mathring{I}\right)}\right]_{0;0}^{j_{r;m_{2}+1}^{\prime}}.\label{eq:consise iterative identity for Xi_=00007Br; m_2+1=00007D^(I, 1) -- 4}
\end{equation}

\noindent Upon comparing Eqs.~\eqref{eq:consise iterative identity for Xi_=00007Br; m_2+1=00007D^(I, 1) -- 3}
and \eqref{eq:consise iterative identity for Xi_=00007Br; m_2+1=00007D^(I, 1) -- 4}
to Eqs.~\eqref{eq:ring I_=00007Br; 0=00007D^(1) -- 1} and \eqref{eq:ring I_=00007Br; m_2+1=00007D^(1) -- 1},
it should be straightforward to see that

\begin{equation}
\left[\Xi_{r;m_{2}}^{\left(\mathring{I},1\right)}\right]_{b_{r;0};0}^{j_{r;0},\ldots,j_{r;m_{2}}}=\mathring{I}_{r;m_{2}}^{\left(1\right)}\left(b_{r;0};j_{r;m\in\left[0,m_{2}\right]}\right).\label{eq:Xi_=00007Br; m_2=00007D^(I, 1) to ring I_=00007Br; m_2=00007D^(1) -- 1}
\end{equation}

\noindent Strictly speaking, the ``$=$'' symbol in Eq.~\eqref{eq:Xi_=00007Br; m_2=00007D^(I, 1) to ring I_=00007Br; m_2=00007D^(1) -- 1}
should be ``$\approx$'' when compression is performed: the larger
the bond dimension, the more accurate the approximation.

The reason why $\Xi_{r;m_{2}}^{\left(\mathring{I},1\right)}$ is expressed
as two MPS's joined together in Eq.~\eqref{eq:Xi_=00007Br; m_2=00007D^(I, 1) -- 1}
is that the left MPS $\Xi_{r;0,\mu_{m_{2}+1;\tau}-1}^{\left(\mathring{I},1,1\right)}$
is not used in the iterative step. This fact will be exploited when
calculating the remainder of the path integral representation of $\rho^{\left(A\right)}\left(\mathbf{j}_{\left(n+1\right)\Delta m};t_{n}\right)$
{[}Eq.~\eqref{eq:rearranging terms in rho^(A) -- 1}{]}, which we
discuss in Secs.~\ref{subsec:Rearranging terms in the path integral expression of rho^(A) further -- 1}-\ref{subsec:Remaining elements of tensor network algorithm -- 1}.

It is worth pointing out that $\Xi_{r;m_{2}}^{\left(\mathring{I},1,2\right)}$
{[}Eq.~\eqref{eq:Xi_=00007Br; m_2=00007D^(I, 1) -- 1}{]}, as an
MPS, can have no more than $N_{\max}^{\left(\mathring{I}\right)}=K_{\tau}\Delta m-1$
nodes by design. At the beginning of the iterative procedure, $\Xi_{r;m_{2}=0}^{\left(\mathring{I},1,2\right)}$
is an MPS of one node. Upon each iterative step, $\Xi_{r;m_{2}}^{\left(\mathring{I},1,2\right)}$
grows in size by one node, until it contains $N_{\max}^{\left(\mathring{I}\right)}$
nodes. The number of nodes then remains fixed with subsequent iterative
steps, although the actual nodes making up $\Xi_{r;m_{2}}^{\left(\mathring{I},1,2\right)}$
will change.

Once $\Xi_{r;n\Delta m-1}^{\left(\mathring{I},1,2\right)}$ has been
calculated using the iterative procedure described above, we can use
it as input to another iterative procedure which is used to calculate
$\mathring{I}_{r;n}\left(b_{r;0};j_{r;m\in\left[0,\left(n+1\right)\Delta m\right]}\right)$
{[}Eq.~\eqref{eq:ring I_=00007Br; n=00007D -- 1}{]}. Below we describe
this second iterative procedure: First, we introduce the following
MPS, which itself is two MPS's joined together:

\begin{align}
\left[\Xi_{r;n;m_{2}}^{\left(\mathring{I},\dashv\right)}\right]_{b_{r;\mu_{n\Delta m;\tau}};b_{r;m_{2}+1}=0}^{j_{r;n\Delta m},\ldots,j_{r;m_{2}}} & =\sum_{b_{r;\mu_{m_{2}+1;\tau}}^{\prime}}\left[\Xi_{r;n;\mu_{n\Delta m;\tau},\mu_{m_{2}+1;\tau}-1}^{\left(\mathring{I},\dashv,1\right)}\right]_{b_{r;\mu_{n\Delta m;\tau}};b_{r;\mu_{m_{2}+1;\tau}}^{\prime}}^{j_{r;\mu_{n\Delta m;\tau}},\ldots,j_{r;\mu_{m_{2}+1;\tau}-1}}\nonumber \\
 & \mathrel{\phantom{=}}\mathop{\times}\left[\Xi_{r;n;m_{2}}^{\left(\mathring{I},\dashv,2\right)}\right]_{b_{r;\mu_{m_{2}+1;\tau}}^{\prime};b_{r;m_{2}+1}=0}^{j_{r;\mu_{m_{2}+1;\tau}},\ldots,j_{r;m_{2}}},\mathrel{\phantom{=}}\quad\text{for }n\Delta m-1\le m_{2}\le\left(n+1\right)\Delta m,\label{eq:Xi_=00007Br; n; m_2=00007D^(I, dashv) -- 1}
\end{align}

\noindent where $\Xi_{r;n;m_{1},m_{2}}^{\left(\mathring{I},\dashv,1\right)}$
is the first MPS:

\begin{equation}
\left[\Xi_{r;n;m_{1},m_{2}}^{\left(\mathring{I},\dashv,1\right)}\right]_{b_{r;m_{1}};b_{r;m_{2}+1}}^{j_{r;m_{1}},\ldots,j_{r;m_{2}}}=\begin{cases}
\prod_{m=m_{1}+1}^{m_{2}}\left\{ \sum_{b_{r;m}}\right\} \prod_{m=m_{1}}^{m_{2}}\left\{ \left[M_{r;n;m}^{\left(\mathring{I},\dashv,1\right)}\right]_{b_{r;m};b_{r;m+1}}^{j_{r;m}}\right\} , & \text{if }m_{1}\le m_{2},\\
\delta_{b_{r;m_{1}},b_{r;m_{2}+1}}, & \text{otherwise},
\end{cases}\label{eq:Xi_=00007Br; n; m_1, m_2=00007D^(I, dashv, 1) -- 1}
\end{equation}

\noindent with the rank-$3$ tensors $M_{r;n;m}^{\left(\mathring{I},\dashv,1\right)}$
determined from previous steps in the iteration procedure; and $\Xi_{r;m_{2}}^{\left(\mathring{I},\dashv,2\right)}$
is the second MPS:

\begin{equation}
\left[\Xi_{r;n;m_{2}}^{\left(\mathring{I},\dashv,2\right)}\right]_{b_{r;\mu_{m_{2}+1;\tau}};b_{r;m_{2}+1}=0}^{j_{r;\mu_{m_{2}+1;\tau}},\ldots,j_{r;m_{2}}}=\prod_{m_{1}=\mu_{m_{2}+1;\tau}+1}^{m_{2}}\left\{ \sum_{b_{r;m_{1}}}\right\} \prod_{m_{1}=\mu_{m_{2}+1;\tau}}^{m_{2}}\left\{ \left[M_{r;n;m_{1},m_{2}}^{\left(\mathring{I},\dashv,2\right)}\right]_{b_{r;m_{1}};b_{r;m_{1}+1}}^{j_{r;m_{1}}}\right\} ,\label{eq:Xi_=00007Br; n; m_2=00007D^(I, dashv, 2) -- 1}
\end{equation}

\noindent with the rank-$3$ tensors $M_{r;n;m_{1},m_{2}}^{\left(\mathring{I},\dashv,2\right)}$
determined from previous steps in the iteration procedure. The initial
step to our iterative procedure is to set:

\begin{equation}
\left[M_{r;n;m_{1},n\Delta m-1}^{\left(\mathring{I},\dashv,2\right)}\right]_{b_{r;m_{1}};b_{r;m_{1}+1}}^{j_{r;m_{1}}}=\left[M_{r;m_{1},n\Delta m-1}^{\left(\mathring{I},1,2\right)}\right]_{b_{r;m_{1}};b_{r;m_{1}+1}}^{j_{r;m_{1}}},\quad\text{for }\mu_{n\Delta m;\tau}\le m_{1}\le n\Delta m-1,\label{eq:M_=00007Br; n; m_1, m_=00007By+z; n-1=00007D-1=00007D^(I, dashv, 2) -- 1}
\end{equation}

\noindent where the $M_{r;m_{1},n\Delta m-1}^{\left(\mathring{I},1,2\right)}$
nodes are those of $\Xi_{r;n\Delta m-1}^{\left(\mathring{I},1,2\right)}$
{[}see Eq.~\eqref{eq:Xi_=00007Br; m_1, m_2=00007D^(I, 1, 2) -- 1}{]}.
The iterative step consists of calculating first:

\begin{align}
\left[\Xi_{r;n;m_{2}+1}^{\left(\mathring{I},\dashv,2,1\right)}\right]_{b_{r;\mu_{m_{2}+1;\tau}};b_{r;m_{2}+2}=0}^{j_{r;\mu_{m_{2}+1;\tau}},\ldots,j_{r;m_{2}+1}} & =\prod_{m_{1}=\mu_{m_{2}+1;\tau}}^{m_{2}+1}\left\{ \sum_{j_{r;m}^{\prime}}\right\} \left[\Omega_{r;n;m_{2}+1}^{\left(\mathring{I}\right)}\right]_{0;b_{r;m_{2}+2}^{\prime\prime}}^{j_{r;\mu_{m_{2}+1;\tau}},\ldots,j_{r;m_{2}+1};j_{r;\mu_{m_{2}+1;\tau}}^{\prime},\ldots,j_{r;m_{2}+1}^{\prime}}\nonumber \\
 & \mathrel{\phantom{=}}\mathop{\times}\left[\Xi_{r;n;m_{2}}^{\left(\mathring{I},\dashv,2\right)}\right]_{b_{r;\mu_{m_{2}+1;\tau}};0}^{j_{r;\mu_{m_{2}+1;\tau}}^{\prime},\ldots,j_{r;m_{2}}^{\prime}}\left[M^{\left(\mathring{I}\right)}\right]_{0;0}^{j_{r;m_{2}+1}^{\prime}}\nonumber \\
 & \quad=\prod_{m_{1}=\mu_{m_{2}+1;\tau}+1}^{m_{2}+1}\left\{ \sum_{b_{r;m_{1}}}\right\} \prod_{m_{1}=\mu_{m_{2}+1;\tau}}^{m_{2}+1}\left\{ \left[M_{r;n;m_{1},m_{2}+1}^{\left(\mathring{I},\dashv,2,1\right)}\right]_{b_{r;m_{1}};b_{r;m_{1}+1}}^{j_{r;m_{1}}}\right\} ,\label{eq:Xi_=00007Br; n; m_2+1=00007D^(I, dashv, 2, 1) -- 1}
\end{align}

\noindent where

\begin{align}
 & \left[M_{r;n;m_{1}=\mu_{m_{2}+1;\tau},m_{2}+1}^{\left(\mathring{I},\dashv,2,1\right)}\right]_{b_{r;\mu_{m_{2}+1;\tau}}=b_{r;\mu_{m_{2}+1;\tau}}^{\prime};b_{r;\mu_{m_{2}+1;\tau}+1}=\left(b_{r;\mu_{m_{2}+1;\tau}+1}^{\prime\prime},b_{r;\mu_{m_{2}+1;\tau}+1}^{\prime}\right)}^{j_{r;\mu_{m_{2}+1;\tau}}}\nonumber \\
 & \quad=\sum_{j_{r;\mu_{m_{2}+1;\tau}}^{\prime}}\left[W_{r;n;\mu_{m_{2}+1;\tau},m_{2}+1}^{\left(\mathring{I}\right)}\right]_{0;b_{r;\mu_{m_{2}+1;\tau}+1}^{\prime\prime}}^{j_{r;\mu_{m_{2}+1;\tau}};j_{r;\mu_{m_{2}+1;\tau}}^{\prime}}\left[M_{r;n;\mu_{m_{2}+1;\tau},m_{2}}^{\left(\mathring{I},\dashv,2\right)}\right]_{b_{r;\mu_{m_{2}+1;\tau}}^{\prime};b_{r;\mu_{m_{2}+1;\tau}+1}^{\prime}}^{j_{r;\mu_{m_{2}+1;\tau}}^{\prime}},\label{eq:M_=00007Br; n; m_1, m_2+1=00007D^(I, dashv, 2, 1) -- 1}
\end{align}

\begin{align}
 & \left[M_{r;n;m_{1},m_{2}+1}^{\left(\mathring{I},\dashv,2,1\right)}\right]_{b_{r;m_{1}}=\left(b_{r;m_{1}}^{\prime\prime},b_{r;m_{1}}^{\prime}\right);b_{r;m_{1}+1}=\left(b_{r;m_{1}+1}^{\prime\prime},b_{r;m_{1}+1}^{\prime}\right)}^{j_{r;m_{1}}}\nonumber \\
 & \quad=\sum_{j_{r;m_{1}}^{\prime}}\left[W_{r;n;m_{1},m_{2}+1}^{\left(\mathring{I}\right)}\right]_{b_{r;m_{1}}^{\prime\prime};b_{r;m_{1}+1}^{\prime\prime}}^{j_{r;m_{1}};j_{r;m_{1}}^{\prime}}\left[M_{r;n;m_{1},m_{2}}^{\left(\mathring{I},\dashv,2\right)}\right]_{b_{r;m_{1}}^{\prime};b_{r;m_{1}+1}^{\prime}}^{j_{r;m_{1}}^{\prime}},\quad\text{for }\mu_{m_{2}+1;\tau}+1\le m_{1}\le m_{2}-1,\label{eq:M_=00007Br; n; m_1, m_2+1=00007D^(I, dashv, 2, 1) -- 2}
\end{align}

\begin{equation}
\left[M_{r;n;m_{1}=m_{2},m_{2}+1}^{\left(\mathring{I},\dashv,2,1\right)}\right]_{b_{r;m_{2}}=\left(b_{r;m_{2}}^{\prime\prime},b_{r;m_{2}}^{\prime}\right);b_{r;m_{2}+1}=b_{r;m_{2}+1}^{\prime\prime}}^{j_{r;m_{2}}}=\sum_{j_{r;m_{2}}^{\prime}}\left[W_{r;n;m_{2},m_{2}+1}^{\left(\mathring{I}\right)}\right]_{b_{r;m_{2}}^{\prime\prime};b_{r;m_{2}+1}^{\prime\prime}}^{j_{r;m_{2}};j_{r;m_{2}}^{\prime}}\left[M_{r;n;m_{2},m_{2}}^{\left(\mathring{I},\dashv,2\right)}\right]_{b_{r;m_{2}}^{\prime};0}^{j_{r;m_{2}}^{\prime}},\label{eq:M_=00007Br; n; m_1, m_2+1=00007D^(I, dashv, 2, 1) -- 3}
\end{equation}

\noindent and

\begin{equation}
\left[M_{r;n;m_{1}=m_{2}+1,m_{2}+1}^{\left(\mathring{I},\dashv,2,1\right)}\right]_{b_{r;m_{2}+1}=b_{r;m_{2}+1}^{\prime\prime};b_{r;m_{2}+2}=b_{r;m_{2}+2}^{\prime\prime}}^{j_{r;m_{2}+1}}=\sum_{j_{r;m_{2}+1}^{\prime}}\left[W_{r;n;m_{2}+1,m_{2}+1}^{\left(\mathring{I}\right)}\right]_{b_{r;m_{2}+1}^{\prime\prime};b_{r;m_{2}+2}^{\prime\prime}}^{j_{r;m_{2}+1};j_{r;m_{2}+1}^{\prime}}\left[M^{\left(\mathring{I}\right)}\right]_{0;0}^{j_{r;m_{2}+1}^{\prime}},\label{eq:M_=00007Br; n; m_1, m_2+1=00007D^(I, dashv, 2, 1) -- 4}
\end{equation}

\noindent with

\begin{equation}
b_{r;m_{2}+2}^{\prime\prime}\in\begin{cases}
\left\{ 0\right\} , & \text{if }m_{2}+1<\left(n+1\right)\Delta m,\\
\left\{ 0,1,2,3\right\} , & \text{if }m_{2}+1=\left(n+1\right)\Delta m.
\end{cases}\label{eq:b range -- 1}
\end{equation}

\noindent Lastly, we compress the MPS $\Xi_{r;n;m_{2}+1}^{\left(\mathring{I},\dashv,2,1\right)}$
and set

\begin{equation}
\left[M_{r;n;m_{1},m_{2}+1}^{\left(\mathring{I},\dashv,2\right)}\right]_{b_{r;m_{1}};b_{r;m_{1}+1}}^{j_{r;m_{1}}}=\left[M_{r;n;m_{1},m_{2}+1}^{\left(\mathring{I},\dashv,2,1\right)}\right]_{b_{r;m_{1}};b_{r;m_{1}+1}}^{j_{r;m_{1}}},\quad\text{for }\mu_{m_{2}+2;\tau}\le m_{1}\le m_{2}+1,\label{eq:M_=00007Br; n; m_1, m_2+1=00007D^(I, dashv, 2) -- 1}
\end{equation}

\noindent and

\begin{equation}
\left[M_{r;n;\mu_{m_{2}+1;\tau}}^{\left(\mathring{I},\dashv,1\right)}\right]_{b_{r;\mu_{m_{2}+1;\tau}};b_{r;\mu_{m_{2}+1;\tau}+1}}^{j_{r;\mu_{m_{2}+1;\tau}}}=\left[M_{r;n;\mu_{m_{2}+1;\tau},m_{2}+1}^{\left(\mathring{I},\dashv,2,1\right)}\right]_{b_{r;\mu_{m_{2}+1;\tau}};b_{r;\mu_{m_{2}+1;\tau}+1}}^{j_{r;\mu_{m_{2}+1;\tau}}},\quad\text{if }\mu_{m_{2}+2;\tau}\ge1,\label{eq:M_=00007Br; n; mu=00007D^(I, dashv, 1) -- 1}
\end{equation}

\noindent which completes the iterative step. 

One can show, using similar arguments that lead to Eq.~\eqref{eq:Xi_=00007Br; m_2=00007D^(I, 1) to ring I_=00007Br; m_2=00007D^(1) -- 1},
that

\begin{align}
 & \mathring{I}_{r;n}\left(b_{r;0};j_{r;m\in\left[0,\left(n+1\right)\Delta m\right]}\right)\nonumber \\
 & \quad=\sum_{b_{r;\mu_{n\Delta m;\tau}}^{\prime}}\sum_{b_{r;\left(n+1\right)\Delta m+1}}\left[\Xi_{r;0,\mu_{n\Delta m;\tau}-1}^{\left(\mathring{I},1,1\right)}\right]_{b_{r;0};b_{r;\mu_{n\Delta m;\tau}}^{\prime}}^{j_{r;0},\ldots,j_{r;\mu_{n\Delta m;\tau}-1}}\left[\Xi_{r;n;\left(n+1\right)\Delta m}^{\left(\mathring{I},\dashv\right)}\right]_{b_{r;\mu_{n\Delta m;\tau}}^{\prime};b_{r;\left(n+1\right)\Delta m+1}}^{j_{r;\mu_{n\Delta m;\tau}},\ldots,j_{r;\left(n+1\right)\Delta m}},\label{eq:Obtaining ring I_=00007Br; n=00007D from iterating MPS procedure -- 1}
\end{align}

\noindent where $\mathring{I}_{r;n}\left(b_{r;0};j_{r;m\in\left[0,\left(n+1\right)\Delta m\right]}\right)$
is the entire influence functional for site $r$ at time $t_{n}=n\Delta t$
{[}given by Eqs.~\eqref{eq:rearranging terms in rho^(A) -- 1}, \eqref{eq:ring I_=00007Br; n=00007D -- 1}-\eqref{eq:ring I^(y<->z, 2) -- 1}{]}.
For later convenience, we also express $\Xi_{r;n;\left(n+1\right)\Delta m}^{\left(\mathring{I},\dashv\right)}$
{[}Eqs.~\eqref{eq:Xi_=00007Br; n; m_2=00007D^(I, dashv) -- 1}-\eqref{eq:Xi_=00007Br; n; m_2=00007D^(I, dashv, 2) -- 1}{]}
as

\begin{align}
\left[\Xi_{r;n;m_{2}}^{\left(\mathring{I},\dashv\right)}\right]_{b_{r;\mu_{n\Delta m;\tau}};b_{r;m_{2}+1}=0}^{j_{r;\mu_{n\Delta m;\tau}},\ldots,j_{r;m_{2}}} & =\prod_{m_{1}=\mu_{n\Delta m;\tau}+1}^{m_{2}}\left\{ \sum_{b_{r;m_{1}}}\right\} \prod_{m_{1}=\mu_{n\Delta m;\tau}}^{m_{2}}\left\{ \left[M_{r;n;m_{1},m_{2}}^{\left(\mathring{I},\dashv,\right)}\right]_{b_{r;m_{1}};b_{r;m_{1}+1}}^{j_{r;m_{1}}}\right\} ,\nonumber \\
 & \quad\text{for }n\Delta m-1\le m_{2}\le\left(n+1\right)\Delta m,\label{eq:Xi_=00007Br; n; m_2=00007D^(I, dashv) -- 2}
\end{align}

\begin{equation}
\left[M_{r;n;m_{1},m_{2}}^{\left(\mathring{I},\dashv,\right)}\right]_{b_{r;m_{1}};b_{r;m_{1}+1}}^{j_{r;m_{1}}}=\begin{cases}
\left[M_{r;n;m_{1}}^{\left(\mathring{I},\dashv,1\right)}\right]_{b_{r;m_{1}};b_{r;m_{1}+1}}^{j_{r;m_{1}}}, & \text{if }\mu_{n\Delta m;\tau}\le m_{1}\le\mu_{m_{2}+1;\tau}-1,\\
\left[M_{r;n;m_{1},m_{2}}^{\left(\mathring{I},\dashv,2\right)}\right]_{b_{r;m_{1}};b_{r;m_{1}+1}}^{j_{r;m_{1}}} & \text{if }\mu_{m_{2}+1;\tau}\le m_{1}\le m_{2},
\end{cases}\label{eq:M_=00007Br; n; m_1, m_2=00007D^(I, dashv) -- 1}
\end{equation}

\noindent where the rank-$3$ tensors $M_{r;n;m_{1}}^{\left(\mathring{I},\dashv,1\right)}$
and $M_{r;n;m_{1},m_{2}}^{\left(\mathring{I},\dashv,2\right)}$ are
determined from the iterative procedures described above.

In $\texttt{spinbosonchain}$, influence paths/functionals $\mathring{I}_{r;n}\left(b_{r;0};j_{r;m\in\left[0,\left(n+1\right)\Delta m\right]}\right)$
are represented by \\
$\texttt{spinbosonchain.\_influence.path}$ objects.

\subsection{Rearranging terms in the path integral expression of $\hat{\rho}^{\left(A\right)}\left(t\right)$
further\label{subsec:Rearranging terms in the path integral expression of rho^(A) further -- 1}}

We now return to the path integral expression of $\hat{\rho}^{\left(A\right)}\left(t\right)$
that we introduced in Eq.~\eqref{eq:rearranging terms in rho^(A) -- 1}
and rearrange terms in that expression further to facilitate our subsequent
discussion of the remainder of the tensor network algorithm. Using
Eqs.~\eqref{eq:rearranging terms in rho^(A) -- 1}, \eqref{eq:Xi_=00007Br; m_1, m_2=00007D^(I, 1, 1) -- 1},
\eqref{eq:Obtaining ring I_=00007Br; n=00007D from iterating MPS procedure -- 1}-\eqref{eq:M_=00007Br; n; m_1, m_2=00007D^(I, dashv) -- 1},
we can rearrange terms in the path integral expression of $\hat{\rho}^{\left(A\right)}\left(t\right)$
as follows:

\begin{align}
\mathring{\rho}_{n}^{\left(A\right)}\left(\mathbf{j}_{\left(n+1\right)\Delta m}\right) & =\left\langle g_{1}\left(\mathbf{j}_{\left(n+1\right)\Delta m}\right)z\right|\hat{\rho}^{\left(A\right)}\left(t_{n}\right)\left|g_{-1}\left(\mathbf{j}_{\left(n+1\right)\Delta m}\right)z\right\rangle \nonumber \\
 & =\prod_{u=-N}^{N}\prod_{r=0}^{L-1}\left[\sum_{b_{r+uL;\mu_{n\Delta m;\tau}}=0}^{3}\prod_{m=\mu_{n\Delta m;\tau}}^{\left(n+1\right)\Delta m-1}\left\{ \sum_{j_{r+uL;m}=0}^{3}\right\} \right]\nonumber \\
 & \mathrel{\phantom{=}}\quad\mathop{\times}e^{i\mathring{\phi}_{n}^{\left(\text{lcafc},\dashv\right)}\left(\mathbf{j}_{m\in\left\{ k\Delta m\right\} _{k=q_{n;\tau}}^{q_{z;n}}}\right)}\mathring{I}_{n}^{\left(\dashv\right)}\left(\mathbf{b}_{\mu_{n\Delta m;\tau}};\mathbf{j}_{m\in\left[\mu_{n\Delta m;\tau},\left(n+1\right)\Delta m\right]}\right)\nonumber \\
 & \mathrel{\phantom{=}}\quad\mathop{\times}\mathring{\rho}_{\mu_{n\Delta m;\tau}}^{\left(\vdash\right)}\left(\mathbf{b}_{\mu_{n\Delta m;\tau}}\right)+\mathcal{O}\left[\Delta t^{2}\right],\label{eq:rearranging terms in rho^(A) -- 2}
\end{align}

\noindent where

\begin{equation}
q_{n;\tau}=\max\left(0,n-K_{\tau}+1\right),\label{eq:q_=00007Bn; tau=00007D -- 1}
\end{equation}

\noindent with $K_{\tau}$ given by Eq.~\eqref{eq:K_=00007Btau=00007D in main manuscript -- 1},
and $\tau$ being the system's memory; $\mu_{m;\tau}$ is given by
Eq.~\eqref{eq:mu_=00007Bm; tau=00007D -- 1}; $\Delta m$ is given
by Eq.~\eqref{eq:Delta m -- 1};

\begin{equation}
\mathring{\phi}_{n}^{\left(\text{lcafc},\dashv\right)}\left(\mathbf{j}_{m\in\left\{ k\Delta m\right\} _{k=q_{n;\tau}}^{q_{z;n}}}\right)=\sum_{k=q_{n;\tau}}^{q_{z;n}}\mathring{\phi}_{n;k}^{\left(\text{lcafc}\right)}\left(\mathbf{j}_{k\Delta m}\right),\label{eq:ring phi_n^=00007Blcafc, dashv=00007D -- 1}
\end{equation}

\noindent with

\begin{equation}
\mathring{\phi}_{n;k}^{\left(\text{lcafc}\right)}\left(\mathbf{j}_{k\Delta m}\right)=\sum_{\alpha=\pm1}\phi_{\alpha;n;k}^{\left(\text{lcafc}\right)}\left(g_{\alpha}\left(\mathbf{j}_{k\Delta m}\right)\right),\label{eq:ring phi_=00007Bn, k=00007D^=00007Blcafc, dashv=00007D -- 1}
\end{equation}

\noindent $\phi_{\alpha;n;k}^{\left(\text{lcafc}\right)}\left(g_{\alpha}\left(\mathbf{j}_{k\Delta m}\right)\right)$
given by Eq.~\eqref{eq:phi_=00007Balpha; n; k=00007D^(lcafc) in main manuscript -- 1};

\begin{equation}
\mathring{I}_{n}^{\left(\dashv\right)}\left(\mathbf{b}_{\mu_{n\Delta m;\tau}};\mathbf{j}_{m\in\left[\mu_{n\Delta m;\tau},\left(n+1\right)\Delta m\right]}\right)=\prod_{u=-N}^{N}\prod_{r=0}^{L-1}\left\{ \left[\Xi_{r+uL;n;\left(n+1\right)\Delta m}^{\left(\mathring{I},\dashv\right)}\right]_{b_{r;\mu_{n\Delta m;\tau}};b_{r;\left(n+1\right)\Delta m+1}=0}^{j_{r;\mu_{n\Delta m;\tau}},\ldots,j_{r;\left(n+1\right)\Delta m}}\right\} ,\label{eq:ring I_n^=00007Bdashv=00007D -- 1}
\end{equation}

\noindent with the tensor $\Xi_{r;n;\left(n+1\right)\Delta m}^{\left(\mathring{I},\dashv\right)}$
being determined by the iteration procedures described in Sec.~\ref{subsec:Tensor network representation of influence functional -- 1};

\begin{align}
\mathring{\rho}_{\left(k+1\right)\Delta m+1}^{\left(\vdash\right)}\left(\mathbf{b}_{\left(k+1\right)\Delta m+1}\right) & =\prod_{u=-N}^{N}\prod_{r=0}^{L-1}\left\{ \sum_{b_{r+uL;k\Delta m}=0}^{3}\right\} \nonumber \\
 & \mathrel{\phantom{=}}\quad\mathop{\times}\mathring{U}_{k}^{\left(\vdash\right)}\left(\mathbf{b}_{k\Delta m+1},\mathbf{b}_{\left(k+1\right)\Delta m+1}\right)\mathring{\rho}_{k\Delta m+1}^{\left(\vdash\right)}\left(\mathbf{b}_{k\Delta m+1}\right),\quad\text{for }k\ge0,\label{eq:ring rho_m^(vdash) recursive relation -- 1}
\end{align}

\noindent with

\begin{equation}
\mathring{\rho}_{0}^{\left(\vdash\right)}\left(\mathbf{b}_{0}\right)=\mathring{\rho}^{\left(i,A\right)}\left(\mathbf{b}_{0}\right),\label{eq:ring rho_0^(vdash) -- 1}
\end{equation}

\noindent $\mathring{\rho}^{\left(i,A\right)}\left(\mathbf{b}_{0}\right)$
given by Eq.~\eqref{eq:ring rho^(i, A) -- 1};

\begin{equation}
\mathring{\rho}_{1}^{\left(\vdash\right)}\left(\mathbf{b}_{1}\right)=\prod_{u=-N}^{N}\prod_{r=0}^{L-1}\left\{ \sum_{b_{r+uL;0}=0}^{3}\right\} \mathring{U}^{\left(\vdash,i\right)}\left(\mathbf{b}_{0},\mathbf{b}_{1}\right)\mathring{\rho}_{0}^{\left(\vdash\right)}\left(\mathbf{b}_{0}\right),\label{eq:ring rho_1^(vdash)(b_1) -- 1}
\end{equation}

\begin{equation}
\mathring{U}^{\left(\vdash,i\right)}\left(\mathbf{b}_{0},\mathbf{b}_{1}\right)=\prod_{u=-N}^{N}\prod_{r=0}^{L-1}\left\{ \sum_{j_{r+uL;0}=0}^{3}\right\} \prod_{u=-N}^{N}\prod_{r=0}^{L-1}\left\{ \left[M_{r+uL;0}^{\left(\mathring{I},1,1\right)}\right]_{b_{r+uL;0};b_{r+uL;1}}^{j_{r+uL;0}}\right\} e^{i\mathring{\phi}_{n=\infty;0}^{\left(\text{lcafc}\right)}\left(\mathbf{j}_{0}\right)},\label{eq:ring U^(vdash, i) -- 1}
\end{equation}

\begin{align}
 & \mathring{U}_{k}^{\left(\vdash\right)}\left(\mathbf{b}_{k\Delta m+1},\mathbf{b}_{\left(k+1\right)\Delta m+1}\right)\nonumber \\
 & \quad=\prod_{u=-N}^{N}\prod_{r=0}^{L-1}\prod_{p=1}^{\Delta m}\left\{ \sum_{j_{r+uL;k\Delta m+p}=0}^{3}\right\} \prod_{u=-N}^{N}\prod_{r=0}^{L-1}\prod_{p=2}^{\Delta m}\left\{ \sum_{b_{r+uL;k\Delta m+p}=0}^{3}\right\} \nonumber \\
 & \quad\mathrel{\phantom{=}}\mathop{\times}\prod_{u=-N}^{N}\prod_{r=0}^{L-1}\prod_{p=1}^{\Delta m}\left\{ \left[M_{r+uL;k\Delta m+p}^{\left(\mathring{I},1,1\right)}\right]_{b_{r+uL;k\Delta m+p};b_{r+uL;k\Delta m+p+1}}^{j_{r+uL;k\Delta m+p}}\right\} e^{i\mathring{\phi}_{n=\infty;k+1}^{\left(\text{lcafc}\right)}\left(\mathbf{j}_{\left(k+1\right)\Delta m}\right)}.\label{eq:ring U_k^(vdash) -- 1}
\end{align}

\noindent and the rank-$3$ tensors $M_{r;m}^{\left(\mathring{I},1,1\right)}$
being determined from the first iteration procedure described in Sec.~\ref{subsec:Tensor network representation of influence functional -- 1}.

We can continue to rearrange terms as follows:

\begin{align}
\mathring{\rho}_{n}^{\left(A\right)}\left(\mathbf{j}_{\left(n+1\right)\Delta m}\right) & =\left\langle g_{1}\left(\mathbf{j}_{\left(n+1\right)\Delta m}\right)z\right|\hat{\rho}^{\left(A\right)}\left(t_{n}\right)\left|g_{-1}\left(\mathbf{j}_{\left(n+1\right)\Delta m}\right)z\right\rangle \nonumber \\
 & =\prod_{u=-N}^{N}\prod_{r=0}^{L-1}\left\{ \sum_{b_{r+uL;n\Delta m+1}=0}^{3}\right\} \mathring{U}_{n}^{\left(f\right)}\left(\mathbf{b}_{n\Delta m+1},\mathbf{j}_{\left(n+1\right)\Delta m}\right)\mathring{\rho}_{n;n\Delta m+1}\left(\mathbf{b}_{n\Delta m+1}\right)+\mathcal{O}\left[\Delta t^{2}\right],\label{eq:rearranging terms in rho^(A) -- 3}
\end{align}

\noindent where

\begin{align}
\mathring{\rho}_{n;\left(k+1\right)\Delta m+1}\left(\mathbf{b}_{\left(k+1\right)\Delta m+1}\right) & =\prod_{u=-N}^{N}\prod_{r=0}^{L-1}\left\{ \sum_{b_{r+uL;k\Delta m}=0}^{3}\right\} \mathring{U}_{n;k}\left(\mathbf{b}_{k\Delta m+1},\mathbf{b}_{\left(k+1\right)\Delta m+1}\right)\mathring{\rho}_{n;k\Delta m+1}\left(\mathbf{b}_{k\Delta m+1}\right),\nonumber \\
 & \mathrel{\phantom{=}}\quad\text{for }\mu_{n\Delta m;\tau}\le k\Delta m+1\le n\Delta m-2,\label{eq:ring rho_=00007Bn; m=00007D recursive relation -- 1}
\end{align}

\noindent with 

\begin{equation}
\mathring{\rho}_{n;1}\left(\mathbf{b}_{1}\right)=\prod_{u=-N}^{N}\prod_{r=0}^{L-1}\left\{ \sum_{b_{r+uL;0}=0}^{3}\right\} \mathring{U}_{n}^{\left(i\right)}\left(\mathbf{b}_{0},\mathbf{b}_{1}\right)\mathring{\rho}_{0}^{\left(\vdash\right)}\left(\mathbf{b}_{0}\right),\quad\text{for }1\le n\le K_{\tau}-1,\label{eq:ring rho_=00007Bn; 1=00007D -- 1}
\end{equation}

\begin{equation}
\mathring{\rho}_{n;\mu_{n\Delta m;\tau}}\left(\mathbf{b}_{\mu_{n\Delta m;\tau}}\right)=\mathring{\rho}_{\mu_{n\Delta m;\tau}}^{\left(\vdash\right)}\left(\mathbf{b}_{\mu_{n\Delta m;\tau}}\right),\quad\text{for }n\ge K_{\tau},\label{eq:ring rho_=00007Bn; mu=00007D -- 1}
\end{equation}

\noindent $\mathring{\rho}_{m}^{\left(\vdash\right)}\left(\mathbf{b}_{m}\right)$
being determined by Eqs.~\eqref{eq:ring rho_m^(vdash) recursive relation -- 1}-\eqref{eq:ring U_k^(vdash) -- 1},

\begin{equation}
\mathring{U}_{n}^{\left(i\right)}\left(\mathbf{b}_{0},\mathbf{b}_{1}\right)=\prod_{u=-N}^{N}\prod_{r=0}^{L-1}\left\{ \sum_{j_{r+uL;0}=0}^{3}\right\} \prod_{u=-N}^{N}\prod_{r=0}^{L-1}\left\{ \left[M_{r+uL;n;0,\left(n+1\right)\Delta m}^{\left(\mathring{I},\dashv,\right)}\right]_{b_{r+uL;0};b_{r+uL;1}}^{j_{r+uL;0}}\right\} e^{i\mathring{\phi}_{n;0}^{\left(\text{lcafc}\right)}\left(\mathbf{j}_{0}\right)},\label{eq:ring U_n^(i) -- 1}
\end{equation}

\begin{align}
\mathring{U}_{n;k}\left(\mathbf{b}_{k\Delta m+1},\mathbf{b}_{\left(k+1\right)\Delta m+1}\right) & =\prod_{u=-N}^{N}\prod_{r=0}^{L-1}\prod_{p=1}^{\Delta m}\left\{ \sum_{j_{r;k\Delta m+p}=0}^{3}\right\} \prod_{u=-N}^{N}\prod_{r=0}^{L-1}\prod_{p=2}^{\Delta m}\left\{ \sum_{b_{r;k\Delta m+p}=0}^{3}\right\} \nonumber \\
 & \mathrel{\phantom{=}}\mathop{\times}\prod_{u=-N}^{N}\prod_{r=0}^{L-1}\prod_{p=1}^{\Delta m}\left\{ \left[M_{r;n;k\Delta m+p,\left(n+1\right)\Delta m}^{\left(\mathring{I},\dashv,\right)}\right]_{b_{r;k\Delta m+p};b_{r;k\Delta m+p+1}}^{j_{r;k\Delta m+p}}\right\} e^{i\mathring{\phi}_{n;k+1}^{\left(\text{lcafc}\right)}\left(\mathbf{j}_{\left(k+1\right)\Delta m}\right)},\label{eq:ring U_=00007Bn; k=00007D -- 1}
\end{align}

\noindent and

\begin{align}
 & \mathring{U}_{n}^{\left(f\right)}\left(\mathbf{b}_{n\Delta m+1},\mathbf{j}_{\left(n+1\right)\Delta m}\right)\nonumber \\
 & \quad=\prod_{u=-N}^{N}\prod_{r=0}^{L-1}\prod_{p=1}^{\Delta m}\left\{ \sum_{j_{r;n\Delta m+p}^{\prime}=0}^{3}\right\} \prod_{u=-N}^{N}\prod_{r=0}^{L-1}\prod_{p=2}^{\Delta m}\left\{ \sum_{b_{r;n\Delta m+p}=0}^{3}\right\} \nonumber \\
 & \quad\mathrel{\phantom{=}}\mathop{\times}\prod_{u=-N}^{N}\prod_{r=0}^{L-1}\prod_{p=1}^{\Delta m-1}\left\{ \left[M_{r;n;n\Delta m+p,\left(n+1\right)\Delta m}^{\left(\mathring{I},\dashv,\right)}\right]_{b_{r;n\Delta m+p};b_{r;n\Delta m+p+1}}^{j_{r;3n+p}^{\prime}}\right\} \nonumber \\
 & \quad\mathrel{\phantom{=}}\mathop{\times}\left[M_{r;n;\left(n+1\right)\Delta m,\left(n+1\right)\Delta m}^{\left(\mathring{I},\dashv,\right)}\right]_{b_{r;\left(n+1\right)\Delta m};b_{r;\left(n+1\right)\Delta m+1}=j_{r;\left(n+1\right)\Delta m}}^{j_{r;\left(n+1\right)\Delta m}^{\prime}}e^{i\mathring{\phi}_{n;n+1}^{\left(\text{lcafc}\right)}\left(\mathbf{j}_{\left(n+1\right)\Delta m}^{\prime}\right)}.\label{eq:ring U_n^(f) -- 1}
\end{align}

\noindent with the rank-$3$ tensors $M_{r;n;m_{1},m_{2}}^{\left(\mathring{I},\dashv,\right)}$
being determined by the iteration procedures described in Sec.~\ref{subsec:Tensor network representation of influence functional -- 1}.

In the following four sections, we describe the remainder of the tensor
network algorithm while making several references to this current
section.

\subsection{MPS representation of the initial state operator\label{subsec:MPS representation of the initial state operator -- 1}}

Let us now express the system's reduced density matrix at $t=0$ {[}see
Eqs.~\eqref{eq:sigma_=00007Bv; r=00007D eigenvalue equation in main manuscript -- 1},
\eqref{eq:matrix elements of rho^(i, A) in main manuscript -- 1},
\eqref{eq:g_=00007Balpha=00007D -- 1}, \eqref{eq:j vector -- 1},
\eqref{eq:applying g_=00007Balpha=00007D to j vector -- 1}, and \eqref{eq:ring rho^(i, A) -- 1}{]},

\begin{equation}
\mathring{\rho}^{\left(i,A\right)}\left(\mathbf{b}_{0}\right)=\left\langle g_{1}\left(\mathbf{b}_{0}\right)z\vphantom{\Psi^{\left(i,A\right)}}\right.\left|\Psi^{\left(i,A\right)}\right\rangle \left\langle \Psi^{\left(i,A\right)}\right|\left.g_{-1}\left(\mathbf{b}_{0}\right)z\vphantom{\Psi^{\left(i,A\right)}}\right\rangle ,\label{eq:ring rho^(i, A) -- 2}
\end{equation}

\noindent as a MPS. As a reminder, $g_{\alpha}\left(\mathbf{b}_{0}\right)$
is defined by Eqs.~\eqref{eq:g_=00007Balpha=00007D -- 1} and \eqref{eq:applying g_=00007Balpha=00007D to j vector -- 1},
and $\left|\Psi^{\left(i,A\right)}\right\rangle $ is a pure state
vector. As a pure state vector, $\left|\Psi^{\left(i,A\right)}\right\rangle $
can be expressed as a MPS \citep{Schollwock1}:

\begin{align}
\Psi^{\left(i,A\right)}\left(g_{\alpha}\left(\mathbf{b}_{0}\right)\right) & =\left\langle g_{\alpha}\left(\mathbf{b}_{0}\right)z\vphantom{\Psi^{\left(i,A\right)}}\right.\left|\Psi^{\left(i,A\right)}\right\rangle \nonumber \\
 & =\prod_{u=-N}^{N}\left\{ \sum_{a_{\alpha;uL;0}}\sum_{a_{\alpha;\left(u+1\right)L;0}}\right\} \prod_{u=-N}^{N}\left\{ \left[\Xi_{u}^{\left(\rho,i,A\right)}\right]_{a_{\alpha;uL;0};a_{\alpha;\left(u+1\right)L;0}}^{g_{\alpha}\left(b_{uL;0}\right),\ldots,g_{\alpha}\left(b_{\left(u+1\right)L-1;0}\right)}\right\} ,\label{eq:Psi^(i, A) in MPS form -- 1}
\end{align}

\noindent where

\begin{equation}
\left[\Xi_{u}^{\left(\rho,i,A\right)}\right]_{a_{\alpha;uL;0};a_{\alpha;\left(u+1\right)L;0}}^{g_{\alpha}\left(b_{uL;0}\right),\ldots,g_{\alpha}\left(b_{\left(u+1\right)L-1;0}\right)}=\prod_{r=1}^{L-1}\left\{ \sum_{a_{\alpha;r+uL;0}}\right\} \prod_{r=0}^{L-1}\left\{ \left[M_{r+uL}^{\left(\Psi,i,A\right)}\right]_{a_{\alpha;r+uL;0};a_{\alpha;r+uL+1;0}}^{g_{\alpha}\left(b_{r+uL;0}\right)}\right\} ,\label{eq:Xi_u^(rho, i, A) -- 1}
\end{equation}

\noindent and the $a_{\alpha;r;0}$ are bond indices. Note that $a_{\alpha;-NL;0}$
and $a_{\alpha;\left(N+1\right)L;0}$ are trivial bond indices:

\begin{equation}
a_{\alpha;-NL;0},a_{\alpha;\left(N+1\right)L;0}\in\left\{ 0\right\} ,\label{eq:rho^(i, A) trivial bond index -- 1}
\end{equation}

\noindent Typically, $\left|\Psi^{\left(i,A\right)}\right\rangle $
is an instantaneous ground of the system Hamiltonian $\hat{H}^{\left(A\right)}\left(t=0\right)$
{[}Eqs.~\eqref{eq:H_u^(A) -- 1}-\eqref{eq:H_=00007Bu; z=00007D^(A) -- 1}{]},
in which case we can use the density matrix renormalization group
(DMRG) method \citep{Schollwock1} to calculate $\left|\Psi^{\left(i,A\right)}\right\rangle $
in the form of a MPS. Using Eqs.~\eqref{eq:ring rho^(i, A) -- 2}
and \eqref{eq:Psi^(i, A) in MPS form -- 1}, we can write:

\begin{equation}
\mathring{\rho}^{\left(i,A\right)}\left(\mathbf{b}_{0}\right)=\prod_{u=-N}^{N}\left\{ \sum_{c_{uL;0}}\sum_{c_{\left(u+1\right)L;0}}\right\} \prod_{u=-N}^{N}\left\{ \left[\Xi_{u}^{\left(\mathring{\rho},i,A\right)}\right]_{c_{uL;0};c_{\left(u+1\right)L;0}}^{b_{uL;0},\ldots,b_{\left(u+1\right)L-1;0}}\right\} ,\label{eq:ring rho^(i, A) in MPS form -- 1}
\end{equation}

\noindent where

\begin{equation}
\left[\Xi_{u}^{\left(\rho,i,A\right)}\right]_{a_{\alpha;uL;0};a_{\alpha;\left(u+1\right)L;0}}^{g_{\alpha}\left(b_{uL;0}\right),\ldots,g_{\alpha}\left(b_{\left(u+1\right)L-1;0}\right)}=\prod_{r=1}^{L-1}\left\{ \sum_{c_{r+uL;0}}\right\} \prod_{r=0}^{L-1}\left\{ \left[M_{r+uL}^{\left(\mathring{\rho},i,A\right)}\right]_{c_{r+uL;0};c_{r+uL+1;0}}^{b_{r+uL;0}}\right\} ,\label{eq:Xi_u^(ring rho, i, A) -- 1}
\end{equation}

\noindent and

\begin{align}
 & \left[M_{r+uL}^{\left(\mathring{\rho},i,A\right)}\right]_{c_{r+uL;0}=\left(a_{1;r+uL;0},a_{-1;r+uL;0}\right);c_{r+uL+1;0}=\left(a_{1;r+uL+1;0},a_{-1;r+uL+1;0}\right)}^{b_{r+uL;0}}\nonumber \\
 & \quad=\left[M_{r+uL}^{\left(\Psi,i,A\right)}\right]_{a_{1;r+uL;0};a_{1;r+uL+1;0}}^{g_{1}\left(b_{r+uL;0}\right)}\left\{ \left[M_{r+uL}^{\left(\Psi,i,A\right)}\right]_{a_{-1;r+uL;0};a_{-1;r+uL+1;0}}^{g_{-1}\left(b_{r+uL;0}\right)}\right\} ^{*}.\label{eq:M_=00007Br+uL=00007D^(rho, i, A) -- 1}
\end{align}

\subsection{MPS representation of $e^{i\mathring{\phi}_{n;k}^{\left(\text{lcafc}\right)}\left(\mathbf{j}_{k\Delta m}\right)}$\label{subsec:MPS representation of e^(i phi) -- 1}}

In this section, we show how to express $e^{i\mathring{\phi}_{n;k}^{\left(\text{lcafc}\right)}\left(\mathbf{j}_{k\Delta m}\right)}$
{[}see Eqs.~\eqref{eq:phi_=00007Balpha; n; k=00007D^(lcafc) in main manuscript -- 1}-\eqref{eq:k composite quadrature rule in main manuscript -- 1},
\eqref{eq:ring phi_=00007Bn, k=00007D^=00007Blcafc, dashv=00007D -- 1}{]}
as a MPS. 

Using Eqs.~\eqref{eq:phi_=00007Balpha; n; k=00007D^(lcafc) in main manuscript -- 1}-\eqref{eq:k composite quadrature rule in main manuscript -- 1},
\eqref{eq:ring phi_=00007Bn, k=00007D^=00007Blcafc, dashv=00007D -- 1},
we express $\mathring{\phi}_{n;k}^{\left(\text{lcafc}\right)}\left(\mathbf{j}_{k\Delta m}\right)$
as:

\begin{equation}
\mathring{\phi}_{n;k}^{\left(\text{lcafc}\right)}\left(\mathbf{j}_{k\Delta m}\right)=\sum_{u=-N}^{N}\sum_{r=0}^{L-1}\left\{ \mathring{\phi}_{r+uL;n;k}^{\left(\text{lfc}\right)}\left(j_{r+uL;k\Delta m}\right)+\mathring{\phi}_{r+uL,r+uL+1;n;k}^{\left(\text{lcc}\right)}\left(j_{r+uL;k\Delta m},j_{r+uL+1;k\Delta m}\right)\right\} ,\label{eq:ring phi_=00007Bn; k=00007D^(lcafc) rewritten -- 1}
\end{equation}

\noindent where

\begin{equation}
\mathring{\phi}_{r;n;k}^{\left(\text{lfc}\right)}\left(j_{r;k\Delta m}\right)=-\frac{\Delta t}{2}\sum_{\alpha=\pm1}\alpha\sum_{k^{\prime}=k-1}^{k}w_{n;k^{\prime}}H_{z;r}^{\left(\text{lfc}\right)}\left(t_{k^{\prime}};g_{\alpha}\left(j_{r;k\Delta m}\right)\right);\label{eq:ring phi_=00007Br; n; k=00007D^(lfc) -- 1}
\end{equation}

\begin{equation}
\mathring{\phi}_{r,r+1;n;k}^{\left(\text{lcc}\right)}\left(j_{r;k\Delta m},j_{r+1;k\Delta m}\right)=-\frac{\Delta t}{2}\sum_{\alpha=\pm1}\alpha\sum_{k^{\prime}=k-1}^{k}w_{n;k^{\prime}}H_{z;r,r+1}^{\left(\text{lcc}\right)}\left(t_{k^{\prime}};g_{\alpha}\left(j_{r;k\Delta m}\right),g_{\alpha}\left(j_{r+1;k\Delta m}\right)\right);\label{eq:ring phi_=00007Br, r+1; n; k=00007D^(lcc) -- 1}
\end{equation}

\begin{equation}
H_{z;r}^{\left(\text{lfc}\right)}\left(t;g_{\alpha}\left(j_{r;k\Delta m}\right)\right)=h_{z;r}\left(t\right)g_{\alpha}\left(j_{r;k\Delta m}\right);\label{eq:H_=00007Bz; r=00007D^(lfc) -- 1}
\end{equation}

\begin{equation}
H_{z;r,r+1}^{\left(\text{lcc}\right)}\left(t;g_{\alpha}\left(j_{r;k\Delta m}\right),g_{\alpha}\left(j_{r+1;k\Delta m}\right)\right)=J_{z,z;r,r+1}\left(t\right)g_{\alpha}\left(j_{r;k\Delta m}\right)g_{\alpha}\left(j_{r+1;k\Delta m}\right);\label{eq:H_=00007Bz; r, r+1=00007D^(lcc) -- 1}
\end{equation}

\noindent $\Delta m$ and $g_{\alpha}\left(\cdots\right)$ are defined
by Eqs.~\eqref{eq:Delta m -- 1} and \eqref{eq:g_=00007Balpha=00007D -- 1}.
Next, we introduce two rank-2 tensors:

\begin{equation}
\left[X_{r,r+1;n;k}^{\left(\text{lcc}\right)}\right]_{j_{r;k\Delta m}^{>};j_{r+1;k\Delta m}^{<}}=e^{i\mathring{\phi}_{r,r+1;n;k}^{\left(\text{lcc}\right)}\left(j_{r;k\Delta m}^{>},j_{r+1;k\Delta m}^{<}\right)},\label{eq:X_=00007Br, r+1; n; k=00007D^(lcc) -- 1}
\end{equation}

\begin{equation}
\left[X_{r;n;k}^{\left(\text{lfc}\right)}\right]_{j_{r;k\Delta m};j_{r;k\Delta m}^{\prime}}=\delta_{j_{r;k\Delta m},j_{r;k\Delta m}^{\prime}}e^{i\mathring{\phi}_{r;n;k}^{\left(\text{lfc}\right)}\left(j_{r;k\Delta m}\right)},\label{eq:X_=00007Br; n; k=00007D^(lfc) -- 1}
\end{equation}

\noindent and split $X_{r,r+1;n;k}^{\left(\text{lcc}\right)}$ into
two rank-2 tensors using e.g. QR decomposition:

\begin{align}
\left[X_{r,r+1;n;k}^{\left(\text{lcc}\right)}\right]_{j_{r;k\Delta m}^{>};j_{r+1;k\Delta m}^{<}} & =\sum_{c_{r+1;k\Delta m}}\left[X_{r,r+1;n;k}^{\left(\text{lcc},>\right)}\right]_{j_{r;k\Delta m}^{>};c_{r+1;k\Delta m}}\left[X_{r,r+1;n;k}^{\left(\text{lcc},<\right)}\right]_{c_{r+1;k\Delta m};j_{r+1;k\Delta m}^{<}},\nonumber \\
 & \mathrel{\phantom{=}}\quad\text{for }-NL\le r\le\left(N+1\right)L-2.\label{eq:X_=00007Br, r+1; n; k=00007D^(lcc) split -- 1}
\end{align}

\noindent where if one used QR decomposition then $X_{r,r+1;n;k}^{\left(\text{lcc},>\right)}$
and $X_{r,r+1;n;k}^{\left(\text{lcc},<\right)}$ are orthogonal and
upper triangular matrices respectively. For convenience, we also define
$X_{r,r+1;n;k}^{\left(\text{lcc},<\right)}$ and $X_{r,r+1;n;k}^{\left(\text{lcc},>\right)}$
for $r=-NL-1$ and $r=\left(N+1\right)L-1$ respectively:

\begin{equation}
\left[X_{-NL-1,-NL;n;k}^{\left(\text{lcc},<\right)}\right]_{c_{-NL;k\Delta m};j_{-NL;k\Delta m}^{<}}=1,\label{eq:trivial X_=00007Br, r+1; n; k=00007D^(lcc, <) -- 1}
\end{equation}

\begin{equation}
\left[X_{\left(N+1\right)L-1,\left(N+1\right)L;n;k}^{\left(\text{lcc},>\right)}\right]_{j_{\left(N+1\right)L-1;k\Delta m}^{>};c_{\left(N+1\right)L;k\Delta m}}=1,\label{eq:trivial X_=00007Br, r+1; n; k=00007D^(lcc, >) -- 1}
\end{equation}

\noindent where

\begin{equation}
j_{r;k\Delta m}^{<}\in\begin{cases}
\left\{ 0\right\} , & \text{if }r=-NL,\\
\left\{ 0,1,2,3\right\} , & \text{if }r>-NL,
\end{cases}\label{eq:j_=00007Br; k Delta m=00007D^(<) -- 1}
\end{equation}

\begin{equation}
j_{r;k\Delta m}\in\left\{ 0,1,2,3\right\} ,\quad\forall r.\label{eq:j_=00007Br; k Delta m=00007D -- 1}
\end{equation}

\begin{equation}
j_{r;k\Delta m}^{>}\in\begin{cases}
\left\{ 0,1,2,3\right\} , & \text{if }r<\left(N+1\right)L-1,\\
\left\{ 0\right\} , & \text{if }r=\left(N+1\right)L-1,
\end{cases}\label{eq:j_=00007Br; k Delta m=00007D^(>) -- 1}
\end{equation}

\noindent and

\begin{equation}
c_{r=-NL;k\Delta m},c_{r=\left(N+1\right)L;k\Delta m}\in\left\{ 0\right\} .\label{eq:trivial c_=00007Br; k Delta m=00007D -- 1}
\end{equation}

\noindent We also introduce a rank-3 tensor:

\begin{equation}
\left[M_{r}^{\left(\delta\right)}\right]_{j_{r;k\Delta m}^{<};j_{r;k\Delta m}^{>}}^{j_{r;k\Delta m}}=\begin{cases}
1, & \text{if }L+N=1,\\
\delta_{j_{r;k\Delta m},j_{r;k\Delta m}^{>}}, & \text{if }r=-NL\text{ and }L+N>1,\\
\delta_{j_{r;k\Delta m}^{<},j_{r;k\Delta m}}, & \text{if }r=\left(N+1\right)L-1\text{ and }L+N>1,\\
\delta_{j_{r;k\Delta m}^{<},j_{r;k\Delta m}}\delta_{j_{r;k\Delta m},j_{r;k\Delta m}^{>}}, & \text{otherwise}.
\end{cases}\label{eq:M_r^(delta) -- 1}
\end{equation}

\noindent In $\texttt{spinbosonchain}$, the tensors $X_{r;n;k}^{\left(\text{lfc}\right)}$
and $X_{r,r+1;n;k}^{\left(\text{lcc}\right)}$ are built by instances
of the ``factory'' classes \\
$\texttt{spinbosonchain.\_phasefactor.tensorfactory.ZFieldPhaseFactorNodeRank2}$
and \\
$\texttt{spinbosonchain.\_phasefactor.tensorfactory.ZZCouplerPhaseFactorNodeRank2}$.

Next, we construct a new rank-3 tensor for the tensors introduced
in Eqs.~\eqref{eq:X_=00007Br, r+1; n; k=00007D^(lcc) split -- 1}-\eqref{eq:M_r^(delta) -- 1}:

\begin{align}
\left[M_{r;n;k}^{\left(\text{lacfc}\right)}\right]_{c_{r;k\Delta m};c_{r+1;k\Delta m}}^{j_{r;k\Delta m}} & =\sum_{j_{r;k\Delta m}^{<}=0}^{3}\sum_{j_{r;k\Delta m}^{\prime}=0}^{3}\sum_{j_{r;k\Delta m}^{>}=0}^{3}\nonumber \\
 & \mathrel{\phantom{=}}\mathop{\times}\left[X_{r-1,r;n;k}^{\left(\text{lcc},<\right)}\right]_{c_{r;k\Delta m};j_{r;k\Delta m}^{<}}\nonumber \\
 & \mathrel{\phantom{=}}\mathop{\times}\left[X_{r;n;k}^{\left(\text{lfc}\right)}\right]_{j_{r;k\Delta m};j_{r;k\Delta m}^{\prime}}\left[M_{r}^{\left(\delta\right)}\right]_{j_{r;k\Delta m}^{<};j_{r;k\Delta m}^{>}}^{j_{r;k\Delta m}^{\prime}}\nonumber \\
 & \mathrel{\phantom{=}}\mathop{\times}\left[X_{r,r+1;n;k}^{\left(\text{lcc},>\right)}\right]_{j_{r;k\Delta m}^{>};c_{r+1;k\Delta m}}.\label{eq:M_=00007Br; n; k=00007D^(lacfc) -- 1}
\end{align}

\noindent Using the $M_{r;n;k}^{\left(\text{lacfc}\right)}$ tensors
we can express $e^{i\mathring{\phi}_{n;k}^{\left(\text{lcafc}\right)}\left(\mathbf{j}_{k\Delta m}\right)}$
as a MPS:

\begin{equation}
e^{i\mathring{\phi}_{n;k}^{\left(\text{lcafc}\right)}\left(\mathbf{j}_{k\Delta m}\right)}=\prod_{u=-N}^{N}\left\{ \sum_{c_{uL;k\Delta m}}\sum_{c_{\left(u+1\right)L;k\Delta m}}\right\} \prod_{u=-N}^{N}\left\{ \left[\Xi_{u;n;k}^{\left(\text{lcafc}\right)}\right]_{c_{uL;k\Delta m};c_{\left(u+1\right)L;k\Delta m}}^{j_{uL;k\Delta m},\ldots,j_{\left(u+1\right)L-1;k\Delta m}}\right\} ,\label{eq:expression lcafc phase factor as MPS -- 1}
\end{equation}

\noindent where

\begin{equation}
\left[\Xi_{u;n;k}^{\left(\text{lcafc}\right)}\right]_{c_{uL;k\Delta m};c_{\left(u+1\right)L;k\Delta m}}^{j_{uL;k\Delta m},\ldots,j_{\left(u+1\right)L-1;k\Delta m}}=\prod_{r=1}^{L-1}\left\{ \sum_{c_{r+uL;k\Delta m}}\right\} \prod_{r=0}^{L-1}\left\{ \left[M_{r;n;k}^{\left(\text{lacfc}\right)}\right]_{c_{r+uL;k\Delta m};c_{r+uL+1;k\Delta m}}^{j_{r+uL;k\Delta m}}\right\} .\label{eq:Xi_=00007Bu; n; k=00007D^(lcafc) -- 1}
\end{equation}

\subsection{MPO representations of $\mathring{U}^{\left(\vdash,i\right)}$, $\mathring{U}_{k}^{\left(\vdash\right)}$,
$\mathring{U}_{n}^{\left(i\right)}$,$\mathring{U}_{n;k}$, and $\mathring{U}_{n}^{\left(f\right)}$\label{subsec:MPO representations of U quantities -- 1}}

In this section, we show how to express $\mathring{U}^{\left(\vdash,i\right)}\left(\mathbf{b}_{0},\mathbf{b}_{1}\right)$
{[}Eq.~\eqref{eq:ring U^(vdash, i) -- 1}{]}, $\mathring{U}_{k}^{\left(\vdash\right)}\left(\mathbf{b}_{3k+1},\mathbf{b}_{3k+4}\right)$
{[}Eq.~\eqref{eq:ring U_k^(vdash) -- 1}{]}, $\mathring{U}_{n}^{\left(i\right)}\left(\mathbf{b}_{0},\mathbf{b}_{1}\right)$
{[}Eq.~\eqref{eq:ring U_n^(i) -- 1}{]}, $\mathring{U}_{n;k}\left(\mathbf{b}_{3k+1},\mathbf{b}_{3k+4}\right)$
{[}Eq.~\eqref{eq:ring U_=00007Bn; k=00007D -- 1}{]}, and $\mathring{U}_{n}^{\left(f\right)}\left(\mathbf{b}_{3n+1},\mathbf{j}_{3n+3}\right)$
{[}Eq.~\eqref{eq:ring U_n^(f) -- 1}{]} as MPO's.

Using Eqs.~\eqref{eq:ring U^(vdash, i) -- 1}, \eqref{eq:expression lcafc phase factor as MPS -- 1},
and \eqref{eq:Xi_=00007Bu; n; k=00007D^(lcafc) -- 1}, we can express
$\mathring{U}^{\left(\vdash,i\right)}\left(\mathbf{b}_{0},\mathbf{b}_{1}\right)$
by the following MPO:

\begin{equation}
\mathring{U}^{\left(\vdash,i\right)}\left(\mathbf{b}_{0},\mathbf{b}_{1}\right)=\prod_{u=-N}^{N}\left\{ \sum_{c_{uL;0}}\sum_{c_{\left(u+1\right)L;0}}\right\} \prod_{u=-N}^{N}\left\{ \left[\Omega_{u}^{\left(\mathring{U},\vdash,i\right)}\right]_{c_{uL;0};c_{\left(u+1\right)L;0}}^{b_{uL;1},\ldots,b_{\left(u+1\right)L-1;1};b_{uL;0},\ldots,b_{\left(u+1\right)L-1;0}}\right\} ,\label{eq:Ohm^(U, vdash, i) -- 1}
\end{equation}

\noindent where

\begin{equation}
\left[\Omega_{u}^{\left(\mathring{U},\vdash,i\right)}\right]_{c_{uL;0};c_{\left(u+1\right)L;0}}^{b_{uL;1},\ldots,b_{\left(u+1\right)L-1;1};b_{uL;0},\ldots,b_{\left(u+1\right)L-1;0}}=\prod_{r=1}^{L-1}\left\{ \sum_{c_{r+uL;0}}\right\} \prod_{r=0}^{L-1}\left\{ \left[W_{r+uL}^{\left(\mathring{U},\vdash,i\right)}\right]_{c_{r+uL;0};c_{r+uL+1;0}}^{b_{r+uL;1};b_{r+uL;0}}\right\} ,\label{eq:Ohm_u^(ring U, vdash, i) -- 1}
\end{equation}

\noindent with

\begin{equation}
\left[W_{r}^{\left(\mathring{U},\vdash,i\right)}\right]_{c_{r;0};c_{r+1;0}}^{b_{r;1};b_{r;0}}=\sum_{j_{r;0}=0}^{3}\left[M_{r;n=\infty;0}^{\left(\text{lacfc}\right)}\right]_{c_{r;0};c_{r+1;0}}^{j_{r;0}}\left[M_{r;0}^{\left(\mathring{I},1,1\right)}\right]_{b_{r;0};b_{r;1}}^{j_{r;0}},\label{eq:W_r^(U, vdash, i) -- 1}
\end{equation}

\noindent the rank-$3$ tensors $M_{r;m}^{\left(\mathring{I},1,1\right)}$
being determined from the first iteration procedure described in Sec.~\ref{subsec:Tensor network representation of influence functional -- 1},
and the rank-$3$ tensors $M_{r;n;k}^{\left(\text{lacfc}\right)}$
being determined from the procedure described in Sec.~\ref{subsec:MPS representation of e^(i phi) -- 1}.

Using Eqs.~\eqref{eq:ring U_k^(vdash) -- 1} and \eqref{eq:Xi_=00007Bu; n; k=00007D^(lcafc) -- 1},
we can express $\mathring{U}_{k}^{\left(\vdash\right)}\left(\mathbf{b}_{3k+1},\mathbf{b}_{3k+4}\right)$
by the following MPO:

\begin{align}
 & \mathring{U}_{k}^{\left(\vdash\right)}\left(\mathbf{b}_{k\Delta m+1},\mathbf{b}_{\left(k+1\right)\Delta m+1}\right)\nonumber \\
 & \quad=\prod_{u=-N}^{N}\left\{ \sum_{c_{uL;\left(k+1\right)\Delta m}}\sum_{c_{\left(u+1\right)L;\left(k+1\right)\Delta m}}\right\} \nonumber \\
 & \quad\mathrel{\phantom{=}}\mathop{\times}\prod_{u=-N}^{N}\left\{ \left[\Omega_{u;k}^{\left(\mathring{U},\vdash\right)}\right]_{c_{uL;\left(k+1\right)\Delta m};c_{\left(u+1\right)L;\left(k+1\right)\Delta m}}^{b_{uL;\left(k+1\right)\Delta m+1},\ldots,b_{\left(u+1\right)L-1;\left(k+1\right)\Delta m+1};b_{uL;k\Delta m+1},\ldots,b_{\left(u+1\right)L-1;k\Delta m+1}}\right\} ,\label{eq:Ohm_k^(U, vdash) -- 1}
\end{align}

\noindent where

\begin{align}
 & \left[\Omega_{u;k}^{\left(\mathring{U},\vdash\right)}\right]_{c_{uL;\left(k+1\right)\Delta m};c_{\left(u+1\right)L;\left(k+1\right)\Delta m}}^{b_{uL;\left(k+1\right)\Delta m+1},\ldots,b_{\left(u+1\right)L-1;\left(k+1\right)\Delta m+1};b_{uL;k\Delta m+1},\ldots,b_{\left(u+1\right)L-1;k\Delta m+1}}\nonumber \\
 & \quad=\prod_{r=1}^{L-1}\left\{ \sum_{c_{r+uL;\left(k+1\right)\Delta m}}\right\} \prod_{r=0}^{L-1}\left\{ \left[W_{r+uL;k}^{\left(\mathring{U},\vdash\right)}\right]_{c_{r+uL;\left(k+1\right)\Delta m};c_{r+uL+1;\left(k+1\right)\Delta m}}^{b_{r+uL;\left(k+1\right)\Delta m+1};b_{r+uL;k\Delta m+1}}\right\} ,\label{eq:Ohm_=00007Bu; k=00007D^(ring U, vdash) -- 1}
\end{align}

\noindent with

\begin{align}
\left[W_{r;k}^{\left(\mathring{U},\vdash\right)}\right]_{c_{r;\left(k+1\right)\Delta m};c_{r+1;\left(k+1\right)\Delta m}}^{b_{r;\left(k+1\right)\Delta m+1};b_{r;k\Delta m+1}} & =\prod_{p=1}^{\Delta m}\left\{ \sum_{j_{r;k\Delta m+p}=0}^{3}\right\} \prod_{p=2}^{\Delta m}\left\{ \sum_{b_{r;k\Delta m+p}=0}^{3}\right\} \nonumber \\
 & \mathrel{\phantom{=}}\mathop{\times}\prod_{p=1}^{\Delta m}\left\{ \left[M_{r;k\Delta m+p}^{\left(\mathring{I},1,1\right)}\right]_{b_{r;k\Delta m+p};b_{r;k\Delta m+p+1}}^{j_{r;k\Delta m+p}}\right\} \nonumber \\
 & \mathrel{\phantom{=}}\mathop{\times}\left[M_{r;n=\infty;k+1}^{\left(\text{lacfc}\right)}\right]_{c_{r;\left(k+1\right)\Delta m};c_{r+1;\left(k+1\right)\Delta m}}^{j_{r;\left(k+1\right)\Delta m}},\label{eq:W_=00007Br; k=00007D^(U, vdash) -- 1}
\end{align}

\noindent and $\Delta m$ being defined by Eq.~\eqref{eq:Delta m -- 1}.

Using Eqs.~\eqref{eq:ring U_n^(i) -- 1} and \eqref{eq:Xi_=00007Bu; n; k=00007D^(lcafc) -- 1},
we can express $\mathring{U}_{n}^{\left(i\right)}\left(\mathbf{b}_{0},\mathbf{b}_{1}\right)$
by the following MPO:

\begin{equation}
\mathring{U}_{n}^{\left(i\right)}\left(\mathbf{b}_{0},\mathbf{b}_{1}\right)=\prod_{u=-N}^{N}\left\{ \sum_{c_{uL;0}}\sum_{c_{\left(u+1\right)L;0}}\right\} \prod_{u=-N}^{N}\left\{ \left[\Omega_{u;n}^{\left(\mathring{U},i\right)}\right]_{c_{uL;0};c_{\left(u+1\right)L;0}}^{b_{uL;1},\ldots,b_{\left(u+1\right)L-1;1};b_{uL;0},\ldots,b_{\left(u+1\right)L-1;0}}\right\} ,\label{eq:Ohm_n^(U, i) -- 1}
\end{equation}

\noindent where

\begin{equation}
\left[\Omega_{u;n}^{\left(\mathring{U},i\right)}\right]_{c_{uL;0};c_{\left(u+1\right)L;0}}^{b_{uL;1},\ldots,b_{\left(u+1\right)L-1;1};b_{uL;0},\ldots,b_{\left(u+1\right)L-1;0}}=\prod_{r=1}^{L-1}\left\{ \sum_{c_{r+uL;0}}\right\} \prod_{r=0}^{L-1}\left\{ \left[W_{r+uL;n}^{\left(\mathring{U},i\right)}\right]_{c_{r+uL;0};c_{r+uL+1;0}}^{b_{r+uL;1};b_{r+uL;0}}\right\} ,\label{eq:Ohm_=00007Bu; n=00007D^(ring U, i) -- 1}
\end{equation}

\noindent with

\begin{equation}
\left[W_{r;n}^{\left(\mathring{U},i\right)}\right]_{c_{r;0};c_{r+1;0}}^{b_{r;1};b_{r;0}}=\sum_{j_{r;0}=0}^{3}\left[M_{r;n;0}^{\left(\text{lacfc}\right)}\right]_{c_{r;0};c_{r+1;0}}^{j_{r;0}}\left[M_{r;n;0,\left(n+1\right)\Delta m}^{\left(\mathring{I},\dashv\right)}\right]_{b_{r;0};b_{r;1}}^{j_{r;0}},\label{eq:W_=00007Br; n=00007D^(U, i) -- 1}
\end{equation}

\noindent and the rank-$3$ tensors $M_{r;n;m_{1},m_{2}}^{\left(\mathring{I},\dashv,\right)}$
being determined by the iteration procedures described in Sec.~\ref{subsec:Tensor network representation of influence functional -- 1}.

Using Eqs.~\eqref{eq:ring U_=00007Bn; k=00007D -- 1} and \eqref{eq:Xi_=00007Bu; n; k=00007D^(lcafc) -- 1},
we can express $\mathring{U}_{n;k}\left(\mathbf{b}_{k\Delta m+1},\mathbf{b}_{\left(k+1\right)\Delta m+1}\right)$
by the following MPO:

\begin{align}
 & \mathring{U}_{n;k}\left(\mathbf{b}_{k\Delta m+1},\mathbf{b}_{\left(k+1\right)\Delta m+1}\right)\nonumber \\
 & \quad=\prod_{u=-N}^{N}\left\{ \sum_{c_{uL;\left(k+1\right)\Delta m}}\sum_{c_{\left(u+1\right)L;\left(k+1\right)\Delta m}}\right\} \nonumber \\
 & \quad\mathrel{\phantom{=}}\mathop{\times}\prod_{u=-N}^{N}\left\{ \left[\Omega_{u;n;k}^{\left(\mathring{U}\right)}\right]_{c_{uL;\left(k+1\right)\Delta m};c_{\left(u+1\right)L;\left(k+1\right)\Delta m}}^{b_{uL;\left(k+1\right)\Delta m+1},\ldots,b_{\left(u+1\right)L-1;\left(k+1\right)\Delta m+1};b_{uL;k\Delta m+1},\ldots,b_{\left(u+1\right)L-1;k\Delta m+1}}\right\} ,\label{eq:Ohm_=00007Bn; k=00007D^(U) -- 1}
\end{align}

\noindent where

\begin{align}
 & \left[\Omega_{u;n;k}^{\left(\mathring{U}\right)}\right]_{c_{uL;\left(k+1\right)\Delta m};c_{\left(u+1\right)L;\left(k+1\right)\Delta m}}^{b_{uL;\left(k+1\right)\Delta m+1},\ldots,b_{\left(u+1\right)L-1;\left(k+1\right)\Delta m+1};b_{uL;k\Delta m+1},\ldots,b_{\left(u+1\right)L-1;k\Delta m+1}}\nonumber \\
 & \quad=\prod_{r=1}^{L-1}\left\{ \sum_{c_{r+uL;\left(k+1\right)\Delta m}}\right\} \prod_{r=0}^{L-1}\left\{ \left[W_{r+uL;n;k}^{\left(\mathring{U}\right)}\right]_{c_{r+uL;\left(k+1\right)\Delta m};c_{r+uL+1;\left(k+1\right)\Delta m}}^{b_{r+uL;\left(k+1\right)\Delta m+1};b_{r+uL;k\Delta m+1}}\right\} ,\label{eq:Ohm_=00007Bu; n; k=00007D^(ring U) -- 1}
\end{align}

\noindent with

\begin{align}
\left[W_{r;n;k}^{\left(\mathring{U}\right)}\right]_{c_{r+uL;\left(k+1\right)\Delta m};c_{r+uL+1;\left(k+1\right)\Delta m}}^{b_{r+uL;\left(k+1\right)\Delta m+1};b_{r+uL;k\Delta m+1}} & =\prod_{p=1}^{\Delta m}\left\{ \sum_{j_{r;k\Delta m+p}=0}^{3}\right\} \prod_{p=2}^{\Delta m}\left\{ \sum_{b_{r;k\Delta m+p}=0}^{3}\right\} \nonumber \\
 & \mathrel{\phantom{=}}\mathop{\times}\prod_{p=1}^{\Delta m}\left\{ \left[M_{r;n;k\Delta m+p,\left(n+1\right)\Delta m}^{\left(\mathring{I},\dashv,\right)}\right]_{b_{r;k\Delta m+p};b_{r;k\Delta m+p+1}}^{j_{r;k\Delta m+p}}\right\} \nonumber \\
 & \mathrel{\phantom{=}}\mathop{\times}\left[M_{r;n;k+1}^{\left(\text{lacfc}\right)}\right]_{c_{r;\left(k+1\right)\Delta m};c_{r+1;\left(k+1\right)\Delta m}}^{j_{r;\left(k+1\right)\Delta m}}.\label{eq:W_=00007Br; n; k=00007D^(U) -- 1}
\end{align}

Using Eqs.~\eqref{eq:ring U_n^(f) -- 1} and \eqref{eq:Xi_=00007Bu; n; k=00007D^(lcafc) -- 1},
we can express $\mathring{U}_{n}^{\left(f\right)}\left(\mathbf{b}_{3n+1},\mathbf{j}_{3n+3}\right)$
by the following MPO:

\begin{align}
 & \mathring{U}_{n}^{\left(f\right)}\left(\mathbf{b}_{n\Delta m+1},\mathbf{j}_{\left(n+1\right)\Delta m}\right)\nonumber \\
 & \quad=\prod_{u=-N}^{N}\left\{ \sum_{c_{uL;\left(n+1\right)\Delta m}}\sum_{c_{\left(u+1\right)L;\left(n+1\right)\Delta m}}\right\} \nonumber \\
 & \quad\mathrel{\phantom{=}}\mathop{\times}\prod_{u=-N}^{N}\left\{ \left[\Omega_{u;n}^{\left(\mathring{U},f\right)}\right]_{c_{uL;\left(n+1\right)\Delta m};c_{\left(u+1\right)L;\left(n+1\right)\Delta m}}^{j_{uL;\left(n+1\right)\Delta m},\ldots,j_{\left(u+1\right)L-1;\left(n+1\right)\Delta m};b_{uL;n\Delta m+1},\ldots,b_{\left(u+1\right)L-1;n\Delta m+1}}\right\} ,\label{eq:Ohm_=00007Bn=00007D^(U, f) -- 1}
\end{align}

\noindent where

\begin{align}
 & \left[\Omega_{u;n}^{\left(\mathring{U},f\right)}\right]_{c_{uL;\left(n+1\right)\Delta m};c_{\left(u+1\right)L;\left(n+1\right)\Delta m}}^{j_{uL;\left(n+1\right)\Delta m},\ldots,j_{\left(u+1\right)L-1;\left(n+1\right)\Delta m};b_{uL;n\Delta m+1},\ldots,b_{\left(u+1\right)L-1;n\Delta m+1}}\nonumber \\
 & \quad=\prod_{r=1}^{L-1}\left\{ \sum_{c_{r+uL;\left(n+1\right)\Delta m}}\right\} \prod_{r=0}^{L-1}\left\{ \left[W_{r+uL;n}^{\left(\mathring{U},f\right)}\right]_{c_{r+uL;\left(n+1\right)\Delta m};c_{r+uL+1;\left(n+1\right)\Delta m}}^{j_{r+uL;\left(n+1\right)\Delta m};b_{r+uL;n\Delta m+1}}\right\} ,\label{eq:Ohm_=00007Bu; n=00007D^(ring U, f) -- 1}
\end{align}

\noindent with

\begin{align}
\left[W_{r;n}^{\left(\mathring{U},f\right)}\right]_{c_{r;\left(n+1\right)\Delta m};c_{r+1;\left(n+1\right)\Delta m}}^{b_{r;\left(n+1\right)\Delta m};b_{r;n\Delta m+1}} & =\prod_{p=1}^{\Delta m}\left\{ \sum_{j_{r;n\Delta m+p}^{\prime}=0}^{3}\right\} \prod_{p=2}^{\Delta m}\left\{ \sum_{b_{r;n\Delta m+p}=0}^{3}\right\} \nonumber \\
 & \mathrel{\phantom{=}}\mathop{\times}\prod_{p=1}^{\Delta m}\left\{ \left[M_{r;n;n\Delta m+p,\left(n+1\right)\Delta m}^{\left(\mathring{I},\dashv,\right)}\right]_{b_{r;n\Delta m+p};b_{r;n\Delta m+p+1}}^{j_{r;n\Delta m+p}^{\prime}}\right\} \nonumber \\
 & \mathrel{\phantom{=}}\mathop{\times}\left[M_{r;n;n+1}^{\left(\text{lacfc}\right)}\right]_{c_{r;\left(n+1\right)\Delta m};c_{r+1;\left(n+1\right)\Delta m}}^{j_{r;\left(n+1\right)\Delta m}^{\prime}}.\label{eq:W_=00007Br; n=00007D^(U, f) -- 1}
\end{align}

\noindent Note that $b_{r;\left(n+1\right)\Delta m}$ is set to $j_{r;\left(n+1\right)\Delta m}$
in calculating $W_{r;n}^{\left(\mathring{U},f\right)}$ and $\mathring{U}_{n}^{\left(f\right)}\left(\mathbf{b}_{n\Delta m+1},\mathbf{j}_{\left(n+1\right)\Delta m}\right)$.

\subsection{Remaining elements of tensor network algorithm\label{subsec:Remaining elements of tensor network algorithm -- 1}}

All that remains is to show how we implement Eqs.~\eqref{eq:ring rho_m^(vdash) recursive relation -- 1}-\eqref{eq:ring rho_1^(vdash)(b_1) -- 1}
and \eqref{eq:rearranging terms in rho^(A) -- 3}-\eqref{eq:ring rho_=00007Bn; mu=00007D -- 1}
using tensor networks. In MPS notation, we can express $\mathring{\rho}_{m}^{\left(\vdash\right)}\left(\mathbf{b}_{m}\right)$,
$\mathring{\rho}_{n;m}\left(\mathbf{b}_{m}\right)$, and $\mathring{\rho}_{n}^{\left(A\right)}\left(\mathbf{j}_{\left(n+1\right)\Delta m}\right)$
as:

\begin{equation}
\mathring{\rho}_{m}^{\left(\vdash\right)}\left(\mathbf{b}_{m}\right)=\prod_{u=-N}^{N}\left\{ \sum_{c_{uL;m}}\sum_{c_{\left(u+1\right)L;m}}\right\} \prod_{u=-N}^{N}\left\{ \left[\Xi_{u;m}^{\left(\mathring{\rho},\vdash\right)}\right]_{c_{uL;m};c_{\left(u+1\right)L;m}}^{b_{uL;m},\ldots,b_{\left(u+1\right)L-1;m}}\right\} ,\label{eq:Xi_=00007Bu; m=00007D^(rho, vdash) -- 1}
\end{equation}

\begin{equation}
\mathring{\rho}_{n;m}\left(\mathbf{b}_{m}\right)=\prod_{u=-N}^{N}\left\{ \sum_{c_{uL;m}}\sum_{c_{\left(u+1\right)L;m}}\right\} \prod_{u=-N}^{N}\left\{ \left[\Xi_{u;n;m}^{\left(\mathring{\rho}\right)}\right]_{c_{uL;m};c_{\left(u+1\right)L;m}}^{b_{uL;m},\ldots,b_{\left(u+1\right)L-1;m}}\right\} ,\label{eq:Xi_=00007Bu; n; m=00007D^(rho) -- 1}
\end{equation}

\begin{equation}
\mathring{\rho}_{n}^{\left(A\right)}\left(\mathbf{j}_{\left(n+1\right)\Delta m}\right)=\prod_{u=-N}^{N}\left\{ \sum_{c_{uL;\left(n+1\right)\Delta m}}\sum_{c_{\left(u+1\right)L;\left(n+1\right)\Delta m}}\right\} \prod_{u=-N}^{N}\left\{ \left[\Xi_{u;n}^{\left(\mathring{\rho},A\right)}\right]_{c_{uL;\left(n+1\right)\Delta m};c_{\left(u+1\right)L;\left(n+1\right)\Delta m}}^{j_{uL;\left(n+1\right)\Delta m},\ldots,j_{\left(u+1\right)L-1;\left(n+1\right)\Delta m}}\right\} .\label{eq:Xi_=00007Bu; n=00007D^(rho, A) -- 1}
\end{equation}

Equation~\eqref{eq:ring rho_0^(vdash) -- 1} is equivalent to

\begin{equation}
\left[\Xi_{u;0}^{\left(\mathring{\rho},\vdash\right)}\right]_{c_{uL;0};c_{\left(u+1\right)L;0}}^{b_{uL;0},\ldots,b_{\left(u+1\right)L-1;0}}=\left[\Xi_{u}^{\left(\mathring{\rho},i,A\right)}\right]_{c_{uL;0};c_{\left(u+1\right)L;0}}^{b_{uL;0},\ldots,b_{\left(u+1\right)L-1;0}},\label{eq:Xi_0^(rho, vdash) -- 1}
\end{equation}

\noindent where $\Xi^{\left(\mathring{\rho},i,A\right)}$ is the MPS
representing $\mathring{\rho}^{\left(i,A\right)}\left(\mathbf{b}_{0}\right)$,
which is determined by the procedure described in Sec.~\ref{subsec:MPS representation of the initial state operator -- 1}. 

Equation~\eqref{eq:ring rho_1^(vdash)(b_1) -- 1} is equivalent to
the following MPO application to a MPS:

\begin{align}
\left[\Xi_{u;1}^{\left(\mathring{\rho},\vdash\right)}\right]_{c_{uL;1};c_{\left(u+1\right)L;1}}^{b_{uL;1},\ldots,b_{\left(u+1\right)L-1;1}} & =\prod_{r=0}^{L-1}\left\{ \sum_{b_{r+uL;0}=0}^{3}\right\} \left[\Omega_{u}^{\left(\mathring{U},\vdash,i\right)}\right]_{c_{uL;0}^{\prime};c_{\left(u+1\right)L;0}^{\prime}}^{b_{uL;1},\ldots,b_{\left(u+1\right)L-1;1};b_{uL;0},\ldots,b_{\left(u+1\right)L-1;0}}\nonumber \\
 & \mathrel{\phantom{=}}\mathop{\times}\left[\Xi_{u;0}^{\left(\mathring{\rho},\vdash\right)}\right]_{c_{uL;0};c_{\left(u+1\right)L;0}}^{b_{uL;0},\ldots,b_{\left(u+1\right)L-1;0}},\nonumber \\
 & \quad\text{*Perform MPS compression}\nonumber \\
 & =\prod_{r=1}^{L-1}\left\{ \sum_{c_{r+uL;1}}\right\} \prod_{r=0}^{L-1}\left\{ \left[M_{r+uL;1}^{\left(\mathring{\rho},\vdash\right)}\right]_{c_{r+uL;1};c_{r+uL+1;1}}^{b_{r+uL;1}}\right\} ,\label{eq:Xi_1^(rho, vdash) -- 1}
\end{align}

\noindent where the $M_{r;1}^{\left(\mathring{\rho},\vdash\right)}$
are the rank-$3$ tensors resulting from the application of the MPO
$\Omega_{u}^{\left(\mathring{U},\vdash,i\right)}$ {[}see Eqs.~\eqref{eq:Ohm^(U, vdash, i) -- 1}-\eqref{eq:W_=00007Br; k=00007D^(U, vdash) -- 1}{]}
to the MPS $\Xi_{u;0}^{\left(\mathring{\rho},\vdash\right)}$ followed
by MPS compression {[}see Appendix~\ref{sec:MPS compression techniques used in sbc}
for references on compression techniques used in $\texttt{spinbosonchain}${]}.

Equation~\eqref{eq:ring rho_m^(vdash) recursive relation -- 1} is
equivalent to the following MPO application to a MPS:

\begin{align}
 & \left[\Xi_{u;\left(k+1\right)\Delta m+1}^{\left(\mathring{\rho},\vdash\right)}\right]_{c_{uL;\left(k+1\right)\Delta m+1};c_{\left(u+1\right)L;\left(k+1\right)\Delta m+1}}^{b_{uL;\left(k+1\right)\Delta m+1},\ldots,b_{\left(u+1\right)L-1;\left(k+1\right)\Delta m+1}}\nonumber \\
 & \quad=\prod_{r=0}^{L-1}\left\{ \sum_{b_{r+uL;k\Delta m+1}=0}^{3}\right\} \left[\Omega_{u;k}^{\left(\mathring{U},\vdash\right)}\right]_{c_{uL;\left(k+1\right)\Delta m};c_{\left(u+1\right)L;\left(k+1\right)\Delta m}}^{b_{uL;\left(k+1\right)\Delta m+1},\ldots,b_{\left(u+1\right)L-1;\left(k+1\right)\Delta m+1};b_{uL;k\Delta m+1},\ldots,b_{\left(u+1\right)L-1;k\Delta m+1}}\nonumber \\
 & \quad\mathrel{\phantom{=}}\mathop{\times}\left[\Xi_{u;k\Delta m+1}^{\left(\mathring{\rho},\vdash\right)}\right]_{c_{uL;k\Delta m+1};c_{\left(u+1\right)L;k\Delta m+1}}^{b_{uL;k\Delta m+1},\ldots,b_{\left(u+1\right)L-1;k\Delta m+1}},\nonumber \\
 & \quad\mathrel{\phantom{=}}\text{*Perform MPS compression}\nonumber \\
 & \quad=\prod_{r=1}^{L-1}\left\{ \sum_{c_{r+uL;\left(k+1\right)\Delta m+1}}\right\} \prod_{r=0}^{L-1}\left\{ \left[M_{r+uL;\left(k+1\right)\Delta m+1}^{\left(\mathring{\rho},\vdash\right)}\right]_{c_{uL;\left(k+1\right)\Delta m+1};c_{\left(u+1\right)L;\left(k+1\right)\Delta m+1}}^{b_{r+uL;\left(k+1\right)\Delta m+1}}\right\} ,\quad\text{for }k\ge0,\label{eq:Xi_=00007B3k+4=00007D^(rho, vdash) -- 1}
\end{align}

\noindent where the $M_{r;\left(k+1\right)\Delta m+1}^{\left(\mathring{\rho},\vdash\right)}$
are the rank-$3$ tensors resulting from the application of the MPO
$\Omega_{u;k}^{\left(\mathring{U},\vdash\right)}$ {[}see Eqs.~\eqref{eq:Ohm_k^(U, vdash) -- 1}-\eqref{eq:W_=00007Br; k=00007D^(U, vdash) -- 1}{]}
to the MPS $\Xi_{u;k\Delta m+1}^{\left(\mathring{\rho},\vdash\right)}$
followed MPS compression.

Equation~\eqref{eq:ring rho_m^(vdash) recursive relation -- 1} is
equivalent to the following MPO application to a MPS:

\begin{align}
\left[\Xi_{u;n;1}^{\left(\mathring{\rho}\right)}\right]_{c_{uL;1};c_{\left(u+1\right)L;1}}^{b_{uL;1},\ldots,b_{\left(u+1\right)L-1;1}} & =\prod_{r=0}^{L-1}\left\{ \sum_{b_{r+uL;0}=0}^{3}\right\} \left[\Omega_{u;n}^{\left(\mathring{U},i\right)}\right]_{c_{uL;0}^{\prime};c_{\left(u+1\right)L;0}^{\prime}}^{b_{uL;1},\ldots,b_{\left(u+1\right)L-1;1};b_{uL;0},\ldots,b_{\left(u+1\right)L-1;0}}\nonumber \\
 & \mathrel{\phantom{=}}\mathop{\times}\left[\Xi_{u;0}^{\left(\mathring{\rho},\vdash\right)}\right]_{c_{uL;0};c_{\left(u+1\right)L;0}}^{b_{uL;0},\ldots,b_{\left(u+1\right)L-1;0}},\nonumber \\
 & \quad\text{*Perform MPS compression}\nonumber \\
 & =\prod_{r=1}^{L-1}\left\{ \sum_{c_{r+uL;1}}\right\} \prod_{r=0}^{L-1}\left\{ \left[M_{r+uL;n;1}^{\left(\mathring{\rho}\right)}\right]_{c_{r+uL;1};c_{r+uL+1;1}}^{b_{r+uL;1}}\right\} ,\quad\text{for }1\le n\le K_{\tau}-1,\label{eq:Xi_=00007Bn; 1=00007D^(rho) -- 1}
\end{align}

\noindent where $K_{\tau}$ is given by Eq.~\eqref{eq:K_=00007Btau=00007D in main manuscript -- 1},
$\tau$ is the system's memory, and the $M_{r;n;1}^{\left(\mathring{\rho}\right)}$
are the rank-$3$ tensors resulting from the application of the MPO
$\Omega_{n}^{\left(\mathring{U},i\right)}$ {[}see Eqs.~\eqref{eq:Ohm_n^(U, i) -- 1}-\eqref{eq:W_=00007Br; n=00007D^(U, i) -- 1}{]}
to the MPS $\Xi_{u;0}^{\left(\mathring{\rho},\vdash\right)}$ followed
by MPS compression.

Equation~\eqref{eq:ring rho_=00007Bn; mu=00007D -- 1} is equivalent
to

\begin{equation}
\left[\Xi_{u;n;\mu_{n\Delta m;\tau}}^{\left(\mathring{\rho}\right)}\right]_{c_{uL;\mu_{n\Delta m;\tau}};c_{\left(u+1\right)L;\mu_{n\Delta m;\tau}}}^{b_{uL;\mu_{n\Delta m;\tau}},\ldots,b_{\left(u+1\right)L-1;\mu_{n\Delta m;\tau}}}=\left[\Xi_{u;\mu_{n\Delta m;\tau}}^{\left(\mathring{\rho},\vdash\right)}\right]_{c_{uL;\mu_{n\Delta m;\tau}};c_{\left(u+1\right)L;\mu_{n\Delta m;\tau}}}^{b_{uL;\mu_{n\Delta m;\tau}},\ldots,b_{\left(u+1\right)L-1;\mu_{n\Delta m;\tau}}},\label{eq:Xi_=00007Bn; mu=00007D^(rho) -- 1}
\end{equation}

\noindent where $\mu_{m;\tau}$ is given by Eq.~\eqref{eq:mu_=00007Bm; tau=00007D -- 1}.

Equation~\eqref{eq:ring rho_m^(vdash) recursive relation -- 1} is
equivalent to the following MPO application to a MPS:

\begin{align}
 & \left[\Xi_{u;n;\left(k+1\right)\Delta m+1}^{\left(\mathring{\rho}\right)}\right]_{c_{uL;\left(k+1\right)\Delta m+1};c_{\left(u+1\right)L;\left(k+1\right)\Delta m+1}}^{b_{uL;\left(k+1\right)\Delta m+1},\ldots,b_{\left(u+1\right)L-1;\left(k+1\right)\Delta m+1}}\nonumber \\
 & \quad=\prod_{r=0}^{L-1}\left\{ \sum_{b_{r+uL;k\Delta m+1}=0}^{3}\right\} \left[\Omega_{u;n;k}^{\left(\mathring{U}\right)}\right]_{c_{uL;\left(k+1\right)\Delta m};c_{\left(u+1\right)L;\left(k+1\right)\Delta m}}^{b_{uL;\left(k+1\right)\Delta m+1},\ldots,b_{\left(u+1\right)L-1;\left(k+1\right)\Delta m+1};b_{uL;k\Delta m+1},\ldots,b_{\left(u+1\right)L-1;k\Delta m+1}}\nonumber \\
 & \quad\mathrel{\phantom{=}}\mathop{\times}\left[\Xi_{u;n;k\Delta m+1}^{\left(\mathring{\rho}\right)}\right]_{c_{uL;k\Delta m+1};c_{\left(u+1\right)L;k\Delta m+1}}^{b_{uL;k\Delta m+1},\ldots,b_{\left(u+1\right)L-1;k\Delta m+1}},\nonumber \\
 & \quad\mathrel{\phantom{=}}\text{*Perform MPS compression}\nonumber \\
 & \quad=\prod_{r=1}^{L-1}\left\{ \sum_{c_{r+uL;\left(k+1\right)\Delta m+1}}\right\} \prod_{r=0}^{L-1}\left\{ \left[M_{r+uL;n;\left(k+1\right)\Delta m+1}^{\left(\mathring{\rho}\right)}\right]_{c_{uL;\left(k+1\right)\Delta m+1};c_{\left(u+1\right)L;\left(k+1\right)\Delta m+1}}^{b_{r+uL;\left(k+1\right)\Delta m+1}}\right\} ,\nonumber \\
 & \quad\mathrel{\phantom{=}}\text{for }\mu_{n\Delta m;\tau}\le k\Delta m+1\le n\Delta m-2,\label{eq:Xi_=00007Bn; 3k+4=00007D^(rho) -- 1}
\end{align}

\noindent where the $M_{r;n;\left(k+1\right)\Delta m+1}^{\left(\mathring{\rho}\right)}$
are the rank-$3$ tensors resulting from the application of the MPO
$\Omega_{u;n;k}^{\left(\mathring{U}\right)}$ {[}see Eqs.~\eqref{eq:Ohm_=00007Bn; k=00007D^(U) -- 1}-\eqref{eq:W_=00007Br; n; k=00007D^(U) -- 1}{]}
to the MPS $\Xi_{u;n;k\Delta m+1}^{\left(\mathring{\rho}\right)}$
followed MPS compression.

Lastly, Eq.~\eqref{eq:rearranging terms in rho^(A) -- 3} is equivalent
to the following MPO application to a MPS:

\begin{align}
 & \left[\Xi_{u;n}^{\left(\mathring{\rho},A\right)}\right]_{c_{uL;\left(n+1\right)\Delta m};c_{\left(u+1\right)L;\left(n+1\right)\Delta m}}^{j_{uL;\left(n+1\right)\Delta m},\ldots,j_{\left(u+1\right)L-1;\left(n+1\right)\Delta m}}\nonumber \\
 & \quad=\prod_{r=0}^{L-1}\left\{ \sum_{b_{r+uL;n\Delta m+1}=0}^{3}\right\} \left[\Omega_{u;n}^{\left(\mathring{U},f\right)}\right]_{c_{uL;\left(n+1\right)\Delta m}^{\prime};c_{\left(u+1\right)L;\left(n+1\right)\Delta m}^{\prime}}^{j_{uL;\left(n+1\right)\Delta m+1},\ldots,j_{\left(u+1\right)L-1;\left(n+1\right)\Delta m+1};b_{uL;n\Delta m+1},\ldots,b_{\left(u+1\right)L-1;n\Delta m+1}}\nonumber \\
 & \quad\mathrel{\phantom{=}}\mathop{\times}\left[\Xi_{u;n;n\Delta m+1}^{\left(\mathring{\rho}\right)}\right]_{c_{uL;n\Delta m+1};c_{\left(u+1\right)L;n\Delta m+1}}^{b_{uL;n\Delta m+1},\ldots,b_{\left(u+1\right)L-1;n\Delta m+1}},\nonumber \\
 & \quad\mathrel{\phantom{=}}\text{*Perform MPS compression}\nonumber \\
 & \quad=\prod_{r=1}^{L-1}\left\{ \sum_{c_{r+uL;\left(n+1\right)\Delta m}}\right\} \prod_{r=0}^{L-1}\left\{ \left[M_{r+uL;n}^{\left(\mathring{\rho},A\right)}\right]_{c_{uL;\left(n+1\right)\Delta m};c_{\left(u+1\right)L;\left(n+1\right)\Delta m}}^{j_{r+uL;\left(n+1\right)\Delta m}}\right\} ,\label{eq:Xi_=00007Bn=00007D^(rho, A) -- 2}
\end{align}

\noindent where the $M_{r;n}^{\left(\mathring{\rho},A\right)}$ are
the rank-$3$ tensors resulting from the application of the MPO $\Omega_{u;n}^{\left(\mathring{U},f\right)}$
{[}see Eqs.~\eqref{eq:Ohm_=00007Bn=00007D^(U, f) -- 1}-\eqref{eq:W_=00007Br; n=00007D^(U, f) -- 1}{]}
to the MPS $\Xi_{u;n;n\Delta m+1}^{\left(\mathring{\rho}\right)}$
followed by MPS compression.

In $\texttt{spinbosonchain}$, a given system's reduced density matrix
$\mathring{\rho}_{n}^{\left(A\right)}\left(\mathbf{j}_{\left(n+1\right)\Delta m}\right)$
is represented by an instance of the $\texttt{spinbosonchain.state.SystemState}$
class. This class handles the evolution of $\mathring{\rho}_{n}^{\left(A\right)}\left(\mathbf{j}_{\left(n+1\right)\Delta m}\right)$
via the $\texttt{spinbosonchain.state.SystemState.evolve}$ method.

\subsection{Computable quantities\label{subsec:Computable quantities -- 1}}

There are some useful quantities that can be calculated straightforwardly
using the MPS algorithm described in the previous sections. The first
quantity we consider is the trace of the system's reduced state operator
$\hat{\rho}^{\left(A\right)}\left(t_{n}\right)$. Stricly speaking,
due to the way MPS's are normalized upon truncations or applications
of MPO's in $\texttt{spinbosonchain}$, $\hat{\rho}^{\left(A\right)}\left(t_{n}\right)$
might not have a unit trace. Therefore it is important that we normalize
$\hat{\rho}^{\left(A\right)}\left(t_{n}\right)$ by its trace upon
calculating any expectation values of observables. Using Eqs.~\eqref{eq:mapping spin pairs to j -- 1},
\eqref{eq:g_=00007Balpha=00007D -- 1}, \eqref{eq:j vector -- 1},
\eqref{eq:applying g_=00007Balpha=00007D to j vector -- 1}, \eqref{eq:rearranging terms in rho^(A) -- 1},
and \eqref{eq:Xi_=00007Bu; n=00007D^(rho, A) -- 1}, the trace can
be expressed as

\begin{equation}
\text{Tr}^{\left(A\right)}\left\{ \hat{\rho}^{\left(A\right)}\left(t_{n}\right)\right\} =\prod_{u=-N}^{N}\left\{ \sum_{c_{uL}}\sum_{c_{\left(u+1\right)L}}\left[X_{u;n}^{\left(\mathring{\rho},A\right)}\right]_{c_{uL};c_{\left(u+1\right)L}}\right\} ,\label{eq:trace of rho^(A) -- 1}
\end{equation}

\noindent where

\begin{equation}
\left[X_{u;n}^{\left(\mathring{\rho},A\right)}\right]_{c_{uL};c_{\left(u+1\right)L}}=\prod_{r=0}^{L-1}\left\{ \sum_{j_{r+uL}=0,3}\right\} \left[\Xi_{u;n}^{\left(\mathring{\rho},A\right)}\right]_{c_{uL};c_{\left(u+1\right)L}}^{j_{uL},\ldots,j_{\left(u+1\right)L-1}},\label{eq:X_=00007Bu; n=00007D^(ring rho, A) -- 1}
\end{equation}

\noindent with $\Delta m$ is given by Eq.~\eqref{eq:Delta m -- 1}.
We can rewrite Eq.~\eqref{eq:trace of rho^(A) -- 1} in the following
form:

\begin{equation}
\text{Tr}^{\left(A\right)}\left\{ \hat{\rho}^{\left(A\right)}\left(t_{n}\right)\right\} =\sum_{c_{0}}\sum_{c_{L}}\left[v_{n;N;N}^{\left(\mathring{\rho},A,\vdash\right)}\right]_{c_{0}}\left[X_{0;n}^{\left(\mathring{\rho},A\right)}\right]_{c_{0};c_{L}}\left[v_{n;N;N}^{\left(\mathring{\rho},A,\dashv\right)}\right]_{c_{L}}.\label{eq:trace of rho^(A) -- 2}
\end{equation}

\noindent where

\begin{equation}
\left[v_{n;N;\tilde{N}}^{\left(\mathring{\rho},A,\vdash\right)}\right]_{c_{0}}=\prod_{u=-N}^{-N+\tilde{N}-1}\left\{ \sum_{c_{uL}}\sum_{c_{\left(u+1\right)L}}\left[X_{u;n}^{\left(\mathring{\rho},A\right)}\right]_{c_{uL};c_{\left(u+1\right)L}}\right\} ,\label{eq:v_=00007Bn; N; tilde N=00007D^(ring rho, A, vdash) -- 1}
\end{equation}

\begin{equation}
\left[v_{n;N;\tilde{N}}^{\left(\mathring{\rho},A,\dashv\right)}\right]_{c_{L}}=\prod_{u=N-\tilde{N}+1}^{N}\left\{ \sum_{c_{uL}}\sum_{c_{\left(u+1\right)L}}\left[X_{u;n}^{\left(\mathring{\rho},A\right)}\right]_{c_{uL};c_{\left(u+1\right)L}}\right\} ,\label{eq:v_=00007Bn; N; tilde N=00007D^(ring rho, A, dashv) -- 1}
\end{equation}

\noindent and as a reminder we are using the product convention given
by Eq.~\eqref{eq:product convention -- 1}. It is useful to rewrite
Eqs.~\eqref{eq:v_=00007Bn; N; tilde N=00007D^(ring rho, A, vdash) -- 1}
and \eqref{eq:v_=00007Bn; N; tilde N=00007D^(ring rho, A, dashv) -- 1}
as:

\begin{equation}
\left[v_{n;N;\tilde{N}}^{\left(\mathring{\rho},A,\vdash\right)}\right]_{c_{0}}=C_{n;N;\tilde{N}}^{\left(\mathring{\rho},A,\vdash\right)}x_{n;N;0}^{\tilde{N}}\left[\tilde{v}_{n;N;0}^{\left(\mathring{\rho},A,\vdash\right)}\right]_{c_{0}},\label{eq:v_=00007Bn; N; tilde N=00007D^(ring rho, A, vdash) -- 2}
\end{equation}

\begin{equation}
\left[v_{n;N;\tilde{N}}^{\left(\mathring{\rho},A,\dashv\right)}\right]_{c_{L}}=C_{n;N;\tilde{N}}^{\left(\mathring{\rho},A,\dashv\right)}x_{n;N;0}^{\tilde{N}}\left[\tilde{v}_{n;N;0}^{\left(\mathring{\rho},A,\dashv\right)}\right]_{c_{L}},\label{eq:v_=00007Bn; N; tilde N=00007D^(ring rho, A, dashv) -- 2}
\end{equation}

\noindent where

\begin{equation}
\lim_{\tilde{N}\to\infty}C_{n;N\ge\tilde{N};\tilde{N}}^{\left(\mathring{\rho},A,\vdash\right)}=C_{n}^{\left(\mathring{\rho},A,\vdash\right)},\label{eq:limit of C_=00007Bn; N; tilde N=00007D^(ring rho, A, vdash) -- 1}
\end{equation}

\begin{equation}
\lim_{\tilde{N}\to\infty}C_{n;N\ge\tilde{N};\tilde{N}}^{\left(\mathring{\rho},A,\dashv\right)}=C_{n}^{\left(\mathring{\rho},A,\dashv\right)},\label{eq:limit of C_=00007Bn; N; tilde N=00007D^(ring rho, A, dashv) -- 1}
\end{equation}

\noindent with $\tilde{v}_{n;N;0}^{\left(\mathring{\rho},A,\vdash\right)}$,
$\tilde{v}_{n;N;0}^{\left(\mathring{\rho},A,\dashv\right)}$, $C_{n;N;\tilde{N}}^{\left(\mathring{\rho},A,\vdash\right)}$,
$C_{n;N;\tilde{N}}^{\left(\mathring{\rho},A,\dashv\right)}$, and
$x_{n;N;0}$ to be determined. Recall from Sec.~\ref{subsec:Model}
that in $\texttt{spinbosonchain}$, we set $N=0$ for finite systems
and take $N\to\infty$ for infinite systems. For finite systems, it
is clear from Eq.~\eqref{eq:product convention -- 1} that

\begin{equation}
\left[\tilde{v}_{n;0;0}^{\left(\mathring{\rho},A,\vdash\right)}\right]_{c_{0}}=\left[\tilde{v}_{n;0;0}^{\left(\mathring{\rho},A,\dashv\right)}\right]_{c_{L}}=C_{n;0;0}^{\left(\mathring{\rho},A,\vdash\right)}=C_{n;0;0}^{\left(\mathring{\rho},A,\dashv\right)}=1,\label{eq:tilde v_n^(ring rho, A, vdash or dashv) for N=00003D0 -- 1}
\end{equation}

\noindent where

\begin{equation}
c_{0},c_{L}\in\left\{ 0\right\} ,\quad\text{for }N=0.\label{eq:c_=00007B0 or L=00007D for N=00003D0 -- 1}
\end{equation}

\noindent For infinite systems, Eqs.~\eqref{eq:v_=00007Bn; N; tilde N=00007D^(ring rho, A, vdash) -- 1}
and \eqref{eq:v_=00007Bn; N; tilde N=00007D^(ring rho, A, dashv) -- 1}
are just instances of unnormalized left and right power iterations
respectively, as should be apparent from the fact that 

\begin{equation}
\lim_{N\to\infty}\left[X_{u;n}^{\left(\mathring{\rho},A\right)}\right]_{c_{0};c_{L}}=\left[X_{0;n}^{\left(\mathring{\rho},A\right)}\right]_{c_{0};c_{L}},\quad\text{for finite }u.\label{eq:bulk property for infinite chains -- 1}
\end{equation}

\noindent If we assume that there is only one dominant eigenvalue
of $X_{0;n}^{\left(\mathring{\rho},A\right)}$ {[}i.e. that there
is only one eigenvalue that is largest in magnitude{]}, then as $N\to\infty$,
$x_{n;N;0}$ converges to the dominant eigenvalue of $X_{0;n}^{\left(\mathring{\rho},A\right)}$
whereas $\tilde{v}_{n;N;0}^{\left(\mathring{\rho},A,\vdash\right)}$
and $\tilde{v}_{n;N;0}^{\left(\mathring{\rho},A,\dashv\right)}$ converge
to left and right dominant eigenvectors respectively:

\begin{equation}
\sum_{c_{0}}\left[\tilde{v}_{n;N;0}^{\left(\mathring{\rho},A,\vdash\right)}\right]_{c_{0}}\left[X_{0;n}^{\left(\mathring{\rho},A\right)}\right]_{c_{0};c_{L}}=x_{n;N;0}\left[\tilde{v}_{n;N;0}^{\left(\mathring{\rho},A,\vdash\right)}\right]_{c_{L}},\quad\text{as }N\to\infty,\label{eq:X_=00007B0; n=00007D^(ring rho, A) left eigval equation -- 1}
\end{equation}

\begin{equation}
\sum_{c_{L}}\left[X_{0;n}^{\left(\mathring{\rho},A\right)}\right]_{c_{0};c_{L}}\left[\tilde{v}_{n;N;0}^{\left(\mathring{\rho},A,\dashv\right)}\right]_{c_{L}}=x_{n;N;0}\left[\tilde{v}_{n;N;0}^{\left(\mathring{\rho},A,\dashv\right)}\right]_{c_{0}},\quad\text{as }N\to\infty.\label{eq:X_=00007B0; n=00007D^(ring rho, A) right eigval equation -- 1}
\end{equation}

\noindent The assumption that the dominant eigenvalue is non-degenerate
is typically valid, however strictly speaking there is no guarantee
that this assumption is valid \citep{Schollwock1,Orus2}. Nevertheless,
we assume this in $\texttt{spinbosonchain}$, as is often done in
practice \citep{Schollwock1}. We can normalize $\tilde{v}_{n;N;0}^{\left(\mathring{\rho},A,\vdash\right)}$
and $\tilde{v}_{n;N;0}^{\left(\mathring{\rho},A,\dashv\right)}$ such
that 

\begin{equation}
\sum_{c}\left[\tilde{v}_{n;N;0}^{\left(\mathring{\rho},A,\vdash\right)}\right]_{c}\left[\tilde{v}_{n;N;0}^{\left(\mathring{\rho},A,\dashv\right)}\right]_{c}=1.\label{eq:renormalizing tilde v vectors -- 1}
\end{equation}

\noindent As we show below, we do not need to determine either $C_{n;N;\tilde{N}}^{\left(\mathring{\rho},A,\vdash\right)}$
or $C_{n;N;\tilde{N}}^{\left(\mathring{\rho},A,\dashv\right)}$ in
order to calculate any expectation values. From the above discussion,
it follows that

\begin{equation}
\text{Tr}^{\left(A\right)}\left\{ \hat{\rho}^{\left(A\right)}\left(t_{n}\right)\right\} =C_{n;N;N}^{\left(\mathring{\rho},A,\vdash\right)}C_{n;N;N}^{\left(\mathring{\rho},A,\dashv\right)}x_{n;N;0}^{2N+1},\quad\text{for }N=0\text{ or }N\to\infty.\label{eq:trace of rho^(A) -- 3}
\end{equation}

The next quantity we consider is the expectation value of an operator
$\hat{O}\left(t\right)$ at time $t=t_{n}$, where $\hat{O}\left(t\right)$
is an operator in the Schrödinger picture {[}implying possible intrinsic
time dependence{]}. Using Eqs.~\eqref{eq:mapping spin pairs to j -- 1},
\eqref{eq:g_=00007Balpha=00007D -- 1}, \eqref{eq:j vector -- 1},
\eqref{eq:applying g_=00007Balpha=00007D to j vector -- 1}, \eqref{eq:rearranging terms in rho^(A) -- 1},
and \eqref{eq:Xi_=00007Bu; n=00007D^(rho, A) -- 1}, we can write

\begin{align}
\left\langle \hat{O}\left(t_{n}\right)\right\rangle _{t_{n}} & \equiv\frac{1}{\text{Tr}^{\left(A\right)}\left\{ \hat{\rho}^{\left(A\right)}\left(t_{n}\right)\right\} }\text{Tr}\left\{ \hat{\rho}^{\left(A\right)}\left(t_{n}\right)\hat{O}\left(t_{n}\right)\right\} \nonumber \\
 & =\frac{1}{\text{Tr}^{\left(A\right)}\left\{ \hat{\rho}^{\left(A\right)}\left(t_{n}\right)\right\} }\prod_{u=-N}^{N}\left\{ \sum_{c_{uL}}\sum_{c_{\left(u+1\right)L}}\prod_{r=0}^{L-1}\left\{ \sum_{j_{r+uL}=0}^{3}\right\} \left[\Xi_{u;n}^{\left(\mathring{\rho},A\right)}\right]_{c_{uL};c_{\left(u+1\right)L}}^{j_{uL},\ldots,j_{\left(u+1\right)L-1}}\right\} \nonumber \\
 & \mathrel{\phantom{=}}\mathop{\times}O\left(t_{n};\overline{j_{-NL}},\ldots,\overline{j_{\left(N+1\right)L-1}}\right),\label{eq:expectation value of an operator -- 1}
\end{align}

\noindent where

\begin{equation}
\overline{j}=\begin{cases}
0, & \text{if }j_{r}=0,\\
2, & \text{if }j_{r}=1,\\
1, & \text{if }j_{r}=2,\\
3, & \text{if }j_{r}=3,
\end{cases}\label{eq:overbar j -- 1}
\end{equation}

\noindent and

\begin{align}
 & O\left(t_{n};j_{-NL},\ldots,j_{\left(N+1\right)L-1}\right)\nonumber \\
 & \quad=\left\langle g_{1}\left(j_{-NL}\right)z;\ldots;g_{1}\left(j_{\left(N+1\right)L-1}\right)z\right|\hat{O}\left(t_{n}\right)\left|g_{-1}\left(j_{-NL}\right)z;\ldots;g_{-1}\left(j_{\left(N+1\right)L-1}\right)z\right\rangle ,\label{eq:O tensor -- 1}
\end{align}

\noindent with $g_{\alpha}\left(j\right)$ defined by Eq.~\eqref{eq:g_=00007Balpha=00007D -- 1},

\begin{equation}
\left|g_{\alpha}\left(j_{-NL}\right)z;\ldots;g_{\alpha}\left(j_{\left(N+1\right)L-1}\right)z\right\rangle =\bigotimes_{u=-N}^{N}\bigotimes_{r=0}^{L-1}\left|g_{\alpha}\left(j_{r+uL}\right)z\right\rangle ,\label{eq:spin vector -- 1}
\end{equation}

\noindent and 

\begin{equation}
\hat{\sigma}_{z;r}\left|g_{\alpha}\left(j_{r}\right)z\right\rangle =g_{\alpha}\left(j_{r}\right)\left|g_{\alpha}\left(j_{r}\right)z\right\rangle .\label{eq:acting sigma_=00007Bz; r=00007D on |g_=00007Balpha=00007D(j_=00007Br=00007D)z> -- 1}
\end{equation}

\noindent If $\hat{O}$ is a finite string of single-spin operators:

\begin{equation}
\hat{O}\left(t\right)=\bigotimes_{u=u_{i}}^{u_{f}}\bigotimes_{r=0}^{L-1}\hat{O}_{r+uL}\left(t\right),\label{eq:a string of single-spin operators -- 1}
\end{equation}

\noindent where

\begin{equation}
-N\le u_{i}\le0\le u_{f}\le N,\label{eq:u_i and u_f -- 1}
\end{equation}

\noindent then Eq.~\eqref{eq:expectation value of an operator -- 1}
becomes:

\begin{align}
\left\langle \hat{O}\left(t_{n}\right)\right\rangle _{t_{n}} & =\frac{1}{C_{n;N;N}^{\left(\mathring{\rho},A,\vdash\right)}C_{n;N;N}^{\left(\mathring{\rho},A,\dashv\right)}x_{n;N;0}^{2N+1}}\prod_{u=u_{i}}^{u_{f}}\left\{ \sum_{c_{uL}}\sum_{c_{\left(u+1\right)L}}\right\} \left[v_{n;N;N+u_{i}}^{\left(\mathring{\rho},A,\vdash\right)}\right]_{c_{u_{i}L}}\nonumber \\
 & \mathrel{\phantom{=}}\mathop{\times}\prod_{u=u_{i}}^{u_{f}}\left\{ \prod_{r=0}^{L-1}\left\{ \sum_{j_{r+uL}=0}^{3}O_{r+uL}\left(t_{n};\overline{j_{r+uL}}\right)\right\} \left[\Xi_{u;n}^{\left(\mathring{\rho},A\right)}\right]_{c_{uL};c_{\left(u+1\right)L}}^{j_{uL},\ldots,j_{\left(u+1\right)L-1}}\right\} \left[v_{n;N;N-u_{f}}^{\left(\mathring{\rho},A,\dashv\right)}\right]_{c_{\left(u_{f}+1\right)L}},\label{eq:expectation value of an operator -- 2}
\end{align}

\noindent where

\begin{equation}
O_{r}\left(t_{n};j_{r}\right)=\left\langle g_{1}\left(j_{r}\right)z\right|\hat{O}_{r}\left|g_{-1}\left(j_{r}\right)z\right\rangle ,\label{eq:O_r(t_n; j_=00007Br=00007D) -- 1}
\end{equation}

\noindent and we used Eq.~\eqref{eq:trace of rho^(A) -- 3}. From
Eqs.~\eqref{eq:v_=00007Bn; N; tilde N=00007D^(ring rho, A, vdash) -- 2},
\eqref{eq:v_=00007Bn; N; tilde N=00007D^(ring rho, A, dashv) -- 2},
\eqref{eq:tilde v_n^(ring rho, A, vdash or dashv) for N=00003D0 -- 1},
\eqref{eq:c_=00007B0 or L=00007D for N=00003D0 -- 1}, and \eqref{eq:u_i and u_f -- 1},
it should be clear that for finite systems Eq.~\eqref{eq:expectation value of an operator -- 2}
reduces to:

\begin{equation}
\left\langle \hat{O}\left(t_{n}\right)\right\rangle _{t_{n}}=\frac{1}{x_{n;0;0}}\prod_{r=0}^{L-1}\left\{ \sum_{j_{r}=0}^{3}O_{r}\left(t_{n};\overline{j_{r}}\right)\right\} \left[\Xi_{0;n}^{\left(\mathring{\rho},A\right)}\right]_{0;0}^{j_{0},\ldots,j_{L-1}},\quad\text{for }N=0.\label{eq:expectation value of an operator -- 3}
\end{equation}

\noindent For infinite system, we can use Eqs.~\eqref{eq:u_i and u_f -- 1}-\eqref{eq:u_i and u_f -- 1},
\eqref{eq:bulk property for infinite chains -- 1}-\eqref{eq:renormalizing tilde v vectors -- 1},
and \eqref{eq:renormalizing tilde v vectors -- 1} to show that Eq.~\eqref{eq:expectation value of an operator -- 2}
reduces as follows:

\begin{align}
\left\langle \hat{O}\left(t_{n}\right)\right\rangle _{t_{n}} & =\frac{1}{\cancel{C_{n}^{\left(\mathring{\rho},A,\vdash\right)}}\cancel{C_{n}^{\left(\mathring{\rho},A,\dashv\right)}}x_{n;N;0}^{2N+1}}\prod_{u=u_{i}}^{u_{f}}\left\{ \sum_{c_{uL}}\sum_{c_{\left(u+1\right)L}}\right\} \left\{ \cancel{C_{n}^{\left(\mathring{\rho},A,\vdash\right)}}x_{n;N;0}^{N+u_{i}}\left[\tilde{v}_{n;N;0}^{\left(\mathring{\rho},A,\vdash\right)}\right]_{c_{u_{i}L}}\right\} \nonumber \\
 & \mathrel{\phantom{=}}\mathop{\times}\prod_{u=u_{i}}^{u_{f}}\left\{ \prod_{r=0}^{L-1}\left\{ \sum_{j_{r+uL}=0}^{3}O_{r+uL}\left(t_{n};\overline{j_{r+uL}}\right)\right\} \left[\Xi_{u;n}^{\left(\mathring{\rho},A\right)}\right]_{c_{uL};c_{\left(u+1\right)L}}^{j_{uL},\ldots,j_{\left(u+1\right)L-1}}\right\} \nonumber \\
 & \mathrel{\phantom{=}}\mathop{\times}\left\{ \cancel{C_{n}^{\left(\mathring{\rho},A,\dashv\right)}}x_{n;N;0}^{N-u_{f}}\left[\tilde{v}_{n;N;0}^{\left(\mathring{\rho},A,\dashv\right)}\right]_{c_{\left(u_{f}+1\right)L}}\right\} \nonumber \\
 & =\frac{1}{x_{n;N;0}^{u_{f}-u_{i}+1}}\prod_{u=u_{i}}^{u_{f}}\left\{ \sum_{c_{uL}}\sum_{c_{\left(u+1\right)L}}\right\} \left[\tilde{v}_{n;N;0}^{\left(\mathring{\rho},A,\vdash\right)}\right]_{c_{u_{i}L}}\nonumber \\
 & \mathrel{\phantom{=}}\mathop{\times}\prod_{u=u_{i}}^{u_{f}}\left\{ \prod_{r=0}^{L-1}\left\{ \sum_{j_{r+uL}=0}^{3}O_{r+uL}\left(t_{n};\overline{j_{r+uL}}\right)\right\} \left[\Xi_{u;n}^{\left(\mathring{\rho},A\right)}\right]_{c_{uL};c_{\left(u+1\right)L}}^{j_{uL},\ldots,j_{\left(u+1\right)L-1}}\right\} \left[\tilde{v}_{n;N;0}^{\left(\mathring{\rho},A,\dashv\right)}\right]_{c_{\left(u_{f}+1\right)L}},\nonumber \\
 & \mathrel{\phantom{=}}\quad\text{for }N\to\infty.\label{eq:expectation value of an operator -- 4}
\end{align}

\noindent The result of Eq.~\eqref{eq:expectation value of an operator -- 3}
is a special case of Eq.~\eqref{eq:expectation value of an operator -- 4},
hence we may write:

\begin{align}
\left\langle \hat{O}\left(t_{n}\right)\right\rangle _{t_{n}} & =\frac{1}{x_{n;N;0}^{u_{f}-u_{i}+1}}\prod_{u=u_{i}}^{u_{f}}\left\{ \sum_{c_{uL}}\sum_{c_{\left(u+1\right)L}}\right\} \left[\tilde{v}_{n;N;0}^{\left(\mathring{\rho},A,\vdash\right)}\right]_{c_{u_{i}L}}\nonumber \\
 & \mathrel{\phantom{=}}\mathop{\times}\prod_{u=u_{i}}^{u_{f}}\left\{ \prod_{r=0}^{L-1}\left\{ \sum_{j_{r+uL}=0}^{3}O_{r+uL}\left(t_{n};\overline{j_{r+uL}}\right)\right\} \left[\Xi_{u;n}^{\left(\mathring{\rho},A\right)}\right]_{c_{uL};c_{\left(u+1\right)L}}^{j_{uL},\ldots,j_{\left(u+1\right)L-1}}\right\} \left[\tilde{v}_{n;N;0}^{\left(\mathring{\rho},A,\dashv\right)}\right]_{c_{\left(u_{f}+1\right)L}},\nonumber \\
 & \mathrel{\phantom{=}}\quad\text{for }N=0\text{ or }N\to\infty.\label{eq:expectation value of an operator -- 5}
\end{align}

\noindent In $\texttt{spinbosonchain}$, the function $\texttt{spinbosonchain.ev.multi\_site\_spin\_op}$
calculates $\left\langle \hat{O}\left(t_{n}\right)\right\rangle _{t_{n}}$
for a given $\hat{O}\left(t_{n}\right)$ of the form given by Eq.~\eqref{eq:a string of single-spin operators -- 1}.
The implementation of the function $\texttt{spinbosonchain.ev.multi\_site\_spin\_op}$
is based on Eq.~\eqref{eq:expectation value of an operator -- 5}.
This function can be used to calculate the expectation value of strings
of Pauli spin operators $\hat{\sigma}_{\nu;r}$ and the identity operator.

Another application of Eq.~\eqref{eq:expectation value of an operator -- 5}
is to calculate classical spin probabilities. Given a classical spin
configuration of a subset of the system's spins:

\begin{equation}
\left(\sigma_{z;u_{i}L},\ldots\sigma_{z;\left(u_{f}+1\right)L-1}\right),\label{eq:classical spin configuration -- 1}
\end{equation}

\noindent with

\begin{equation}
\sigma_{z;r}=\pm1,\label{eq:sigma_=00007Bz; r=00007D -- 1}
\end{equation}

\noindent we can calculate the probability of measuring the $z$-components
this subset of spin to be in the given configuration $\boldsymbol{\sigma}_{z;u_{i},u_{f}}$
at time $t_{n}$ by setting 

\begin{equation}
\hat{O}\left(t_{n}\right)=\bigotimes_{u=u_{i}}^{u_{f}}\bigotimes_{r=0}^{L-1}\left\{ \frac{1+\sigma_{z;r+uL}\hat{\sigma}_{z;r+uL}}{2}\right\} ,\label{eq:O(t_n) for spin config prob -- 1}
\end{equation}

\noindent and using Eq.~\eqref{eq:expectation value of an operator -- 5}
to get

\begin{align}
 & \text{Prob}\left(t_{n};\sigma_{z;u_{i}L},\ldots\sigma_{z;\left(u_{f}+1\right)L-1}\right)\nonumber \\
 & \quad=\frac{1}{x_{n;N;0}^{u_{f}-u_{i}+1}}\prod_{u=u_{i}}^{u_{f}}\left\{ \sum_{c_{uL}}\sum_{c_{\left(u+1\right)L}}\right\} \left[\tilde{v}_{n;N;0}^{\left(\mathring{\rho},A,\vdash\right)}\right]_{c_{u_{i}L}}\nonumber \\
 & \quad\mathrel{\phantom{=}}\mathop{\times}\prod_{u=u_{i}}^{u_{f}}\left\{ \prod_{r=0}^{L-1}\left\{ \sum_{j_{r+uL}=0}^{3}O_{r+uL}\left(t_{n};\overline{j_{r+uL}}\right)\right\} \left[\Xi_{u;n}^{\left(\mathring{\rho},A\right)}\right]_{c_{uL};c_{\left(u+1\right)L}}^{\frac{3}{2}\left(1-\sigma_{z;uL}\right),\ldots,\frac{3}{2}\left(1-\sigma_{z;\left(u+1\right)L-1}\right)}\right\} \left[\tilde{v}_{n;N;0}^{\left(\mathring{\rho},A,\dashv\right)}\right]_{c_{\left(u_{f}+1\right)L}},\nonumber \\
 & \quad\mathrel{\phantom{=}}\quad\text{for }N=0\text{ or }N\to\infty.\label{eq:spin configuration probability -- 1}
\end{align}

\noindent In $\texttt{spinbosonchain}$, the function $\texttt{spinbosonchain.state.spin\_config\_prob}$
calculates \\
$\text{Prob}\left(t_{n};\sigma_{z;u_{i}L},\ldots\sigma_{z;\left(u_{f}+1\right)L-1}\right)$.

We can also use Eq.~\eqref{eq:expectation value of an operator -- 5}
to calculate the expectation value of the system's energy per unit
cell. To see how this can be done, first note that

\begin{equation}
\left\langle \hat{\varepsilon}\left(t_{n}\right)\right\rangle _{t_{n}}\equiv\frac{1}{2N+1}\left\langle \hat{H}^{\left(A\right)}\left(t\right)\right\rangle _{t_{n}}=\left\langle \hat{H}_{u=0}^{\left(A\right)}\left(t\right)\right\rangle _{t_{n}},\quad\text{for }N=0\text{ or }N\to\infty,\label{eq:average energy -- 1}
\end{equation}

\noindent where we have used Eqs\@.~\eqref{eq:total Hamiltonian -- 1}
and \eqref{eq:total Hamiltonian -- 2}. Using Eqs.~\eqref{eq:H_u^(A) -- 1}-\eqref{eq:H_=00007Bu; z=00007D^(A) -- 1},
we can rewrite Eq.~\eqref{eq:average energy -- 1} as

\begin{align}
\left\langle \hat{\varepsilon}\left(t_{n}\right)\right\rangle _{t_{n}} & =\sum_{r=0}^{L-1}\left\{ h_{x;r}\left(t_{n}\right)\left\langle \hat{\sigma}_{x;r}\right\rangle _{t_{n}}+h_{z;r}\left(t_{n}\right)\left\langle \hat{\sigma}_{z;r}\right\rangle _{t_{n}}+J_{z,z;r,r+1}\left(t_{n}\right)\left\langle \hat{\sigma}_{z;r}\hat{\sigma}_{z;r+1}\right\rangle _{t_{n}}\right\} ,\nonumber \\
 & \mathrel{\phantom{=}}\text{for }N=0\text{ or }N\to\infty.\label{eq:average energy -- 2}
\end{align}

\noindent From Eq.~\eqref{eq:average energy -- 2}, it should be
clear that we can apply Eq.~\eqref{eq:expectation value of an operator -- 5}
to calculate $\left\langle \hat{\varepsilon}\left(t_{n}\right)\right\rangle _{t_{n}}$.
In $\texttt{spinbosonchain}$, the function $\texttt{spinbosonchain.ev.energy}$
calculates $\left\langle \hat{\varepsilon}\left(t_{n}\right)\right\rangle _{t_{n}}$,
doing so by calling $\texttt{spinbosonchain.ev.single\_site\_spin\_op}$
and $\texttt{spinbosonchain.ev.nn\_two\_site\_spin\_op}$: the former
calculates expectation values of single-site spin operators $\hat{\sigma}_{\nu;r}$,
and the latter calculates the expectation values of nearest-neighbour
two-site spin operators $\hat{\sigma}_{\nu_{1};r}\hat{\sigma}_{\nu_{2};r+1}$.

Next, we use Eq.~\eqref{eq:expectation value of an operator -- 5}
to derive an expression for characteristic correlation lengths of
two-point correlators for infinite systems. Let us consider the following
coarse-grained two-point correlation function:

\begin{equation}
G_{u_{i},u_{f}}\left(t_{n}\right)=\left\langle \hat{O}_{u_{i}}^{\left(1\right)}\hat{O}_{u_{f}}^{\left(2\right)}\right\rangle _{t_{n}},\label{eq:coarse grain two-point correlator -- 1}
\end{equation}

\noindent where $\hat{O}_{u_{i}L}^{\left(1\right)}$ and $\hat{O}_{u_{f}L}^{\left(2\right)}$
are multi-site operators that operate on the unit cells $u=u_{i}$
and $u=u_{f}$ respectively. From Eq.~\eqref{eq:expectation value of an operator -- 5},
we can see that $G_{u_{i},u_{f}}\left(t_{n}\right)$ can be expressed
as

\begin{align}
G_{u_{i},u_{f}}\left(t_{n}\right) & =\frac{1}{x_{n;N;0}^{u_{f}-u_{i}+1}}\prod_{u=u_{i}}^{u_{f}}\left\{ \sum_{c_{uL}}\sum_{c_{\left(u+1\right)L}}\right\} \left[\tilde{v}_{n;N;0}^{\left(\mathring{\rho},A,\vdash\right)}\right]_{c_{u_{i}L}}\nonumber \\
 & \mathrel{\phantom{=}}\mathop{\times}\prod_{r=0}^{L-1}\left\{ \sum_{j_{r+u_{i}L}=0}^{3}\right\} O_{u_{i}}^{\left(1\right)}\left(t_{n};\overline{j_{u_{i}L}},\ldots,\overline{j_{\left(u_{i}+1\right)L-1}}\right)\left[\Xi_{0;n}^{\left(\mathring{\rho},A\right)}\right]_{c_{u_{i}L};c_{\left(u_{i}+1\right)L}}^{j_{u_{i}L},\ldots,j_{\left(u_{i}+1\right)L-1}}\nonumber \\
 & \mathrel{\phantom{=}}\mathop{\times}\prod_{u=u_{i}+1}^{u_{f}-1}\left\{ \prod_{r=0}^{L-1}\left\{ \sum_{j_{r+uL}=0,3}\right\} \left[\Xi_{0;n}^{\left(\mathring{\rho},A\right)}\right]_{c_{uL};c_{\left(u+1\right)L}}^{j_{uL},\ldots,j_{\left(u+1\right)L-1}}\right\} \nonumber \\
 & \mathrel{\phantom{=}}\mathop{\times}\prod_{r=0}^{L-1}\left\{ \sum_{j_{r+u_{f}L}=0}^{3}\right\} O_{u_{f}}^{\left(2\right)}\left(t_{n};\overline{j_{u_{f}L}},\ldots,\overline{j_{\left(u_{f}+1\right)L-1}}\right)\left[\Xi_{0;n}^{\left(\mathring{\rho},A\right)}\right]_{c_{u_{f}L};c_{\left(u_{f}+1\right)L}}^{j_{u_{f}L},\ldots,j_{\left(u_{f}+1\right)L-1}}\nonumber \\
 & \mathrel{\phantom{=}}\mathop{\times}\left[\tilde{v}_{n;N;0}^{\left(\mathring{\rho},A,\dashv\right)}\right]_{c_{\left(u_{f}+1\right)L}}\nonumber \\
 & =x_{n;N;0}^{-u_{f}+u_{i}+1}\prod_{u=u_{i}+1}^{u_{f}-1}\left\{ \sum_{c_{uL}}\sum_{c_{\left(u+1\right)L}}\right\} \left[v_{n;N}^{\left(O,1\right)}\right]_{c_{\left(u_{i}+1\right)L}}\prod_{u=u_{i}+1}^{u_{f}-1}\left\{ \left[X_{0;n}^{\left(\mathring{\rho},A\right)}\right]_{c_{uL};c_{\left(u+1\right)L}}\right\} \left[v_{n;N}^{\left(O,2\right)}\right]_{c_{\left(u_{f}\right)L}},\nonumber \\
 & \mathrel{\phantom{=}}\text{for }N\to\infty,\label{eq:coarse grain two-point correlator -- 2}
\end{align}

\noindent where

\begin{align}
\left[v_{n;N}^{\left(O,1\right)}\right]_{c_{\left(u_{i}+1\right)L}} & =\frac{1}{x_{n;N;0}}\prod_{r=0}^{L-1}\left\{ \sum_{j_{r+u_{i}L}=0}^{3}\right\} O_{u_{i}}^{\left(1\right)}\left(t_{n};\overline{j_{u_{i}L}},\ldots,\overline{j_{\left(u_{i}+1\right)L-1}}\right)\nonumber \\
 & \mathrel{\phantom{=}}\mathop{\times}\sum_{c_{u_{i}L}}\left[\tilde{v}_{n;N;0}^{\left(\mathring{\rho},A,\vdash\right)}\right]_{c_{u_{i}L}}\left[\Xi_{0;n}^{\left(\mathring{\rho},A\right)}\right]_{c_{u_{i}L};c_{\left(u_{i}+1\right)L}}^{j_{u_{i}L},\ldots,j_{\left(u_{i}+1\right)L-1}},\label{eq:v_=00007Bn; N=00007D^(O, 1) -- 1}
\end{align}

\begin{align}
\left[v_{n;N}^{\left(O,2\right)}\right]_{c_{\left(u_{f}\right)L}} & =\frac{1}{x_{n;N;0}}\prod_{r=0}^{L-1}\left\{ \sum_{j_{r+u_{f}L}=0}^{3}\right\} O_{u_{f}}^{\left(2\right)}\left(t_{n};\overline{j_{u_{f}L}},\ldots,\overline{j_{\left(u_{f}+1\right)L-1}}\right)\nonumber \\
 & \mathrel{\phantom{=}}\mathop{\times}\sum_{c_{\left(u_{f}+1\right)L}}\left[\Xi_{0;n}^{\left(\mathring{\rho},A\right)}\right]_{c_{u_{f}L};c_{\left(u_{f}+1\right)L}}^{j_{u_{f}L},\ldots,j_{\left(u_{f}+1\right)L-1}}\left[\tilde{v}_{n;N;0}^{\left(\mathring{\rho},A,\dashv\right)}\right]_{c_{\left(u_{f}+1\right)L}}.\label{eq:v_=00007Bn; N=00007D^(O, 2) -- 1}
\end{align}

\noindent If we assume that $X_{0;n}^{\left(\mathring{\rho},A\right)}$
is not a defective matrix, then we can expand $v_{n;N}^{\left(O,1\right)}$
and $v_{n;N}^{\left(O,2\right)}$ in terms of left and right eigenbases
of $X_{0;n}^{\left(\mathring{\rho},A\right)}$ respectively such that
for each left eigenvector from the left eigenbasis, there is exactly
one right eigenvector from the right eigenbasis to which it is not
orthogonal. Moreover, this right eigenvector shares the same eigenvalue
as the given left eigenvector and their dot product is unity. Mathematically,
if we denote $\left\{ \tilde{v}_{n;N;l}^{\left(\mathring{\rho},A,\vdash\right)}\right\} _{l=0}^{\chi_{s}-1}$
and $\left\{ \tilde{v}_{n;N;l}^{\left(\mathring{\rho},A,\dashv\right)}\right\} _{l=0}^{\chi_{s}-1}$
as the aforementioned left and right eigenbases respectively, then
we have:

\begin{equation}
\left[v_{n;N}^{\left(O,1\right)}\right]_{c_{\left(u_{i}+1\right)L}}=\sum_{l}\psi_{n;N;l}^{\left(O,1\right)}\left[\tilde{v}_{n;N;l}^{\left(\mathring{\rho},A,\vdash\right)}\right]_{c_{\left(u_{i}+1\right)L}},\label{eq:v_=00007Bn; N=00007D^(O, 1) expansion -- 1}
\end{equation}

\begin{equation}
\left[v_{n;N}^{\left(O,2\right)}\right]_{c_{\left(u_{f}\right)L}}=\sum_{l}\psi_{n;N;l}^{\left(O,2\right)}\left[\tilde{v}_{n;N;l}^{\left(\mathring{\rho},A,\dashv\right)}\right]_{c_{\left(u_{f}\right)L}},\label{eq:v_=00007Bn; N=00007D^(O, 2) expansion -- 1}
\end{equation}

\noindent where $\left(\chi_{s}-1\right)\times\left(\chi_{s}-1\right)$
are the dimensions of the matrix $X_{0;n}^{\left(\mathring{\rho},A\right)}$;
$\psi_{n;N;l}^{\left(O,1\right)}$ and $\psi_{n;N;l}^{\left(O,2\right)}$
are expansion coefficients which we need not determine for the purposes
of this discussion; and 

\begin{equation}
\sum_{c_{0}}\left[\tilde{v}_{n;N;l}^{\left(\mathring{\rho},A,\vdash\right)}\right]_{c_{0}}\left[X_{0;n}^{\left(\mathring{\rho},A\right)}\right]_{c_{0};c_{L}}=x_{n;N;l}\left[\tilde{v}_{n;N;l}^{\left(\mathring{\rho},A,\vdash\right)}\right]_{c_{L}},\label{eq:left eigenvalue equation of X_=00007B0; n=00007D^(ring rho, A) -- 1}
\end{equation}

\begin{equation}
\sum_{c_{L}}\left[X_{0;n}^{\left(\mathring{\rho},A\right)}\right]_{c_{0};c_{L}}\left[\tilde{v}_{n;N;l}^{\left(\mathring{\rho},A,\dashv\right)}\right]_{c_{L}}=x_{n;N;l}\left[\tilde{v}_{n;N;0}^{\left(\mathring{\rho},A,\dashv\right)}\right]_{c_{0}},\label{eq:right eigenvalue equation of X_=00007B0; n=00007D^(ring rho, A) -- 1}
\end{equation}

\begin{equation}
\sum_{c}\left[\tilde{v}_{n;N;l_{1}}^{\left(\mathring{\rho},A,\vdash\right)}\right]_{c}\left[\tilde{v}_{n;N;l_{2}}^{\left(\mathring{\rho},A,\dashv\right)}\right]_{c}=\delta_{l_{1},l_{2}},\label{eq:orthonormality condition of left and right eigenbases -- 1}
\end{equation}

\noindent with $\left\{ x_{n;N;l}\right\} _{l=0}^{\chi_{s}-1}$ being
the eigenspectrum of $X_{0;n}^{\left(\mathring{\rho},A\right)}$ {[}note
that we have already denoted $x_{n;N;l=0}$ as the dominant eigenvalue
of $X_{0;n}^{\left(\mathring{\rho},A\right)}${]}. The assumption
that $X_{0;n}^{\left(\mathring{\rho},A\right)}$ is not a defective
matrix is typically valid, however stricly speaking there is no guarantee
that this assumption is valid. Nevertheless, we assume this in $\texttt{spinbosonchain}$
for practicality. Using Eqs.~\eqref{eq:v_=00007Bn; N=00007D^(O, 1) expansion -- 1}-\eqref{eq:orthonormality condition of left and right eigenbases -- 1},
we can rewrite Eq.~\eqref{eq:coarse grain two-point correlator -- 2}
as

\begin{align}
G_{u_{i},u_{f}}\left(t_{n}\right) & =\sum_{l=0}^{\chi_{s}-1}\psi_{n;N;l}^{\left(O,1\right)}\psi_{n;N;l}^{\left(O,2\right)}\left\{ \frac{x_{n;N;l}}{x_{n;N;0}}\right\} ^{u_{f}-u_{i}-1}\nonumber \\
 & =\psi_{n;N;0}^{\left(O,1\right)}\psi_{n;N;0}^{\left(O,2\right)}+\sum_{l=1}^{\chi_{s}-1}e^{i\varphi_{l}\left\{ u_{f}-u_{i}-1\right\} }\psi_{n;N;l}^{\left(O,1\right)}\psi_{n;N;l}^{\left(O,2\right)}e^{-\frac{L}{\xi_{l}}\left\{ u_{f}-u_{i}-1\right\} },\quad\text{for }N\to\infty,\label{eq:coarse grain two-point correlator -- 3}
\end{align}

\noindent where

\begin{equation}
\varphi_{l}=\arg\left(\frac{x_{n;N;l}}{x_{n;N;0}}\right),\label{eq:arg of x ration -- 1}
\end{equation}

\noindent and

\begin{equation}
\xi_{l}=-\frac{L}{\ln\left(\left|x_{n;N;l}/x_{n;N;0}\right|\right)},\quad\text{for }1\le l\le\chi_{s}-1\text{ and }N\to\infty.\label{eq:correlation lengths -- 1}
\end{equation}

\noindent From Eq.~\eqref{eq:coarse grain two-point correlator -- 3},
we see that two-point correlators can be long-ranged if $\psi_{n;N;0}^{\left(O,1\right)}\psi_{n;N;0}^{\left(O,2\right)}\neq0$
and/or be made up of a superposition of exponentials with decay lengths
$\xi_{l}$. These $\xi_{l}$ are the correlation lengths mentioned
above. For infinite systems, the $\xi_{l}$ are calculated automatically
upon evolving the system in $\texttt{spinbosonchain}$, and are stored
as the attribute $\texttt{correlation\_lengths}$ of the $\texttt{spinbosonchain.state.SystemState}$
class.

Lastly, we consider entanglement measures. Even though the system
is prepared in a pure state, i.e. $\hat{\rho}^{\left(i,A\right)}$
is a pure state operator {[}see Eq.~\eqref{eq:rho^(i, A) -- 1}{]},
generally speaking $\hat{\rho}^{\left(A\right)}\left(t_{n}\right)$
will be a mixed-state operator since the system is coupled to an environment.
As a result, the bi-partite von Neumann entanglement entropy is not
a good entanglement measure for it cannot distinguish between classical
and quantum correlations, and a mixed-state can possess both, unlike
a pure state which can only possess quantum correlations. Below we
describe an alternative approach that works for mixed states. First,
we express $\Xi_{n}^{\left(\mathring{\rho},A\right)}$, the MPS representing
$\hat{\rho}^{\left(A\right)}\left(t_{n}\right)$, as a mixed-canonical
MPS:

\begin{align}
\mathring{\rho}_{n}^{\left(A\right)}\left(\mathbf{j}_{\left(n+1\right)\Delta m}\right) & =\sum_{c_{r;\left(n+1\right)\Delta m}}\left[\Xi_{r;n}^{\left(\mathring{\rho},A,\vdash\right)}\right]_{0;c_{r;\left(n+1\right)\Delta m}}^{j_{-NL;\left(n+1\right)\Delta m},\ldots,j_{r-1;\left(n+1\right)\Delta m}}\nonumber \\
 & \mathrel{\phantom{=}}\mathop{\times}\tilde{\Lambda}_{r;c_{r;\left(n+1\right)\Delta m}}\left[\Xi_{r;n}^{\left(\mathring{\rho},A,\dashv\right)}\right]_{c_{r;\left(n+1\right)\Delta m};0}^{j_{r;\left(n+1\right)\Delta m},\ldots,j_{\left(N+1\right)L-1;\left(n+1\right)\Delta m}},\quad\text{for }0\le r\le L-1,\label{eq:mixed canonical form of MPS -- 1}
\end{align}

\noindent where $\Lambda_{r;c_{r;\left(n+1\right)\Delta m}}$ is the
Schmidt spectrum associated with the bond between sites $r-1$ and
$r$, and $\Xi_{r;n}^{\left(\mathring{\rho},A,\vdash\right)}$ and
$\Xi_{r;n}^{\left(\mathring{\rho},A,\dashv\right)}$ satisfy

\begin{align}
 & \prod_{r^{\prime}=-NL}^{r-1}\left\{ \sum_{j_{r^{\prime};\left(n+1\right)\Delta m}=0}^{3}\right\} \left\{ \left[\Xi_{r;n}^{\left(\mathring{\rho},A,\vdash\right)}\right]_{0;c_{r;\left(n+1\right)\Delta m}}^{j_{-NL;\left(n+1\right)\Delta m},\ldots,j_{r-1;\left(n+1\right)\Delta m}}\right\} ^{*}\nonumber \\
 & \mathrel{\phantom{=}}\mathop{\times}\left[\Xi_{r;n}^{\left(\mathring{\rho},A,\vdash\right)}\right]_{0;c_{r;\left(n+1\right)\Delta m}^{\prime}}^{j_{-NL;\left(n+1\right)\Delta m},\ldots,j_{r-1;\left(n+1\right)\Delta m}}=\delta_{c_{r;\left(n+1\right)\Delta m},c_{r;\left(n+1\right)\Delta m}^{\prime}},\label{eq:condition for left tensor of mixed canonical form -- 1}
\end{align}

\begin{align}
 & \prod_{r^{\prime}=r}^{\left(N+1\right)L-1}\left\{ \sum_{j_{r^{\prime};\left(n+1\right)\Delta m}=0}^{3}\right\} \left\{ \left[\Xi_{r;n}^{\left(\mathring{\rho},A,\dashv\right)}\right]_{c_{r;\left(n+1\right)\Delta m};0}^{j_{r;\left(n+1\right)\Delta m},\ldots,j_{\left(N+1\right)L-1;\left(n+1\right)\Delta m}}\right\} ^{*}\nonumber \\
 & \mathrel{\phantom{=}}\mathop{\times}\left[\Xi_{r;n}^{\left(\mathring{\rho},A,\dashv\right)}\right]_{c_{r;\left(n+1\right)\Delta m};0}^{j_{r;\left(n+1\right)\Delta m},\ldots,j_{\left(N+1\right)L-1;\left(n+1\right)\Delta m}}=\delta_{c_{r;\left(n+1\right)\Delta m},c_{r;\left(n+1\right)\Delta m}^{\prime}}.\label{eq:condition for right tensor of mixed canonical form -- 1}
\end{align}

\noindent For detailed discussions on the mixed-canonical form of
a finite MPS see Ref.~\citep{Schollwock1}, and for infinite systems
see Refs.~\citep{Orus2,McCulloch1,Schollwock1}. Let us normalize
the Schmidt spectra by the trace of the system's reduced density matrix:

\begin{equation}
\Lambda_{r;c}=\frac{\tilde{\Lambda}_{r;c}}{\text{Tr}^{\left(A\right)}\left\{ \hat{\rho}^{\left(A\right)}\left(t_{n}\right)\right\} }.\label{eq:renormalized Schmidt spectra -- 1}
\end{equation}

\noindent According to the realignment criterion \citep{Chen1,Rudolph1},
if $\sum_{c}\Lambda_{r;c}>1$ then the system is in a bipartite entangled
state for the bipartition formed by splitting the chain in two between
sites $r-1$ and $r$. It is important to note that $\sum_{c}\Lambda_{r;c}\le1$
is a necessary condition for a state to be separable {[}i.e. not entangled{]}
for the aforementioned bipartition, however it is not \emph{sufficient}.
Therefore, for a given bipartition of the chain, there exist bipartite
entangled states that violate $\sum_{c}\Lambda_{r;c}>1$. That being
said $\sum_{c}\Lambda_{r;c}\le1$ is considered to be a strong condition
for separability \citep{Chen1,Rudolph1}, hence the criterion $\sum_{c}\Lambda_{r;c}>1$
should detect most bipartite entangled states. Unfortunately, we cannot
apply the realignment criterion to infinite chains since we cannot
determine $\text{Tr}^{\left(A\right)}\left\{ \hat{\rho}^{\left(A\right)}\left(t_{n}\right)\right\} $.

In $\texttt{spinbosonchain}$ for finite systems, one can determine
if a given system state possesses any bipartite entanglement using
the $\texttt{spinbosonchain.state.realignment\_criterion}$ function.
The function checks for bipartite entanglement for every bipartition.

\newpage{}

\section{Conclusion\label{sec:Conclusion}}

In this manuscript, we have extended the work of Refs.~\citep{Strathearn1,Suzuki1,Oshiyama1}
by developing a generalized QUAPI formalism that is capable of describing
the dynamics of a generalized spin-boson chain model, where both the
$z$- and $y$-components of the quantum Ising spins are coupled to
bosonic baths, rather than only the $z$-components. We have derived
a path integral expression for the system's reduced density matrix
and have developed a tensor network technique that enables us to evaluate
the path integral in a computationally efficient manner. Our formalism
is flexible in that it can handle both Markovian and non-Markovian
types of noise, which is particularly useful when studying hybrid
types of noise consisting of e.g. a $1/f$ component and an ohmic
component. We have implemented our approach in a Python library called
$\texttt{spinbosonchain}$ \citep{sbc1}.

One possible application of $\texttt{spinbosonchain}$ is to study
the noisy dynamics of flux qubit chains subject to both charge and
flux noise. In this context, charge and flux noise are modelled by
$y$- and $z$-noise respectively in our generalized spin-boson model.
In future work, we plan on using $\texttt{spinbosonchain}$ to study
the Kibble-Zurek mechanism in flux-qubit chains in the presence of
charge and flux noise, and to connect with experimental results obtained
by D-Wave's flux-qubit quantum annealers. Another possible future
direction of this work is to generalized our formalism to different
topologies, e.g. ladders, square lattices, chains with next-nearest
couplings, and so on. The formalism would only need to be modified
in Secs.~\ref{subsec:MPS representation of e^(i phi) -- 1} and \ref{subsec:MPO representations of U quantities -- 1},
where the various $\mathring{U}$-tensors would have to be expressed
as MPO's with bond dimensions greater than or equal to $4$, in contrast
to chain topologies where all MPO bond dimensions are equal to $4$.
Therefore, the simulation of more complicated topologies will be more
computationally expensive as compared to that of chain topologies.
\begin{acknowledgments}
This work was supported by Mitacs through the Mitacs Accelerate program.
\end{acknowledgments}

\appendix
\newpage{}

\section{Coherent states\label{sec:Coherent states -- 1}}

Let $\hat{b}_{\nu;r;\epsilon}^{\vphantom{\dagger}}$ be the bosonic
annihilation operator for a harmonic oscillator at site $r$ in mode
$\epsilon$ with angular frequency $\omega_{\nu;\epsilon}$, coupled
to the $\nu^{\text{th}}$ component of the spin at the same site.
We define the bosonic coherent states $\left|\mathbf{b}_{\nu;\varphi}^{\vphantom{\prime}}\right\rangle $
as the set of normalized kets that are common eigenstates of the annihilation
operators $\hat{b}_{\nu;r;\epsilon}^{\vphantom{\dagger}}$. These
bosonic coherent states can be expressed as \citep{Negele1}

\begin{equation}
\left|\mathbf{b}_{\nu;\varphi}^{\vphantom{\prime}}\right\rangle \equiv\sum_{\epsilon}\sum_{u=-N}^{N}\sum_{r=0}^{L-1}\sum_{n=0}^{\infty}\frac{b_{\nu;r+uL;\epsilon;\varphi}^{n\vphantom{\dagger}}}{n!}\left(\hat{b}_{\nu;r+uL;\epsilon}^{\dagger}\right)^{n}\left|\mathbf{0}_{\nu}^{\vphantom{\prime}}\right\rangle ,\label{eq:defining coherent state in appendix -- 1}
\end{equation}

\noindent where $2N+1$ is the number of unit cells in the system
{[}see Sec.~\ref{subsec:Model} for a brief discussion on unit cells{]},
$L$ is the number of sites in every unit cell, $\varphi$ is a set
of labels {[}e.g. $\varphi=\{\},\varphi=\{j;\alpha;k\}${]} used to
distinguish between different instances of the resolution of identity
{[}see below{]}, $\left|\mathbf{0}_{\nu}^{\vphantom{\prime}}\right\rangle $
is the vacuum state, i.e. a state with no bosons, and the complex
numbers $b_{\nu;r;\epsilon;\varphi}^{\vphantom{\dagger}}$ make up
the components of the complex vector $\mathbf{b}_{\nu;\varphi}^{\vphantom{\prime}}$,
which parameterizes the coherent state $\left|\mathbf{b}_{\nu;\varphi}^{\vphantom{\prime}}\right\rangle $.
Using Eq.~\eqref{eq:defining coherent state in appendix -- 1}, we
can verify that the coherent states $\left|\mathbf{b}_{\nu;\varphi}^{\vphantom{\prime}}\right\rangle $
are right-handed eigenvectors of the annihilation operator $\hat{b}_{\nu;r;\epsilon}^{\vphantom{\dagger}}$:

\begin{equation}
\hat{b}_{\nu;r;\epsilon}^{\vphantom{\dagger}}\left|\mathbf{b}_{\nu;\varphi}^{\vphantom{\prime}}\right\rangle =b_{\nu;r;\epsilon;\varphi}^{\vphantom{\dagger}}\left|\mathbf{b}_{\nu;\varphi}^{\vphantom{\prime}}\right\rangle .\label{eq:annihilation op right-hand eigenvalue equation -- 1}
\end{equation}

\noindent The dual equation to Eq.~\eqref{eq:annihilation op right-hand eigenvalue equation -- 1}
is

\begin{equation}
\left\langle \mathbf{b}_{\nu}^{\vphantom{\prime}}\right|\hat{b}_{\nu;r;\epsilon}^{\dagger}=\left\langle \mathbf{b}_{\nu}^{\vphantom{\prime}}\right|b_{\nu;r;\epsilon;\varphi}^{*\vphantom{\dagger}}.\label{eq:creation op left-hand eigenvalue equation -- 1}
\end{equation}

\noindent Equations~\eqref{eq:annihilation op right-hand eigenvalue equation -- 1}
and \eqref{eq:creation op left-hand eigenvalue equation -- 1} further
imply that

\begin{equation}
f\left(\hat{b}_{\nu;r;\epsilon}^{\vphantom{\dagger}}\right)\left|\mathbf{b}_{\nu;\varphi}^{\vphantom{\prime}}\right\rangle =f\left(b_{\nu;r;\epsilon;\varphi}^{\vphantom{\dagger}}\right)\left|\mathbf{b}_{\nu;\varphi}^{\vphantom{\prime}}\right\rangle ,\label{eq:annihilation op right-hand eigenvalue equation -- 2}
\end{equation}

\noindent 
\begin{equation}
\left\langle \mathbf{b}_{\nu;\varphi}^{\vphantom{\prime}}\right|\hat{b}_{\nu;r;\epsilon}^{\dagger}=\left\langle \mathbf{b}_{\nu;\varphi}^{\vphantom{\prime}}\right|f\left(b_{\nu;r;\epsilon;\varphi}^{*\vphantom{\dagger}}\right),\label{eq:creation op left-hand eigenvalue equation -- 2}
\end{equation}

\noindent where $f\left(\ldots\right)$ denotes an arbitrary function.

Coherent states $\left|\mathbf{b}_{\nu;\varphi}^{\vphantom{\prime}}\right\rangle $
can be used as basis vectors, but they are not orthonormal, as can
be seen by taking the inner product between two arbitrary coherent
states $\left|\mathbf{b}_{\nu;\varphi}^{\vphantom{\prime}}\right\rangle $
and $\left|\mathbf{b}_{\nu;\varphi}^{\prime}\right\rangle $

\begin{equation}
\left\langle \mathbf{b}_{\nu;\varphi^{\vphantom{\varphi}}}^{\vphantom{\prime}}\right|\left.\mathbf{b}_{\nu;\varphi^{\prime}}^{\prime}\right\rangle =e^{\mathbf{b}_{\nu;\varphi^{\vphantom{\varphi}}}^{\vphantom{\prime}}\cdot\mathbf{b}_{\nu;\varphi^{\prime}}^{\prime}}\equiv\prod_{\epsilon}\prod_{u=-N}^{N}\prod_{r=0}^{L-1}\left\{ e^{b_{\nu;r+uL;\epsilon;\varphi^{\vphantom{\prime}}}^{*\vphantom{\prime}}b_{\nu;r+uL;\epsilon;\varphi^{\prime}}^{\vphantom{*}\prime}}\right\} ,\label{eq:coherent states inner product -- 1}
\end{equation}

\noindent The coherent state vectors obey the following completeness
relation

\begin{equation}
\int d^{2}\mathbf{b}_{\nu;\varphi}^{\vphantom{\prime}}\,e^{-\mathbf{b}_{\nu;\varphi}^{\vphantom{\prime}}\cdot\mathbf{b}_{\nu;\varphi}^{\vphantom{\prime}}}\left|\mathbf{b}_{\nu;\varphi}^{\vphantom{\prime}}\right\rangle \left\langle \mathbf{b}_{\nu;\varphi}^{\vphantom{\prime}}\right|=\prod_{\epsilon}\prod_{u=-N}^{N}\prod_{r=0}^{L-1}\left\{ \hat{1}_{\nu;r+uL;\epsilon}^{\left(\mathcal{B}\right)}\right\} ,\label{eq:coherent state completeness relation -- 1}
\end{equation}

\noindent where

\begin{align}
d^{2}\mathbf{b}_{\nu;\varphi}^{\vphantom{\prime}} & \equiv\prod_{\epsilon}\prod_{u=-N}^{N}\prod_{r=0}^{L-1}\left\{ d^{2}b_{\nu;r+uL;\epsilon;\varphi}^{\vphantom{\prime}}\right\} \nonumber \\
 & \equiv\prod_{\epsilon}\prod_{u=-N}^{N}\prod_{r=0}^{L-1}\left\{ \frac{d\left(\text{Re}\left[b_{\nu;r+uL;\epsilon;\varphi}^{\vphantom{*}}\right]\right)d\left(\text{Im}\left[b_{\nu;r+uL;\epsilon;\varphi}^{\vphantom{*}}\right]\right)}{\pi}\right\} ,\label{eq:coherent state differentials -- 1}
\end{align}

\noindent and $\hat{1}_{\nu;r;\epsilon}^{\left(\mathcal{B}\right)}$
is the identity operator of the Hilbert space associated with the
harmonic oscillator at site $r$ in mode $\epsilon$ with angular
frequency $\omega_{\nu;\epsilon}$, coupled to the $\nu^{\text{th}}$
component of the spin at the same site. The integration is over the
entire complex plane $\mathbb{C}$. The trace of an operator $\hat{A}$
in the coherent state basis is given by

\begin{equation}
\text{Tr}^{\left(B\right)}\left\{ \hat{O}\right\} =\int d^{2}\mathbf{b}_{\nu;\varphi}^{\vphantom{\prime}}\,e^{-\mathbf{b}_{\nu;\varphi}^{\vphantom{\prime}}\cdot\mathbf{b}_{\nu;\varphi}^{\vphantom{\prime}}}\left\langle \mathbf{b}_{\nu;\varphi}^{\vphantom{\prime}}\right|\hat{O}\left|\mathbf{b}_{\nu;\varphi}^{\vphantom{\prime}}\right\rangle .\label{eq:coherent state trace -- 1}
\end{equation}

\newpage{}

\section{Deriving the path integral form of the system's reduced density matrix\label{sec:Path integral representation derivation -- 1}}

In this appendix, we present a derivation of one particular coherent
state path integral representation of the device's reduced state operator
$\hat{\rho}^{\left(A\right)}\left(t\right)$, drawing from references
Refs.~\citep{Strathearn1,Suzuki1}, but in certain respects we are
extending the work therein. Recall from Sec.~\ref{sec:QUAPI formalism}
that $\hat{\rho}^{\left(A\right)}\left(t\right)$ can be expressed
as

\begin{equation}
\hat{\rho}^{\left(A\right)}\left(t\right)=\text{Tr}^{\left(B\right)}\left\{ \hat{\rho}\left(t\right)\right\} =\text{Tr}^{\left(B\right)}\left\{ \hat{U}\left(t,0\right)\hat{\rho}^{\left(i\right)}\hat{U}\left(0,t\right)\right\} ,\label{eq:rho^(A) in path integral appendix -- 1}
\end{equation}

\noindent where $\hat{H}\left(t\right)$ is given by Eqs.~\eqref{eq:total Hamiltonian -- 1}-\eqref{eq:Q_=00007Bv; r=00007D -- 1},$\hat{\rho}^{\left(i\right)}$
is given by Eqs.~\eqref{eq:rho^(i) -- 1}-\eqref{eq:rho^(i, B) -- 1},
\eqref{eq:rho_u^(i, B) -- 1},\eqref{eq:rho_=00007Bu; v=00007D^(i, B) -- 1},
and \eqref{eq:Z_=00007Bu; v=00007D^(B) -- 2}, and $\hat{U}\left(t,0\right)$
is given by Eq.~\eqref{eq:U -- 1}.

Shortly down below, the time integrals in Eq.~\eqref{eq:U -- 1}
will be evaluated using the trapezoidal quadrature scheme on a discretized
uniform time grid $t_{k}$ defined by

\begin{equation}
t_{k}=k\Delta t,\label{eq:discrete time grid -- 1}
\end{equation}

\noindent where $\Delta t$ is the time step size, and $k$ is an
integer. The quadrature weights $\left\{ \tilde{w}_{l}\right\} _{l=0}^{1}$
over a single quadrature interval are

\begin{equation}
\tilde{w}_{l=0,1}=\frac{1}{2}.\label{eq:weights over single interval -- 1}
\end{equation}

\noindent The error of the numerical integration over a single interval
is third order in $\Delta t$.

Recall that $\hat{U}\left(t,t^{\prime}\right)$, being a time evolution
operator, satisfies

\begin{equation}
\hat{U}\left(t,t^{\prime}\right)=\hat{U}\left(t,t^{\prime\prime}\right)\hat{U}\left(t^{\prime\prime},t^{\prime}\right).\label{eq:time evolution operator property -- 1}
\end{equation}

\noindent This allows us to rewrite, for instance, $\hat{U}\left(t,0\right)$
as

\begin{equation}
\hat{U}\left(t_{n},0\right)=\prod_{k=0}^{n-1}\left[\hat{U}\left(t_{n-k},t_{n-k-1}\right)\right],\label{eq:rewritting U(t_n, 0) -- 1}
\end{equation}

\noindent where we have introduced the notation

\begin{equation}
\prod_{l=l_{i}}^{l_{f}}\left\{ x_{l_{\vphantom{i}}}\right\} =x_{l_{i}}x_{l_{i}+1}\cdots x_{l_{f}-1}x_{l_{f}}.\label{eq:capital Pi notation -- 2}
\end{equation}

\noindent Next, let us approximate $\hat{U}\left(t_{n-k},t_{n-k-1}\right)$
using the aforementioned quadrature scheme:

\begin{align}
\hat{U}\left(t_{n-k},t_{n-k-1}\right) & =T\left\{ e^{-i\int_{t_{n-k-1}}^{t_{n-k}}dt^{\prime\prime}\hat{H}\left(t^{\prime\prime}\right)}\right\} \nonumber \\
 & =\sum_{M=0}^{\infty}\frac{\left(-i\right)^{M}}{M!}\prod_{m=1}^{M}\left\{ \int_{t_{n-k-1}}^{t_{n-k}}dt^{\left(m\right)}\right\} T\left\{ \prod_{m=1}^{M}\left[\hat{H}\left(t^{\left(m\right)}\right)\right]\right\} \nonumber \\
 & =\sum_{M=0}^{\infty}\frac{\left(-i\right)^{M}}{M!}\prod_{m=1}^{M}\left\{ \sum_{l_{m}=0}^{1}\left[\Delta t\tilde{w}_{l_{m}}\right]\right\} T\left\{ \prod_{m=1}^{M}\left[\hat{H}\left(t_{n-k-l_{m}}\right)\right]\right\} +\mathcal{O}\left[\Delta t^{3}\right].\label{eq:approximating U over single step -- 1}
\end{align}

\noindent We note that we can rewrite the following term in Eq.~\eqref{eq:approximating U over single step -- 1}
as

\begin{align}
 & \prod_{m=1}^{M}\left\{ \sum_{l_{m}=0}^{1}\left[\Delta t\tilde{w}_{l_{m}}\right]\right\} T\left\{ \prod_{m=1}^{M}\left[\hat{H}\left(t_{n-k-l_{m}}\right)\right]\right\} \nonumber \\
 & \quad=\prod_{l=0}^{1}\left\{ \sum_{M_{n-k-l}=0}^{M}\right\} \delta_{M_{n-k}+M_{n-k-1},M}\left[\frac{M!}{\left(M_{n-k}\right)!\left(M_{n-k-1}\right)!}\right]\prod_{l=0}^{1}\left\{ \left[\Delta t\tilde{w}_{l}\hat{H}\left(t_{n-k-l}\right)\right]^{M_{n-k-l}}\right\} .\label{eq:approximating U over single step -- 2}
\end{align}

\noindent where $\delta_{i,j}$ is the kronecker delta function. Substituting
Eq.~\eqref{eq:approximating U over single step -- 2} into \eqref{eq:approximating U over single step -- 1}
yields

\begin{align}
\hat{U}\left(t_{n-k},t_{n-k-1}\right) & =\sum_{M=0}^{\infty}\frac{\left\{ -i\Delta t\right\} ^{M}}{M!}\prod_{m=1}^{M}\left\{ \sum_{l_{m}=0}^{1}\left[\tilde{w}_{l_{m}}\right]\right\} T\left\{ \prod_{m=1}^{M}\left[\hat{H}\left(t_{n-k-l}\right)\right]\right\} +\mathcal{O}\left[\Delta t^{3}\right]\nonumber \\
 & =\sum_{M=0}^{\infty}\frac{\left\{ -i\Delta t\right\} ^{M}}{\cancel{M!}}\prod_{l=0}^{1}\left\{ \sum_{M_{n-k-l}=0}^{M}\right\} \delta_{M_{n-k}+M_{n-k-1},M}\left[\frac{M!}{\left(M_{n-k}\right)!\left(M_{n-k-1}\right)!}\right]\nonumber \\
 & \phantom{\mathrel{=}\sum_{M=0}^{\infty}\frac{\left\{ -i\Delta t\right\} ^{M}}{\cancel{M!}}\prod_{l=0}^{1}\left\{ \sum_{M_{n-k-l}=0}^{M}\right\} }\mathord{\times}\prod_{l=0}^{1}\left\{ \left[\tilde{w}_{l}\hat{H}\left(t_{n-k-l}\right)\right]^{M_{n-k-l}}\right\} +\mathcal{O}\left[\Delta t^{3}\right]\nonumber \\
 & =\sum_{M=0}^{\infty}\prod_{l=0}^{1}\left\{ \sum_{M_{n-k-l}=0}^{M}\right\} \delta_{M_{n-k}+M_{n-k-1},M}\prod_{l=0}^{1}\left\{ \frac{\left[-i\Delta t\tilde{w}_{l}\hat{H}\left(t_{n-k-l}\right)\right]^{M_{n-k-l}}}{\left(M_{n-k-l}\right)!}\right\} \nonumber \\
 & \mathrel{\phantom{=}}\mathord{+}\mathcal{O}\left[\Delta t^{3}\right]\nonumber \\
 & =\prod_{l=0}^{1}\left\{ \sum_{M_{n-k-l}=0}^{M}\frac{\left[-i\Delta t\tilde{w}_{l}\hat{H}\left(t_{n-k-l}\right)\right]^{M_{n-k-l}}}{\left(M_{n-k-l}\right)!}\right\} +\mathcal{O}\left[\Delta t^{3}\right]\nonumber \\
 & =\prod_{l=0}^{1}\left\{ e^{-i\Delta t\tilde{w}_{l}\hat{H}\left(t_{n-k-l}\right)}\right\} +\mathcal{O}\left[\Delta t^{3}\right].\label{eq:approximating U over single step -- 3}
\end{align}

\noindent Using Eq.~\eqref{eq:approximating U over single step -- 3},
we can write

\begin{align}
\hat{U}\left(t_{n},0\right) & =\prod_{k=0}^{n-1}\left[\hat{U}_{I}\left(t_{n-k},t_{n-k-1}\right)\right]\nonumber \\
 & =\prod_{k=0}^{n-1}\left\{ \prod_{l=0}^{1}\left[e^{-i\Delta t\tilde{w}_{l}\hat{H}\left(t_{n-k-l}\right)}\right]+\mathcal{O}\left[\Delta t^{3}\right]\right\} \nonumber \\
 & =\prod_{k=0}^{n-1}\left\{ \prod_{l=0}^{1}\left[e^{-i\Delta t\tilde{w}_{l}\hat{H}\left(t_{n-k-l}\right)}\right]\right\} +\mathcal{O}\left[\Delta t^{2}\right]\nonumber \\
 & =\prod_{k=0}^{n}\left\{ e^{-i\Delta tw_{n;k}\hat{H}\left(t_{n-k}\right)}\right\} +\mathcal{O}\left[\Delta t^{2}\right],\label{eq:approximating U over =00005Bt, 0=00005D -- 1}
\end{align}

\noindent where $w_{n;k}$ are the weights for the composite quadrature
scheme:

\begin{equation}
w_{n;k}=\begin{cases}
0, & \text{if }k=-1,n+1,\\
\frac{1}{2}, & \text{if }k=0,n,\\
1, & \text{if }0<k<n.
\end{cases}\label{eq:composite quadrature rule -- 1}
\end{equation}

\noindent with the $k=-1,n+1$ cases added for later convenience.
Similarly, we have

\begin{equation}
\hat{U}\left(0,t_{n}\right)=\prod_{k=0}^{n}\left\{ e^{i\Delta tw_{n;k}\hat{H}\left(t_{k}\right)}\right\} +\mathcal{O}\left[\Delta t^{2}\right].\label{eq:approximating U over =00005B0, t=00005D -- 1}
\end{equation}

Next, we rewrite the total Hamiltonian $\hat{H}\left(t\right)$ as

\begin{equation}
\hat{H}\left(t\right)=\sum_{\nu\in\left\{ x,z\right\} }\hat{H}_{\nu}^{\left(A\right)}\left(t\right)+\sum_{\nu\in\left\{ y,z\right\} }\hat{H}_{\nu}^{\left(AB+B\right)}\left(t\right),\label{eq:total Hamiltonian in path integral appendix -- 1}
\end{equation}

\noindent where

\begin{equation}
\hat{H}_{\nu\in\left\{ x,z\right\} }^{\left(A\right)}\left(t\right)=\sum_{u=-N}^{N}\hat{H}_{u;\nu}^{\left(A\right)}\left(t\right),\label{eq:H_v^(A) in path integral appendix -- 1}
\end{equation}

\begin{equation}
\hat{H}_{\nu\in\left\{ y,z\right\} }^{\left(AB\right)}\left(t\right)=\sum_{u=-N}^{N}\hat{H}_{u;\nu}^{\left(AB\right)}\left(t\right),\label{eq:H_v^(AB) in path integral appendix -- 1}
\end{equation}

\begin{equation}
\hat{H}_{\nu\in\left\{ y,z\right\} }^{\left(B\right)}=\sum_{u=-N}^{N}\hat{H}_{u;\nu}^{\left(B\right)},\label{eq:H_v^(B) in path integral appendix -- 1}
\end{equation}

\begin{equation}
\hat{H}_{\nu\in\left\{ y,z\right\} }^{\left(AB+B\right)}\left(t\right)=\hat{H}_{\nu}^{\left(AB\right)}\left(t\right)+\hat{H}_{\nu}^{\left(B\right)},\label{eq:H_v^(AB+B) in path integral appendix -- 1}
\end{equation}

\noindent and $\hat{H}_{u;x}^{\left(A\right)}\left(t\right)$, $\hat{H}_{u;z}^{\left(A\right)}\left(t\right)$,
$\hat{H}_{u;\nu\in\left\{ y,z\right\} }^{\left(B\right)}$, and $\hat{H}_{u;\nu\in\left\{ y,z\right\} }^{\left(AB\right)}$
are defined by Eqs.~\eqref{eq:H_=00007Bu; x=00007D^(A) -- 1}, \eqref{eq:H_=00007Bu; z=00007D^(A) -- 1},
\eqref{eq:H_=00007Bu; v=00007D^(B) -- 1}, and \eqref{eq:H_=00007Bu; v=00007D^(AB) -- 1}.
Using

\begin{equation}
e^{\Delta t\left\{ \hat{X}+\hat{Y}\right\} }=e^{\frac{\Delta t}{2}\hat{X}}e^{\Delta t\hat{Y}}e^{\frac{\Delta t}{2}\hat{X}}+\mathcal{O}\left[\Delta t^{3}\right],\label{eq:symmetric Suzuki-Trotter decomposition -- 1}
\end{equation}

\noindent Eq.~\eqref{eq:H_v^(AB+B) in path integral appendix -- 1}
we rewrite $e^{i\Delta tw_{n;k}\hat{H}\left(t_{k}\right)}$ as

\begin{align}
e^{i\Delta tw_{n;k}\hat{H}\left(t_{k}\right)} & =e^{i\Delta tw_{n;k}\left\{ \left[\hat{H}_{z}^{\left(A\right)}\left(t_{k}\right)+\hat{H}_{z}^{\left(AB+B\right)}\left(t_{k}\right)\right]+\left[\hat{H}_{x}^{\left(A\right)}\left(t_{k}\right)+\hat{H}_{y}^{\left(AB+B\right)}\left(t_{k}\right)\right]\right\} }\nonumber \\
 & =e^{\frac{i}{2}\Delta tw_{n;k}\left\{ \hat{H}_{z}^{\left(A\right)}\left(t_{k}\right)+\hat{H}_{z}^{\left(AB+B\right)}\left(t_{k}\right)\right\} }\nonumber \\
 & \mathrel{\phantom{=}}\mathop{\times}e^{i\Delta tw_{n;k}\left\{ \hat{H}_{x}^{\left(A\right)}\left(t_{k}\right)+\hat{H}_{y}^{\left(AB+B\right)}\left(t_{k}\right)\right\} }\nonumber \\
 & \mathrel{\phantom{=}}\mathop{\times}e^{\frac{i}{2}\Delta tw_{n;k}\left\{ \hat{H}_{z}^{\left(A\right)}\left(t_{k}\right)+\hat{H}_{z}^{\left(AB+B\right)}\left(t_{k}\right)\right\} }+\mathcal{O}\left[\Delta t^{3}\right]\nonumber \\
 & =e^{\frac{i}{2}\Delta tw_{n;k}\hat{H}_{z}^{\left(A\right)}\left(t_{k}\right)}e^{\frac{i}{2}\Delta tw_{n;k}\hat{H}_{z}^{\left(AB+B\right)}\left(t_{k}\right)}\nonumber \\
 & \mathrel{\phantom{=}}\mathop{\times}e^{\frac{i}{2}\Delta tw_{n;k}\hat{H}_{y}^{\left(AB+B\right)}\left(t_{k}\right)}e^{i\Delta tw_{n;k}\hat{H}_{x}^{\left(A\right)}\left(t_{k}\right)}e^{\frac{i}{2}\Delta tw_{n;k}\hat{H}_{y}^{\left(AB+B\right)}\left(t_{k}\right)}\nonumber \\
 & \mathrel{\phantom{=}}\mathop{\times}e^{\frac{i}{2}\Delta tw_{n;k}\hat{H}_{z}^{\left(A\right)}\left(t_{k}\right)}e^{\frac{i}{2}\Delta tw_{n;k}\hat{H}_{z}^{\left(AB+B\right)}\left(t_{k}\right)}+\mathcal{O}\left[\Delta t^{3}\right],\label{eq:rewritting e^=00007Bi dt w_k H=00007D -- 1}
\end{align}

Next, we substitute Eq.~\eqref{eq:rewritting e^=00007Bi dt w_k H=00007D -- 1}
into Eq.~\eqref{eq:approximating U over =00005B0, t=00005D -- 1},
and insert various resolutions of the identity:

\begin{align}
\hat{U}\left(0,t_{n}\right) & =\prod_{k=0}^{n}\left\{ e^{i\Delta tw_{n;k}\hat{H}\left(t_{k}\right)}\right\} \nonumber \\
 & =\prod_{k=0}^{n}\left\{ \hat{1}_{z;-1;2k}^{\left(\sigma\right)}e^{\frac{i}{2}\Delta tw_{n;k}\hat{H}_{z}^{\left(A\right)}\left(t_{k}\right)}e^{\frac{i}{2}\Delta tw_{n;k}\hat{H}_{z}^{\left(AB+B\right)}\left(t_{k}\right)}\right.\nonumber \\
 & \phantom{=\prod_{k=1}^{n}}\left.\quad\mathord{\times}e^{\frac{i}{2}\Delta tw_{n;k}\hat{H}_{y}^{\left(AB+B\right)}\left(t_{k}\right)}\hat{1}_{y;-1;2k}^{\left(\sigma\right)}e^{i\Delta tw_{n;k}\hat{H}_{x}^{\left(A\right)}\left(t_{k}\right)}\hat{1}_{y;-1;2k+1}^{\left(\sigma\right)}e^{\frac{i}{2}\Delta tw_{n;k}\hat{H}_{y}^{\left(AB+B\right)}\left(t_{k}\right)}\right.\nonumber \\
 & \phantom{=\prod_{k=1}^{n}}\left.\quad\mathord{\times}\hat{1}_{z;-1;2k+1}^{\left(\sigma\right)}e^{\frac{i}{2}\Delta tw_{n;k}\hat{H}_{z}^{\left(A\right)}\left(t_{k}\right)}e^{\frac{i}{2}\Delta tw_{n;k}\hat{H}_{z}^{\left(AB+B\right)}\left(t_{k}\right)}\right\} +\mathcal{O}\left[\Delta t^{2}\right],\label{eq:approximating U over =00005B0, t=00005D -- 2}
\end{align}

\noindent where

\begin{equation}
\hat{1}_{\nu;\alpha;k}^{\left(\sigma\right)}=\prod_{u=-N}^{N}\prod_{r=0}^{L-1}\left\{ \sum_{\sigma_{\nu;r+uL;\alpha;k}=\pm1}\right\} \left|\boldsymbol{\sigma}_{\nu;\alpha;k}\nu\right\rangle \left\langle \boldsymbol{\sigma}_{\nu;\alpha;k}\nu\right|;\label{eq:1_=00007Bv; alpha; k=00007D^(sigma) -- 1}
\end{equation}

\begin{equation}
\hat{\sigma}_{\nu;r}\left|\boldsymbol{\sigma}_{\nu;\alpha;k}\nu\right\rangle =\sigma_{\nu;r;\alpha;k}\left|\boldsymbol{\sigma}_{\nu;\alpha;k}\nu\right\rangle ,\label{eq:sigma_=00007Bv; r=00007D eigenvalue equation -- 1}
\end{equation}

\noindent with

\begin{equation}
\boldsymbol{\sigma}_{\nu;\alpha;k}=\left(\sigma_{\nu;r=-NL;\alpha;k},\ldots,\sigma_{\nu;r=\left(N+1\right)L-1;\alpha;k}\right).\label{eq:sigma vector -- 1}
\end{equation}

\noindent Expanding the resolutions of the identity yields:

\begin{align}
\hat{U}\left(0,t_{n}\right) & =\prod_{u=-N}^{N}\prod_{r=0}^{L-1}\left[\prod_{q=0}^{2n+1}\left\{ \sum_{\sigma_{y;r+uL;-1;q}=\pm1}\right\} \prod_{q=0}^{2n+1}\left\{ \sum_{\sigma_{z;r+uL;-1;q}=\pm1}\right\} \right]\nonumber \\
 & \mathrel{\phantom{=}}\mathop{\times}\prod_{u=-N}^{N}\prod_{r=0}^{L-1}\prod_{l=1}^{n}\left\{ \delta_{\sigma_{z;r+uL;-1;2l-1},\sigma_{z;r+uL;-1;2l}}\right\} e^{i\tilde{\phi}_{-1;n}^{\left(\text{lcafc}\right)}\left(\boldsymbol{\sigma}_{z;-1;q\in\left[0,2n+1\right]}\right)}\tilde{I}_{-1;n}^{\left(\text{tfc}\right)}\left(\boldsymbol{\sigma}_{y;-1;q\in\left[0,2n+1\right]}\right)\nonumber \\
 & \mathrel{\phantom{=}}\mathop{\times}\tilde{I}_{-1;n}^{\left(y\leftrightarrow z\right)}\left(\boldsymbol{\sigma}_{\nu\in\left\{ y,z\right\} ;-1;q\in\left[0,2n+1\right]}\right)e^{i\Delta t\sum_{\nu\in\left\{ y,z\right\} }\sum_{q=0}^{2n+1}\tilde{w}_{n;q}\hat{H}_{\nu;k=\left\lfloor q/2\right\rfloor }^{\left(AB+B\right)}\left(\boldsymbol{\sigma}_{\nu;-1;q}\right)}\nonumber \\
 & \mathrel{\phantom{=}}\mathop{\times}\left|\boldsymbol{\sigma}_{z;-1;0}z\right\rangle \left\langle \boldsymbol{\sigma}_{z;-1;n+1}z\right|+\mathcal{O}\left[\Delta t^{2}\right],\label{eq:approximating U over =00005B0, t=00005D -- 3}
\end{align}

\noindent where

\begin{equation}
\tilde{w}_{n;q}=\begin{cases}
\frac{1}{4}, & \text{if }q=0,1,2n,2n+1,\\
\frac{1}{2}, & \text{if }2\le q\le2n-1;
\end{cases}\label{eq:tilde w_=00007Bn; q=00007D -- 1}
\end{equation}

\begin{equation}
\tilde{\phi}_{\alpha;n}^{\left(\text{lcafc}\right)}\left(\boldsymbol{\sigma}_{z;\alpha;q\in\left[0,2n+1\right]}\right)=\sum_{q=0}^{2n+1}\tilde{\phi}_{\alpha;n;q}^{\left(\text{lcafc}\right)}\left(\boldsymbol{\sigma}_{z;\alpha;q}\right),\label{eq:tilde phi_=00007Balpha; n=00007D^(lcafc) -- 1}
\end{equation}

\noindent with

\begin{equation}
\tilde{\phi}_{\alpha;n;q}^{\left(\text{lcafc}\right)}\left(\boldsymbol{\sigma}_{z;\alpha;q}\right)=-\frac{\alpha\Delta t}{2}\tilde{w}_{n;q}H_{z}^{\left(A\right)}\left(t=\left\lfloor q/2\right\rfloor \Delta t;\boldsymbol{\sigma}_{z;\alpha;q}\right),\label{eq:tilde phi_=00007Balpha; n; q=00007D^(lcafc) -- 1}
\end{equation}

\begin{align}
H_{z}^{\left(A\right)}\left(t;\boldsymbol{\sigma}_{z;\alpha;q}\right) & \equiv\left\langle \boldsymbol{\sigma}_{z;\alpha;q}z\right|\hat{H}_{z}^{\left(A\right)}\left(t\right)\left|\boldsymbol{\sigma}_{z;\alpha;q}z\right\rangle \nonumber \\
 & =\sum_{u=-N}^{N}\sum_{r=0}^{L-1}h_{z;r}\left(t\right)\sigma_{z;r+uL;\alpha;q}\mathop{+}\sum_{u=-N}^{N}\sum_{r=0}^{L-1}J_{z,z;r,r+1}\left(t\right)\sigma_{z;r+uL;\alpha;q}\sigma_{z;r+uL+1;\alpha;q};\label{eq:H_z^(A)(t; sigma) -- 1}
\end{align}

\begin{equation}
\tilde{I}_{\alpha;n}^{\left(\text{tfc}\right)}\left(\boldsymbol{\sigma}_{y;\alpha;q\in\left[0,2n+1\right]}\right)=\prod_{l=0}^{n}\prod_{u=-N}^{N}\prod_{r=0}^{L-1}\left\{ \tilde{I}_{\alpha;r+uL;n;k=l}^{\left(\text{tfc}\right)}\left(\sigma_{y;r+uL;\alpha;2l},\sigma_{y;r+uL;\alpha;2l+1}\right)\right\} ,\label{eq:tilde I_=00007Balpha; n=00007D^(tfc) -- 1}
\end{equation}

\noindent with

\begin{equation}
\tilde{I}_{\alpha;r;n;k}^{\left(\text{tfc}\right)}\left(\sigma_{y;r;\alpha;2l},\sigma_{y;r;\alpha;2l+1}\right)=\begin{cases}
\left\langle \sigma_{y;r;-1;2l}y\right|e^{i\Delta tw_{n;k}h_{x;r}\left(t_{k}\right)\hat{\sigma}_{x;r}}\left|\sigma_{y;r;-1;2l+1}y\right\rangle , & \text{if }\alpha=-1,\\
\left\langle \sigma_{y;r;1;2l+1}y\right|e^{-i\Delta tw_{n;k}h_{x;r}\left(t_{k}\right)\hat{\sigma}_{x;r}}\left|\sigma_{y;r;1;2l}y\right\rangle , & \text{if }\alpha=1;
\end{cases}\label{eq:tilde I_=00007Balpha; r; n; k=00007D^(tfc) -- 1}
\end{equation}

\begin{equation}
h_{x;r+uL}\left(t_{k}\right)=h_{x;r}\left(t_{k}\right),\label{eq:periodicity of transverse field -- 1}
\end{equation}

\begin{align}
\hat{H}_{\nu;k=\left\lfloor q/2\right\rfloor }^{\left(AB+B\right)}\left(\boldsymbol{\sigma}_{\nu;\alpha,q}\right) & \equiv\left\langle \boldsymbol{\sigma}_{\nu;\alpha;q}\nu\right|\hat{H}_{\nu}^{\left(AB+B\right)}\left(t_{k}\right)\left|\boldsymbol{\sigma}_{\nu;\alpha;q}\nu\right\rangle \nonumber \\
 & =\sum_{u=-N}^{N}\sum_{r=0}^{L-1}\sum_{\epsilon}\left[\mathcal{E}_{v;r}^{\left(\lambda\right)}\left(t_{k}\right)\lambda_{\nu;r;\epsilon}\sigma_{\nu;r+uL;\alpha;k}\left\{ \hat{b}_{\nu;r+uL;\epsilon}^{\dagger}+\hat{b}_{\nu;r+uL;\epsilon}^{\vphantom{\dagger}}\right\} \right]\nonumber \\
 & \phantom{=\sum_{u=-N}^{N}\sum_{r=0}^{L-1}\sum_{\epsilon}}\left.\quad\mathop{+}\omega_{\nu;\epsilon}\hat{b}_{\nu;r+uL;\epsilon}^{\dagger}\hat{b}_{\nu;r+uL;\epsilon}^{\vphantom{\dagger}}\right],\label{eq:H_=00007Bv; k=00007D^(AB+B)(sigma) -- 1}
\end{align}

\noindent with

\begin{equation}
\mathcal{E}_{v;r;q}^{\left(\lambda\right)}=\mathcal{E}_{v;r}^{\left(\lambda\right)}\left(t=\left\lfloor q/2\right\rfloor \Delta t\right);\label{eq:mathcal E_=00007Bv; r; q=00007D^(Lambda) -- 1}
\end{equation}

\begin{equation}
\tilde{I}_{\alpha;n}^{\left(y\leftrightarrow z\right)}\left(\boldsymbol{\sigma}_{\nu\in\left\{ y,z\right\} ;\alpha;q\in\left[0,2n+1\right]}\right)=\prod_{u=-N}^{N}\prod_{r=0}^{L-1}\left\{ \tilde{I}_{\alpha;n}^{\left(y\leftrightarrow z\right)}\left(\sigma_{\nu\in\left\{ y,z\right\} ;r+uL;\alpha;q\in\left[0,2n+1\right]}\right)\right\} ,\label{eq:tilde I_=00007Balpha; n=00007D^(y<->z) -- 1}
\end{equation}

\noindent with

\begin{align}
 & \tilde{I}_{\alpha;n}^{\left(y\leftrightarrow z\right)}\left(\sigma_{\nu\in\left\{ y,z\right\} ;r;\alpha;q\in\left[0,2n+1\right]}\right)\nonumber \\
 & \quad=\begin{cases}
\prod_{l=0}^{n}\left\{ I^{\left(y\to z\right)}\left(\sigma_{z;r;-1;2l};\sigma_{y;r;-1;2l}\right)I^{\left(z\to y\right)}\left(\sigma_{y;r;-1;2l+1};\sigma_{z;r;-1;2l+1}\right)\right\} , & \text{if }\alpha=-1,\\
\prod_{k=0}^{n}\left\{ I^{\left(y\to z\right)}\left(\sigma_{z;r;1;2l+1};\sigma_{y;r;1;2l+1}\right)I^{\left(z\to y\right)}\left(\sigma_{y;r;1;2l};\sigma_{z;r;1;2l}\right)\right\} , & \text{if }\alpha=1,
\end{cases}\label{eq:tilde I_=00007Balpha; n=00007D^(y<->z) -- 2}
\end{align}

\noindent and

\begin{align}
I^{\left(z\to y\right)}\left(\sigma_{y;r;\alpha;q^{\vphantom{\prime}}},\sigma_{z;r;\alpha;q^{\prime}}\right) & \equiv\left.\left\langle \sigma_{y;r;\alpha;q^{\vphantom{\prime}}}y\right|\sigma_{z;r;\alpha;q^{\prime}}z\right\rangle \nonumber \\
 & =\frac{1}{\sqrt{2}}\left\{ \frac{\left(1+\sigma_{z;r;\alpha;q^{\prime}}\right)}{2}-i\sigma_{y;r;\alpha;q^{\vphantom{\prime}}}\frac{\left(1-\sigma_{z;r;\alpha;q^{\prime}}\right)}{2}\right\} \nonumber \\
 & =\frac{1}{2\sqrt{2}}\left\{ 1-i\sigma_{y;r;\alpha;q^{\vphantom{\prime}}}+\sigma_{z;r;\alpha;q^{\prime}}+i\sigma_{y;r;\alpha;q^{\vphantom{\prime}}}\sigma_{z;r;\alpha;q^{\prime}}\right\} ,\label{eq:I^(z->y) -- 1}\\
I^{\left(y\to z\right)}\left(\sigma_{z;r;\alpha;q^{\prime}},\sigma_{y;r;\alpha;q^{\vphantom{\prime}}}\right) & =\left\{ I^{\left(z\to y\right)}\left(\sigma_{y;r;\alpha;q^{\vphantom{\prime}}},\sigma_{z;r;\alpha;q^{\prime}}\right)\right\} ^{*}.\label{eq:I^(y->z) -- 1}
\end{align}

\noindent Note that the labels ``lcafc'' and ``tfc'' that appear
in Eq.~\eqref{eq:approximating U over =00005B0, t=00005D -- 3} are
initialisms for ``longitudinal coupler and field components'' and
``transverse field components'' respectively. 

Next, we rewrite the term $\hat{H}_{\nu;k}^{\left(AB+B\right)}\left(\boldsymbol{\sigma}_{\nu;\alpha,q}\right)$
in Eq.~\eqref{eq:approximating U over =00005B0, t=00005D -- 3} as:

\begin{align}
 & \hat{H}_{\nu;k=\left\lfloor q/2\right\rfloor }^{\left(AB+B\right)}\left(\boldsymbol{\sigma}_{\nu;\alpha,q}\right)\nonumber \\
 & \quad=\sum_{u=-N}^{N}\sum_{r=0}^{L-1}\sum_{\epsilon}\left[\mathcal{E}_{v;r;q}^{\left(\lambda\right)}\lambda_{\nu;r;\epsilon}\sigma_{\nu;r+uL;\alpha;q}\left\{ \hat{b}_{\nu;r+uL;\epsilon}^{\dagger}+\hat{b}_{\nu;r+uL;\epsilon}^{\vphantom{\dagger}}\right\} \right]\nonumber \\
 & \quad\phantom{=\sum_{u=-N}^{N}\sum_{r=0}^{L-1}\sum_{\epsilon}}\left.\mathop{+}\omega_{\nu;\epsilon}\hat{b}_{\nu;r+uL;\epsilon}^{\dagger}\hat{b}_{\nu;r+uL;\epsilon}^{\vphantom{\dagger}}\right],\nonumber \\
 & \quad=\sum_{u=-N}^{N}\sum_{r=0}^{L-1}\sum_{\epsilon}\left[\omega_{\nu;\epsilon}\left\{ \hat{b}_{\nu;r+uL;\epsilon}^{\dagger}+\mathcal{E}_{v;r;q}^{\left(\lambda\right)}\frac{\lambda_{\nu;r;\epsilon}}{\omega_{\nu;\epsilon}}\sigma_{\nu;r+uL;\alpha;q}\right\} \vphantom{-\frac{\left\{ \mathcal{E}_{v;r;q}^{\left(\lambda\right)}\lambda_{\nu;r;\epsilon}\right\} ^{2}}{\omega_{\nu;\epsilon}}}\right]\nonumber \\
 & \quad\phantom{=\sum_{u=-N}^{N}\sum_{r=0}^{L-1}\sum_{\epsilon}}\left.\quad\mathop{\times}\left\{ \hat{b}_{\nu;r+uL;\epsilon}^{\vphantom{\dagger}}+\mathcal{E}_{v;r;q}^{\left(\lambda\right)}\frac{\lambda_{\nu;r;\epsilon}}{\omega_{\nu;\epsilon}}\sigma_{\nu;r+uL;\alpha;q}\right\} -\frac{\left\{ \mathcal{E}_{v;r;q}^{\left(\lambda\right)}\lambda_{\nu;r;\epsilon}\right\} ^{2}}{\omega_{\nu;\epsilon}}\right]\nonumber \\
 & \quad=\prod_{u=-N}^{N}\prod_{r=0}^{L-1}\prod_{\epsilon}\left\{ \hat{D}_{\nu;r+uL;\epsilon}^{\dagger}\left(\mathcal{E}_{v;r;q}^{\left(\lambda\right)}\frac{\lambda_{\nu;r;\epsilon}}{\omega_{\nu;\epsilon}}\sigma_{\nu;r+uL;\alpha;q}\right)\right\} \hat{H}_{\nu}^{\left(B\right)}\nonumber \\
 & \quad\mathrel{\phantom{=}}\mathop{\times}\prod_{u=-N}^{N}\prod_{r=0}^{L-1}\prod_{\epsilon}\left\{ \hat{D}_{\nu;r+uL;\epsilon}^{\vphantom{\dagger}}\left(\mathcal{E}_{v;r;q}^{\left(\lambda\right)}\frac{\lambda_{\nu;r;\epsilon}}{\omega_{\nu;\epsilon}}\sigma_{\nu;r+uL;\alpha;q}\right)\right\} \nonumber \\
 & \quad\mathrel{\phantom{=}}\mathop{-}\sum_{u=-N}^{N}\sum_{r=0}^{L-1}\sum_{\epsilon}\frac{\left\{ \mathcal{E}_{v;r;q}^{\left(\lambda\right)}\lambda_{\nu;r;\epsilon}\right\} ^{2}}{\omega_{\nu;\epsilon}},\label{eq:rewritting H_v^(AB+B) -- 1}
\end{align}

\noindent where $\hat{H}_{\nu\in\left\{ y,z\right\} }^{\left(B\right)}$
is given by Eq.~\eqref{eq:H_v^(B) in path integral appendix -- 1},
and $\hat{D}_{\nu;r;\epsilon}^{\vphantom{\dagger}}\left(z\right)$
is the displacement operator \citep{Ballentine}:

\begin{equation}
\hat{D}_{\nu\in\left\{ y,z\right\} ;r;\epsilon}^{\vphantom{\dagger}}\left(z\right)=e^{z\hat{b}_{\nu;r;\epsilon}^{\dagger}-z^{*}\hat{b}_{\nu;r;\epsilon}^{\vphantom{\dagger}}}.\label{eq:displacement operator in path integral appendix -- 1}
\end{equation}

\noindent Next, we use the fact that the displacement operator $\hat{D}_{\nu;r;\epsilon}^{\vphantom{\dagger}}\left(z\right)$
is unitary to write:

\begin{align}
 & e^{i\Delta t\tilde{w}_{n;q}\hat{H}_{\nu;k=\left\lfloor q/2\right\rfloor }^{\left(AB+B\right)}\left(\boldsymbol{\sigma}_{\nu;-1;q}\right)}\nonumber \\
 & \quad=e^{-i\Delta t\tilde{w}_{n;q}\sum_{u,r,\epsilon}\frac{\left\{ \mathcal{E}_{v;r;q}^{\left(\lambda\right)}\lambda_{\nu;r;\epsilon}\right\} ^{2}}{\omega_{\nu;\epsilon}}}\nonumber \\
 & \quad\mathrel{\phantom{=}}\mathop{\times}\prod_{u=-N}^{N}\prod_{r=0}^{L-1}\prod_{\epsilon}\left\{ \hat{D}_{\nu;r+uL;\epsilon}^{\dagger}\left(\mathcal{E}_{v;r;q}^{\left(\lambda\right)}\frac{\lambda_{\nu;r;\epsilon}}{\omega_{\nu;\epsilon}}\sigma_{\nu;r+uL;-1;q}\right)\right\} e^{i\Delta t\tilde{w}_{n;q}\hat{H}_{\nu}^{\left(B\right)}}\nonumber \\
 & \quad\mathrel{\phantom{=}}\mathop{\times}\prod_{u=-N}^{N}\prod_{r=0}^{L-1}\prod_{\epsilon}\left\{ \hat{D}_{\nu;r+uL;\epsilon}^{\vphantom{\dagger}}\left(\mathcal{E}_{v;r;q}^{\left(\lambda\right)}\frac{\lambda_{\nu;r;\epsilon}}{\omega_{\nu;\epsilon}}\sigma_{\nu;r+uL;-1;q}\right)\right\} ,\label{eq:rewritting exp term with H_v^=00007BAB+B=00007D -- 1}
\end{align}

\noindent where

\begin{equation}
\sum_{u,r,\epsilon}f\left(u,r,\epsilon,\ldots\right)=\sum_{u=-N}^{N}\sum_{r=0}^{L-1}\sum_{\epsilon}f\left(u,r,\epsilon,\ldots\right),\label{eq:u r epsilon sum notation -- 1}
\end{equation}

\noindent with $f\left(u,r,\epsilon,\ldots\right)$ being an arbitrary
function. Using the Baker\textendash Campbell\textendash Hausdorff
formula:

\begin{equation}
e^{\hat{X}}e^{\hat{Y}}=e^{\hat{Z}},\label{eq:BCH formula -- 1}
\end{equation}

\noindent where

\begin{equation}
\hat{Z}=\hat{X}+\hat{Y}+\frac{1}{2}\left[\hat{X},\hat{Y}\right]+\frac{1}{12}\left[\hat{X},\left[\hat{X},\hat{Y}\right]\right]-\ldots,\label{eq:BCH formula -- 2}
\end{equation}

\noindent and the fact that

\begin{equation}
\left[\hat{b}_{\nu_{1};r_{1};\epsilon_{1}}^{\vphantom{\dagger}},\hat{b}_{\nu_{2};r_{2};\epsilon_{2}}^{\dagger}\right]=\delta_{\nu_{1},\nu_{2}}\delta_{r_{1},r_{2}}\delta_{\epsilon_{1},\epsilon_{2}}\hat{1}_{\nu_{1};r_{1};\epsilon_{1}}^{\left(\mathcal{B}\right)}\hat{1}_{\nu_{2};r_{2};\epsilon_{2}}^{\left(\mathcal{B}\right)},\label{eq:bosonic commutation relation in path integral appendix -- 1}
\end{equation}

\noindent we can write

\begin{align}
 & \prod_{u=-N}^{N}\prod_{r=0}^{L-1}\prod_{\epsilon}\left\{ \hat{D}_{\nu;r+uL;\epsilon}^{\vphantom{\dagger}}\left(\mathcal{E}_{v;r;q}^{\left(\lambda\right)}\frac{\lambda_{\nu;r;\epsilon}}{\omega_{\nu;\epsilon}}\sigma_{\nu;r+uL;-1;q}\right)\right\} \nonumber \\
 & \quad=e^{\frac{1}{2}\sum_{u,r,\epsilon}\left\{ \mathcal{E}_{v;r;q}^{\left(\lambda\right)}\frac{\lambda_{\nu;r;\epsilon}}{\omega_{\nu;\epsilon}}\right\} ^{2}}e^{-\sum_{u,r,\epsilon}\left\{ \mathcal{E}_{v;r;q}^{\left(\lambda\right)}\frac{\lambda_{\nu;r;\epsilon}}{\omega_{\nu;\epsilon}}\sigma_{\nu;r+uL;-1;q}\right\} \hat{b}_{\nu;r+uL;\epsilon}^{\vphantom{\dagger}}}e^{\sum_{u,r,\epsilon}\left\{ \mathcal{E}_{v;r;q}^{\left(\lambda\right)}\frac{\lambda_{\nu;r;\epsilon}}{\omega_{\nu;\epsilon}}\sigma_{\nu;r+uL;-1;q}\right\} \hat{b}_{\nu;r+uL;\epsilon}^{\dagger}},\label{eq:rewritting displacement operator -- 1}
\end{align}

\begin{align}
 & \prod_{u=-N}^{N}\prod_{r=0}^{L-1}\prod_{\epsilon}\left\{ \hat{D}_{\nu;r+uL;\epsilon}^{\dagger}\left(\mathcal{E}_{v;r;q}^{\left(\lambda\right)}\frac{\lambda_{\nu;r;\epsilon}}{\omega_{\nu;\epsilon}}\sigma_{\nu;r+uL;-1;q}\right)\right\} \nonumber \\
 & \quad=e^{\frac{1}{2}\sum_{u,r,\epsilon}\left\{ \mathcal{E}_{v;r;q}^{\left(\lambda\right)}\frac{\lambda_{\nu;r;\epsilon}}{\omega_{\nu;\epsilon}}\right\} ^{2}}e^{\sum_{u,r,\epsilon}\left\{ \mathcal{E}_{v;r;q}^{\left(\lambda\right)}\frac{\lambda_{\nu;r;\epsilon}}{\omega_{\nu;\epsilon}}\sigma_{\nu;r+uL;-1;q}\right\} \hat{b}_{\nu;r+uL;\epsilon}^{\vphantom{\dagger}}}e^{-\sum_{u,r,\epsilon}\left\{ \mathcal{E}_{v;r;q}^{\left(\lambda\right)}\frac{\lambda_{\nu;r;\epsilon}}{\omega_{\nu;\epsilon}}\sigma_{\nu;r+uL;-1;q}\right\} \hat{b}_{\nu;r+uL;\epsilon}^{\dagger}}.\label{eq:rewritting displacement operator -- 2}
\end{align}

\noindent We can go further and insert a coherent state resolution
of the identity {[}see Eq.~\eqref{eq:coherent state completeness relation -- 1}
in Appendix~\ref{sec:Coherent states -- 1}{]}:

\begin{equation}
\int d^{2}\mathbf{b}_{\nu;j;\alpha;q}\,e^{-\mathbf{b}_{\nu;j;\alpha;q}\cdot\mathbf{b}_{\nu;j;\alpha;q}}\left|\mathbf{b}_{\nu;j;\alpha;q}\right\rangle \left\langle \mathbf{b}_{\nu;j;\alpha;q}\right|=\prod_{u=-N}^{N}\prod_{r=0}^{L-1}\prod_{\epsilon}\left\{ \hat{1}_{\nu;r;\epsilon}^{\left(\mathcal{B}\right)}\right\} ,\quad\text{with }j\in\left\{ 0,1\right\} ,\label{eq:coherent state completeness relation in path integral appendix -- 1}
\end{equation}

\noindent in between the second and third exponential terms in both
Eq.~\eqref{eq:rewritting displacement operator -- 1} and \eqref{eq:rewritting displacement operator -- 2}
to get

\begin{align}
 & \prod_{u=-N}^{N}\prod_{r=0}^{L-1}\prod_{\epsilon}\left\{ \hat{D}_{\nu;r+uL;\epsilon}^{\vphantom{\dagger}}\left(\mathcal{E}_{v;r;q}^{\left(\lambda\right)}\frac{\lambda_{\nu;r;\epsilon}}{\omega_{\nu;\epsilon}}\sigma_{\nu;r+uL;-1;q}\right)\right\} \nonumber \\
 & \quad=\int d^{2}\mathbf{b}_{\nu;1;\alpha;q}\,e^{-\mathbf{b}_{\nu;1;\alpha;q}\cdot\mathbf{b}_{\nu;1;\alpha;q}}e^{\frac{1}{2}\sum_{u,r,\epsilon}\left\{ \mathcal{E}_{v;r;q}^{\left(\lambda\right)}\frac{\lambda_{\nu;r;\epsilon}}{\omega_{\nu;\epsilon}}\right\} ^{2}}\nonumber \\
 & \quad\mathrel{\phantom{=}}\mathop{\times}e^{-\sum_{u,r,\epsilon}\left\{ \mathcal{E}_{v;r;q}^{\left(\lambda\right)}\frac{\lambda_{\nu;r;\epsilon}}{\omega_{\nu;\epsilon}}\sigma_{\nu;r+uL;-1;q}\right\} \left\{ b_{\nu;r+uL;\epsilon;1;-1;q}^{\vphantom{*}}-b_{\nu;r+uL;\epsilon;1;-1;q}^{*}\right\} }\left|\mathbf{b}_{\nu;1;\alpha;q}\right\rangle \left\langle \mathbf{b}_{\nu;1;\alpha;q}\right|,\label{eq:rewritting displacement operator -- 3}
\end{align}

\begin{align}
 & \prod_{u=-N}^{N}\prod_{r=0}^{L-1}\prod_{\epsilon}\left\{ \hat{D}_{\nu;r+uL;\epsilon}^{\dagger}\left(\mathcal{E}_{v;r;q}^{\left(\lambda\right)}\frac{\lambda_{\nu;r;\epsilon}}{\omega_{\nu;\epsilon}}\sigma_{\nu;r+uL;-1;q}\right)\right\} \nonumber \\
 & \quad=\int d^{2}\mathbf{b}_{\nu;0;\alpha;q}\,e^{-\mathbf{b}_{\nu;0;\alpha;q}\cdot\mathbf{b}_{\nu;0;\alpha;q}}e^{\frac{1}{2}\sum_{u,r,\epsilon}\left\{ \mathcal{E}_{v;r;q}^{\left(\lambda\right)}\frac{\lambda_{\nu;r;\epsilon}}{\omega_{\nu;\epsilon}}\right\} ^{2}}\nonumber \\
 & \quad\mathrel{\phantom{=}}\mathop{\times}e^{\sum_{u,r,\epsilon}\left\{ \mathcal{E}_{v;r;q}^{\left(\lambda\right)}\frac{\lambda_{\nu;r;\epsilon}}{\omega_{\nu;\epsilon}}\sigma_{\nu;r+uL;-1;q}\right\} \left\{ b_{\nu;r+uL;\epsilon;0;-1;q}^{\vphantom{*}}-b_{\nu;r+uL;\epsilon;0;-1;q}^{*}\right\} }\left|\mathbf{b}_{\nu;0;\alpha;q}\right\rangle \left\langle \mathbf{b}_{\nu;0;\alpha;q}\right|,\label{eq:rewritting displacement operator -- 4}
\end{align}

\noindent where the set of labels $\left\{ j;\alpha;q\right\} $ is
used to distinguish between the different instances of the resolution
of the identity. Applying Eqs.~\eqref{eq:rewritting displacement operator -- 3}
and \eqref{eq:rewritting displacement operator -- 4} to Eq.~\eqref{eq:rewritting exp term with H_v^=00007BAB+B=00007D -- 1}
yields:

\begin{align}
e^{i\Delta t\tilde{w}_{n;q}\hat{H}_{\nu;k=\left\lfloor q/2\right\rfloor }^{\left(AB+B\right)}\left(\boldsymbol{\sigma}_{\nu;-1;q}\right)} & =\prod_{j=0,1}\left\{ \int d^{2}\mathbf{b}_{\nu;j;-1;q}\,e^{-\mathbf{b}_{\nu;j;-1;q}\cdot\mathbf{b}_{\nu;j;-1;q}}\right\} \nonumber \\
 & \mathrel{\phantom{=}}\mathop{\times}e^{\sum_{u,r,\epsilon}\left\{ \mathcal{E}_{v;r;q}^{\left(\lambda\right)}\frac{\lambda_{\nu;r;\epsilon}}{\omega_{\nu;\epsilon}}\right\} ^{2}}e^{-i\Delta t\tilde{w}_{n;q}\sum_{u,r,\epsilon}\frac{\left\{ \mathcal{E}_{v;r;q}^{\left(\lambda\right)}\lambda_{\nu;r;\epsilon}\right\} ^{2}}{\omega_{\nu;\epsilon}}}\nonumber \\
 & \mathrel{\phantom{=}}\mathop{\times}e^{\sum_{u,r,\epsilon}\left\{ \mathcal{E}_{v;r;q}^{\left(\lambda\right)}\frac{\lambda_{\nu;r;\epsilon}}{\omega_{\nu;\epsilon}}\sigma_{\nu;r+uL;-1;q}\right\} \left[b_{\nu;r+uL;\epsilon;0;-1;q}^{\vphantom{*}}-b_{\nu;r+uL;\epsilon;0;-1;q}^{*}\right]}\nonumber \\
 & \mathrel{\phantom{=}}\mathop{\times}e^{-\sum_{u,r,\epsilon}\left\{ \mathcal{E}_{v;r;q}^{\left(\lambda\right)}\frac{\lambda_{\nu;r;\epsilon}}{\omega_{\nu;\epsilon}}\sigma_{\nu;r+uL;-1;q}\right\} \left[b_{\nu;r+uL;\epsilon;1;-1;q}^{\vphantom{*}}-b_{\nu;r+uL;\epsilon;1;-1;q}^{*}\right]}\nonumber \\
 & \mathrel{\phantom{=}}\mathop{\times}\left\langle \mathbf{b}_{\nu;0;-1;q}\right|e^{i\Delta t\tilde{w}_{n;q}\hat{H}_{\nu}^{\left(B\right)}}\left|\mathbf{b}_{\nu;1;-1;q}\right\rangle \left|\mathbf{b}_{\nu;0;-1;q}\right\rangle \left\langle \mathbf{b}_{\nu;1;-1;q}\right|.\label{eq:rewritting exp term with H_v^=00007BAB+B=00007D -- 2}
\end{align}

To calculate the matrix elements of $e^{i\Delta t\tilde{w}_{n;q}\hat{H}_{\nu}^{\left(B\right)}}$,
we use the following identity

\begin{equation}
f_{b^{\vphantom{\prime}}b^{\prime}}\left(a\right)\equiv\left\langle b^{\vphantom{\prime}}\right|a^{\hat{b}^{\dagger}\hat{b}}\left|b^{\prime}\right\rangle =e^{b^{*\vphantom{\prime}}b^{\prime}a},\label{eq:exponent coherent state identity -- 1}
\end{equation}

\noindent where $\left|b^{\vphantom{\prime}}\right\rangle $ and $\left|b^{\prime}\right\rangle $
are coherent states and $a$ is a complex number. Equation~\eqref{eq:exponent coherent state identity -- 1}
can be proven from the following operator relation:

\begin{equation}
g\left(\hat{b}^{\dagger}\hat{b}\right)\hat{b}=\hat{b}g\left(\hat{b}^{\dagger}\hat{b}-\hat{1}\right),\label{eq:bosonic field operator relation -- 1}
\end{equation}

\noindent where $g$ is an arbitrary function. Equation~\eqref{eq:bosonic field operator relation -- 1}
can be proven by acting both sides of Eq.~\eqref{eq:bosonic field operator relation -- 1}
by an arbitrary occupation number state $\left|n\right\rangle $:

\begin{align}
 & g\left(\hat{b}^{\dagger}\hat{b}\right)\hat{b}\left|n\right\rangle =\hat{b}g\left(\hat{b}^{\dagger}\hat{b}-\hat{1}\right)\left|n\right\rangle \nonumber \\
\Rightarrow & g\left(\hat{b}^{\dagger}\hat{b}\right)\left\{ \hat{b}\left|n\right\rangle \right\} =\hat{b}\left\{ g\left(\hat{b}^{\dagger}\hat{b}-\hat{1}\right)\left|n\right\rangle \right\} \nonumber \\
\Rightarrow & g\left(\hat{b}^{\dagger}\hat{b}\right)\left\{ \sqrt{n}\left|n-1\right\rangle \right\} =\hat{b}\left\{ g\left(n-1\right)\left|n\right\rangle \right\} \nonumber \\
\Rightarrow & \sqrt{n}\left\{ g\left(\hat{b}^{\dagger}\hat{b}\right)\left|n-1\right\rangle \right\} =g\left(n-1\right)\left\{ \hat{b}\left|n\right\rangle \right\} \nonumber \\
\Rightarrow & \sqrt{n}\left\{ g\left(n-1\right)\left|n-1\right\rangle \right\} =g\left(n-1\right)\left\{ \sqrt{n}\left|n-1\right\rangle \right\} \nonumber \\
\Rightarrow & \sqrt{n}g\left(n-1\right)\left|n-1\right\rangle =\sqrt{n}g\left(n-1\right)\left|n-1\right\rangle .\label{eq:proving bosonic field operator relation -- 1}
\end{align}

\noindent Since the set of occupation number states form a complete
basis, Eq.~\eqref{eq:proving bosonic field operator relation -- 1}
implies that Eq.~\eqref{eq:bosonic field operator relation -- 1}
is valid for all states in the single-boson Hilbert space. By differentiating
$f_{b^{\vphantom{\prime}}b^{\prime}}\left(a\right)$ with respect
to $a$, and using Eqs.~\eqref{eq:annihilation op right-hand eigenvalue equation -- 1},
\eqref{eq:creation op left-hand eigenvalue equation -- 1}, and \eqref{eq:bosonic field operator relation -- 1},
one obtains

\begin{align}
\frac{\partial}{\partial a}\left\{ f_{b^{\vphantom{\prime}}b^{\prime}}\left(a\right)\right\}  & =\left\langle b^{\vphantom{\prime}}\right|\frac{\partial}{\partial a}\left\{ a^{\hat{b}^{\dagger}\hat{b}}\right\} \left|b^{\prime}\right\rangle \nonumber \\
 & =\left\langle b^{\vphantom{\prime}}\right|\hat{b}^{\dagger}\hat{b}a^{\hat{b}^{\dagger}\hat{b}-\hat{1}}\left|b^{\prime}\right\rangle \nonumber \\
 & =\left\langle b^{\vphantom{\prime}}\right|\hat{b}^{\dagger}a^{\hat{b}^{\dagger}\hat{b}}\hat{b}\left|b^{\prime}\right\rangle \nonumber \\
 & =b^{*\vphantom{\prime}}b^{\prime}\left\langle b^{\vphantom{\prime}}\right|a^{\hat{b}^{\dagger}\hat{b}}\left|b^{\prime}\right\rangle \nonumber \\
 & =b^{*\vphantom{\prime}}b^{\prime}f_{b^{\vphantom{\prime}}b^{\prime}}\left(a\right).\label{eq:proving exponent coherent state identity -- 1}
\end{align}

\noindent We can then solve the ordinary differential equation as
follows:

\begin{align}
 & \frac{\partial}{\partial a}\left\{ f_{b^{\vphantom{\prime}}b^{\prime}}\left(a\right)\right\} =b^{*\vphantom{\prime}}b^{\prime}f_{b^{\vphantom{\prime}}b^{\prime}}\left(a\right)\nonumber \\
\Rightarrow & \frac{1}{f_{b^{\vphantom{\prime}}b^{\prime}}\left(a\right)}\frac{\partial}{\partial a}\left\{ f_{b^{\vphantom{\prime}}b^{\prime}}\left(a\right)\right\} =b^{*\vphantom{\prime}}b^{\prime}\nonumber \\
\Rightarrow & \int_{1}^{a}da^{\prime}\,\left[\frac{1}{f_{b^{\vphantom{\prime}}b^{\prime}}\left(a\right)}\frac{\partial}{\partial a}\left\{ f_{b^{\vphantom{\prime}}b^{\prime}}\left(a\right)\right\} \right]=\int_{1}^{a}da^{\prime}\,b^{*\vphantom{\prime}}b^{\prime}\nonumber \\
\Rightarrow & \int_{1}^{a}\left[\left(da^{\prime}\,\frac{\partial}{\partial a}\left\{ f_{b^{\vphantom{\prime}}b^{\prime}}\left(a\right)\right\} \right)\frac{1}{f_{b^{\vphantom{\prime}}b^{\prime}}\left(a\right)}\right]=\left(a-1\right)b^{*\vphantom{\prime}}b^{\prime}\nonumber \\
\Rightarrow & \int_{f_{b^{\vphantom{\prime}}b^{\prime}}\left(1\right)}^{f_{b^{\vphantom{\prime}}b^{\prime}}\left(a\right)}\left[df\,\frac{1}{f}\right]=\left(a-1\right)b^{*\vphantom{\prime}}b^{\prime}\nonumber \\
\Rightarrow & \ln\left(\frac{f_{b^{\vphantom{\prime}}b^{\prime}}\left(a\right)}{f_{b^{\vphantom{\prime}}b^{\prime}}\left(1\right)}\right)=\left(a-1\right)b^{*\vphantom{\prime}}b^{\prime}\nonumber \\
\Rightarrow & f_{b^{\vphantom{\prime}}b^{\prime}}\left(a\right)=f_{b^{\vphantom{\prime}}b^{\prime}}\left(1\right)e^{\left(a-1\right)b^{*\vphantom{\prime}}b^{\prime}}\nonumber \\
\Rightarrow & f_{b^{\vphantom{\prime}}b^{\prime}}\left(a\right)=\left\langle b^{\vphantom{\prime}}\right|1^{\hat{b}^{\dagger}\hat{b}}\left|b^{\prime}\right\rangle e^{\left(a-1\right)b^{*\vphantom{\prime}}b^{\prime}}\nonumber \\
\Rightarrow & f_{b^{\vphantom{\prime}}b^{\prime}}\left(a\right)=\left\langle b^{\vphantom{\prime}}\right|\hat{1}\left|b^{\prime}\right\rangle e^{\left(a-1\right)b^{*\vphantom{\prime}}b^{\prime}}\nonumber \\
\Rightarrow & f_{b^{\vphantom{\prime}}b^{\prime}}\left(a\right)=e^{b^{*\vphantom{\prime}}b^{\prime}}e^{\left(a-1\right)b^{*\vphantom{\prime}}b^{\prime}}\nonumber \\
\Rightarrow & f_{b^{\vphantom{\prime}}b^{\prime}}\left(a\right)=e^{b^{*\vphantom{\prime}}b^{\prime}a},\label{eq:proving exponent coherent state identity -- 2}
\end{align}

\noindent where we used Eq.~\eqref{eq:coherent states inner product -- 1}
in going from the third last to the second last line. This completes
the proof of Eq.~\eqref{eq:exponent coherent state identity -- 1}.
Using Eqs.~\eqref{eq:H_v^(B) in path integral appendix -- 1} and
\eqref{eq:exponent coherent state identity -- 1}, we can express
the matrix elements of $e^{i\Delta t\tilde{w}_{n;q}\hat{H}_{\nu}^{\left(B\right)}}$
as

\begin{equation}
\left\langle \mathbf{b}_{\nu;0;-1;q}\right|e^{i\Delta t\tilde{w}_{n;q}\hat{H}_{\nu}^{\left(B\right)}}\left|\mathbf{b}_{\nu;1;-1;q}\right\rangle =e^{\sum_{u,r,\epsilon}b_{\nu;r+uL;\epsilon;0;-1;k}^{*}b_{\nu;r+uL;\epsilon;1;-1;k}^{\vphantom{*}}e^{i\Delta t\tilde{w}_{n;q}\omega_{\nu;\epsilon}}}.\label{eq:matrix elements of e^=00007Bi dt w H_v^(B)=00007D -- 1}
\end{equation}

\noindent Substituting Eq.~\eqref{eq:matrix elements of e^=00007Bi dt w H_v^(B)=00007D -- 1}
into Eq.~\eqref{eq:rewritting exp term with H_v^=00007BAB+B=00007D -- 2}
yields

\begin{align}
e^{i\Delta t\tilde{w}_{n;q}\hat{H}_{\nu;k=\left\lfloor q/2\right\rfloor }^{\left(AB+B\right)}\left(\boldsymbol{\sigma}_{\nu;-1;q}\right)} & =\prod_{j=0,1}\left\{ \int d^{2}\mathbf{b}_{\nu;j;-1;q}\,e^{-\mathbf{b}_{\nu;j;-1;q}\cdot\mathbf{b}_{\nu;j;-1;q}}\right\} \nonumber \\
 & \mathrel{\phantom{=}}\mathop{\times}e^{\sum_{u,r,\epsilon}\left\{ \mathcal{E}_{v;r;q}^{\left(\lambda\right)}\frac{\lambda_{\nu;r;\epsilon}}{\omega_{\nu;\epsilon}}\right\} ^{2}}e^{-i\Delta t\tilde{w}_{n;q}\sum_{u,r,\epsilon}\frac{\left\{ \mathcal{E}_{v;r;q}^{\left(\lambda\right)}\lambda_{\nu;r;\epsilon}\right\} ^{2}}{\omega_{\nu;\epsilon}}}\nonumber \\
 & \mathrel{\phantom{=}}\mathop{\times}e^{\sum_{u,r,\epsilon}\left\{ \mathcal{E}_{v;r;q}^{\left(\lambda\right)}\frac{\lambda_{\nu;r;\epsilon}}{\omega_{\nu;\epsilon}}\sigma_{\nu;r+uL;-1;q}\right\} \left[b_{\nu;r+uL;\epsilon;0;-1;q}^{\vphantom{*}}-b_{\nu;r+uL;\epsilon;0;-1;q}^{*}\right]}\nonumber \\
 & \mathrel{\phantom{=}}\mathop{\times}e^{-\sum_{u,r,\epsilon}\left\{ \mathcal{E}_{v;r;q}^{\left(\lambda\right)}\frac{\lambda_{\nu;r;\epsilon}}{\omega_{\nu;\epsilon}}\sigma_{\nu;r+uL;-1;q}\right\} \left[b_{\nu;r+uL;\epsilon;1;-1;q}^{\vphantom{*}}-b_{\nu;r+uL;\epsilon;1;-1;q}^{*}\right]}\nonumber \\
 & \mathrel{\phantom{=}}\mathop{\times}e^{\sum_{u,r,\epsilon}b_{\nu;r+uL;\epsilon;0;-1;k}^{*}b_{\nu;r+uL;\epsilon;1;-1;k}^{\vphantom{*}}e^{i\Delta t\tilde{w}_{n;q}\omega_{\nu;\epsilon}}}\left|\mathbf{b}_{\nu;0;-1;q}\right\rangle \left\langle \mathbf{b}_{\nu;1;-1;q}\right|.\label{eq:rewritting exp term with H_v^=00007BAB+B=00007D -- 3}
\end{align}

Next, we return to Eq.~\eqref{eq:rewritting exp term with H_v^=00007BAB+B=00007D -- 2}
and rewrite it as follows:

\begin{equation}
\hat{\rho}^{\left(A\right)}\left(t_{n}\right)=\text{Tr}^{\left(B\right)}\left\{ \hat{U}\left(t_{n},0\right)\hat{\rho}^{\left(i\right)}\hat{U}\left(0,t_{n}\right)\right\} =\text{Tr}^{\left(B\right)}\left\{ \hat{U}^{\dagger}\left(0,t_{n}\right)\hat{\rho}^{\left(i\right)}\hat{U}\left(0,t_{n}\right)\right\} .\label{eq:rho^(A) in path integral appendix -- 2}
\end{equation}

\noindent Next, we evaluate the trace $\text{Tr}^{\left(B\right)}\left\{ \cdots\right\} $
using the coherent states $\left|\mathbf{b}_{y;0;0};\mathbf{b}_{z;0;0}\right\rangle $
as basis states {[}see Eq.~\eqref{eq:coherent state trace -- 1}
for expression for trace in coherent state basis{]}:

\begin{equation}
\hat{\rho}^{\left(A\right)}\left(t_{n}\right)=\int d^{2}\mathbf{b}_{y;0;0}\,\int d^{2}\mathbf{b}_{z;0;0}\,e^{-\mathbf{b}_{y;0;0}\cdot\mathbf{b}_{y;0;0}-\mathbf{b}_{z;0;0}\cdot\mathbf{b}_{z;0;0}}\left\langle \mathbf{b}_{y;0;0};\mathbf{b}_{z;0;0}\right|\hat{U}^{\dagger}\left(0,t_{n}\right)\hat{\rho}^{\left(i\right)}\hat{U}\left(0,t_{n}\right)\left|\mathbf{b}_{y;0;0};\mathbf{b}_{z;0;0}\right\rangle ,\label{eq:rho^(A) in path integral appendix -- 3}
\end{equation}

\noindent Next, using Eqs.~\eqref{eq:rho^(i) -- 1}-\eqref{eq:rho^(i, B) -- 1},
\eqref{eq:rho_u^(i, B) -- 1},\eqref{eq:rho_=00007Bu; v=00007D^(i, B) -- 1},
\eqref{eq:Z_=00007Bu; v=00007D^(B) -- 2}, \eqref{eq:coherent states inner product -- 1},
\eqref{eq:approximating U over =00005B0, t=00005D -- 3}, \eqref{eq:exponent coherent state identity -- 1},
\eqref{eq:rewritting exp term with H_v^=00007BAB+B=00007D -- 3},
and \eqref{eq:rho^(A) in path integral appendix -- 3} we get:

\begin{align}
 & \rho^{\left(A\right)}\left(\boldsymbol{\sigma}_{z;1;2n+1},\boldsymbol{\sigma}_{z;-1;2n+1}\right)\nonumber \\
 & \quad\equiv\left\langle \boldsymbol{\sigma}_{z;1;2n+1}z\right|\hat{\rho}^{\left(A\right)}\left(t_{n}\right)\left|\boldsymbol{\sigma}_{z;-1;2n+1}z\right\rangle \nonumber \\
 & \quad=\prod_{\alpha=\pm1}\prod_{u=-N}^{N}\prod_{r=0}^{L-1}\left[\prod_{q=0}^{2n+1}\left\{ \sum_{\sigma_{y;r+uL;\alpha;q}=\pm1}\right\} \prod_{q=0}^{2n}\left\{ \sum_{\sigma_{z;r+uL;\alpha;q}=\pm1}\right\} \right]\nonumber \\
 & \quad\mathrel{\phantom{=}}\mathop{\times}\prod_{\alpha=\pm1}\prod_{u=-N}^{N}\prod_{r=0}^{L-1}\prod_{l=1}^{n}\left\{ \delta_{\sigma_{z;r+uL;\alpha;2l-1},\sigma_{z;r+uL;\alpha;2l}}\right\} \nonumber \\
 & \quad\mathrel{\phantom{=}}\mathop{\times}e^{i\sum_{\alpha=\pm1}\tilde{\phi}_{\alpha;n}^{\left(\text{lcafc}\right)}\left(\boldsymbol{\sigma}_{z;\alpha;q\in\left[0,2n+1\right]}\right)}e^{2\sum_{\nu\in\left\{ y,z\right\} }\sum_{q=0}^{2n+1}\sum_{u,r,\epsilon}\left\{ \mathcal{E}_{v;r;q}^{\left(\lambda\right)}\frac{\lambda_{\nu;r;\epsilon}}{\omega_{\nu;\epsilon}}\right\} ^{2}}\nonumber \\
 & \quad\mathrel{\phantom{=}}\mathop{\times}\prod_{\alpha=\pm1}\left\{ \tilde{I}_{\alpha;n}^{\left(\text{tfc}\right)}\left(\boldsymbol{\sigma}_{y;\alpha;q\in\left[0,2n+1\right]}\right)\tilde{I}_{\alpha;n}^{\left(y\leftrightarrow z\right)}\left(\boldsymbol{\sigma}_{\nu\in\left\{ y,z\right\} ;\alpha;q\in\left[0,2n+1\right]}\right)\right\} \nonumber \\
 & \quad\mathrel{\phantom{=}}\mathop{\times}\int\mathcal{D}b\,F_{n\vphantom{\nu;r;\epsilon}}\left(\boldsymbol{\sigma}_{\nu\in\left\{ y,z\right\} ;\alpha=\pm1;q\in\left[0,2n+1\right]};b\right)\prod_{u=-N}^{N}\left\{ \left[\mathcal{Z}_{u;y}^{\left(B\right)}\mathcal{Z}_{u;z}^{\left(B\right)}\right]^{-1}\right\} \rho^{\left(i,A\right)}\left(\boldsymbol{\sigma}_{z;1;0},\boldsymbol{\sigma}_{z;-1;0}\right)+\mathcal{O}\left[\Delta t^{2}\right],\label{eq:rho^(A) in path integral appendix -- 4}
\end{align}

\noindent where

\begin{equation}
\rho^{\left(i,A\right)}\left(\boldsymbol{\sigma}_{z;1;0},\boldsymbol{\sigma}_{z;-1;0}\right)=\left\langle \boldsymbol{\sigma}_{z;1;0}z\right|\hat{\rho}^{\left(i,A\right)}\left|\boldsymbol{\sigma}_{z;-1;0}z\right\rangle =\left\langle \boldsymbol{\sigma}_{z;1;0}z\vphantom{\Psi^{\left(i,A\right)}}\right.\left|\Psi^{\left(i,A\right)}\right\rangle \left\langle \Psi^{\left(i,A\right)}\right|\left.\boldsymbol{\sigma}_{z;-1;0}z\vphantom{\Psi^{\left(i,A\right)}}\right\rangle ;\label{eq:matrix elements of rho^(i, A) -- 1}
\end{equation}

\noindent $\int\mathcal{D}b$ is a bosonic measure:

\begin{equation}
\int\mathcal{D}b=\prod_{\nu\in\left\{ y,z\right\} }\prod_{u=-N}^{N}\prod_{r=0}^{L-1}\prod_{\epsilon}\left\{ \int\mathcal{D}b_{\nu;r+uL;\epsilon}\right\} ,\label{eq:defining b measure -- 1}
\end{equation}

\noindent with

\begin{equation}
\int\mathcal{D}b_{\nu;r;\epsilon}=\int d^{2}b_{\nu;r;\epsilon;j=0;\alpha=0}\,\prod_{q=0}^{2n+1}\prod_{\alpha=\pm1}\prod_{j=0,1}\left\{ \int d^{2}b_{\nu;r;\epsilon;j;\alpha;q}\right\} ;\label{eq:defining  b_=00007Bv; r; epsilon=00007D measure -- 1}
\end{equation}

\noindent $F_{n\vphantom{\nu;r;\epsilon}}\left(\boldsymbol{\sigma}_{\nu\in\left\{ y,z\right\} ;\alpha=\pm1;q\in\left[0,2n+1\right]};b\right)$
is an integrand that is Gaussian in the bosonic fields with:

\begin{align}
F_{n\vphantom{\nu;r;\epsilon}}\left(\boldsymbol{\sigma}_{\nu\in\left\{ y,z\right\} ;\alpha=\pm1;q\in\left[0,2n+1\right]};b\right) & =\prod_{\nu\in\left\{ y,z\right\} }\prod_{u=-N}^{N}\prod_{r=0}^{L-1}\prod_{\epsilon}\left\{ F_{\nu;r+uL;\epsilon;n}\left(\sigma_{\nu;r+uL;\alpha=\pm1;q\in\left[0,2n+1\right]};\right.\right.\nonumber \\
 & \phantom{=\prod_{\nu\in\left\{ y,z\right\} }\prod_{u=-N}^{N}\prod_{r=0}^{L-1}\prod_{\epsilon}F_{\nu;r+uL;\epsilon;n}}\left.\left.\quad b_{\nu;r+uL;\epsilon;j=0,1;\alpha=\pm1;q\in\left[0,2n+1\right]};\right.\right.\nonumber \\
 & \phantom{=\prod_{\nu\in\left\{ y,z\right\} }\prod_{u=-N}^{N}\prod_{r=0}^{L-1}\prod_{\epsilon}F_{\nu;r+uL;\epsilon;n}}\left.\quad\vphantom{F_{\nu;r+uL;\epsilon;n}}\left.b_{\nu;r+uL;\epsilon;j=0;\alpha=0}\right)\right\} ,\label{eq:defining F_n -- 1}
\end{align}

\begin{align}
 & F_{\nu;r;\epsilon;n}\left(\sigma_{\nu;r;\alpha=\pm1;q\in\left[0,2n+1\right]};b_{\nu;r;\epsilon;j=0,1;\alpha=\pm1;q\in\left[0,2n+1\right]};b_{\nu;r;\epsilon;j=0;\alpha=0}\right)\nonumber \\
 & \quad=e^{-\left(\boldsymbol{\zeta}_{\nu;r;\epsilon;n}\right)^{\dagger}\mathbf{A}_{\nu;r;\epsilon;n}\boldsymbol{\zeta}_{\nu;r;\epsilon;n}+\left(\boldsymbol{\xi}_{\nu;r;\epsilon;n}\right)^{\dagger}\boldsymbol{\zeta}_{\nu;r;\epsilon;n}+\left(\boldsymbol{\zeta}_{\nu;r;\epsilon;n}\right)^{\dagger}\boldsymbol{\psi}_{\nu;r;\epsilon;n}},\label{eq:defining F_=00007Bv; r; epsilon; n=00007D -- 1}
\end{align}

\begin{equation}
\boldsymbol{\zeta}_{\nu;r;\epsilon;n}=\begin{pmatrix}\boldsymbol{\zeta}_{\nu;r;\epsilon;n;j=0}\\
\boldsymbol{\zeta}_{\nu;r;\epsilon;n;j=1}
\end{pmatrix},\label{eq:zeta_=00007Bv; r; epsilon; n=00007D -- 1}
\end{equation}

\begin{equation}
\boldsymbol{\zeta}_{\nu;r;\epsilon;n;j=0}=\begin{pmatrix}\boldsymbol{\zeta}_{\nu;r;\epsilon;n;j;\alpha=1}\\
\boldsymbol{\zeta}_{\nu;r;\epsilon;n;j;\alpha=0}\\
\boldsymbol{\zeta}_{\nu;r;\epsilon;n;j;\alpha=-1}
\end{pmatrix},\label{eq:zeta_=00007Bv; r; epsilon; n; j=00003D0=00007D -- 1}
\end{equation}

\begin{equation}
\boldsymbol{\zeta}_{\nu;r;\epsilon;n;j=1}=\begin{pmatrix}\boldsymbol{\zeta}_{\nu;r;\epsilon;n;j;\alpha=1}\\
\boldsymbol{\zeta}_{\nu;r;\epsilon;n;j;\alpha=-1}
\end{pmatrix},\label{eq:zeta_=00007Bv; r; epsilon; n; j=00003D1=00007D -- 1}
\end{equation}

\begin{equation}
\boldsymbol{\zeta}_{\nu;r;\epsilon;n;j;\alpha=\pm1}=\begin{pmatrix}b_{\nu;r;\epsilon;j;\alpha;q=0}\\
\vdots\\
b_{\nu;r;\epsilon;j;\alpha;q=2n+1}
\end{pmatrix},\label{eq:zeta_=00007Bv; r; epsilon; n; j; alpha=00003D+/-1=00007D -- 1}
\end{equation}

\begin{equation}
\boldsymbol{\zeta}_{\nu;r;\epsilon;n;j=0;\alpha=0}=\begin{pmatrix}b_{\nu;r;\epsilon}\end{pmatrix},\label{eq:zeta_=00007Bv; r; epsilon; n; j=00003D0; alpha=00003D0=00007D -- 1}
\end{equation}

\begin{equation}
\boldsymbol{\xi}_{\nu;r;\epsilon;n}=\begin{pmatrix}\boldsymbol{\xi}_{\nu;r;\epsilon;n;j=0}\\
\boldsymbol{\xi}_{\nu;r;\epsilon;n;j=1}
\end{pmatrix},\label{eq:xi_=00007Bv; r; epsilon; n=00007D -- 1}
\end{equation}

\begin{equation}
\boldsymbol{\xi}_{\nu;r;\epsilon;n;j=0}=\begin{pmatrix}\boldsymbol{\xi}_{\nu;r;\epsilon;n;j;\alpha=1}\\
\boldsymbol{\xi}_{\nu;r;\epsilon;n;j;\alpha=0}\\
\boldsymbol{\xi}_{\nu;r;\epsilon;n;j;\alpha=-1}
\end{pmatrix},\label{eq:xi_=00007Bv; r; epsilon; n; j=00003D0=00007D -- 1}
\end{equation}

\begin{equation}
\boldsymbol{\xi}_{\nu;r;\epsilon;n;j=1}=\begin{pmatrix}\boldsymbol{\xi}_{\nu;r;\epsilon;n;j;\alpha=1}\\
\boldsymbol{\xi}_{\nu;r;\epsilon;n;j;\alpha=-1}
\end{pmatrix},\label{eq:xi_=00007Bv; r; epsilon; n; j=00003D1=00007D -- 1}
\end{equation}

\begin{equation}
\boldsymbol{\xi}_{\nu;r;\epsilon;n;j=0,1;\alpha=\pm1}=-\alpha\left(-1\right)^{j}\frac{\lambda_{\nu;r;\epsilon}}{\omega_{\nu;\epsilon}}\begin{pmatrix}\sigma_{\nu;r;\alpha;q=0}\mathcal{E}_{v;r;q=0}^{\left(\lambda\right)}\\
\vdots\\
\sigma_{\nu;r;\alpha;q=2n+1}\mathcal{E}_{v;r;q=2n+1}^{\left(\lambda\right)}
\end{pmatrix},\label{eq:xi_=00007Bv; r; epsilon; n; j=00003D0, 1; alpha=00003D+/-1=00007D -- 1}
\end{equation}

\begin{equation}
\lambda_{\nu;r+uL;\epsilon}=\lambda_{\nu;r;\epsilon},\label{eq:periodicity of Lambda_=00007Bv; r; epsilon=00007D -- 1}
\end{equation}

\begin{equation}
\mathcal{E}_{v;r+uL;q}^{\left(\lambda\right)}=\mathcal{E}_{v;r;q}^{\left(\lambda\right)},\label{eq:periodicity of mathcal E_=00007Bv; r=00007D^(Lambda) -- 1}
\end{equation}

\begin{equation}
\boldsymbol{\xi}_{\nu;r;\epsilon;n;j=0;\alpha=0}=\begin{pmatrix}0\end{pmatrix},\label{eq:xi_=00007Bv; r; epsilon; n; j=00003D0; alpha=00003D0=00007D -- 1}
\end{equation}

\begin{equation}
\boldsymbol{\psi}_{\nu;r;\epsilon;n}=-\boldsymbol{\xi}_{\nu;r;\epsilon;n},\label{eq:psi_=00007Bv; r; epsilon; n=00007D -- 1}
\end{equation}

\begin{equation}
\mathbf{A}_{\nu;r;\epsilon;n}=\begin{pmatrix}\mathbf{A}_{\nu;r;\epsilon;n;0,0} & \mathbf{A}_{\nu;r;\epsilon;n;0,1}\\
\mathbf{A}_{\nu;r;\epsilon;n;1,0} & \mathbf{A}_{\nu;r;\epsilon;n;1,1}
\end{pmatrix},\label{eq:A_=00007Bv; epsilon; n=00007D -- 1}
\end{equation}

\begin{align}
\mathbf{A}_{\nu;r;\epsilon;n;0,0} & =\begin{pmatrix}\mathbf{A}_{\nu;r;\epsilon;n;0,0;1,1} & \mathbf{A}_{\nu;r;\epsilon;n;0,0;1,0} & \mathbf{A}_{\nu;r;\epsilon;n;0,0;1,-1}\\
\mathbf{A}_{\nu;r;\epsilon;n;0,0;0,1} & \mathbf{A}_{\nu;r;\epsilon;n;0,0;0,0} & \mathbf{A}_{\nu;r;\epsilon;n;0,0;0,-1}\\
\mathbf{A}_{\nu;r;\epsilon;n;0,0;-1,1} & \mathbf{A}_{\nu;r;\epsilon;n;0,0;-1,0} & \mathbf{A}_{\nu;r;\epsilon;n;0,0;-1,-1}
\end{pmatrix}\nonumber \\
 & =\begin{pmatrix}\mathbf{1}_{\left(2n+2\right)\times\left(2n+2\right)} & \mathbf{0}_{\left(2n+2\right)\times\left(1\right)} & \mathbf{A}_{\nu;r;\epsilon;n;0,0;1,-1}\\
\mathbf{0}_{\left(1\right)\times\left(2n+2\right)} & \mathbf{1}_{\left(1\right)\times\left(1\right)} & \mathbf{0}_{\left(1\right)\times\left(2n+2\right)}\\
\mathbf{0}_{\left(2n+2\right)\times\left(2n+2\right)} & \mathbf{0}_{\left(2n+2\right)\times\left(1\right)} & \mathbf{1}_{\left(2n+2\right)\times\left(2n+2\right)}
\end{pmatrix},\label{eq:A_=00007Bv; epsilon; n; 0, 0=00007D -- 1}
\end{align}

\begin{equation}
\mathbf{1}_{\left(N\right)\times\left(N\right)}=\underbrace{\begin{pmatrix}1 & 0 & \cdots & \cdots & 0\\
0 & 1 & \ddots &  & \vdots\\
\vdots & \ddots & \ddots & \ddots & \vdots\\
\vdots &  & \ddots & 1 & 0\\
0 & \cdots & \cdots & 0 & 1
\end{pmatrix}}_{N\text{ columns}}\left.\vphantom{\begin{pmatrix}1\\
0\\
\vdots\\
\vdots\\
0
\end{pmatrix}}\right\} N\text{ rows},\label{eq:1_=00007B(N)x(N)=00007D -- 1}
\end{equation}

\begin{equation}
\mathbf{0}_{\left(N_{1}\right)\times\left(N_{1}\right)}=\underbrace{\begin{pmatrix}0 & \cdots & \cdots & \cdots & 0\\
\vdots &  &  &  & \vdots\\
\vdots &  &  &  & \vdots\\
\vdots &  &  &  & \vdots\\
0 & \cdots & \cdots & \cdots & 0
\end{pmatrix}}_{N_{2}\text{ columns}}\left.\vphantom{\begin{pmatrix}1\\
0\\
\vdots\\
\vdots\\
0
\end{pmatrix}}\right\} N_{1}\text{ rows},\label{eq:0_=00007B(N_1)x(N_2)=00007D -- 1}
\end{equation}

\begin{equation}
\left[\mathbf{A}_{\nu;r;\epsilon;n;0,0;1,-1}\right]_{q_{1},q_{2}}=-\delta_{q_{1},0}\delta_{0,q_{2}}g_{\nu;\epsilon},\label{eq:A_=00007Bv; epsilon; n; 0, 0; 1, -1=00007D -- 1}
\end{equation}

\begin{equation}
\delta_{q,q^{\prime}}=\begin{cases}
1, & \text{if }q=q^{\prime},\\
0, & \text{if }q\neq q^{\prime},
\end{cases}\label{eq:introduce kronecker delta -- 1}
\end{equation}

\begin{equation}
g_{\nu;\epsilon}=e^{-\beta\omega_{\nu;\epsilon}},\label{eq:g_=00007Bv; epsilon=00007D -- 1}
\end{equation}

\begin{equation}
\mathbf{A}_{\nu;r;\epsilon;n;0,1}=\begin{pmatrix}\mathbf{A}_{\nu;r;\epsilon;n;0,1;1,1} & \mathbf{0}_{\left(2n+2\right)\times\left(2n+2\right)}\\
\mathbf{A}_{\nu;r;\epsilon;n;0,1;0,1} & \mathbf{0}_{\left(1\right)\times\left(2n+2\right)}\\
\mathbf{0}_{\left(2n+2\right)\times\left(2n+2\right)} & \mathbf{A}_{\nu;r;\epsilon;n;0,1;-1,-1}
\end{pmatrix},\label{eq:A_=00007Bv; epsilon; n; 0, 1=00007D --1}
\end{equation}

\begin{equation}
\left[\mathbf{A}_{\nu;r;\epsilon;n;0,1;1,1}\right]_{q_{1},q_{2}}=-\delta_{q_{1},q_{2}+1},\label{eq:A_=00007Bv; epsilon; n; 0, 1; 1, 1=00007D -- 1}
\end{equation}

\begin{equation}
\left[\mathbf{A}_{\nu;r;\epsilon;n;0,1;0,1}\right]_{q_{1},q_{2}}=-\delta_{2n+1,q_{2}},\label{eq:A_=00007Bv; epsilon; n; 0, 1; 0, 1=00007D -- 1}
\end{equation}

\begin{equation}
\left[\mathbf{A}_{\nu;r;\epsilon;n;0,1;-1,-1}\right]_{q_{1},q_{2}}=-\delta_{q_{1},q_{2}}f_{\nu;\epsilon;n;-1;q_{1}},\label{eq:A_=00007Bv; epsilon; n; 0, 1; -1, -1=00007D -- 1}
\end{equation}

\begin{equation}
f_{\nu;r;\epsilon;n;\alpha;q}=e^{-\alpha i\Delta t\tilde{w}_{n;q}\omega_{\nu;\epsilon}},\label{eq:f_=00007Bv; epsilon; n; alpha; k=00007D -- 1}
\end{equation}

\begin{equation}
\mathbf{A}_{\nu;r;\epsilon;n;1,0}=\begin{pmatrix}\mathbf{A}_{\nu;r;\epsilon;n;1,0;1,1} & \mathbf{0}_{\left(2n+2\right)\times\left(1\right)} & \mathbf{0}_{\left(2n+2\right)\times\left(2n+2\right)}\\
\mathbf{0}_{\left(2n+2\right)\times\left(2n+2\right)} & \mathbf{A}_{\nu;r;\epsilon;n;1,0;-1,0} & \mathbf{A}_{\nu;r;\epsilon;n;1,0;-1,-1}
\end{pmatrix},\label{eq:A_=00007Bv; epsilon; n; 1, 0=00007D -- 1}
\end{equation}

\begin{equation}
\left[\mathbf{A}_{\nu;r;\epsilon;n;1,0;1,1}\right]_{q_{1},q_{2}}=-\delta_{q_{1},q_{2}}f_{\nu;\epsilon;n;1;q_{1}},\label{eq:A_=00007Bv; epsilon; n; 1, 0; 1, 1=00007D -- 1}
\end{equation}

\begin{equation}
\left[\mathbf{A}_{\nu;r;\epsilon;n;1,0;-1,0}\right]_{q_{1},q_{2}}=-\delta_{q_{1},2n+1},\label{eq:A_=00007Bv; epsilon; n; 1, 0; -1, 0=00007D -- 1}
\end{equation}

\begin{equation}
\left[\mathbf{A}_{\nu;r;\epsilon;n;1,0;-1,-1}\right]_{q_{1},q_{2}}=-\delta_{q_{1}+1,q_{2}},\label{eq:A_=00007Bv; epsilon; n; 1, 0; -1, -1=00007D -- 1}
\end{equation}

\begin{equation}
\mathbf{A}_{\nu;r;\epsilon;n;1,1}=\begin{pmatrix}\mathbf{1}_{\left(2n+2\right)\times\left(2n+2\right)} & \mathbf{0}_{\left(2n+2\right)\times\left(2n+2\right)}\\
\mathbf{0}_{\left(2n+2\right)\times\left(2n+2\right)} & \mathbf{1}_{\left(2n+2\right)\times\left(2n+2\right)}
\end{pmatrix}.\label{eq:A_=00007Bv; epsilon; n; 1, 1=00007D -- 1}
\end{equation}

The product of Gaussian integrals in Eq.~\eqref{eq:rho^(A) in path integral appendix -- 4}
can be evaluated using the identity \citep{Altland1}:

\begin{equation}
\mathcal{D}b_{\nu;r;\epsilon}e^{-\left(\boldsymbol{\zeta}_{\nu;r;\epsilon;n}\right)^{\dagger}\mathbf{A}_{\nu;r;\epsilon;n}\boldsymbol{\zeta}_{\nu;r;\epsilon;n}+\left(\boldsymbol{\xi}_{\nu;r;\epsilon;n}\right)^{\dagger}\boldsymbol{\zeta}_{\nu;r;\epsilon;n}+\left(\boldsymbol{\zeta}_{\nu;r;\epsilon;n}\right)^{\dagger}\boldsymbol{\psi}_{\nu;r;\epsilon;n}}=\frac{e^{\left(\boldsymbol{\xi}_{\nu;r;\epsilon;n}\right)^{\dagger}\left(\mathbf{A}_{\nu;r;\epsilon;n}\right)^{-1}\mathbf{\boldsymbol{\psi}}_{\nu;r;\epsilon;n}}}{\det\left[\mathbf{A}_{\nu;r;\epsilon;n}\right]},\label{eq:Gaussian integral identity -- 1}
\end{equation}

\noindent where the calculations of $\left(\mathbf{A}_{\nu;r;\epsilon;n}\right)^{-1}$
and $\det\left[\mathbf{A}_{\nu;r;\epsilon;n}\right]$ are very long
and cumbersome. The matrix $\mathbf{A}_{\nu;r;\epsilon;n}$ was deliberately
cast in the above block form such that we could exploit certain block
matrix identities. Let $\mathbf{M}$ be a block matrix of the form

\begin{equation}
\mathbf{M}=\begin{pmatrix}\mathbf{M}_{1,1} & \mathbf{M}_{1,2}\\
\mathbf{M}_{2,1} & \mathbf{M}_{2,2,}
\end{pmatrix}.\label{eq:M block matrix -- 1}
\end{equation}

\noindent Assuming that $\mathbf{M}_{1,1}$ is invertible, then we
have the following identities:

\begin{equation}
\det\left[\mathbf{M}\right]=\det\left[\mathbf{M}_{1,1}\right]\det\left[\mathbf{M}_{2,2}-\mathbf{M}_{2,1}\left(\mathbf{M}_{1,1}\right)^{-1}\mathbf{M}_{1,2}\right],\label{eq:determinant of a block matrix -- 1}
\end{equation}

\begin{equation}
\mathbf{M}=\begin{pmatrix}\mathbf{M}_{1,1} & \mathbf{M}_{1,2}\\
\mathbf{M}_{2,1} & \mathbf{M}_{2,2,}
\end{pmatrix}.\label{eq:inverse of a block matrix -- 1}
\end{equation}

\begin{equation}
\left[\mathbf{M}^{-1}\right]_{1,1}=\left(\mathbf{M}_{1,1}\right)^{-1}-\left(\mathbf{M}_{1,1}\right)^{-1}\mathbf{M}_{1,2}\left[\mathbf{M}^{-1}\right]_{2,1},\label{eq:=00005BM^=00007B-1=00007D=00005D_=00007B1, 1=00007D -- 1}
\end{equation}

\begin{equation}
\left[\mathbf{M}^{-1}\right]_{1,2}=-\left(\mathbf{M}_{1,1}\right)^{-1}\mathbf{M}_{1,2}\left[\mathbf{M}^{-1}\right]_{2,2},\label{eq:=00005BM^=00007B-1=00007D=00005D_=00007B1, 2=00007D -- 1}
\end{equation}

\begin{equation}
\left[\mathbf{M}^{-1}\right]_{2,1}=-\left[\mathbf{M}^{-1}\right]_{2,2}\mathbf{M}_{2,1}\left(\mathbf{M}_{1,1}\right)^{-1},\label{eq:=00005BM^=00007B-1=00007D=00005D_=00007B2, 1=00007D -- 1}
\end{equation}

\begin{equation}
\left[\mathbf{M}^{-1}\right]_{2,2}=\left(\mathbf{M}_{2,2}-\mathbf{M}_{2,1}\left(\mathbf{M}_{1,1}\right)^{-1}\mathbf{M}_{1,2}\right)^{-1}.\label{eq:=00005BM^=00007B-1=00007D=00005D_=00007B2, 2=00007D -- 1}
\end{equation}

In what follows, we use Eqs.~\eqref{eq:determinant of a block matrix -- 1}-\eqref{eq:=00005BM^=00007B-1=00007D=00005D_=00007B2, 2=00007D -- 1}
to calculate $\left(\mathbf{A}_{\nu;r;\epsilon;n}\right)^{-1}$ and
$\det\left[\mathbf{A}_{\nu;r;\epsilon;n}\right]$. If we let 

\begin{equation}
\mathbf{C}_{\nu;r;\epsilon;n}=\left(\mathbf{A}_{\nu;r;\epsilon;n}\right)^{-1},\label{eq:introduce C_=00007Bv; epsilon; n=00007D -- 1}
\end{equation}

\noindent then according to Eqs.~\eqref{eq:inverse of a block matrix -- 1}\eqref{eq:=00005BM^=00007B-1=00007D=00005D_=00007B2, 2=00007D -- 1},
we can write

\begin{equation}
\mathbf{C}_{\nu;r;\epsilon;n}=\begin{pmatrix}\mathbf{C}_{\nu;r;\epsilon;n;0,0} & \mathbf{C}_{\nu;r;\epsilon;n;0,1}\\
\mathbf{C}_{\nu;r;\epsilon;n;1,0} & \mathbf{C}_{\nu;r;\epsilon;n;1,1}
\end{pmatrix},\label{eq:C_=00007Bv; epsilon; n=00007D block structure -- 2}
\end{equation}

\noindent with

\begin{equation}
\mathbf{C}_{\nu;r;\epsilon;n;0,0}=\left(\mathbf{A}_{\nu;r;\epsilon;n;0,0}\right)^{-1}-\left(\mathbf{A}_{\nu;r;\epsilon;n;0,0}\right)^{-1}\mathbf{A}_{\nu;r;\epsilon;n;0,1}\mathbf{C}_{\nu;r;\epsilon;n;1,0},\label{eq:introduce C_=00007Bv; epsilon; n; 0, 0=00007D -- 1}
\end{equation}

\begin{equation}
\mathbf{C}_{\nu;r;\epsilon;n;0,1}=-\left(\mathbf{A}_{\nu;r;\epsilon;n;0,0}\right)^{-1}\mathbf{A}_{\nu;r;\epsilon;n;0,1}\mathbf{C}_{\nu;r;\epsilon;n;1,1},\label{eq:introduce C_=00007Bv; epsilon; n; 0, 1=00007D -- 1}
\end{equation}

\begin{equation}
\mathbf{C}_{\nu;r;\epsilon;n;1,0}=-\mathbf{C}_{\nu;r;\epsilon;n;1,1}\mathbf{A}_{\nu;r;\epsilon;n;1,0}\left(\mathbf{A}_{\nu;r;\epsilon;n;0,0}\right)^{-1},\label{eq:introduce C_=00007Bv; epsilon; n; 1, 0=00007D -- 1}
\end{equation}

\begin{equation}
\mathbf{C}_{\nu;r;\epsilon;n;1,1}=\left(\mathbf{B}_{\nu;r;\epsilon;n;1,1}\right)^{-1},\label{eq:introduce C_=00007Bv; epsilon; n; 1, 1=00007D -- 1}
\end{equation}

\begin{equation}
\mathbf{B}_{\nu;r;\epsilon;n;1,1}=\mathbf{A}_{\nu;r;\epsilon;n;1,1}-\mathbf{A}_{\nu;r;\epsilon;n;1,0}\left(\mathbf{A}_{\nu;r;\epsilon;n;0,0}\right)^{-1}\mathbf{A}_{\nu;r;\epsilon;n;0,1}.\label{eq:B_=00007Bv; epsilon; n; 1, 1=00007D -- 1}
\end{equation}

\noindent Let us first calculate $\left(\mathbf{A}_{\nu;r;\epsilon;n;0,0}\right)^{-1}$.
To calculate $\left(\mathbf{A}_{\nu;r;\epsilon;n;0,0}\right)^{-1}$,
we recast $\mathbf{A}_{\nu;r;\epsilon;n;0,0}$ as a $2\times2$ block
matrix so that Eqs.~\eqref{eq:determinant of a block matrix -- 1}-\eqref{eq:=00005BM^=00007B-1=00007D=00005D_=00007B2, 2=00007D -- 1}
are applicable:

\begin{equation}
\mathbf{A}_{\nu;r;\epsilon;n;0,0}=\begin{pmatrix}\mathbf{1}_{\left(2n+3\right)\times\left(2n+3\right)} & \mathbf{A}_{\nu;r;\epsilon;n;0,0;0,1}^{\prime}\\
\mathbf{0}_{\left(2n+2\right)\times\left(2n+3\right)} & \mathbf{1}_{\left(2n+2\right)\times\left(2n+2\right)}
\end{pmatrix},\label{eq:A_=00007Bv; epsilon; n; 0, 0=00007D as 2x2 block matrix -- 1}
\end{equation}

\begin{equation}
\mathbf{A}_{\nu;r;\epsilon;n;0,0;0,1}^{\prime}=\begin{pmatrix}\mathbf{A}_{\nu;r;\epsilon;n;0,0;1,-1}\\
\mathbf{0}_{\left(1\right)\times\left(2n+2\right)}
\end{pmatrix}.\label{eq:A_=00007Bv; epsilon; 0, 0; 0, 1=00007D^=00007Bprime=00007D -- 1}
\end{equation}

\noindent Because $\mathbf{A}_{\nu;r;\epsilon;n;0,0}$ in Eq.~\eqref{eq:A_=00007Bv; epsilon; n; 0, 0=00007D as 2x2 block matrix -- 1}
has a simple $2\times2$ block matrix structure, it is straightforward
to see from Eqs.~\eqref{eq:=00005BM^=00007B-1=00007D=00005D_=00007B1, 1=00007D -- 1}-\eqref{eq:=00005BM^=00007B-1=00007D=00005D_=00007B2, 2=00007D -- 1}
that

\begin{align}
\left(\mathbf{A}_{\nu;r;\epsilon;n;0,0}\right)^{-1} & =\begin{pmatrix}\mathbf{1}_{\left(2n+3\right)\times\left(2n+3\right)} & -\mathbf{A}_{\nu;r;\epsilon;n;0,0;0,1}^{\prime}\\
\mathbf{0}_{\left(2n+2\right)\times\left(2n+3\right)} & \mathbf{1}_{\left(2n+2\right)\times\left(2n+2\right)}
\end{pmatrix}\nonumber \\
 & =\begin{pmatrix}\mathbf{1}_{\left(2n+2\right)\times\left(2n+2\right)} & \mathbf{0}_{\left(2n+2\right)\times\left(1\right)} & -\mathbf{A}_{\nu;r;\epsilon;n;0,0;1,-1}\\
\mathbf{0}_{\left(1\right)\times\left(2n+2\right)} & \mathbf{1}_{\left(1\right)\times\left(1\right)} & \mathbf{0}_{\left(1\right)\times\left(2n+2\right)}\\
\mathbf{0}_{\left(2n+2\right)\times\left(2n+2\right)} & \mathbf{0}_{\left(2n+2\right)\times\left(1\right)} & \mathbf{1}_{\left(2n+2\right)\times\left(2n+2\right)}
\end{pmatrix}.\label{eq:(A_=00007Bv; epsilon; n; 0, 0=00007D)^=00007B-1=00007D -- 1}
\end{align}

\noindent Next, we calculate $\mathbf{A}_{\nu;r;\epsilon;n;1,0}\left(\mathbf{A}_{\nu;r;\epsilon;n;0,0}\right)^{-1}$.
Using Eqs.~\eqref{eq:A_=00007Bv; epsilon; n; 0, 0; 1, -1=00007D -- 1},
\eqref{eq:A_=00007Bv; epsilon; n; 1, 0=00007D -- 1}, \eqref{eq:A_=00007Bv; epsilon; n; 1, 0; 1, 1=00007D -- 1},
and \eqref{eq:(A_=00007Bv; epsilon; n; 0, 0=00007D)^=00007B-1=00007D -- 1}
we get

\begin{equation}
\mathbf{A}_{\nu;r;\epsilon;n;1,0}\left(\mathbf{A}_{\nu;r;\epsilon;n;0,0}\right)^{-1}=\begin{pmatrix}\mathbf{A}_{\nu;r;\epsilon;n;1,0;1,1} & \mathbf{0}_{\left(2n+2\right)\times\left(1\right)} & -\mathbf{A}_{\nu;r;\epsilon;n;1,0;1,1}\mathbf{A}_{\nu;r;\epsilon;n;0,0;1,-1}\\
\mathbf{0}_{\left(2n+2\right)\times\left(2n+2\right)} & \mathbf{A}_{\nu;r;\epsilon;n;1,0;-1,0} & \mathbf{A}_{\nu;r;\epsilon;n;1,0;-1,-1}
\end{pmatrix},\label{eq:A_=00007B1, 0=00007D (A_=00007B0, 0=00007D)^=00007B-1=00007D -- 1}
\end{equation}

\noindent with

\begin{equation}
\left[\mathbf{A}_{\nu;r;\epsilon;n;1,0;1,1}\mathbf{A}_{\nu;r;\epsilon;n;0,0;1,-1}\right]_{q_{1},q_{2}}=\delta_{q_{1},0}\delta_{0,q_{2}}f_{\nu;\epsilon;n;1;0}g_{\nu;\epsilon}.\label{eq:A_=00007B1, 0; 1, 1=00007D A_=00007B0, 0; 1, -1=00007D -- 1}
\end{equation}

\noindent Next, we calculate $\left(\mathbf{A}_{\nu;r;\epsilon;n;0,0}\right)^{-1}\mathbf{A}_{\nu;r;\epsilon;n;0,1}$.
Using Eqs.~\eqref{eq:A_=00007Bv; epsilon; n; 0, 0; 1, -1=00007D -- 1},
\eqref{eq:A_=00007Bv; epsilon; n; 0, 1=00007D --1}, \eqref{eq:A_=00007Bv; epsilon; n; 0, 1; -1, -1=00007D -- 1},
and \eqref{eq:(A_=00007Bv; epsilon; n; 0, 0=00007D)^=00007B-1=00007D -- 1}
we get

\begin{equation}
\left(\mathbf{A}_{\nu;r;\epsilon;n;0,0}\right)^{-1}\mathbf{A}_{\nu;r;\epsilon;n;0,1}=\begin{pmatrix}\mathbf{A}_{\nu;r;\epsilon;n;0,1;1,1} & -\mathbf{A}_{\nu;r;\epsilon;n;0,0;1,-1}\mathbf{A}_{\nu;r;\epsilon;n;0,1;-1,-1}\\
\mathbf{A}_{\nu;r;\epsilon;n;0,1;0,1} & \mathbf{0}_{\left(1\right)\times\left(2n+2\right)}\\
\mathbf{0}_{\left(2n+2\right)\times\left(2n+2\right)} & \mathbf{A}_{\nu;r;\epsilon;n;0,1;-1,-1}
\end{pmatrix},\label{eq:(A_=00007B0, 0=00007D)^=00007B-1=00007D A_=00007B0, 1=00007D -- 1}
\end{equation}

\noindent with

\begin{equation}
\left[\mathbf{A}_{\nu;r;\epsilon;n;0,0;1,-1}\mathbf{A}_{\nu;r;\epsilon;n;0,1;-1,-1}\right]_{q_{1},q_{2}}=\delta_{q_{1},0}\delta_{0,q_{2}}f_{\nu;\epsilon;n;-1;0}g_{\nu;\epsilon}.\label{eq:A_=00007B0, 0; 1, -1=00007D A_=00007B0, 1; -1, -1=00007D -- 1}
\end{equation}

\noindent Next, we calculate $\mathbf{A}_{\nu;r;\epsilon;n;1,0}\left(\mathbf{A}_{\nu;r;\epsilon;n;0,0}\right)^{-1}\mathbf{A}_{\nu;r;\epsilon;n;0,1}$.
Using Eqs.~\eqref{eq:A_=00007Bv; epsilon; n; 0, 0; 1, -1=00007D -- 1},
\eqref{eq:A_=00007Bv; epsilon; n; 0, 1; 1, 1=00007D -- 1}-\eqref{eq:A_=00007Bv; epsilon; n; 0, 1; -1, -1=00007D -- 1},
\eqref{eq:A_=00007Bv; epsilon; n; 1, 0=00007D -- 1}-\eqref{eq:A_=00007Bv; epsilon; n; 1, 0; -1, -1=00007D -- 1},
and \eqref{eq:(A_=00007B0, 0=00007D)^=00007B-1=00007D A_=00007B0, 1=00007D -- 1}
we get

\begin{align}
 & \mathbf{A}_{\nu;r;\epsilon;n;1,0}\left(\mathbf{A}_{\nu;r;\epsilon;n;0,0}\right)^{-1}\mathbf{A}_{\nu;r;\epsilon;n;0,1}\nonumber \\
 & \quad=\begin{pmatrix}\mathbf{A}_{\nu;r;\epsilon;n;1,0;1,1}\mathbf{A}_{\nu;r;\epsilon;n;0,1;1,1} & -\mathbf{A}_{\nu;r;\epsilon;n;1,0;1,1}\mathbf{A}_{\nu;r;\epsilon;n;0,0;1,-1}\mathbf{A}_{\nu;r;\epsilon;n;0,1;-1,-1}\\
\mathbf{A}_{\nu;r;\epsilon;n;1,0;-1,0}\mathbf{A}_{\nu;r;\epsilon;n;0,1;0,1} & \mathbf{A}_{\nu;r;\epsilon;n;1,0;-1,-1}\mathbf{A}_{\nu;r;\epsilon;n;0,1;-1,-1}
\end{pmatrix},\label{eq:A_=00007B1, 0=00007D (A_=00007B0, 0=00007D)^=00007B-1=00007D A_=00007B0, 1=00007D -- 1}
\end{align}

\noindent with

\begin{equation}
\left[\mathbf{A}_{\nu;r;\epsilon;n;1,0;1,1}\mathbf{A}_{\nu;r;\epsilon;n;0,1;1,1}\right]_{q_{1},q_{2}}=\delta_{q_{1},q_{2}+1}f_{\nu;\epsilon;n;1;q_{1}},\label{eq:A_=00007B1, 0; 1, 1=00007D A_=00007B0, 1; 1, 1=00007D -- 1}
\end{equation}

\begin{equation}
\left[\mathbf{A}_{\nu;r;\epsilon;n;1,0;1,1}\mathbf{A}_{\nu;r;\epsilon;n;0,0;1,-1}\mathbf{A}_{\nu;r;\epsilon;n;0,1;-1,-1}\right]_{q_{1},q_{2}}=-\delta_{q_{1},0}\delta_{0,q_{2}}g_{\nu;\epsilon},\label{eq:A_=00007B1, 0; 1, 1=00007D A_=00007B0, 0; 1, -1=00007D A_=00007B0, 1; -1, -1=00007D -- 1}
\end{equation}

\begin{equation}
\left[\mathbf{A}_{\nu;r;\epsilon;n;1,0;-1,0}\mathbf{A}_{\nu;r;\epsilon;n;0,1;0,1}\right]_{q_{1},q_{2}}=\delta_{q_{1},2n+1}\delta_{2n+1,q_{2}},\label{eq:A_=00007B1, 0; -1, 0=00007D A_=00007B0, 1; 0, 1=00007D -- 1}
\end{equation}

\begin{equation}
\left[\mathbf{A}_{\nu;r;\epsilon;n;1,0;-1,-1}\mathbf{A}_{\nu;r;\epsilon;n;0,1;-1,-1}\right]_{q_{1},q_{2}}=\delta_{q_{1}+1,q_{2}}f_{\nu;\epsilon;n;-1;q_{2}}.\label{eq:A_=00007B1, 0; -1, -1=00007D A_=00007B0, 1; -1, -1=00007D -- 1}
\end{equation}

\noindent Next, we calculate $\mathbf{B}_{\nu;r;\epsilon;n;1,1}$.
Using Eqs.~\eqref{eq:A_=00007Bv; epsilon; n; 1, 1=00007D -- 1},
\eqref{eq:B_=00007Bv; epsilon; n; 1, 1=00007D -- 1}, and \eqref{eq:A_=00007B1, 0=00007D (A_=00007B0, 0=00007D)^=00007B-1=00007D A_=00007B0, 1=00007D -- 1}-\eqref{eq:A_=00007B1, 0; -1, -1=00007D A_=00007B0, 1; -1, -1=00007D -- 1}
we get

\begin{equation}
\mathbf{B}_{\nu;r;\epsilon;n;1,1}=\begin{pmatrix}\mathbf{B}_{\nu;r;\epsilon;n;1,1;1,1} & \mathbf{B}_{\nu;r;\epsilon;n;1,1;1,-1}\\
\mathbf{B}_{\nu;r;\epsilon;n;1,1;-1,1} & \mathbf{B}_{\nu;r;\epsilon;n;1,1;-1,-1}
\end{pmatrix},\label{eq:B_=00007Bv; epsilon; n; 1, 1=00007D -- 2}
\end{equation}

\noindent with

\begin{equation}
\left[\mathbf{B}_{\nu;r;\epsilon;n;1,1;1,1}\right]_{q_{1},q_{2}}=\delta_{q_{1},q_{2}}-\delta_{q_{1},q_{2}+1}f_{\nu;\epsilon;n;1;q_{1}},\label{eq:B_=00007Bv; epsilon; n; 1, 1; 1, 1=00007D -- 1}
\end{equation}

\begin{equation}
\left[\mathbf{B}_{\nu;r;\epsilon;n;1,1;1,-1}\right]_{q_{1},q_{2}}=-\delta_{q_{1},0}\delta_{0,q_{2}}g_{\nu;\epsilon},\label{eq:B_=00007Bv; epsilon; n; 1, 1; 1, -1=00007D -- 1}
\end{equation}

\begin{equation}
\left[\mathbf{B}_{\nu;r;\epsilon;n;1,1;-1,1}\right]_{q_{1},q_{2}}=-\delta_{q_{1},2n+1}\delta_{2n+1,q_{2}},\label{eq:B_=00007Bv; epsilon; n; 1, 1; -1, 1=00007D -- 1}
\end{equation}

\begin{equation}
\left[\mathbf{B}_{\nu;r;\epsilon;n;1,1;-1,-1}\right]_{q_{1},q_{2}}=\delta_{q_{1},q_{2}}-\delta_{q_{1}+1,q_{2}}f_{\nu;\epsilon;n;-1;q_{2}}.\label{eq:B_=00007Bv; epsilon; n; 1, 1; -1, -1=00007D -- 1}
\end{equation}

Now that we have calculated $\mathbf{B}_{\nu;r;\epsilon;n;1,1}$,
we need to calculate its inverse, i.e. $\mathbf{C}_{\nu;r;\epsilon;n;1,1}$
{[}see Eq.~\eqref{eq:introduce C_=00007Bv; epsilon; n; 1, 1=00007D -- 1}{]}.
Using Eqs.~\eqref{eq:inverse of a block matrix -- 1}-\eqref{eq:=00005BM^=00007B-1=00007D=00005D_=00007B2, 2=00007D -- 1},
and \eqref{eq:introduce C_=00007Bv; epsilon; n; 1, 1=00007D -- 1},
we can write

\begin{equation}
\mathbf{C}_{\nu;r;\epsilon;n;1,1}=\begin{pmatrix}\mathbf{C}_{\nu;r;\epsilon;n;1,1;1,1} & \mathbf{C}_{\nu;r;\epsilon;n;1,1;1,-1}\\
\mathbf{C}_{\nu;r;\epsilon;n;1,1;-1,1} & \mathbf{C}_{\nu;r;\epsilon;n;1,1;-1,-1}
\end{pmatrix},\label{eq:C_=00007Bv; epsilon; n; 1, 1=00007D block structure -- 1}
\end{equation}

\noindent with

\begin{equation}
\mathbf{C}_{\nu;r;\epsilon;n;1,1;1,1}=\left(\mathbf{B}_{\nu;r;\epsilon;n;1,1;1,1}\right)^{-1}-\left(\mathbf{B}_{\nu;r;\epsilon;n;1,1;1,1}\right)^{-1}\mathbf{B}_{\nu;r;\epsilon;n;1,1;1,-1}\mathbf{C}_{\nu;r;\epsilon;n;1,1;-1,1},\label{eq:C_=00007Bv; epsilon; n; 1, 1; 1, 1=00007D -- 1}
\end{equation}

\begin{equation}
\mathbf{C}_{\nu;r;\epsilon;n;1,1;1,-1}=-\left(\mathbf{B}_{\nu;r;\epsilon;n;1,1;1,1}\right)^{-1}\mathbf{B}_{\nu;r;\epsilon;n;1,1;1,-1}\mathbf{C}_{\nu;r;\epsilon;n;1,1;-1,-1},\label{eq:C_=00007Bv; epsilon; n; 1, 1; 1, -1=00007D -- 1}
\end{equation}

\begin{equation}
\mathbf{C}_{\nu;r;\epsilon;n;1,1;-1,1}=-\mathbf{C}_{\nu;r;\epsilon;n;1,1;-1,-1}\mathbf{B}_{\nu;r;\epsilon;n;1,1;-1,1}\left(\mathbf{B}_{\nu;r;\epsilon;n;1,1;1,1}\right)^{-1},\label{eq:C_=00007Bv; epsilon; n; 1, 1; -1, 1=00007D -- 1}
\end{equation}

\begin{equation}
\mathbf{C}_{\nu;r;\epsilon;n;1,1;-1,-1}=\left(\mathbf{B}_{\nu;r;\epsilon;n;1,1;-1,-1}-\mathbf{B}_{\nu;r;\epsilon;n;1,1;-1,1}\left(\mathbf{B}_{\nu;r;\epsilon;n;1,1;1,1}\right)^{-1}\mathbf{B}_{\nu;r;\epsilon;n;1,1;1,-1}\right)^{-1}.\label{eq:C_=00007Bv; epsilon; n; 1, 1; -1, -1=00007D -- 1}
\end{equation}

\noindent Let us first calculate $\left(\mathbf{B}_{\nu;r;\epsilon;n;1,1;1,1}\right)^{-1}$.
For an invertible matrix $\mathbf{M}$, we can calculate its inverse
using the following formula:

\begin{equation}
\mathbf{M}^{-1}=\frac{1}{\det\left[\mathbf{M}\right]}\left(\mathbf{q}\right)^{\text{T}},\label{eq:inverse from cofactor matrix -- 1}
\end{equation}

\noindent where $\mathbf{q}$ is the cofactor matrix: 

\begin{equation}
\left[\mathbf{q}\right]_{q_{1},q_{2}}=\left(-1\right)^{q_{1}+q_{2}}\left[\mathbf{D}\right]_{q_{1},q_{2}},\label{eq:cofactor matrix -- 1}
\end{equation}

\noindent with $\left[\mathbf{D}\right]_{q_{1},q_{2}}$ being the
determinant of the matrix that results from deleting row $q_{1}$
and column $q_{2}$ of $\mathbf{M}$. By inspection, we applied the
above formula to $\mathbf{B}_{\nu;r;\epsilon;n;1,1;1,1}$ and obtained

\begin{equation}
\left[\left(\mathbf{B}_{\nu;r;\epsilon;n;1,1;1,1}\right)^{-1}\right]_{q_{1},q_{2}}=\delta_{q_{1},q_{2}}+\Theta_{q_{1},q_{2}+1}\prod_{q=q_{2}+1}^{q_{1}}\left\{ f_{\nu;\epsilon;n;1;q}\right\} .\label{eq:(B_=00007Bv; epsilon; n; 1, 1; 1, 1=00007D)^=00007B-1=00007D -- 1}
\end{equation}

\noindent with

\begin{equation}
\Theta_{q,q^{\prime}}=\begin{cases}
1, & \text{if }k\ge k^{\prime},\\
0, & \text{if }k<k^{\prime}.
\end{cases}\label{eq:introduce step function -- 1}
\end{equation}

\noindent The most straightforward way for the reader to convince
themselves of this is to simply evaluate the matrix multiplication
of $\mathbf{B}_{\nu;r;\epsilon;n;1,1;1,1}$ with its inverse using
Eqs.~\eqref{eq:B_=00007Bv; epsilon; n; 1, 1; 1, 1=00007D -- 1} and
\eqref{eq:(B_=00007Bv; epsilon; n; 1, 1; 1, 1=00007D)^=00007B-1=00007D -- 1}:

\begin{align*}
 & \left[\mathbf{B}_{\nu;r;\epsilon;n;1,1;1,1}\left(\mathbf{B}_{\nu;r;\epsilon;n;1,1;1,1}\right)^{-1}\right]_{q_{1},q_{2}}\\
 & \quad=\sum_{q_{3}}\left[\mathbf{B}_{\nu;r;\epsilon;n;1,1;1,1}\right]_{q_{1},q_{3}}\left[\left(\mathbf{B}_{\nu;r;\epsilon;n;1,1;1,1}\right)^{-1}\right]_{q_{3},q_{2}}\\
 & \quad=\sum_{q_{3}}\left(\delta_{q_{1},q_{3}}-\delta_{q_{1},q_{3}+1}f_{\nu;\epsilon;n;1;q_{1}}\right)\left(\delta_{q_{3},q_{2}}+\Theta_{q_{3},q_{2}+1}\prod_{q=q_{2}+1}^{q_{3}}\left\{ f_{\nu;\epsilon;n;1;q}\right\} \right)\\
 & \quad=\sum_{q_{3}}\delta_{q_{1},q_{3}}\delta_{q_{3},q_{2}}+\sum_{q_{3}}\delta_{q_{1},q_{3}}\Theta_{q_{3},q_{2}+1}\prod_{q=q_{2}+1}^{q_{3}}\left\{ f_{\nu;\epsilon;n;1;q}\right\} \\
 & \quad\phantom{=}\mathord{-}\sum_{q_{3}}\delta_{q_{1},q_{3}+1}\delta_{q_{3},q_{2}}f_{\nu;\epsilon;n;1;q_{1}}-\sum_{q_{3}}\delta_{q_{1},q_{3}+1}\Theta_{q_{3},q_{2}+1}f_{\nu;\epsilon;n;1;q_{1}}\prod_{q=q_{2}+1}^{q_{3}}\left\{ f_{\nu;\epsilon;n;1;q}\right\} \\
 & \quad=\delta_{q_{1},q_{2}}+\Theta_{q_{1},q_{2}+1}\prod_{q=q_{2}+1}^{q_{1}}\left\{ f_{\nu;\epsilon;n;1;q}\right\} \hphantom{f_{\nu;\epsilon;n;1;q_{1}}\prod_{q=q_{2}+1}^{q_{1}-1}\left\{ f_{\nu;\epsilon;n;1;q}\right\} \prod_{q=q_{2}+1}^{q_{3}}\left\{ f_{\nu;\epsilon;n;1;q}\right\} }
\end{align*}

\begin{align}
 & \quad\phantom{=}\mathord{-}\delta_{q_{1},q_{2}+1}f_{\nu;\epsilon;n;1;q_{1}}-\Theta_{q_{1}-1,q_{2}+1}f_{\nu;\epsilon;n;1;q_{1}}\prod_{q=q_{2}+1}^{q_{1}-1}\left\{ f_{\nu;\epsilon;n;1;q}\right\} \nonumber \\
 & \quad=\delta_{q_{1},q_{2}}+\Theta_{q_{1},q_{2}+1}\prod_{q=q_{2}+1}^{q_{1}}\left\{ f_{\nu;\epsilon;n;1;q}\right\} \hphantom{f_{\nu;\epsilon;n;1;q_{1}}\prod_{q=q_{2}+1}^{q_{1}-1}\left\{ f_{\nu;\epsilon;n;1;q}\right\} \prod_{q=q_{2}+1}^{q_{3}}\left\{ f_{\nu;\epsilon;n;1;q}\right\} }\nonumber \\
 & \quad\phantom{=}\mathord{-}\delta_{q_{1},q_{2}+1}f_{\nu;\epsilon;n;1;q_{1}}-\Theta_{q_{1}-1,q_{2}+1}\prod_{q=q_{2}+1}^{q_{1}}\left\{ f_{\nu;\epsilon;n;1;q}\right\} \nonumber \\
 & \quad=\delta_{q_{1},q_{2}}+\Theta_{q_{1},q_{2}+1}\prod_{q=q_{2}+1}^{q_{1}}\left\{ f_{\nu;\epsilon;n;1;q}\right\} -\Theta_{q_{1},q_{2}+1}\prod_{q=q_{2}+1}^{q_{1}}\left\{ f_{\nu;\epsilon;n;1;q}\right\} \nonumber \\
 & \quad=\delta_{q_{1},q_{2}}.\label{eq:checking (B_=00007Bv; epsilon; n; 1, 1; 1, 1=00007D)^=00007B-1=00007D -- 1}
\end{align}

\noindent Next, we calculate $\mathbf{B}_{\nu;r;\epsilon;n;1,1;-1,1}\left(\mathbf{B}_{\nu;r;\epsilon;n;1,1;1,1}\right)^{-1}$.
Using Eqs.~\eqref{eq:B_=00007Bv; epsilon; n; 1, 1; -1, 1=00007D -- 1}
and \eqref{eq:(B_=00007Bv; epsilon; n; 1, 1; 1, 1=00007D)^=00007B-1=00007D -- 1}
we get

\begin{align}
\left[\mathbf{B}_{\nu;r;\epsilon;n;1,1;-1,1}\left(\mathbf{B}_{\nu;r;\epsilon;n;1,1;1,1}\right)^{-1}\right]_{q_{1},q_{2}} & =\sum_{q_{3}}\left(-\delta_{q_{1},2n+1}\delta_{2n+1,q_{3}}\right)\left(\delta_{q_{3},q_{2}}+\Theta_{q_{3},q_{2}+1}\prod_{q=q_{2}+1}^{q_{3}}\left\{ f_{\nu;\epsilon;n;1;q}\right\} \right)\nonumber \\
 & =-\delta_{q_{1},2n+1}\left(\Theta_{2n+1,q_{2}+1}\prod_{q=q_{2}+1}^{2n+1}\left\{ f_{\nu;\epsilon;n;1;q}\right\} +\delta_{2n+1,q_{2}}\right).\label{eq:B_=00007B1, 1; -1, 1=00007D (B_=00007B1, 1; 1, 1=00007D)^=00007B-1=00007D -- 1}
\end{align}

\noindent Next, we calculate $\left(\mathbf{B}_{\nu;r;\epsilon;n;1,1;1,1}\right)^{-1}\mathbf{B}_{\nu;r;\epsilon;n;1,1;1,-1}$.
Using Eqs.~\eqref{eq:B_=00007Bv; epsilon; n; 1, 1; 1, -1=00007D -- 1}
and \eqref{eq:(B_=00007Bv; epsilon; n; 1, 1; 1, 1=00007D)^=00007B-1=00007D -- 1}
we get

\begin{align}
\left[\left(\mathbf{B}_{\nu;r;\epsilon;n;1,1;1,1}\right)^{-1}\mathbf{B}_{\nu;r;\epsilon;n;1,1;1,-1}\right]_{q_{1},q_{2}} & =\sum_{q_{3}}\left(\delta_{q_{1},q_{3}}+\Theta_{q_{1},q_{3}+1}\prod_{q=q_{3}+1}^{q_{1}}\left\{ f_{\nu;\epsilon;n;1;q}\right\} \right)\left(-\delta_{q_{3},0}\delta_{0,q_{2}}g_{\nu;\epsilon}\right)\nonumber \\
 & =-g_{\nu;\epsilon}\left(\delta_{q_{1},0}+\Theta_{q_{1},1}\prod_{q=1}^{q_{1}}\left\{ f_{\nu;\epsilon;n;1;q}\right\} \right)\delta_{0,q_{2}}.\label{eq:(B_=00007B1, 1; 1, 1=00007D)^=00007B-1=00007D B_=00007B1, 1; 1, -1=00007D -- 1}
\end{align}

\noindent Next, we calculate $\mathbf{B}_{\nu;r;\epsilon;n;1,1;-1,1}\left(\mathbf{B}_{\nu;r;\epsilon;n;1,1;1,1}\right)^{-1}\mathbf{B}_{\nu;r;\epsilon;n;1,1;1,-1}$.
Using Eqs.~\eqref{eq:B_=00007Bv; epsilon; n; 1, 1; -1, 1=00007D -- 1}
and \eqref{eq:(B_=00007B1, 1; 1, 1=00007D)^=00007B-1=00007D B_=00007B1, 1; 1, -1=00007D -- 1}
we get

\begin{align}
 & \left[\mathbf{B}_{\nu;r;\epsilon;n;1,1;-1,1}\left(\mathbf{B}_{\nu;r;\epsilon;n;1,1;1,1}\right)^{-1}\mathbf{B}_{\nu;r;\epsilon;n;1,1;1,-1}\right]_{q_{1},q_{2}}\nonumber \\
 & \quad=\sum_{q_{3}}\left(-\delta_{q_{1},2n+1}\delta_{2n+1,q_{3}}\right)\left(-g_{\nu;\epsilon}\left(\delta_{q_{3},0}+\Theta_{q_{3},1}\prod_{q=1}^{q_{3}}\left\{ f_{\nu;\epsilon;n;1;q}\right\} \right)\delta_{0,q_{2}}\right)\nonumber \\
 & \quad=g_{\nu;\epsilon}\prod_{q=1}^{2n+1}\left\{ f_{\nu;\epsilon;n;1;q}\right\} \delta_{q_{1},2n+1}\delta_{0,q_{2}}.\label{eq:B_=00007B1, 1; -1, 1=00007D (B_=00007B1, 1; 1, 1=00007D)^=00007B-1=00007D B_=00007B1, 1; 1, -1=00007D -- 1}
\end{align}

\noindent Next, we calculate $\left(\mathbf{C}_{\nu;r;\epsilon;n;1,1;-1,-1}\right)^{-1}$.
Using Eqs.~\eqref{eq:B_=00007Bv; epsilon; n; 1, 1; -1, -1=00007D -- 1},
\eqref{eq:C_=00007Bv; epsilon; n; 1, 1; -1, -1=00007D -- 1}, and
\eqref{eq:B_=00007B1, 1; -1, 1=00007D (B_=00007B1, 1; 1, 1=00007D)^=00007B-1=00007D B_=00007B1, 1; 1, -1=00007D -- 1}
we get

\begin{align}
\left[\left(\mathbf{C}_{\nu;r;\epsilon;n;1,1;-1,-1}\right)^{-1}\right]_{q_{1},q_{2}} & =\left[\mathbf{B}_{\nu;r;\epsilon;n;1,1;-1,-1}-\mathbf{B}_{\nu;r;\epsilon;n;1,1;-1,1}\left(\mathbf{B}_{\nu;r;\epsilon;n;1,1;1,1}\right)^{-1}\mathbf{B}_{\nu;r;\epsilon;n;1,1;1,-1}\right]_{q_{1},q_{2}}\nonumber \\
 & =\delta_{q_{1},q_{2}}-\delta_{q_{1}+1,q_{2}}f_{\nu;\epsilon;n;-1;q_{2}}-g_{\nu;\epsilon}\prod_{q=1}^{2n+1}\left\{ f_{\nu;\epsilon;n;1;q}\right\} \delta_{q_{1},2n+1}\delta_{0,q_{2}}.\label{eq:(C_=00007Bv; epsilon; n; 1, 1; -1, -1=00007D)^=00007B-1=00007D -- 1}
\end{align}

\noindent To calculate $\mathbf{C}_{\nu;r;\epsilon;n;1,1;-1,-1}$,
we calculate the inverse of $\left(\mathbf{C}_{\nu;r;\epsilon;n;1,1;-1,-1}\right)^{-1}$
as expressed in Eq.~\eqref{eq:(C_=00007Bv; epsilon; n; 1, 1; -1, -1=00007D)^=00007B-1=00007D -- 1}.
We use Eqs\@.~\eqref{eq:inverse from cofactor matrix -- 1} and
\eqref{eq:cofactor matrix -- 1} to calculate the inverse:

\begin{equation}
\left[\mathbf{C}_{\nu;r;\epsilon;n;1,1;-1,-1}\right]_{q_{1},q_{2}}=\frac{1}{1-g_{\nu;\epsilon}}\left(\delta_{q_{1},q_{2}}+\Theta_{q_{1},q_{2}+1}g_{\nu;\epsilon}\prod_{q=q_{2}+1}^{q_{1}}\left\{ f_{\nu;\epsilon;n;1;q}\right\} +\Theta_{q_{2},q_{1}+1}\prod_{q=q_{1}+1}^{q_{2}}\left\{ f_{\nu;\epsilon;n;-1;q}\right\} \right).\label{eq:C_=00007Bv; epsilon; n; 1, 1; -1, -1=00007D -- 2}
\end{equation}

\noindent Let us verify this result:

\begin{align}
 & \left[\left(\mathbf{C}_{\nu;r;\epsilon;n;1,1;-1,-1}\right)^{-1}\right]_{q_{1},q_{3}}\left[\mathbf{C}_{\nu;r;\epsilon;n;1,1;-1,-1}\right]_{q_{3},q_{2}}\nonumber \\
 & \quad=\sum_{q_{3}}\left(\delta_{q_{1},q_{3}}-\delta_{q_{1}+1,q_{3}}f_{\nu;\epsilon;n;-1;q_{3}}-g_{\nu;\epsilon}\prod_{q=1}^{2n+1}\left\{ f_{\nu;\epsilon;n;1;q}\right\} \delta_{q_{1},2n+1}\delta_{0,q_{3}}\right)\left(\frac{1}{1-g_{\nu;\epsilon}}\right)\nonumber \\
 & \quad\phantom{=\sum_{q_{3}}}\quad\times\left(\delta_{q_{3},q_{2}}+\Theta_{q_{3},q_{2}+1}g_{\nu;\epsilon}\prod_{q=q_{2}+1}^{q_{3}}\left\{ f_{\nu;\epsilon;n;1;q}\right\} +\Theta_{q_{2},q_{3}+1}\prod_{q=q_{3}+1}^{q_{2}}\left\{ f_{\nu;\epsilon;n;-1;q}\right\} \right)\nonumber \\
 & \quad=\left(\frac{1}{1-g_{\nu;\epsilon}}\right)\left(\delta_{q_{1},q_{2}}+\Theta_{q_{1},q_{2}+1}g_{\nu;\epsilon}\prod_{q=q_{2}+1}^{q_{1}}\left\{ f_{\nu;\epsilon;n;1;q}\right\} \right.\nonumber \\
 & \quad\phantom{=\left(\frac{1}{1-g_{\nu;\epsilon}}\right)}\left.\quad\mathord{+}\cancel{\Theta_{q_{2},q_{1}+1}\prod_{q=q_{1}+1}^{q_{2}}\left\{ f_{\nu;\epsilon;n;-1;q}\right\} }-\cancel{\delta_{q_{1}+1,q_{2}}f_{\nu;\epsilon;n;-1;q_{2}}}\right.\nonumber \\
 & \quad\phantom{=\left(\frac{1}{1-g_{\nu;\epsilon}}\right)}\left.\quad\mathord{-}\Theta_{2n+1,q_{1}}\Theta_{q_{1}+1,q_{2}+1}g_{\nu;\epsilon}f_{\nu;\epsilon;n;-1;q_{1}+1}\prod_{q=q_{2}+1}^{q_{1}+1}\left\{ f_{\nu;\epsilon;n;1;q}\right\} \right.\nonumber \\
 & \quad\phantom{=\left(\frac{1}{1-g_{\nu;\epsilon}}\right)}\left.\quad\mathord{-}\cancel{\Theta_{q_{2},q_{1}+2}\prod_{q=q_{1}+1}^{q_{2}}\left\{ f_{\nu;\epsilon;n;-1;q}\right\} }-\delta_{q_{1},2n+1}\delta_{0,q_{2}}g_{\nu;\epsilon}\prod_{q=1}^{2n+1}\left\{ f_{\nu;\epsilon;n;1;q}\right\} \right.\nonumber \\
 & \quad\phantom{=\left(\frac{1}{1-g_{\nu;\epsilon}}\right)}\left.\quad\mathord{-}\delta_{q_{1},2n+1}\Theta_{q_{2},1}g_{\nu;\epsilon}\prod_{q=1}^{2n+1}\left\{ f_{\nu;\epsilon;n;1;q}\right\} \prod_{q=1}^{q_{2}}\left\{ f_{\nu;\epsilon;n;-1;q}\right\} \right)\nonumber \\
 & \quad=\left(\frac{1}{1-g_{\nu;\epsilon}}\right)\left(\delta_{q_{1},q_{2}}+\Theta_{q_{1},q_{2}+1}g_{\nu;\epsilon}\prod_{q=q_{2}+1}^{q_{1}}\left\{ f_{\nu;\epsilon;n;1;q}\right\} \right.\nonumber \\
 & \quad\phantom{=\left(\frac{1}{1-g_{\nu;\epsilon}}\right)}\left.\quad\mathord{-}\left\{ \Theta_{2n+1,q_{1}}\delta_{q_{1},q_{2}}g_{\nu;\epsilon}+\Theta_{2n+1,q_{1}}\Theta_{q_{1},q_{2}+1}g_{\nu;\epsilon}\prod_{q=q_{2}+1}^{q_{1}}\left\{ f_{\nu;\epsilon;n;1;q}\right\} \right\} \right.\nonumber \\
 & \quad\phantom{=\left(\frac{1}{1-g_{\nu;\epsilon}}\right)}\left.\quad\mathord{-}\left\{ \delta_{q_{1},2n+1}\delta_{0,q_{2}}g_{\nu;\epsilon}\prod_{q=1}^{2n+1}\left\{ f_{\nu;\epsilon;n;1;q}\right\} \right.\right.\nonumber \\
 & \quad\phantom{=\left(\frac{1}{1-g_{\nu;\epsilon}}\right)}\left.\phantom{\quad\mathord{-}}\left.\quad\mathord{+}\delta_{q_{1},2n+1}\Theta_{q_{2},1}g_{\nu;\epsilon}\prod_{q=1}^{2n+1}\left\{ f_{\nu;\epsilon;n;1;q}\right\} \prod_{q=1}^{q_{2}}\left\{ f_{\nu;\epsilon;n;-1;q}\right\} \right\} \right)\nonumber \\
 & \quad=\left(\frac{1}{1-g_{\nu;\epsilon}}\right)\left(\delta_{q_{1},q_{2}}+\cancel{\Theta_{q_{1},q_{2}+1}g_{\nu;\epsilon}\prod_{q=q_{2}+1}^{q_{1}}\left\{ f_{\nu;\epsilon;n;1;q}\right\} }\right.\nonumber \\
 & \quad\phantom{=\left(\frac{1}{1-g_{\nu;\epsilon}}\right)}\left.\quad\mathord{-}\left\{ \Theta_{2n+1,q_{1}}\delta_{q_{1},q_{2}}g_{\nu;\epsilon}+\Theta_{2n+1,q_{1}}\Theta_{q_{1},q_{2}+1}g_{\nu;\epsilon}\prod_{q=q_{2}+1}^{q_{1}}\left\{ f_{\nu;\epsilon;n;1;q}\right\} \right\} \right.\nonumber \\
 & \quad\phantom{=\left(\frac{1}{1-g_{\nu;\epsilon}}\right)}\left.\quad\mathord{-}\left\{ \delta_{q_{1},2n+1}\delta_{q_{1},q_{2}}g_{\nu;\epsilon}+\cancel{\delta_{q_{1},2n+1}\Theta_{q_{1},q_{2}+1}g_{\nu;\epsilon}\prod_{q=q_{2}+1}^{q_{1}}\left\{ f_{\nu;\epsilon;n;1;q}\right\} }\right\} \right.\nonumber \\
 & \quad=\left(\frac{1}{1-g_{\nu;\epsilon}}\right)\left(\delta_{q_{1},q_{2}}-g_{\nu;\epsilon}\delta_{q_{1},q_{2}}\right)\nonumber \\
 & \quad=\delta_{q_{1},q_{2}}.\label{eq:checking (C_=00007Bv; epsilon; n; 1, 1; -1, -1=00007D)^=00007B-1=00007D -- 1}
\end{align}

\noindent Next, we calculate $\mathbf{C}_{\nu;r;\epsilon;n;1,1;1,-1}$.
Using Eqs.~\eqref{eq:C_=00007Bv; epsilon; n; 1, 1; 1, -1=00007D -- 1},
\eqref{eq:(B_=00007B1, 1; 1, 1=00007D)^=00007B-1=00007D B_=00007B1, 1; 1, -1=00007D -- 1},
and \eqref{eq:C_=00007Bv; epsilon; n; 1, 1; -1, -1=00007D -- 2} we
get

\begin{align}
 & \left[\mathbf{C}_{\nu;r;\epsilon;n;1,1;1,-1}\right]_{q_{1},q_{2}}\nonumber \\
 & \quad=-\sum_{q_{3}}\left[\left(\mathbf{B}_{\nu;r;\epsilon;n;1,1;1,1}\right)^{-1}\mathbf{B}_{\nu;r;\epsilon;n;1,1;1,-1}\right]_{q_{1},q_{3}}\left[\mathbf{C}_{\nu;r;\epsilon;n;1,1;-1,-1}\right]_{q_{3},q_{2}}\nonumber \\
 & \quad=\sum_{q_{3}}g_{\nu;\epsilon}\left(\delta_{q_{1},0}+\Theta_{q_{1},1}\prod_{q=1}^{q_{1}}\left\{ f_{\nu;\epsilon;n;1;q}\right\} \right)\delta_{0,q_{3}}\nonumber \\
 & \quad\phantom{=\sum_{q_{3}}}\quad\mathord{\times}\frac{1}{1-g_{\nu;\epsilon}}\left(\delta_{q_{3},q_{2}}+\Theta_{q_{3},q_{2}+1}g_{\nu;\epsilon}\prod_{q=q_{2}+1}^{q_{3}}\left\{ f_{\nu;\epsilon;n;1;q}\right\} +\Theta_{q_{2},q_{3}+1}\prod_{q=q_{3}+1}^{q_{2}}\left\{ f_{\nu;\epsilon;n;-1;q}\right\} \right)\nonumber \\
 & \quad=\frac{g_{\nu;\epsilon}}{1-g_{\nu;\epsilon}}\left(\delta_{q_{1},0}+\Theta_{q_{1},1}\prod_{q=1}^{q_{1}}\left\{ f_{\nu;\epsilon;n;1;q}\right\} \right)\nonumber \\
 & \quad\phantom{=\frac{g_{\nu;\epsilon}}{1-g_{\nu;\epsilon}}}\mathord{\times}\left(\sum_{q_{3}}\delta_{0,q_{3}}\delta_{q_{3},q_{2}}+\sum_{q_{3}}\delta_{0,q_{3}}\Theta_{q_{3},q_{2}+1}g_{\nu;\epsilon}\prod_{q=q_{2}+1}^{q_{3}}\left\{ f_{\nu;\epsilon;n;1;q}\right\} \right.\nonumber \\
 & \quad\phantom{=\frac{g_{\nu;\epsilon}}{1-g_{\nu;\epsilon}}\mathord{\times}}\left.\quad\mathord{+}\sum_{q_{3}}\delta_{0,q_{3}}\Theta_{q_{2},q_{3}+1}\prod_{q=q_{3}+1}^{q_{2}}\left\{ f_{\nu;\epsilon;n;-1;q}\right\} \right)\nonumber \\
 & \quad=\frac{g_{\nu;\epsilon}}{1-g_{\nu;\epsilon}}\left(\delta_{q_{1},0}+\Theta_{q_{1},1}\prod_{q=1}^{q_{1}}\left\{ f_{\nu;\epsilon;n;1;q}\right\} \right)\left(\delta_{0,q_{2}}+\Theta_{q_{2},1}\prod_{q=1}^{q_{2}}\left\{ f_{\nu;\epsilon;n;-1;q}\right\} \right)\nonumber \\
 & \quad=\frac{g_{\nu;\epsilon}}{1-g_{\nu;\epsilon}}\left(\delta_{q_{1},q_{2}}+\Theta_{q_{1},q_{2}+1}\prod_{q=q_{2}+1}^{q_{1}}\left\{ f_{\nu;\epsilon;n;1;q}\right\} +\Theta_{q_{2},q_{1}+1}\prod_{q=q_{1}+1}^{q_{2}}\left\{ f_{\nu;\epsilon;n;-1;q}\right\} \right).\label{eq:C_=00007Bv; epsilon; n; 1, 1; 1, -1=00007D -- 2}
\end{align}

\noindent Next, we calculate $\mathbf{C}_{\nu;r;\epsilon;n;1,1;-1,1}$.
Using Eqs.~\eqref{eq:C_=00007Bv; epsilon; n; 1, 1; -1, 1=00007D -- 1},
\eqref{eq:B_=00007B1, 1; -1, 1=00007D (B_=00007B1, 1; 1, 1=00007D)^=00007B-1=00007D -- 1},
and \eqref{eq:C_=00007Bv; epsilon; n; 1, 1; -1, -1=00007D -- 2} we
get

\begin{align}
 & \left[\mathbf{C}_{\nu;r;\epsilon;n;1,1;-1,1}\right]_{q_{1},q_{2}}\nonumber \\
 & \quad=-\sum_{q_{3}}\left[\mathbf{C}_{\nu;r;\epsilon;n;1,1;-1,-1}\right]_{q_{1},q_{3}}\left[\mathbf{B}_{\nu;r;\epsilon;n;1,1;-1,1}\left(\mathbf{B}_{\nu;r;\epsilon;n;1,1;1,1}\right)^{-1}\right]_{q_{3},q_{2}}\nonumber \\
 & \quad=\sum_{q_{3}}\frac{1}{1-g_{\nu;\epsilon}}\left(\delta_{q_{1},q_{3}}+\Theta_{q_{1},q_{3}+1}g_{\nu;\epsilon}\prod_{q=q_{3}+1}^{q_{1}}\left\{ f_{\nu;\epsilon;n;1;q}\right\} +\Theta_{q_{3},q_{1}+1}\prod_{q=q_{1}+1}^{q_{3}}\left\{ f_{\nu;\epsilon;n;-1;q}\right\} \right)\nonumber \\
 & \quad\phantom{\mathrel{=}\sum_{q_{3}}}\quad\mathop{\times}\left(\Theta_{2n+1,q_{2}+1}\prod_{q=q_{2}+1}^{2n+1}\left\{ f_{\nu;\epsilon;n;1;q}\right\} +\delta_{2n+1,q_{2}}\right)\delta_{q_{3},2n+1}\nonumber \\
 & \quad=\frac{1}{1-g_{\nu;\epsilon}}\left(\Theta_{2n+1,q_{2}+1}\prod_{q=q_{2}+1}^{2n+1}\left\{ f_{\nu;\epsilon;n;1;q}\right\} +\delta_{2n+1,q_{2}}\right)\nonumber \\
 & \quad\phantom{\mathrel{=}\frac{1}{1-g_{\nu;\epsilon}}}\quad\mathop{\times}\left(\sum_{q_{3}}\delta_{q_{1},q_{3}}\delta_{q_{3},2n+1}+\sum_{q_{3}}\Theta_{q_{1},q_{3}+1}\delta_{q_{3},2n+1}g_{\nu;\epsilon}\prod_{q=q_{3}+1}^{q_{1}}\left\{ f_{\nu;\epsilon;n;1;q}\right\} \right.\nonumber \\
 & \quad\phantom{\mathrel{=}\frac{1}{1-g_{\nu;\epsilon}}}\quad\mathop{\phantom{\times}}\left.\quad\mathop{+}\sum_{q_{3}}\Theta_{q_{3},q_{1}+1}\delta_{q_{3},2n+1}\prod_{q=q_{1}+1}^{q_{3}}\left\{ f_{\nu;\epsilon;n;-1;q}\right\} \right)\nonumber \\
 & \quad=\frac{1}{1-g_{\nu;\epsilon}}\left(\Theta_{2n+1,q_{1}+1}\prod_{q=q_{1}+1}^{2n+1}\left\{ f_{\nu;\epsilon;n;-1;q}\right\} +\delta_{q_{1},2n+1}\right)\nonumber \\
 & \quad\phantom{\mathrel{=}\frac{1}{1-g_{\nu;\epsilon}}}\quad\mathop{\times}\left(\Theta_{2n+1,q_{2}+1}\prod_{q=q_{2}+1}^{2n+1}\left\{ f_{\nu;\epsilon;n;1;q}\right\} +\delta_{2n+1,q_{2}}\right)\nonumber \\
 & \quad=\frac{1}{1-g_{\nu;\epsilon}}\left(\delta_{q_{1},q_{2}}+\Theta_{q_{1},q_{2}+1}\prod_{q=q_{2}+1}^{q_{1}}\left\{ f_{\nu;\epsilon;n;1;q}\right\} +\Theta_{q_{2},q_{1}+1}\prod_{q=q_{1}+1}^{q_{2}}\left\{ f_{\nu;\epsilon;n;-1;q}\right\} \right).\label{eq:C_=00007Bv; epsilon; n; 1, 1; -1, 1=00007D -- 2}
\end{align}

\noindent Next, we calculate $\mathbf{C}_{\nu;r;\epsilon;n;1,1;1,1}$.
Using Eqs.~\eqref{eq:C_=00007Bv; epsilon; n; 1, 1; 1, 1=00007D -- 1},
\eqref{eq:(B_=00007Bv; epsilon; n; 1, 1; 1, 1=00007D)^=00007B-1=00007D -- 1},
\eqref{eq:(B_=00007B1, 1; 1, 1=00007D)^=00007B-1=00007D B_=00007B1, 1; 1, -1=00007D -- 1},
and \eqref{eq:C_=00007Bv; epsilon; n; 1, 1; -1, -1=00007D -- 2} we
get

\begin{align}
 & \left[\mathbf{C}_{\nu;r;\epsilon;n;1,1;1,1}\right]_{q_{1},q_{2}}\nonumber \\
 & \quad=\left[\left(\mathbf{B}_{\nu;r;\epsilon;n;1,1;1,1}\right)^{-1}\right]_{q_{1},q_{2}}\nonumber \\
 & \quad\mathrel{\phantom{=}}\mathop{-}\sum_{q_{3}}\left[\left(\mathbf{B}_{\nu;r;\epsilon;n;1,1;1,1}\right)^{-1}\mathbf{B}_{\nu;r;\epsilon;n;1,1;1,-1}\right]_{q_{1},q_{3}}\left[\mathbf{C}_{\nu;r;\epsilon;n;1,1;-1,1}\right]_{q_{3},q_{2}}\nonumber \\
 & \quad=\delta_{q_{1},q_{2}}+\Theta_{q_{1},q_{2}+1}\prod_{q=q_{2}+1}^{q_{1}}\left\{ f_{\nu;\epsilon;n;1;q}\right\} \nonumber \\
 & \quad\mathrel{\phantom{=}}\mathop{+}\frac{g_{\nu;\epsilon}}{1-g_{\nu;\epsilon}}\sum_{q_{3}}\left(\delta_{q_{1},0}+\Theta_{q_{1},1}\prod_{q=1}^{q_{1}}\left\{ f_{\nu;\epsilon;n;1;q}\right\} \right)\delta_{0,q_{3}}\nonumber \\
 & \quad\phantom{\mathrel{\phantom{=}}\mathop{+}\frac{g_{\nu;\epsilon}}{1-g_{\nu;\epsilon}}\sum_{q_{3}}}\quad\mathop{\times}\left(\Theta_{2n+1,q_{3}+1}\prod_{q=q_{3}+1}^{2n+1}\left\{ f_{\nu;\epsilon;n;-1;q}\right\} +\delta_{q_{3},2n+1}\right)\nonumber \\
 & \quad\phantom{\mathrel{\phantom{=}}\mathop{+}\frac{g_{\nu;\epsilon}}{1-g_{\nu;\epsilon}}\sum_{q_{3}}}\quad\mathop{\times}\left(\Theta_{2n+1,q_{2}+1}\prod_{q=q_{2}+1}^{2n+1}\left\{ f_{\nu;\epsilon;n;1;q}\right\} +\delta_{2n+1,q_{2}}\right)\nonumber \\
 & \quad=\delta_{q_{1},q_{2}}+\Theta_{q_{1},q_{2}+1}\prod_{q=q_{2}+1}^{q_{1}}\left\{ f_{\nu;\epsilon;n;1;q}\right\} \nonumber \\
 & \quad\mathrel{\phantom{=}}\mathop{+}\frac{g_{\nu;\epsilon}}{1-g_{\nu;\epsilon}}\left(\delta_{q_{1},0}+\Theta_{q_{1},1}\prod_{q=1}^{q_{1}}\left\{ f_{\nu;\epsilon;n;1;q}\right\} \right)\prod_{q=1}^{2n+1}\left\{ f_{\nu;\epsilon;n;-1;q}\right\} \nonumber \\
 & \quad\phantom{\mathrel{\phantom{=}}\mathop{+}}\quad\mathop{\times}\left(\Theta_{2n+1,q_{2}+1}\prod_{q=q_{2}+1}^{2n+1}\left\{ f_{\nu;\epsilon;n;1;q}\right\} +\delta_{2n+1,q_{2}}\right)\nonumber \\
 & \quad=\left(1+\frac{g_{\nu;\epsilon}}{1-g_{\nu;\epsilon}}\right)\delta_{q_{1},q_{2}}\nonumber \\
 & \quad\mathrel{\phantom{=}}\mathop{+}\Theta_{q_{1},q_{2}+1}\left(1+\frac{g_{\nu;\epsilon}}{1-g_{\nu;\epsilon}}\right)\prod_{q=q_{2}+1}^{q_{1}}\left\{ f_{\nu;\epsilon;n;1;q}\right\} \nonumber \\
 & \quad\mathrel{\phantom{=}}\mathop{+}\Theta_{q_{2},q_{1}+1}\frac{g_{\nu;\epsilon}}{1-g_{\nu;\epsilon}}\prod_{q=q_{1}+1}^{q_{2}}\left\{ f_{\nu;\epsilon;n;-1;q}\right\} \nonumber \\
 & \quad=\frac{1}{1-g_{\nu;\epsilon}}\left(\delta_{q_{1},q_{2}}+\Theta_{q_{1},q_{2}+1}\prod_{q=q_{2}+1}^{q_{1}}\left\{ f_{\nu;\epsilon;n;1;q}\right\} +\Theta_{q_{2},q_{1}+1}g_{\nu;\epsilon}\prod_{q=q_{1}+1}^{q_{2}}\left\{ f_{\nu;\epsilon;n;-1;q}\right\} \right).\label{eq:C_=00007Bv; epsilon; n; 1, 1; 1, 1=00007D -- 2}
\end{align}

Next, we calculate $\mathbf{C}_{\nu;r;\epsilon;n;0,1}$. This is done
in several steps: first we cast $\mathbf{C}_{\nu;r;\epsilon;n;0,1}$
in block form, and then we calculate those elements of the block matrix
that we ultimately require in order to calculate the right-hand-side
of Eq.~\eqref{eq:Gaussian integral identity -- 1}. Using Eqs.~\eqref{eq:introduce C_=00007Bv; epsilon; n; 0, 1=00007D -- 1},
\eqref{eq:(A_=00007B0, 0=00007D)^=00007B-1=00007D A_=00007B0, 1=00007D -- 1},
and \eqref{eq:C_=00007Bv; epsilon; n; 1, 1=00007D block structure -- 1}
we get

\begin{equation}
\mathbf{C}_{\nu;r;\epsilon;n;0,1}=\begin{pmatrix}\mathbf{C}_{\nu;r;\epsilon;n;0,1;1,1} & \mathbf{C}_{\nu;r;\epsilon;n;0,1;1,-1}\\
\mathbf{C}_{\nu;r;\epsilon;n;0,1;0,1} & \mathbf{C}_{\nu;r;\epsilon;n;0,1;0,-1}\\
\mathbf{C}_{\nu;r;\epsilon;n;0,1;-1,1} & \mathbf{C}_{\nu;r;\epsilon;n;0,1;-1,-1}
\end{pmatrix},\label{eq:C_=00007Bv; epsilon; n; 0, 1=00007D block structure -- 1}
\end{equation}

\noindent with

\begin{align}
\mathbf{C}_{\nu;r;\epsilon;n;0,1;1,1} & =-\mathbf{A}_{\nu;r;\epsilon;n;0,1;1,1}\mathbf{C}_{\nu;r;\epsilon;n;1,1;1,1}\nonumber \\
 & \mathrel{\phantom{=}}\mathop{+}\mathbf{A}_{\nu;r;\epsilon;n;0,0;1,-1}\mathbf{A}_{\nu;r;\epsilon;n;0,1;-1,-1}\mathbf{C}_{\nu;r;\epsilon;n;1,1;-1,1},\label{eq:C_=00007Bv; epsilon; n; 0, 1; 1, 1=00007D -- 1}
\end{align}

\begin{align}
\mathbf{C}_{\nu;r;\epsilon;n;0,1;1,-1} & =-\mathbf{A}_{\nu;r;\epsilon;n;0,1;1,1}\mathbf{C}_{\nu;r;\epsilon;n;1,1;1,-1}\nonumber \\
 & \mathrel{\phantom{=}}\mathop{+}\mathbf{A}_{\nu;r;\epsilon;n;0,0;1,-1}\mathbf{A}_{\nu;r;\epsilon;n;0,1;-1,-1}\mathbf{C}_{\nu;r;\epsilon;n;1,1;-1,-1},\label{eq:C_=00007Bv; epsilon; n; 0, 1; 1, -1=00007D -- 1}
\end{align}

\begin{equation}
\mathbf{C}_{\nu;r;\epsilon;n;0,1;0,1}=-\mathbf{A}_{\nu;r;\epsilon;n;0,1;0,1}\mathbf{C}_{\nu;r;\epsilon;n;1,1;1,1},\label{eq:C_=00007Bv; epsilon; n; 0, 1; 0, 1=00007D -- 1}
\end{equation}

\begin{equation}
\mathbf{C}_{\nu;r;\epsilon;n;0,1;0,-1}=-\mathbf{A}_{\nu;r;\epsilon;n;0,1;0,1}\mathbf{C}_{\nu;r;\epsilon;n;1,1;1,-1},\label{eq:C_=00007Bv; epsilon; n; 0, 1; 0, -1=00007D -- 1}
\end{equation}

\begin{equation}
\mathbf{C}_{\nu;r;\epsilon;n;0,1;-1,1}=-\mathbf{A}_{\nu;r;\epsilon;n;0,1;-1,-1}\mathbf{C}_{\nu;r;\epsilon;n;1,1;-1,1},\label{eq:C_=00007Bv; epsilon; n; 0, 1; -1, 1=00007D -- 1}
\end{equation}

\noindent and

\begin{equation}
\mathbf{C}_{\nu;r;\epsilon;n;0,1;-1,-1}=-\mathbf{A}_{\nu;r;\epsilon;n;0,1;-1,-1}\mathbf{C}_{\nu;r;\epsilon;n;1,1;-1,-1}.\label{eq:C_=00007Bv; epsilon; n; 0, 1; -1, -1=00007D -- 1}
\end{equation}

\noindent From Eqs.~\eqref{eq:xi_=00007Bv; r; epsilon; n; j=00003D0; alpha=00003D0=00007D -- 1},
\eqref{eq:Gaussian integral identity -- 1}, \eqref{eq:introduce C_=00007Bv; epsilon; n=00007D -- 1},
\eqref{eq:C_=00007Bv; epsilon; n=00007D block structure -- 2}, and
\eqref{eq:C_=00007Bv; epsilon; n; 0, 1=00007D block structure -- 1},
we can see that the only matrix elements of Eq.~\eqref{eq:C_=00007Bv; epsilon; n; 0, 1=00007D block structure -- 1}
that we need to calculate explicitly are $\mathbf{C}_{\nu;r;\epsilon;n;0,1;1,1}$,
$\mathbf{C}_{\nu;r;\epsilon;n;0,1;1,-1}$, $\mathbf{C}_{\nu;r;\epsilon;n;0,1;-1,1}$,
and $\mathbf{C}_{\nu;r;\epsilon;n;0,1;-1,-1}$. First, we calculate
$\mathbf{C}_{\nu;r;\epsilon;n;0,1;1,1}$. Using Eqs.~\eqref{eq:A_=00007Bv; epsilon; n; 0, 1; 1, 1=00007D -- 1},
\eqref{eq:A_=00007B0, 0; 1, -1=00007D A_=00007B0, 1; -1, -1=00007D -- 1},
\eqref{eq:C_=00007Bv; epsilon; n; 1, 1; -1, 1=00007D -- 2}, \eqref{eq:C_=00007Bv; epsilon; n; 1, 1; 1, 1=00007D -- 2},
and \eqref{eq:C_=00007Bv; epsilon; n; 0, 1; 1, 1=00007D -- 1} we
get

\begin{align}
 & \left[\mathbf{C}_{\nu;r;\epsilon;n;0,1;1,1}\right]_{q_{1},q_{2}}\nonumber \\
 & \quad=-\sum_{q_{3}}\left[\mathbf{A}_{\nu;r;\epsilon;n;0,1;1,1}\right]_{q_{1},q_{3}}\left[\mathbf{C}_{\nu;r;\epsilon;n;1,1;1,1}\right]_{q_{3},q_{2}}\nonumber \\
 & \quad\mathrel{\phantom{=}}\mathop{+}\sum_{q_{3}}\left[\mathbf{A}_{\nu;r;\epsilon;n;0,0;1,-1}\mathbf{A}_{\nu;r;\epsilon;n;0,1;-1,-1}\right]_{q_{1},q_{3}}\left[\mathbf{C}_{\nu;r;\epsilon;n;1,1;-1,1}\right]_{q_{3},q_{2}}\nonumber \\
 & \quad=\frac{1}{1-g_{\nu;\epsilon}}\delta_{q_{1},q_{2}+1}+\Theta_{q_{1}-1,q_{2}+1}\frac{1}{1-g_{\nu;\epsilon}}\prod_{q=q_{2}+1}^{q_{1}-1}\left\{ f_{\nu;\epsilon;n;1;q}\right\} \nonumber \\
 & \quad\mathrel{\phantom{=}}\mathop{+}\Theta_{q_{2},q_{1}}\Theta_{q_{1},1}\frac{g_{\nu;\epsilon}}{1-g_{\nu;\epsilon}}\prod_{q=q_{1}}^{q_{2}}\left\{ f_{\nu;\epsilon;n;-1;q}\right\} \nonumber \\
 & \quad\mathrel{\phantom{=}}\mathop{+}\delta_{q_{1},0}f_{\nu;\epsilon;n;-1;0}\frac{g_{\nu;\epsilon}}{1-g_{\nu;\epsilon}}\left(\delta_{0,q_{2}}+\Theta_{q_{2},1}\prod_{q=1}^{q_{2}}\left\{ f_{\nu;\epsilon;n;-1;q}\right\} \right)\nonumber \\
 & \quad=\frac{1}{1-g_{\nu;\epsilon}}\left(\delta_{q_{1},q_{2}+1}+\Theta_{q_{1}-1,q_{2}+1}\prod_{q=q_{2}+1}^{q_{1}-1}\left\{ f_{\nu;\epsilon;n;1;q}\right\} +\Theta_{q_{2},q_{1}}g_{\nu;\epsilon}\prod_{q=q_{1}}^{q_{2}}\left\{ f_{\nu;\epsilon;n;-1;q}\right\} \right).\label{eq:C_=00007Bv; epsilon; n; 0, 1; 1, 1=00007D -- 2}
\end{align}

\noindent Next, we calculate $\mathbf{C}_{\nu;r;\epsilon;n;0,1;1,-1}$.
Using Eqs.~\eqref{eq:A_=00007Bv; epsilon; n; 0, 1; 1, 1=00007D -- 1},
\eqref{eq:A_=00007B0, 0; 1, -1=00007D A_=00007B0, 1; -1, -1=00007D -- 1},
\eqref{eq:C_=00007Bv; epsilon; n; 1, 1; -1, -1=00007D -- 2}, \eqref{eq:C_=00007Bv; epsilon; n; 1, 1; 1, -1=00007D -- 2},
and \eqref{eq:C_=00007Bv; epsilon; n; 0, 1; 1, -1=00007D -- 1} we
get

\begin{align}
 & \left[\mathbf{C}_{\nu;r;\epsilon;n;0,1;1,-1}\right]_{q_{1},q_{2}}\nonumber \\
 & \quad=-\sum_{q_{3}}\left[\mathbf{A}_{\nu;r;\epsilon;n;0,1;1,1}\right]_{q_{1},q_{3}}\left[\mathbf{C}_{\nu;r;\epsilon;n;1,1;1,-1}\right]_{q_{3},q_{2}}\nonumber \\
 & \quad\mathrel{\phantom{=}}\mathop{+}\sum_{q_{3}}\left[\mathbf{A}_{\nu;r;\epsilon;n;0,0;1,-1}\mathbf{A}_{\nu;r;\epsilon;n;0,1;-1,-1}\right]_{q_{1},q_{3}}\left[\mathbf{C}_{\nu;r;\epsilon;n;1,1;-1,-1}\right]_{q_{3},q_{2}}\nonumber \\
 & \quad=\frac{g_{\nu;\epsilon}}{1-g_{\nu;\epsilon}}\left(\delta_{q_{1}-1,q_{2}}+\Theta_{q_{1}-1,q_{2}+1}\prod_{q=q_{2}+1}^{q_{1}-1}\left\{ f_{\nu;\epsilon;n;1;q}\right\} +\Theta_{q_{2},q_{1}}\Theta_{q_{1},1}\prod_{q=q_{1}}^{q_{2}}\left\{ f_{\nu;\epsilon;n;-1;q}\right\} \right)\nonumber \\
 & \quad\mathrel{\phantom{=}}\mathop{+}\frac{1}{1-g_{\nu;\epsilon}}\left(\delta_{q_{1},0}\delta_{0,q_{2}}f_{\nu;\epsilon;n;-1;0}g_{\nu;\epsilon}+\Theta_{q_{2},1}\delta_{q_{1},0}g_{\nu;\epsilon}\prod_{q=0}^{q_{2}}\left\{ f_{\nu;\epsilon;n;-1;q}\right\} \right)\nonumber \\
 & \quad\mathrel{=}\frac{g_{\nu;\epsilon}}{1-g_{\nu;\epsilon}}\left(\delta_{q_{1},q_{2}+1}+\Theta_{q_{1}-1,q_{2}+1}\prod_{q=q_{2}+1}^{q_{1}-1}\left\{ f_{\nu;\epsilon;n;1;q}\right\} +\Theta_{q_{2},q_{1}}\prod_{q=q_{1}}^{q_{2}}\left\{ f_{\nu;\epsilon;n;-1;q}\right\} \right).\label{eq:C_=00007Bv; epsilon; n; 0, 1; 1, -1=00007D -- 2}
\end{align}

\noindent Next, we calculate $\mathbf{C}_{\nu;r;\epsilon;n;0,1;-1,1}$.
Using Eqs.~\eqref{eq:A_=00007Bv; epsilon; n; 0, 1; -1, -1=00007D -- 1},
\eqref{eq:C_=00007Bv; epsilon; n; 1, 1; -1, 1=00007D -- 2}, and \eqref{eq:C_=00007Bv; epsilon; n; 0, 1; -1, 1=00007D -- 1}
we get

\begin{align}
 & \left[\mathbf{C}_{\nu;r;\epsilon;n;0,1;-1,1}\right]_{q_{1},q_{2}}\nonumber \\
 & \quad=-\sum_{q_{3}}\left[\mathbf{A}_{\nu;r;\epsilon;n;0,1;-1,-1}\right]_{q_{1},q_{3}}\left[\mathbf{C}_{\nu;r;\epsilon;n;1,1;-1,1}\right]_{q_{3},q_{2}}\nonumber \\
 & \quad=\frac{1}{1-g_{\nu;\epsilon}}\left(\delta_{q_{1},q_{2}}f_{\nu;\epsilon;n;-1;q_{1}}+\Theta_{q_{1},q_{2}+1}f_{\nu;\epsilon;n;-1;q_{1}}\prod_{q=q_{2}+1}^{q_{1}}\left\{ f_{\nu;\epsilon;n;1;q}\right\} +\Theta_{q_{2},q_{1}+1}\prod_{q=q_{1}}^{q_{2}}\left\{ f_{\nu;\epsilon;n;-1;q}\right\} \right)\nonumber \\
 & \quad=\frac{1}{1-g_{\nu;\epsilon}}\left(\delta_{q_{1},q_{2}+1}+\Theta_{q_{1}-1,q_{2}+1}\prod_{q=q_{2}+1}^{q_{1}-1}\left\{ f_{\nu;\epsilon;n;1;q}\right\} +\Theta_{q_{2},q_{1}}\prod_{q=q_{1}}^{q_{2}}\left\{ f_{\nu;\epsilon;n;-1;q}\right\} \right).\label{eq:C_=00007Bv; epsilon; n; 0, 1; -1, 1=00007D -- 2}
\end{align}

\noindent Next, we calculate $\mathbf{C}_{\nu;r;\epsilon;n;0,1;-1,-1}$.
Using Eqs.~\eqref{eq:A_=00007Bv; epsilon; n; 0, 1; -1, -1=00007D -- 1},
\eqref{eq:C_=00007Bv; epsilon; n; 1, 1; -1, -1=00007D -- 2}, and
\eqref{eq:C_=00007Bv; epsilon; n; 0, 1; -1, -1=00007D -- 1} we get

\begin{align}
 & \left[\mathbf{C}_{\nu;r;\epsilon;n;0,1;-1,-1}\right]_{q_{1},q_{2}}\nonumber \\
 & \quad=-\sum_{q_{3}}\left[\mathbf{A}_{\nu;r;\epsilon;n;0,1;-1,-1}\right]_{q_{1},q_{3}}\left[\mathbf{C}_{\nu;r;\epsilon;n;1,1;-1,-1}\right]_{q_{3},q_{2}}\nonumber \\
 & \quad=\frac{1}{1-g_{\nu;\epsilon}}\left(\delta_{q_{1},q_{2}}f_{\nu;\epsilon;n;-1;q_{1}}+\Theta_{q_{1},q_{2}+1}g_{\nu;\epsilon}f_{\nu;\epsilon;n;-1;q_{1}}\prod_{q=q_{2}+1}^{q_{1}}\left\{ f_{\nu;\epsilon;n;1;q}\right\} +\Theta_{q_{2},q_{1}+1}\prod_{q=q_{1}}^{q_{2}}\left\{ f_{\nu;\epsilon;n;-1;q}\right\} \right)\nonumber \\
 & \quad=\frac{1}{1-g_{\nu;\epsilon}}\left(\Theta_{q_{2},q_{1}}\prod_{q=q_{1}}^{q_{2}}\left\{ f_{\nu;\epsilon;n;-1;q}\right\} +g_{\nu;\epsilon}\left[\delta_{q_{1},q_{2}+1}+\Theta_{q_{1}-1,q_{2}+1}\prod_{q=q_{2}+1}^{q_{1}-1}\left\{ f_{\nu;\epsilon;n;1;q}\right\} \right]\right)\nonumber \\
 & \quad=\frac{1}{1-g_{\nu;\epsilon}}\left(\delta_{q_{1},q_{2}+1}g_{\nu;\epsilon}+\Theta_{q_{1}-1,q_{2}+1}g_{\nu;\epsilon}\prod_{q=q_{2}+1}^{q_{1}-1}\left\{ f_{\nu;\epsilon;n;1;q}\right\} +\Theta_{q_{2},q_{1}}\prod_{q=q_{1}}^{q_{2}}\left\{ f_{\nu;\epsilon;n;-1;q}\right\} \right).\label{eq:C_=00007Bv; epsilon; n; 0, 1; -1, -1=00007D -- 2}
\end{align}

Next, we calculate $\mathbf{C}_{\nu;r;\epsilon;n;1,0}$. Using Eqs.~\eqref{eq:introduce C_=00007Bv; epsilon; n; 1, 0=00007D -- 1},
\eqref{eq:A_=00007B1, 0=00007D (A_=00007B0, 0=00007D)^=00007B-1=00007D -- 1},
and \eqref{eq:C_=00007Bv; epsilon; n; 1, 1=00007D block structure -- 1}
we get

\begin{equation}
\mathbf{C}_{\nu;r;\epsilon;n;1,0}=\begin{pmatrix}\mathbf{C}_{\nu;r;\epsilon;n;1,0;1,1} & \mathbf{C}_{\nu;r;\epsilon;n;1,0;1,0} & \mathbf{C}_{\nu;r;\epsilon;n;1,0;1,-1}\\
\mathbf{C}_{\nu;r;\epsilon;n;1,0;-1,1} & \mathbf{C}_{\nu;r;\epsilon;n;1,0;-1,0} & \mathbf{C}_{\nu;r;\epsilon;n;1,0;-1,-1}
\end{pmatrix},\label{eq:C_=00007Bv; epsilon; n; 1, 0=00007D block structure -- 1}
\end{equation}

\noindent with

\begin{equation}
\mathbf{C}_{\nu;r;\epsilon;n;1,0;1,1}=\mathbf{C}_{\nu;r;\epsilon;n;1,1;1,1}\mathbf{A}_{\nu;r;\epsilon;n;1,0;1,1},\label{eq:C_=00007Bv; epsilon; n; 1, 0; 1, 1=00007D -- 1}
\end{equation}

\begin{equation}
\mathbf{C}_{\nu;r;\epsilon;n;1,0;1,0}=\mathbf{C}_{\nu;r;\epsilon;n;1,1;1,-1}\mathbf{A}_{\nu;r;\epsilon;n;1,0;-1,0},\label{eq:C_=00007Bv; epsilon; n; 1, 0; 1, 0=00007D -- 1}
\end{equation}

\begin{align}
\mathbf{C}_{\nu;r;\epsilon;n;1,0;1,-1} & =\mathbf{C}_{\nu;r;\epsilon;n;1,1;1,1}\mathbf{A}_{\nu;r;\epsilon;n;1,0;1,1}\mathbf{A}_{\nu;r;\epsilon;n;0,0;1,-1}\nonumber \\
 & \mathrel{\phantom{=}}\mathop{-}\mathbf{C}_{\nu;r;\epsilon;n;1,1;1,-1}\mathbf{A}_{\nu;r;\epsilon;n;1,0;-1,-1},\label{eq:C_=00007Bv; epsilon; n; 1, 0; 1, -1=00007D -- 1}
\end{align}

\begin{equation}
\mathbf{C}_{\nu;r;\epsilon;n;1,0;-1,1}=\mathbf{C}_{\nu;r;\epsilon;n;1,1;-1,1}\mathbf{A}_{\nu;r;\epsilon;n;1,0;1,1},\label{eq:C_=00007Bv; epsilon; n; 1, 0; -1, 1=00007D -- 1}
\end{equation}

\begin{equation}
\mathbf{C}_{\nu;r;\epsilon;n;1,0;-1,0}=\mathbf{C}_{\nu;r;\epsilon;n;1,1;-1,-1}\mathbf{A}_{\nu;r;\epsilon;n;1,0;-1,0},\label{eq:C_=00007Bv; epsilon; n; 1, 0; -1, 0=00007D -- 1}
\end{equation}

\noindent and

\begin{align}
\mathbf{C}_{\nu;r;\epsilon;n;1,0;-1,-1} & =\mathbf{C}_{\nu;r;\epsilon;n;1,1;-1,1}\mathbf{A}_{\nu;r;\epsilon;n;1,0;1,1}\mathbf{A}_{\nu;r;\epsilon;n;0,0;1,-1}\nonumber \\
 & \mathrel{\phantom{=}}\mathop{-}\mathbf{C}_{\nu;r;\epsilon;n;1,1;-1,-1}\mathbf{A}_{\nu;r;\epsilon;n;1,0;-1,-1}.\label{eq:C_=00007Bv; epsilon; n; 1, 0; -1, -1=00007D -- 1}
\end{align}

\noindent From Eqs.~\eqref{eq:xi_=00007Bv; r; epsilon; n; j=00003D0; alpha=00003D0=00007D -- 1},
\eqref{eq:Gaussian integral identity -- 1}, \eqref{eq:introduce C_=00007Bv; epsilon; n=00007D -- 1},
\eqref{eq:C_=00007Bv; epsilon; n=00007D block structure -- 2}, and
\eqref{eq:C_=00007Bv; epsilon; n; 1, 0=00007D block structure -- 1},
we can see that the only matrix elements of Eq.~\eqref{eq:C_=00007Bv; epsilon; n; 0, 1=00007D block structure -- 1}
that we need to calculate explicitly are $\mathbf{C}_{\nu;r;\epsilon;n;1,0;1,1}$,
$\mathbf{C}_{\nu;r;\epsilon;n;1,0;1,-1}$, $\mathbf{C}_{\nu;r;\epsilon;n;1,0;-1,1}$,
and $\mathbf{C}_{\nu;r;\epsilon;n;1,0;-1,-1}$. First, we calculate
$\mathbf{C}_{\nu;r;\epsilon;n;1,0;1,1}$. Using Eqs.~\eqref{eq:A_=00007Bv; epsilon; n; 1, 0; 1, 1=00007D -- 1},
\eqref{eq:C_=00007Bv; epsilon; n; 1, 1; 1, 1=00007D -- 1}, and \eqref{eq:C_=00007Bv; epsilon; n; 1, 0; 1, 1=00007D -- 1}
we get

\begin{align}
 & \left[\mathbf{C}_{\nu;r;\epsilon;n;1,0;1,1}\right]_{q_{1},q_{2}}\nonumber \\
 & \quad=-\sum_{q_{3}}\left[\mathbf{C}_{\nu;r;\epsilon;n;1,1;1,1}\right]_{q_{1},q_{3}}\left[\mathbf{A}_{\nu;r;\epsilon;n;1,0;1,1}\right]_{q_{3},q_{2}}\nonumber \\
 & \quad=\frac{1}{1-g_{\nu;\epsilon}}\left(\delta_{q_{1},q_{2}}f_{\nu;\epsilon;n;1;q_{2}}+\Theta_{q_{1},q_{2}+1}\prod_{q=q_{2}}^{q_{1}}\left\{ f_{\nu;\epsilon;n;1;q}\right\} +\Theta_{q_{2},q_{1}+1}f_{\nu;\epsilon;n;1;q_{2}}g_{\nu;\epsilon}\prod_{q=q_{1}+1}^{q_{2}}\left\{ f_{\nu;\epsilon;n;-1;q}\right\} \right)\nonumber \\
 & \quad=\frac{1}{1-g_{\nu;\epsilon}}\left(\Theta_{q_{1},q_{2}}\prod_{q=q_{2}}^{q_{1}}\left\{ f_{\nu;\epsilon;n;1;q}\right\} +g_{\nu;\epsilon}\left[\delta_{q_{2},q_{1}+1}+\Theta_{q_{2}-1,q_{1}+1}\prod_{q=q_{1}+1}^{q_{2}-1}\left\{ f_{\nu;\epsilon;n;-1;q}\right\} \right]\right)\nonumber \\
 & \quad=\frac{1}{1-g_{\nu;\epsilon}}\left(\delta_{q_{2},q_{1}+1}g_{\nu;\epsilon}+\Theta_{q_{2}-1,q_{1}+1}g_{\nu;\epsilon}\prod_{q=q_{1}+1}^{q_{2}-1}\left\{ f_{\nu;\epsilon;n;-1;q}\right\} +\Theta_{q_{1},q_{2}}\prod_{q=q_{2}}^{q_{1}}\left\{ f_{\nu;\epsilon;n;1;q}\right\} \right).\label{eq:C_=00007Bv; epsilon; n; 1, 0; 1, 1=00007D -- 2}
\end{align}

\noindent Next, we calculate $\mathbf{C}_{\nu;r;\epsilon;n;1,0;1,-1}$.
Using Eqs.~\eqref{eq:A_=00007Bv; epsilon; n; 1, 0; -1, -1=00007D -- 1},
\eqref{eq:A_=00007B1, 0; 1, 1=00007D A_=00007B0, 0; 1, -1=00007D -- 1},
\eqref{eq:C_=00007Bv; epsilon; n; 1, 1; 1, -1=00007D -- 2}, \eqref{eq:C_=00007Bv; epsilon; n; 1, 1; 1, 1=00007D -- 2},
and \eqref{eq:C_=00007Bv; epsilon; n; 1, 0; 1, -1=00007D -- 1} we
get

\begin{align}
 & \left[\mathbf{C}_{\nu;r;\epsilon;n;1,0;1,-1}\right]_{q_{1},q_{2}}\nonumber \\
 & \quad=\sum_{q_{3}}\left[\mathbf{C}_{\nu;r;\epsilon;n;1,1;1,1}\right]_{q_{1},q_{3}}\left[\mathbf{A}_{\nu;r;\epsilon;n;1,0;1,1}\mathbf{A}_{\nu;r;\epsilon;n;0,0;1,-1}\right]_{q_{3},q_{2}}\nonumber \\
 & \quad\mathrel{\phantom{=}}\mathop{-}\sum_{q_{3}}\left[\mathbf{C}_{\nu;r;\epsilon;n;1,1;1,-1}\right]_{q_{1},q_{3}}\left[\mathbf{A}_{\nu;r;\epsilon;n;1,0;-1,-1}\right]_{q_{3},q_{2}}\nonumber \\
 & \quad=\frac{g_{\nu;\epsilon}}{1-g_{\nu;\epsilon}}\left(\delta_{q_{1},0}\delta_{0,q_{2}}f_{\nu;\epsilon;n;1;0}+\Theta_{q_{1},1}\delta_{0,q_{2}}f_{\nu;\epsilon;n;1;0}\prod_{q=1}^{q_{1}}\left\{ f_{\nu;\epsilon;n;1;q}\right\} \right.\nonumber \\
 & \quad\phantom{\mathrel{=}\frac{g_{\nu;\epsilon}}{1-g_{\nu;\epsilon}}}\left.\quad\mathop{+}\delta_{q_{1},q_{2}-1}+\Theta_{q_{1},q_{2}}\Theta_{q_{2},1}\prod_{q=q_{2}}^{q_{1}}\left\{ f_{\nu;\epsilon;n;1;q}\right\} +\Theta_{q_{2}-1,q_{1}+1}\prod_{q=q_{1}+1}^{q_{2}-1}\left\{ f_{\nu;\epsilon;n;-1;q}\right\} \right)\nonumber \\
 & \quad=\frac{g_{\nu;\epsilon}}{1-g_{\nu;\epsilon}}\left(\delta_{0,q_{2}}\prod_{q=0}^{q_{1}}\left\{ f_{\nu;\epsilon;n;1;q}\right\} +\delta_{q_{1},q_{2}-1}+\Theta_{q_{1},q_{2}}\Theta_{q_{2},1}\prod_{q=q_{2}}^{q_{1}}\left\{ f_{\nu;\epsilon;n;1;q}\right\} +\Theta_{q_{2}-1,q_{1}+1}\prod_{q=q_{1}+1}^{q_{2}-1}\left\{ f_{\nu;\epsilon;n;-1;q}\right\} \right)\nonumber \\
 & \quad=\frac{g_{\nu;\epsilon}}{1-g_{\nu;\epsilon}}\left(\delta_{q_{1},q_{2}-1}+\Theta_{q_{1},q_{2}}\prod_{q=q_{2}}^{q_{1}}\left\{ f_{\nu;\epsilon;n;1;q}\right\} +\Theta_{q_{2}-1,q_{1}+1}\prod_{q=q_{1}+1}^{q_{2}-1}\left\{ f_{\nu;\epsilon;n;-1;q}\right\} \right).\label{eq:C_=00007Bv; epsilon; n; 1, 0; 1, -1=00007D -- 2}
\end{align}

\noindent Next, we calculate $\mathbf{C}_{\nu;r;\epsilon;n;1,0;-1,1}$.
Using Eqs.~\eqref{eq:A_=00007Bv; epsilon; n; 1, 0; 1, 1=00007D -- 1},
\eqref{eq:C_=00007Bv; epsilon; n; 1, 1; -1, 1=00007D -- 2}, and \eqref{eq:C_=00007Bv; epsilon; n; 1, 0; -1, 1=00007D -- 1}
we get

\begin{align}
 & \left[\mathbf{C}_{\nu;r;\epsilon;n;1,0;-1,1}\right]_{q_{1},q_{2}}\nonumber \\
 & \quad=-\sum_{q_{3}}\left[\mathbf{C}_{\nu;r;\epsilon;n;1,1;-1,1}\right]_{q_{1},q_{3}}\left[\mathbf{A}_{\nu;r;\epsilon;n;1,0;1,1}\right]_{q_{3},q_{2}}\nonumber \\
 & \quad=\frac{1}{1-g_{\nu;\epsilon}}\left(\delta_{q_{1},q_{2}}f_{\nu;\epsilon;n;1;q_{2}}+\Theta_{q_{1},q_{2}+1}\prod_{q=q_{2}}^{q_{1}}\left\{ f_{\nu;\epsilon;n;1;q}\right\} +\Theta_{q_{2},q_{1}+1}f_{\nu;\epsilon;n;1;q_{2}}\prod_{q=q_{1}+1}^{q_{2}}\left\{ f_{\nu;\epsilon;n;-1;q}\right\} \right)\nonumber \\
 & \quad=\frac{1}{1-g_{\nu;\epsilon}}\left(\delta_{q_{2},q_{1}+1}+\Theta_{q_{1},q_{2}}\prod_{q=q_{2}}^{q_{1}}\left\{ f_{\nu;\epsilon;n;1;q}\right\} +\Theta_{q_{2}-1,q_{1}+1}\prod_{q=q_{1}+1}^{q_{2}-1}\left\{ f_{\nu;\epsilon;n;-1;q}\right\} \right).\label{eq:C_=00007Bv; epsilon; n; 1, 0; -1, 1=00007D -- 2}
\end{align}

\noindent Next, we calculate $\mathbf{C}_{\nu;r;\epsilon;n;1,0;-1,-1}$.
Using Eqs.~\eqref{eq:A_=00007Bv; epsilon; n; 1, 0; -1, -1=00007D -- 1},
\eqref{eq:A_=00007B1, 0; 1, 1=00007D A_=00007B0, 0; 1, -1=00007D -- 1},
\eqref{eq:C_=00007Bv; epsilon; n; 1, 1; -1, 1=00007D -- 1}, \eqref{eq:C_=00007Bv; epsilon; n; 1, 1; -1, -1=00007D -- 1},
and \eqref{eq:C_=00007Bv; epsilon; n; 1, 0; -1, -1=00007D -- 1} we
get

\begin{align}
 & \left[\mathbf{C}_{\nu;r;\epsilon;n;1,0;-1,-1}\right]_{q_{1},q_{2}}\nonumber \\
 & \quad=\sum_{q_{3}}\left[\mathbf{C}_{\nu;r;\epsilon;n;1,1;-1,1}\right]_{q_{1},q_{3}}\left[\mathbf{A}_{\nu;r;\epsilon;n;1,0;1,1}\mathbf{A}_{\nu;r;\epsilon;n;0,0;1,-1}\right]_{q_{3},q_{2}}\nonumber \\
 & \quad\phantom{=}\mathord{-}\sum_{q_{3}}\left[\mathbf{C}_{\nu;r;\epsilon;n;1,1;-1,-1}\right]_{q_{1},q_{3}}\left[\mathbf{A}_{\nu;r;\epsilon;n;1,0;-1,-1}\right]_{q_{3},q_{2}}\nonumber \\
 & \quad=\frac{1}{1-g_{\nu;\epsilon}}\left(\delta_{q_{1},0}\delta_{0,q_{2}}f_{\nu;\epsilon;n;1;0}g_{\nu;\epsilon}+\Theta_{q_{1},1}\delta_{0,q_{2}}f_{\nu;\epsilon;n;1;0}g_{\nu;\epsilon}\prod_{q=1}^{q_{1}}\left\{ f_{\nu;\epsilon;n;1;q}\right\} \right.\nonumber \\
 & \quad\phantom{=\frac{1}{1-g_{\nu;\epsilon}}}\left(\quad\mathord{+}\delta_{q_{1}+1,q_{2}}+\Theta_{q_{1},q_{2}}\Theta_{q_{2},1}g_{\nu;\epsilon}\prod_{q=q_{2}}^{q_{1}}\left\{ f_{\nu;\epsilon;n;1;q}\right\} +\Theta_{q_{2}-1,q_{1}+1}\prod_{q=q_{1}+1}^{q_{2}-1}\left\{ f_{\nu;\epsilon;n;-1;q}\right\} \right)\nonumber \\
 & \quad=\frac{1}{1-g_{\nu;\epsilon}}\left(\delta_{0,q_{2}}g_{\nu;\epsilon}\prod_{q=0}^{q_{1}}\left\{ f_{\nu;\epsilon;n;1;q}\right\} +\delta_{q_{1}+1,q_{2}}\right.\nonumber \\
 & \quad\phantom{=\frac{1}{1-g_{\nu;\epsilon}}}\left.\quad\mathord{+}\Theta_{q_{1},q_{2}}\Theta_{q_{2},1}g_{\nu;\epsilon}\prod_{q=q_{2}}^{q_{1}}\left\{ f_{\nu;\epsilon;n;1;q}\right\} +\Theta_{q_{2}-1,q_{1}+1}\prod_{q=q_{1}+1}^{q_{2}-1}\left\{ f_{\nu;\epsilon;n;-1;q}\right\} \right)\nonumber \\
 & \quad=\frac{1}{1-g_{\nu;\epsilon}}\left(\delta_{q_{1}+1,q_{2}}+\Theta_{q_{1},q_{2}}g_{\nu;\epsilon}\prod_{q=q_{2}}^{q_{1}}\left\{ f_{\nu;\epsilon;n;1;q}\right\} +\Theta_{q_{2}-1,q_{1}+1}\prod_{q=q_{1}+1}^{q_{2}-1}\left\{ f_{\nu;\epsilon;n;-1;q}\right\} \right).\label{eq:C_=00007Bv; epsilon; n; 1, 0; -1, -1=00007D -- 2}
\end{align}

Next, we calculate $\mathbf{C}_{\nu;r;\epsilon;n;0,0}$. Using Eqs.~\eqref{eq:introduce C_=00007Bv; epsilon; n; 0, 0=00007D -- 1},
\eqref{eq:(A_=00007Bv; epsilon; n; 0, 0=00007D)^=00007B-1=00007D -- 1},
\eqref{eq:(A_=00007B0, 0=00007D)^=00007B-1=00007D A_=00007B0, 1=00007D -- 1},
and \eqref{eq:C_=00007Bv; epsilon; n; 1, 0=00007D block structure -- 1}
we get

\begin{equation}
\mathbf{C}_{\nu;r;\epsilon;n;0,0}=\begin{pmatrix}\mathbf{C}_{\nu;r;\epsilon;n;0,0;1,1} & \mathbf{C}_{\nu;r;\epsilon;n;0,0;1,0} & \mathbf{C}_{\nu;r;\epsilon;n;0,0;1,-1}\\
\mathbf{C}_{\nu;r;\epsilon;n;0,0;0,1} & \mathbf{C}_{\nu;r;\epsilon;n;0,0;0,0} & \mathbf{C}_{\nu;r;\epsilon;n;0,0;0,-1}\\
\mathbf{C}_{\nu;r;\epsilon;n;0,0;-1,1} & \mathbf{C}_{\nu;r;\epsilon;n;0,0;-1,0} & \mathbf{C}_{\nu;r;\epsilon;n;0,0;-1,-1}
\end{pmatrix},\label{eq:C_=00007Bv; epsilon; n; 0, 0=00007D block structure -- 1}
\end{equation}

\noindent with

\begin{align}
\mathbf{C}_{\nu;r;\epsilon;n;0,0;1,1} & =\mathbf{1}_{\left(q_{v;n}+1\right)\times\left(q_{v;n}+1\right)}-\mathbf{A}_{\nu;r;\epsilon;n;0,1;1,1}\mathbf{C}_{\nu;r;\epsilon;n;1,0;1,1}\nonumber \\
 & \phantom{=}\mathord{+}\mathbf{A}_{\nu;r;\epsilon;n;0,0;1,-1}\mathbf{A}_{\nu;r;\epsilon;n;0,1;-1,-1}\mathbf{C}_{\nu;r;\epsilon;n;1,0;-1,1},\label{eq:C_=00007Bv; epsilon; n; 0, 0; 1, 1=00007D -- 1}
\end{align}

\begin{align}
\mathbf{C}_{\nu;r;\epsilon;n;0,0;1,0} & =-\mathbf{A}_{\nu;r;\epsilon;n;0,1;1,1}\mathbf{C}_{\nu;r;\epsilon;n;1,0;1,0}\nonumber \\
 & \phantom{=}\mathord{+}\mathbf{A}_{\nu;r;\epsilon;n;0,0;1,-1}\mathbf{A}_{\nu;r;\epsilon;n;0,1;-1,-1}\mathbf{C}_{\nu;r;\epsilon;n;1,0;-1,0},\label{eq:C_=00007Bv; epsilon; n; 0, 0; 1, 0=00007D -- 1}
\end{align}

\begin{align}
\mathbf{C}_{\nu;r;\epsilon;n;0,0;1,-1} & =-\mathbf{A}_{\nu;r;\epsilon;n;0,0,1,-1}-\mathbf{A}_{\nu;r;\epsilon;n;0,1;1,1}\mathbf{C}_{\nu;r;\epsilon;n;1,0;1,-1}\nonumber \\
 & \phantom{=}\mathord{+}\mathbf{A}_{\nu;r;\epsilon;n;0,0;1,-1}\mathbf{A}_{\nu;r;\epsilon;n;0,1;-1,-1}\mathbf{C}_{\nu;r;\epsilon;n;1,0;-1,-1},\label{eq:C_=00007Bv; epsilon; n; 0, 0; 1, -1=00007D -- 1}
\end{align}

\begin{equation}
\mathbf{C}_{\nu;r;\epsilon;n;0,0;0,1}=-\mathbf{A}_{\nu;r;\epsilon;n;0,1;0,1}\mathbf{C}_{\nu;r;\epsilon;n;1,0;1,1},\label{eq:C_=00007Bv; epsilon; n; 0, 0; 0, 1=00007D -- 1}
\end{equation}

\begin{equation}
\mathbf{C}_{\nu;r;\epsilon;n;0,0;0,0}=\mathbf{1}_{\left(1\right)\times\left(1\right)}-\mathbf{A}_{\nu;r;\epsilon;n;0,1;0,1}\mathbf{C}_{\nu;r;\epsilon;n;1,0;1,0},\label{eq:C_=00007Bv; epsilon; n; 0, 0; 0, 0=00007D -- 1}
\end{equation}

\begin{equation}
\mathbf{C}_{\nu;r;\epsilon;n;0,0;0,-1}=-\mathbf{A}_{\nu;r;\epsilon;n;0,1;0,1}\mathbf{C}_{\nu;r;\epsilon;n;1,0;1,-1},\label{eq:C_=00007Bv; epsilon; n; 0, 0; 0, -1=00007D -- 1}
\end{equation}

\begin{equation}
\mathbf{C}_{\nu;r;\epsilon;n;0,0;-1,1}=-\mathbf{A}_{\nu;r;\epsilon;n;0,1;-1,-1}\mathbf{C}_{\nu;r;\epsilon;n;1,0;-1,1},\label{eq:C_=00007Bv; epsilon; n; 0, 0; -1, 1=00007D -- 1}
\end{equation}

\begin{equation}
\mathbf{C}_{\nu;r;\epsilon;n;0,0;-1,0}=-\mathbf{A}_{\nu;r;\epsilon;n;0,1;-1,-1}\mathbf{C}_{\nu;r;\epsilon;n;1,0;-1,0},\label{eq:C_=00007Bv; epsilon; n; 0, 0; -1, 0=00007D -- 1}
\end{equation}

\noindent and

\begin{equation}
\mathbf{C}_{\nu;r;\epsilon;n;0,0;-1,-1}=\mathbf{1}_{\left(q_{v;n}+1\right)\times\left(q_{v;n}+1\right)}-\mathbf{A}_{\nu;r;\epsilon;n;0,1;-1,-1}\mathbf{C}_{\nu;r;\epsilon;n;1,0;-1,-1}.\label{eq:C_=00007Bv; epsilon; n; 0, 0; -1, -1=00007D -- 1}
\end{equation}

\noindent From Eqs.~\eqref{eq:xi_=00007Bv; r; epsilon; n; j=00003D0; alpha=00003D0=00007D -- 1},
\eqref{eq:Gaussian integral identity -- 1}, \eqref{eq:introduce C_=00007Bv; epsilon; n=00007D -- 1},
\eqref{eq:C_=00007Bv; epsilon; n=00007D block structure -- 2}, and
\eqref{eq:C_=00007Bv; epsilon; n; 0, 0=00007D block structure -- 1},
we can see that the only matrix elements of Eq.~\eqref{eq:C_=00007Bv; epsilon; n; 0, 1=00007D block structure -- 1}
that we need to calculate explicitly are $\mathbf{C}_{\nu;r;\epsilon;n;0,0;1,1}$,
$\mathbf{C}_{\nu;r;\epsilon;n;0,0;1,-1}$, $\mathbf{C}_{\nu;r;\epsilon;n;0,0;-1,1}$,
and $\mathbf{C}_{\nu;r;\epsilon;n;0,0;-1,-1}$. First, we calculate
$\mathbf{C}_{\nu;r;\epsilon;n;0,0;1,1}$. Using Eqs.~\eqref{eq:A_=00007Bv; epsilon; n; 0, 1; 1, 1=00007D -- 1},
\eqref{eq:A_=00007B0, 0; 1, -1=00007D A_=00007B0, 1; -1, -1=00007D -- 1},
\eqref{eq:C_=00007Bv; epsilon; n; 1, 0; 1, 1=00007D -- 2}, \eqref{eq:C_=00007Bv; epsilon; n; 1, 0; -1, 1=00007D -- 2},
and \eqref{eq:C_=00007Bv; epsilon; n; 0, 0; 1, 1=00007D -- 1} we
get

\begin{align}
 & \left[\mathbf{C}_{\nu;r;\epsilon;n;0,0;1,1}\right]_{q_{1},q_{2}}\nonumber \\
 & \quad=\left[\mathbf{1}_{\left(q_{v;n}+1\right)\times\left(q_{v;n}+1\right)}\right]_{q_{1},q_{2}}-\sum_{q_{3}}\left[\mathbf{A}_{\nu;r;\epsilon;n;0,1;1,1}\right]_{q_{1},q_{3}}\left[\mathbf{C}_{\nu;r;\epsilon;n;1,0;1,1}\right]_{q_{3},q_{2}}\nonumber \\
 & \quad\phantom{=}\mathord{+}\sum_{q_{3}}\left[\mathbf{A}_{\nu;r;\epsilon;n;0,0;1,-1}\mathbf{A}_{\nu;r;\epsilon;n;0,1;-1,-1}\right]_{q_{1},q_{3}}\left[\mathbf{C}_{\nu;r;\epsilon;n;1,0;-1,1}\right]_{q_{3},q_{2}}\nonumber \\
 & \quad=\delta_{q_{1},q_{2}}\nonumber \\
 & \quad\phantom{=}\mathord{+}\frac{1}{1-g_{\nu;\epsilon}}\left(\delta_{q_{1},q_{2}}\Theta_{q_{1},1}g_{\nu;\epsilon}+\Theta_{q_{2}-1,q_{1}}\Theta_{q_{1},1}g_{\nu;\epsilon}\prod_{q=q_{1}}^{q_{2}-1}\left\{ f_{\nu;\epsilon;n;-1;q}\right\} +\Theta_{q_{1}-1,q_{2}}\prod_{q=q_{2}}^{q_{1}-1}\left\{ f_{\nu;\epsilon;n;1;q}\right\} \right)\nonumber \\
 & \quad\phantom{=}\mathord{+}\frac{g_{\nu;\epsilon}\delta_{q_{1},0}}{1-g_{\nu;\epsilon}}\left(\delta_{q_{2},1}f_{\nu;\epsilon;n;-1;0}+\delta_{0,q_{2}}+\Theta_{q_{2}-1,1}\prod_{q=0}^{q_{2}-1}\left\{ f_{\nu;\epsilon;n;-1;q}\right\} \right)\nonumber \\
 & \quad=\delta_{q_{1},q_{2}}\nonumber \\
 & \quad\phantom{=}\mathord{+}\frac{1}{1-g_{\nu;\epsilon}}\left(\delta_{q_{1},q_{2}}\Theta_{q_{1},1}g_{\nu;\epsilon}+\Theta_{q_{2}-1,q_{1}}\Theta_{q_{1},1}g_{\nu;\epsilon}\prod_{q=q_{1}}^{q_{2}-1}\left\{ f_{\nu;\epsilon;n;-1;q}\right\} +\Theta_{q_{1}-1,q_{2}}\prod_{q=q_{2}}^{q_{1}-1}\left\{ f_{\nu;\epsilon;n;1;q}\right\} \right)\nonumber \\
 & \quad\phantom{=}\mathord{+}\frac{1}{1-g_{\nu;\epsilon}}\left(\delta_{q_{1},0}\delta_{0,q_{2}}g_{\nu;\epsilon}+\Theta_{q_{2}-1,0}\delta_{q_{1},0}g_{\nu;\epsilon}\prod_{q=0}^{q_{2}-1}\left\{ f_{\nu;\epsilon;n;-1;q}\right\} \right)\nonumber \\
 & \quad=\frac{1}{1-g_{\nu;\epsilon}}\left(\delta_{q_{1},q_{2}}+\Theta_{q_{2}-1,q_{1}}g_{\nu;\epsilon}\prod_{q=q_{1}}^{q_{2}-1}\left\{ f_{\nu;\epsilon;n;-1;q}\right\} +\Theta_{q_{1}-1,q_{2}}\prod_{q=q_{2}}^{q_{1}-1}\left\{ f_{\nu;\epsilon;n;1;q}\right\} \right).\label{eq:C_=00007Bv; epsilon; n; 0, 0; 1, 1=00007D -- 2}
\end{align}

\noindent Next, we calculate $\mathbf{C}_{\nu;r;\epsilon;n;0,0;1,-1}$.
Using Eqs.~\eqref{eq:A_=00007Bv; epsilon; n; 0, 0; 1, -1=00007D -- 1},
\eqref{eq:A_=00007Bv; epsilon; n; 0, 1; 1, 1=00007D -- 1}, \eqref{eq:A_=00007B0, 0; 1, -1=00007D A_=00007B0, 1; -1, -1=00007D -- 1},
\eqref{eq:C_=00007Bv; epsilon; n; 1, 0; 1, -1=00007D -- 2}, \eqref{eq:C_=00007Bv; epsilon; n; 1, 0; -1, -1=00007D -- 2},
and \eqref{eq:C_=00007Bv; epsilon; n; 0, 0; 1, -1=00007D -- 1} we
get

\begin{align}
 & \left[\mathbf{C}_{\nu;r;\epsilon;n;0,0;1,-1}\right]_{q_{1},q_{2}}\nonumber \\
 & \quad=-\left[\mathbf{A}_{\nu;r;\epsilon;n;0,0,1,-1}\right]_{q_{1},q_{2}}-\sum_{q_{3}}\left[\mathbf{A}_{\nu;r;\epsilon;n;0,1;1,1}\right]_{q_{1},q_{3}}\left[\mathbf{C}_{\nu;r;\epsilon;n;1,0;1,-1}\right]_{q_{3},q_{2}}\nonumber \\
 & \quad\phantom{=}\mathord{+}\sum_{q_{3}}\left[\mathbf{A}_{\nu;r;\epsilon;n;0,0;1,-1}\mathbf{A}_{\nu;r;\epsilon;n;0,1;-1,-1}\right]_{q_{1},q_{3}}\left[\mathbf{C}_{\nu;r;\epsilon;n;1,0;-1,-1}\right]_{q_{3},q_{2}}\nonumber \\
 & \quad=\delta_{q_{1},0}\delta_{0,q_{2}}g_{\nu;\epsilon}\nonumber \\
 & \quad\phantom{=}\mathord{+}\frac{g_{\nu;\epsilon}}{1-g_{\nu;\epsilon}}\left(\delta_{q_{1}-1,q_{2}-1}\Theta_{q_{1},1}+\Theta_{q_{1}-1,q_{2}}\prod_{q=q_{2}}^{q_{1}-1}\left\{ f_{\nu;\epsilon;n;1;q}\right\} +\Theta_{q_{2}-1,q_{1}}\Theta_{q_{1},1}\prod_{q=q_{1}}^{q_{2}-1}\left\{ f_{\nu;\epsilon;n;-1;q}\right\} \right)\nonumber \\
 & \quad\phantom{=}\mathord{+}\frac{1}{1-g_{\nu;\epsilon}}\left(\delta_{q_{1},0}\delta_{1,q_{2}}f_{\nu;\epsilon;n;-1;0}g_{\nu;\epsilon}+\delta_{q_{1},0}\delta_{0,q_{2}}g_{\nu;\epsilon}^{2}+\delta_{q_{1},0}\Theta_{q_{2}-1,1}g_{\nu;\epsilon}\prod_{q=0}^{q_{2}-1}\left\{ f_{\nu;\epsilon;n;-1;q}\right\} \right)\nonumber \\
 & \quad=\frac{g_{\nu;\epsilon}}{1-g_{\nu;\epsilon}}\left(\delta_{q_{1}-1,q_{2}-1}\Theta_{q_{1},1}+\Theta_{q_{1}-1,q_{2}}\prod_{q=q_{2}}^{q_{1}-1}\left\{ f_{\nu;\epsilon;n;1;q}\right\} +\Theta_{q_{2}-1,q_{1}}\Theta_{q_{1},1}\prod_{q=q_{1}}^{q_{2}-1}\left\{ f_{\nu;\epsilon;n;-1;q}\right\} \right)\nonumber \\
 & \quad\phantom{=}\mathord{+}\frac{g_{\nu;\epsilon}}{1-g_{\nu;\epsilon}}\left(\delta_{q_{1},0}\delta_{0,q_{2}}+\delta_{q_{1},0}\Theta_{q_{2}-1,0}\prod_{q=0}^{q_{2}-1}\left\{ f_{\nu;\epsilon;n;-1;q}\right\} \right)\nonumber \\
 & \quad=\frac{g_{\nu;\epsilon}}{1-g_{\nu;\epsilon}}\left(\delta_{q_{1},q_{2}}+\Theta_{q_{1}-1,q_{2}}\prod_{q=q_{2}}^{q_{1}-1}\left\{ f_{\nu;\epsilon;n;1;q}\right\} +\Theta_{q_{2}-1,q_{1}}\prod_{q=q_{1}}^{q_{2}-1}\left\{ f_{\nu;\epsilon;n;-1;q}\right\} \right).\label{eq:C_=00007Bv; epsilon; n; 0, 0; 1, -1=00007D -- 2}
\end{align}

\noindent Next, we calculate $\mathbf{C}_{\nu;r;\epsilon;n;0,0;-1,1}$.
Using Eqs.~\eqref{eq:A_=00007Bv; epsilon; n; 0, 1; -1, -1=00007D -- 1},
\eqref{eq:C_=00007Bv; epsilon; n; 1, 0; -1, 1=00007D -- 2}, and \eqref{eq:C_=00007Bv; epsilon; n; 0, 0; -1, 1=00007D -- 1}
we get

\begin{align}
 & \left[\mathbf{C}_{\nu;r;\epsilon;n;0,0;-1,1}\right]_{q_{1},q_{2}}\nonumber \\
 & \quad=-\sum_{q_{3}}\left[\mathbf{A}_{\nu;r;\epsilon;n;0,1;-1,-1}\right]_{q_{1},q_{3}}\left[\mathbf{C}_{\nu;r;\epsilon;n;1,0;-1,1}\right]_{q_{3},q_{2}}\nonumber \\
 & \quad=\frac{1}{1-g_{\nu;\epsilon}}\left(\delta_{q_{2},q_{1}+1}f_{\nu;\epsilon;n;-1;q_{1}}+\Theta_{q_{1},q_{2}}f_{\nu;\epsilon;n;-1;q_{1}}\prod_{q=q_{2}}^{q_{1}}\left\{ f_{\nu;\epsilon;n;1;q}\right\} +\Theta_{q_{2}-1,q_{1}+1}\prod_{q=q_{1}}^{q_{2}-1}\left\{ f_{\nu;\epsilon;n;-1;q}\right\} \right)\nonumber \\
 & \quad=\frac{1}{1-g_{\nu;\epsilon}}\left(\delta_{q_{2}-1,q_{1}}f_{\nu;\epsilon;n;-1;q_{1}}+\delta_{q_{1},q_{2}}+\Theta_{q_{1}-1,q_{2}}\prod_{q=q_{2}}^{q_{1}-1}\left\{ f_{\nu;\epsilon;n;1;q}\right\} +\Theta_{q_{2}-1,q_{1}+1}\prod_{q=q_{1}}^{q_{2}-1}\left\{ f_{\nu;\epsilon;n;-1;q}\right\} \right)\nonumber \\
 & \quad=\frac{1}{1-g_{\nu;\epsilon}}\left(\delta_{q_{1},q_{2}}+\Theta_{q_{1}-1,q_{2}}\prod_{q=q_{2}}^{q_{1}-1}\left\{ f_{\nu;\epsilon;n;1;q}\right\} +\Theta_{q_{2}-1,q_{1}}\prod_{q=q_{1}}^{q_{2}-1}\left\{ f_{\nu;\epsilon;n;-1;q}\right\} \right).\label{eq:C_=00007Bv; epsilon; n; 0, 0; -1, 1=00007D -- 2}
\end{align}

\noindent Next, we calculate $\mathbf{C}_{\nu;r;\epsilon;n;0,0;-1,-1}$.
Using Eqs.~\eqref{eq:A_=00007Bv; epsilon; n; 0, 1; -1, -1=00007D -- 1},
\eqref{eq:C_=00007Bv; epsilon; n; 1, 0; -1, -1=00007D -- 2}, and
\eqref{eq:C_=00007Bv; epsilon; n; 0, 0; -1, -1=00007D -- 1} we get

\begin{align}
 & \left[\mathbf{C}_{\nu;r;\epsilon;n;0,0;-1,-1}\right]_{q_{1},q_{2}}\nonumber \\
 & \quad=\left[\mathbf{1}_{\left(q_{v;n}+1\right)\times\left(q_{v;n}+1\right)}\right]_{q_{1},q_{2}}-\sum_{q_{3}}\left[\mathbf{A}_{\nu;r;\epsilon;n;0,1;-1,-1}\right]_{q_{1},q_{3}}\left[\mathbf{C}_{\nu;r;\epsilon;n;1,0;-1,-1}\right]_{q_{3},q_{2}}\nonumber \\
 & \quad=\delta_{q_{1},q_{2}}\nonumber \\
 & \quad\phantom{=}\mathord{+}\frac{1}{1-g_{\nu;\epsilon}}\left(\delta_{q_{1}+1,q_{2}}f_{\nu;\epsilon;n;-1;q_{1}}+\Theta_{q_{1},q_{2}}f_{\nu;\epsilon;n;-1;q_{1}}g_{\nu;\epsilon}\prod_{q=q_{2}}^{q_{1}}\left\{ f_{\nu;\epsilon;n;1;q}\right\} +\Theta_{q_{2}-1,q_{1}+1}\prod_{q=q_{1}}^{q_{2}-1}\left\{ f_{\nu;\epsilon;n;-1;q}\right\} \right)\nonumber \\
 & \quad=\delta_{q_{1},q_{2}}\nonumber \\
 & \quad\phantom{=}\mathord{+}\frac{1}{1-g_{\nu;\epsilon}}\left(\delta_{q_{1}+1,q_{2}}f_{\nu;\epsilon;n;-1;q_{1}}+\delta_{q_{1},q_{2}}g_{\nu;\epsilon}\right.\nonumber \\
 & \quad\phantom{=\mathord{+}\frac{1}{1-g_{\nu;\epsilon}}}\left.\quad\mathord{+}\Theta_{q_{1}-1,q_{2}}g_{\nu;\epsilon}\prod_{q=q_{2}}^{q_{1}-1}\left\{ f_{\nu;\epsilon;n;1;q}\right\} +\Theta_{q_{2}-1,q_{1}+1}\prod_{q=q_{1}}^{q_{2}-1}\left\{ f_{\nu;\epsilon;n;-1;q}\right\} \right)\nonumber \\
 & \quad=\frac{1}{1-g_{\nu;\epsilon}}\left(\delta_{q_{1},q_{2}}+\Theta_{q_{1}-1,q_{2}}g_{\nu;\epsilon}\prod_{q=q_{2}}^{q_{1}-1}\left\{ f_{\nu;\epsilon;n;1;q}\right\} +\Theta_{q_{2}-1,q_{1}}\prod_{q=q_{1}}^{q_{2}-1}\left\{ f_{\nu;\epsilon;n;-1;q}\right\} \right).\label{eq:C_=00007Bv; epsilon; n; 0, 0; -1, -1=00007D -- 2}
\end{align}

We have now calculated the block elements of $\mathbf{C}_{\nu;r;\epsilon;n}$
that are required to calculate the right-hand-side of Eq.~\eqref{eq:Gaussian integral identity -- 1}.
To summarize, these elements are:

\begin{equation}
\left[\mathbf{C}_{\nu;r;\epsilon;n;1,1;1,1}\right]_{q_{1},q_{2}}=\frac{1}{1-g_{\nu;\epsilon}}\left(\delta_{q_{1},q_{2}}+\Theta_{q_{1},q_{2}+1}\prod_{q=q_{2}+1}^{q_{1}}\left\{ f_{\nu;\epsilon;n;1;q}\right\} +\Theta_{q_{2},q_{1}+1}g_{\nu;\epsilon}\prod_{q=q_{1}+1}^{q_{2}}\left\{ f_{\nu;\epsilon;n;-1;q}\right\} \right),\label{eq:C_=00007Bv; epsilon; n; 1, 1; 1, 1=00007D -- 3}
\end{equation}

\begin{equation}
\left[\mathbf{C}_{\nu;r;\epsilon;n;1,1;-1,1}\right]_{q_{1},q_{2}}=\frac{1}{1-g_{\nu;\epsilon}}\left(\delta_{q_{1},q_{2}}+\Theta_{q_{1},q_{2}+1}\prod_{q=q_{2}+1}^{q_{1}}\left\{ f_{\nu;\epsilon;n;1;q}\right\} +\Theta_{q_{2},q_{1}+1}\prod_{q=q_{1}+1}^{q_{2}}\left\{ f_{\nu;\epsilon;n;-1;q}\right\} \right),\label{eq:C_=00007Bv; epsilon; n; 1, 1; -1, 1=00007D -- 3}
\end{equation}

\begin{equation}
\left[\mathbf{C}_{\nu;r;\epsilon;n;1,1;1,-1}\right]_{q_{1},q_{2}}=g_{\nu;\epsilon}\left[\mathbf{C}_{\nu;r;\epsilon;n;1,1;-1,1}\right]_{q_{1},q_{2}},\label{eq:C_=00007Bv; epsilon; n; 1, 1; 1, -1=00007D -- 3}
\end{equation}

\begin{equation}
\left[\mathbf{C}_{\nu;r;\epsilon;n;1,1;-1,-1}\right]_{q_{1},q_{2}}=\left[\mathbf{C}_{\nu;r;\epsilon;n;1,1;1,1}\right]_{q_{2},q_{1}}^{*},\label{eq:C_=00007Bv; epsilon; n; 1, 1; -1, -1=00007D -- 3}
\end{equation}

\begin{equation}
\left[\mathbf{C}_{\nu;r;\epsilon;n;0,1;1,1}\right]_{q_{1},q_{2}}=\frac{1}{1-g_{\nu;\epsilon}}\left(\delta_{q_{1},q_{2}+1}+\Theta_{q_{1}-1,q_{2}+1}\prod_{q=q_{2}+1}^{q_{1}-1}\left\{ f_{\nu;\epsilon;n;1;q}\right\} +\Theta_{q_{2},q_{1}}g_{\nu;\epsilon}\prod_{q=q_{1}}^{q_{2}}\left\{ f_{\nu;\epsilon;n;-1;q}\right\} \right),\label{eq:C_=00007Bv; epsilon; n; 0, 1; 1, 1=00007D -- 3}
\end{equation}

\begin{equation}
\left[\mathbf{C}_{\nu;r;\epsilon;n;0,1;-1,1}\right]_{q_{1},q_{2}}=\frac{1}{1-g_{\nu;\epsilon}}\left(\delta_{q_{1},q_{2}+1}+\Theta_{q_{1}-1,q_{2}+1}\prod_{q=q_{2}+1}^{q_{1}-1}\left\{ f_{\nu;\epsilon;n;1;q}\right\} +\Theta_{q_{2},q_{1}}\prod_{q=q_{1}}^{q_{2}}\left\{ f_{\nu;\epsilon;n;-1;q}\right\} \right),\label{eq:C_=00007Bv; epsilon; n; 0, 1; -1, 1=00007D -- 3}
\end{equation}

\begin{equation}
\left[\mathbf{C}_{\nu;r;\epsilon;n;0,1;1,-1}\right]_{q_{1},q_{2}}=g_{\nu;\epsilon}\left[\mathbf{C}_{\nu;r;\epsilon;n;0,1;-1,1}\right]_{q_{1},q_{2}},\label{eq:C_=00007Bv; epsilon; n; 0, 1; 1, -1=00007D -- 3}
\end{equation}

\begin{equation}
\left[\mathbf{C}_{\nu;r;\epsilon;n;0,1;-1,-1}\right]_{q_{1},q_{2}}=\left[\mathbf{C}_{\nu;r;\epsilon;n;1,0;1,1}\right]_{q_{2},q_{1}}^{*},\label{eq:C_=00007Bv; epsilon; n; 0, 1; -1, -1=00007D -- 3}
\end{equation}

\begin{align}
 & \left[\mathbf{C}_{\nu;r;\epsilon;n;1,0;1,1}\right]_{q_{1},q_{2}}\nonumber \\
 & \quad=\frac{1}{1-g_{\nu;\epsilon}}\left(\delta_{q_{2},q_{1}+1}g_{\nu;\epsilon}+\Theta_{q_{2}-1,q_{1}+1}g_{\nu;\epsilon}\prod_{q=q_{1}+1}^{q_{2}-1}\left\{ f_{\nu;\epsilon;n;-1;q}\right\} +\Theta_{q_{1},q_{2}}\prod_{q=q_{2}}^{q_{1}}\left\{ f_{\nu;\epsilon;n;1;q}\right\} \right),\label{eq:C_=00007Bv; epsilon; n; 1, 0; 1, 1=00007D -- 3}
\end{align}

\begin{equation}
\left[\mathbf{C}_{\nu;r;\epsilon;n;1,0;-1,1}\right]_{q_{1},q_{2}}=\frac{1}{1-g_{\nu;\epsilon}}\left(\delta_{q_{2},q_{1}+1}+\Theta_{q_{2}-1,q_{1}+1}\prod_{q=q_{1}+1}^{q_{2}-1}\left\{ f_{\nu;\epsilon;n;-1;q}\right\} +\Theta_{q_{1},q_{2}}\prod_{q=q_{2}}^{q_{1}}\left\{ f_{\nu;\epsilon;n;1;q}\right\} \right),\label{eq:C_=00007Bv; epsilon; n; 1, 0; -1, 1=00007D -- 3}
\end{equation}

\begin{equation}
\left[\mathbf{C}_{\nu;r;\epsilon;n;1,0;1,-1}\right]_{q_{1},q_{2}}=g_{\nu;\epsilon}\left[\mathbf{C}_{\nu;r;\epsilon;n;1,0;-1,1}\right]_{q_{1},q_{2}},\label{eq:C_=00007Bv; epsilon; n; 1, 0; 1, -1=00007D -- 3}
\end{equation}

\begin{equation}
\left[\mathbf{C}_{\nu;r;\epsilon;n;1,0;-1,-1}\right]_{q_{1},q_{2}}=\left[\mathbf{C}_{\nu;r;\epsilon;n;0,1;1,1}\right]_{q_{2},q_{1}}^{*},\label{eq:C_=00007Bv; epsilon; n; 1, 0; -1, -1=00007D -- 3}
\end{equation}

\begin{equation}
\left[\mathbf{C}_{\nu;r;\epsilon;n;0,0;1,1}\right]_{q_{1},q_{2}}=\frac{1}{1-g_{\nu;\epsilon}}\left(\delta_{q_{1},q_{2}}+\Theta_{q_{2}-1,q_{1}}g_{\nu;\epsilon}\prod_{q=q_{1}}^{q_{2}-1}\left\{ f_{\nu;\epsilon;n;-1;q}\right\} +\Theta_{q_{1}-1,q_{2}}\prod_{q=q_{2}}^{q_{1}-1}\left\{ f_{\nu;\epsilon;n;1;q}\right\} \right),\label{eq:C_=00007Bv; epsilon; n; 0, 0; 1, 1=00007D -- 3}
\end{equation}

\begin{equation}
\left[\mathbf{C}_{\nu;r;\epsilon;n;0,0;-1,1}\right]_{q_{1},q_{2}}=\frac{1}{1-g_{\nu;\epsilon}}\left(\delta_{q_{1},q_{2}}+\Theta_{q_{2}-1,q_{1}}\prod_{q=q_{1}}^{q_{2}-1}\left\{ f_{\nu;\epsilon;n;-1;q}\right\} +\Theta_{q_{1}-1,q_{2}}\prod_{q=q_{2}}^{q_{1}-1}\left\{ f_{\nu;\epsilon;n;1;q}\right\} \right),\label{eq:C_=00007Bv; epsilon; n; 0, 0; -1, 1=00007D -- 3}
\end{equation}

\begin{equation}
\left[\mathbf{C}_{\nu;r;\epsilon;n;0,0;1,-1}\right]_{q_{1},q_{2}}=g_{\nu;\epsilon}\left[\mathbf{C}_{\nu;r;\epsilon;n;0,0;-1,1}\right]_{q_{1},q_{2}},\label{eq:C_=00007Bv; epsilon; n; 0, 0; 1, -1=00007D -- 3}
\end{equation}

\noindent and

\begin{equation}
\left[\mathbf{C}_{\nu;r;\epsilon;n;0,0;-1,-1}\right]_{q_{1},q_{2}}=\left[\mathbf{C}_{\nu;r;\epsilon;n;0,0;1,1}\right]_{q_{2},q_{1}}^{*}.\label{eq:C_=00007Bv; epsilon; n; 0, 0; -1, -1=00007D -- 3}
\end{equation}

The next task is to calculate $\det\left[\mathbf{A}_{\nu;r;\epsilon;n}\right]$,
which appears on the right-hand-side of Eq.~\eqref{eq:Gaussian integral identity -- 1}.
This is done in several steps. First, using Eqs.~\eqref{eq:A_=00007Bv; epsilon; n=00007D -- 1},
\eqref{eq:M block matrix -- 1}, and \eqref{eq:determinant of a block matrix -- 1}
we can write

\begin{align}
\det\left[\mathbf{A}_{\nu;r;\epsilon;n}\right] & =\det\left[\mathbf{A}_{\nu;r;\epsilon;n;0,0}\right]\det\left[\mathbf{A}_{\nu;r;\epsilon;n;1,1}-\mathbf{A}_{\nu;r;\epsilon;n;1,0}\left(\mathbf{A}_{\nu;r;\epsilon;n;0,0}\right)^{-1}\mathbf{A}_{\nu;r;\epsilon;n;0,1}\right]\nonumber \\
 & =\det\left[\mathbf{A}_{\nu;r;\epsilon;n;0,0}\right]\det\left[\mathbf{B}_{\nu;r;\epsilon;n;1,1}\right].\label{eq:det=00005BA_=00007Bv; epsilon; n=00007D=00005D -- 1}
\end{align}

\noindent Next, using Eqs.~\eqref{eq:M block matrix -- 1}, \eqref{eq:determinant of a block matrix -- 1},
and \eqref{eq:A_=00007Bv; epsilon; n; 0, 0=00007D as 2x2 block matrix -- 1}
we can write

\begin{align}
 & \det\left[\mathbf{A}_{\nu;r;\epsilon;n;0,0}\right]\nonumber \\
 & \quad=\det\left[\mathbf{1}_{\left(2n+3\right)\times\left(2n+3\right)}\right]\det\left[\mathbf{1}_{\left(2n+2\right)\times\left(2n+2\right)}-\mathbf{0}_{\left(2n+2\right)\times\left(2n+3\right)}\left(\mathbf{1}_{\left(2n+3\right)\times\left(2n+3\right)}\right)^{-1}\mathbf{A}_{\nu;r;\epsilon;n;0,0;0,1}^{\prime}\right]\nonumber \\
 & \quad=1.\label{eq:det=00005BA_=00007Bv; epsilon; n; 0, 0=00007D=00005D -- 1}
\end{align}

\noindent Next, using Eqs.~\eqref{eq:f_=00007Bv; epsilon; n; alpha; k=00007D -- 1},
\eqref{eq:M block matrix -- 1}, \eqref{eq:determinant of a block matrix -- 1},
\eqref{eq:B_=00007Bv; epsilon; n; 1, 1=00007D -- 2}, \eqref{eq:B_=00007Bv; epsilon; n; 1, 1; 1, 1=00007D -- 1},
and \eqref{eq:(C_=00007Bv; epsilon; n; 1, 1; -1, -1=00007D)^=00007B-1=00007D -- 1}
we can write

\begin{align}
 & \det\left[\mathbf{B}_{\nu;r;\epsilon;n;1,1}\right]\nonumber \\
 & \quad=\det\left[\mathbf{B}_{\nu;r;\epsilon;n;1,1;1,1}\right]\det\left[\mathbf{B}_{\nu;r;\epsilon;n;1,1;-1,-1}-\mathbf{B}_{\nu;r;\epsilon;n;1,1;-1,1}\left(\mathbf{B}_{\nu;r;\epsilon;n;1,1;1,1}\right)^{-1}\mathbf{B}_{\nu;r;\epsilon;n;1,1;1,-1}\right]\nonumber \\
 & \quad=\det\left[\mathbf{B}_{\nu;r;\epsilon;n;1,1;1,1}\right]\det\left[\left(\mathbf{C}_{\nu;r;\epsilon;n;1,1;-1,-1}\right)^{-1}\right]\nonumber \\
 & \quad=\det\left[\delta_{q_{1},q_{2}}-\delta_{q_{1}+1,q_{2}}f_{\nu;\epsilon;n;-1;q_{2}}-g_{\nu;\epsilon}\prod_{q=1}^{2n+1}\left\{ f_{\nu;\epsilon;n;1;q}\right\} \delta_{q_{1},2n+1}\delta_{0,q_{2}}\right]\nonumber \\
 & \quad=1+\left(-1\right)^{2n+1}\left(-g_{\nu;\epsilon}\prod_{q=1}^{2n+1}\left\{ f_{\nu;\epsilon;n;1;q}\right\} \right)\left(\prod_{q=1}^{2n+1}\left\{ -f_{\nu;\epsilon;n;-1;q}\right\} \right)\nonumber \\
 & \quad=1-g_{\nu;\epsilon}.\label{eq:det=00005BB_=00007Bv; epsilon; n; 1, 1=00007D=00005D -- 1}
\end{align}

\noindent Therefore, from Eqs.~\eqref{eq:det=00005BA_=00007Bv; epsilon; n=00007D=00005D -- 1}-\eqref{eq:det=00005BB_=00007Bv; epsilon; n; 1, 1=00007D=00005D -- 1}
we have

\begin{equation}
\det\left[\mathbf{A}_{\nu;r;\epsilon;n}\right]=1-g_{\nu;\epsilon}=1-e^{-\beta\omega_{\nu;\epsilon}}.\label{eq:det=00005BA_=00007Bv; epsilon; n=00007D=00005D -- 2}
\end{equation}

The next task is to calculate $\left(\boldsymbol{\xi}_{\nu;r;\epsilon;n}\right)^{\dagger}\left(\mathbf{A}_{\nu;r;\epsilon;n}\right)^{-1}\mathbf{\boldsymbol{\psi}}_{\nu;r;\epsilon;n}$,
which appears on the right-hand-side of Eq.~\eqref{eq:Gaussian integral identity -- 1}.
Using Eqs.~\eqref{eq:xi_=00007Bv; r; epsilon; n=00007D -- 1}-\eqref{eq:xi_=00007Bv; r; epsilon; n; j=00003D1=00007D -- 1},
\eqref{eq:xi_=00007Bv; r; epsilon; n; j=00003D0; alpha=00003D0=00007D -- 1},
\eqref{eq:psi_=00007Bv; r; epsilon; n=00007D -- 1}, \eqref{eq:introduce C_=00007Bv; epsilon; n=00007D -- 1},
\eqref{eq:C_=00007Bv; epsilon; n=00007D block structure -- 2}, \eqref{eq:C_=00007Bv; epsilon; n; 1, 1=00007D block structure -- 1},
\eqref{eq:C_=00007Bv; epsilon; n; 0, 1=00007D block structure -- 1},
\eqref{eq:C_=00007Bv; epsilon; n; 1, 0=00007D block structure -- 1},
and \eqref{eq:C_=00007Bv; epsilon; n; 0, 0=00007D block structure -- 1}
we get

\begin{align}
\left(\boldsymbol{\xi}_{\nu;r;\epsilon;n}\right)^{\dagger}\left(\mathbf{A}_{\nu;r;\epsilon;n}\right)^{-1}\mathbf{\boldsymbol{\psi}}_{\nu;r;\epsilon;n} & =-\left(\boldsymbol{\xi}_{\nu;r;\epsilon;n}\right)^{\dagger}\mathbf{C}_{\nu;r;\epsilon;n}\mathbf{\boldsymbol{\xi}}_{\nu;r;\epsilon;n}\nonumber \\
 & =-\tilde{\gamma}_{\nu;r;\epsilon;n;1,1}\left(\sigma_{\nu;r;1;q\in\left[0,2n+1\right]},\sigma_{\nu;r;1;q\in\left[0,2n+1\right]}\right)\nonumber \\
 & \mathrel{\phantom{=}}\mathop{-}\tilde{\gamma}_{\nu;r;\epsilon;n;-1,-1}\left(\sigma_{\nu;r;-1;q\in\left[0,2n+1\right]},\sigma_{\nu;r;-1;q\in\left[0,2n+1\right]}\right)\nonumber \\
 & \mathrel{\phantom{=}}\mathop{-}\tilde{\Gamma}_{\nu;r;\epsilon;n}\left(\sigma_{\nu;r;1;q\in\left[0,2n+1\right]},\sigma_{\nu;r;-1;q\in\left[0,2n+1\right]}\right),\label{eq:xi A psi -- 1}
\end{align}

\noindent where

\begin{align}
\tilde{\gamma}_{\nu;r;\epsilon;n;1,1}\left(\sigma_{\nu;r;1;q\in\left[0,2n+1\right]},\sigma_{\nu;r;1;q\in\left[0,2n+1\right]}\right) & =\left(\boldsymbol{\xi}_{\nu;r;\epsilon;n;0;1}\right)^{\dagger}\mathbf{C}_{\nu;r;\epsilon;n;0,0;1,1}\boldsymbol{\xi}_{\nu;r;\epsilon;n;0;1}\nonumber \\
 & \mathrel{\phantom{=}}\mathop{+}\left(\boldsymbol{\xi}_{\nu;r;\epsilon;n;0;1}\right)^{\dagger}\mathbf{C}_{\nu;r;\epsilon;n;0,1;1,1}\boldsymbol{\xi}_{\nu;r;\epsilon;n;1;1}\nonumber \\
 & \mathrel{\phantom{=}}\mathop{+}\left(\boldsymbol{\xi}_{\nu;r;\epsilon;n;1;1}\right)^{\dagger}\mathbf{C}_{\nu;r;\epsilon;n;1,0;1,1}\boldsymbol{\xi}_{\nu;r;\epsilon;n;0;1}\nonumber \\
 & \mathrel{\phantom{=}}\mathop{+}\left(\boldsymbol{\xi}_{\nu;r;\epsilon;n;1;1}\right)^{\dagger}\mathbf{C}_{\nu;r;\epsilon;n;1,1;1,1}\boldsymbol{\xi}_{\nu;r;\epsilon;n;1;1},\label{eq:tilde gamma_=00007Bv; epsilon; n; 1, 1=00007D -- 1}
\end{align}

\begin{align}
\tilde{\gamma}_{\nu;r;\epsilon;n;-1,-1}\left(\sigma_{\nu;r;-1;q\in\left[0,2n+1\right]},\sigma_{\nu;r;-1;q\in\left[0,2n+1\right]}\right) & =\left(\boldsymbol{\xi}_{\nu;r;\epsilon;n;0;-1}\right)^{\dagger}\mathbf{C}_{\nu;r;\epsilon;n;0,0;-1,-1}\boldsymbol{\xi}_{\nu;r;\epsilon;n;0;-1}\nonumber \\
 & \mathrel{\phantom{=}}\mathop{+}\left(\boldsymbol{\xi}_{\nu;r;\epsilon;n;0;-1}\right)^{\dagger}\mathbf{C}_{\nu;r;\epsilon;n;0,1;-1,-1}\boldsymbol{\xi}_{\nu;r;\epsilon;n;1;-1}\nonumber \\
 & \mathrel{\phantom{=}}\mathop{+}\left(\boldsymbol{\xi}_{\nu;r;\epsilon;n;1;-1}\right)^{\dagger}\mathbf{C}_{\nu;r;\epsilon;n;1,0;-1,-1}\boldsymbol{\xi}_{\nu;r;\epsilon;n;0;-1}\nonumber \\
 & \mathrel{\phantom{=}}\mathop{+}\left(\boldsymbol{\xi}_{\nu;r;\epsilon;n;1;-1}\right)^{\dagger}\mathbf{C}_{\nu;r;\epsilon;n;1,1;-1,-1}\boldsymbol{\xi}_{\nu;r;\epsilon;n;1;-1},\label{eq:tilde gamma_=00007Bv; epsilon; n; -1, -1=00007D -- 1}
\end{align}

\noindent and

\begin{align}
\tilde{\Gamma}_{\nu;r;\epsilon;n}\left(\sigma_{\nu;r;1;q\in\left[0,2n+1\right]},\sigma_{\nu;r;-1;q\in\left[0,2n+1\right]}\right) & =\left(\boldsymbol{\xi}_{\nu;r;\epsilon;n;1;-1}\right)^{\dagger}\mathbf{C}_{\nu;r;\epsilon;n;1,1;-1,1}\boldsymbol{\xi}_{\nu;r;\epsilon;n;1;1}\nonumber \\
 & \mathrel{\phantom{=}}\mathop{+}\left(\boldsymbol{\xi}_{\nu;r;\epsilon;n;1;1}\right)^{\dagger}\mathbf{C}_{\nu;r;\epsilon;n;1,1;1,-1}\boldsymbol{\xi}_{\nu;r;\epsilon;n;1;-1}\nonumber \\
 & \mathrel{\phantom{=}}\mathop{+}\left(\boldsymbol{\xi}_{\nu;r;\epsilon;n;0;-1}\right)^{\dagger}\mathbf{C}_{\nu;r;\epsilon;n;0,0;-1,1}\boldsymbol{\xi}_{\nu;r;\epsilon;n;0;1}\nonumber \\
 & \mathrel{\phantom{=}}\mathop{+}\left(\boldsymbol{\xi}_{\nu;r;\epsilon;n;0;1}\right)^{\dagger}\mathbf{C}_{\nu;r;\epsilon;n;0,0;1,-1}\boldsymbol{\xi}_{\nu;r;\epsilon;n;0;-1}\nonumber \\
 & \mathrel{\phantom{=}}\mathop{+}\left(\boldsymbol{\xi}_{\nu;r;\epsilon;n;0;-1}\right)^{\dagger}\mathbf{C}_{\nu;r;\epsilon;n;0,1;-1,1}\boldsymbol{\xi}_{\nu;r;\epsilon;n;1;1}\nonumber \\
 & \mathrel{\phantom{=}}\mathop{+}\left(\boldsymbol{\xi}_{\nu;r;\epsilon;n;0;1}\right)^{\dagger}\mathbf{C}_{\nu;r;\epsilon;n;0,1;1,-1}\boldsymbol{\xi}_{\nu;r;\epsilon;n;1;-1}\nonumber \\
 & \mathrel{\phantom{=}}\mathop{+}\left(\boldsymbol{\xi}_{\nu;r;\epsilon;n;1;-1}\right)^{\dagger}\mathbf{C}_{\nu;r;\epsilon;n;1,0;-1,1}\boldsymbol{\xi}_{\nu;r;\epsilon;n;0;1}\nonumber \\
 & \mathrel{\phantom{=}}\mathop{+}\left(\boldsymbol{\xi}_{\nu;r;\epsilon;n;1;1}\right)^{\dagger}\mathbf{C}_{\nu;r;\epsilon;n;1,0;1,-1}\boldsymbol{\xi}_{\nu;r;\epsilon;n;0;-1}.\label{eq:tilde Gamma_=00007Bv; epsilon; n=00007D -- 1}
\end{align}

\noindent Towards the goal of calculating $\left(\boldsymbol{\xi}_{\nu;r;\epsilon;n}\right)^{\dagger}\left(\mathbf{A}_{\nu;r;\epsilon;n}\right)^{-1}\mathbf{\boldsymbol{\psi}}_{\nu;r;\epsilon;n}$,
we calculate each individual term on the right-hand-side of Eq.~\eqref{eq:xi A psi -- 1}
separately. Starting with the first term: using Eqs.~\eqref{eq:xi_=00007Bv; r; epsilon; n; j=00003D0, 1; alpha=00003D+/-1=00007D -- 1},
\eqref{eq:f_=00007Bv; epsilon; n; alpha; k=00007D -- 1}, \eqref{eq:C_=00007Bv; epsilon; n; 1, 1; 1, 1=00007D -- 3},
\eqref{eq:C_=00007Bv; epsilon; n; 0, 1; 1, 1=00007D -- 3}, \eqref{eq:C_=00007Bv; epsilon; n; 1, 0; 1, 1=00007D -- 3},
\eqref{eq:C_=00007Bv; epsilon; n; 0, 0; 1, 1=00007D -- 3}, and \eqref{eq:tilde gamma_=00007Bv; epsilon; n; 1, 1=00007D -- 1}
we get

\begin{align}
 & \tilde{\gamma}_{\nu;r;\epsilon;n;1,1}\left(\sigma_{\nu;r;1;q\in\left[0,2n+1\right]},\sigma_{\nu;r;1;q\in\left[0,2n+1\right]}\right)\nonumber \\
 & \quad=\frac{\left(\frac{\lambda_{\nu;r;\epsilon}}{\omega_{\nu;\epsilon}}\right)^{2}}{1-g_{\nu;\epsilon}}\sum_{q_{1}=0}^{2n+1}\sum_{q_{2}=0}^{2n+1}\mathcal{E}_{v;r;q_{1}}^{\left(\lambda\right)}\mathcal{E}_{v;r;q_{2}}^{\left(\lambda\right)}\sigma_{\nu;r;1;q_{1}}\sigma_{\nu;r;1;q_{2}}\nonumber \\
 & \quad\phantom{\mathrel{=}\frac{\left(\frac{\lambda_{\nu;r;\epsilon}}{\omega_{\nu;\epsilon}}\right)^{2}}{1-g_{\nu;\epsilon}}\sum_{q_{1}=0}^{2n+1}\sum_{q_{2}=0}^{2n+1}}\mathop{\times}\left(\delta_{q_{1},q_{2}}+g_{\nu;\epsilon}\Theta_{q_{1}-1,q_{2}}e^{i\Delta t\omega_{\nu;\epsilon}\sum_{q=q_{2}}^{q_{1}-1}\tilde{w}_{n;q}}+\Theta_{q_{1}-1,q_{2}}e^{-i\Delta t\omega_{\nu;\epsilon}\sum_{q=q_{2}}^{q_{1}-1}\tilde{w}_{n;q}}\right.\nonumber \\
 & \quad\phantom{\mathrel{=}\frac{\left(\frac{\lambda_{\nu;r;\epsilon}}{\omega_{\nu;\epsilon}}\right)^{2}}{1-g_{\nu;\epsilon}}\sum_{q_{1}=0}^{2n+1}\sum_{q_{2}=0}^{2n+1}\mathop{\times}}\left.\quad\mathop{-}\delta_{q_{1}-1,q_{2}}-\Theta_{q_{1}-1,q_{2}+1}e^{-i\Delta t\omega_{\nu;\epsilon}\sum_{q=q_{2}+1}^{q_{1}-1}\tilde{w}_{n;q}}-g_{\nu;\epsilon}\Theta_{q_{1},q_{2}}e^{i\Delta t\omega_{\nu;\epsilon}\sum_{q=q_{2}}^{q_{1}}\tilde{w}_{n;q}}\right.\nonumber \\
 & \quad\phantom{\mathrel{=}\frac{\left(\frac{\lambda_{\nu;r;\epsilon}}{\omega_{\nu;\epsilon}}\right)^{2}}{1-g_{\nu;\epsilon}}\sum_{q_{1}=0}^{2n+1}\sum_{q_{2}=0}^{2n+1}\mathop{\times}}\left.\quad\mathop{-}\delta_{q_{1}-1,q_{2}}g_{\nu;\epsilon}-g_{\nu;\epsilon}\Theta_{q_{1}-1,q_{2}+1}e^{i\Delta t\omega_{\nu;\epsilon}\sum_{q=q_{2}+1}^{q_{1}-1}\tilde{w}_{n;q}}-\Theta_{q_{1},q_{2}}e^{-i\Delta t\omega_{\nu;\epsilon}\sum_{q=q_{2}}^{q_{1}}\tilde{w}_{n;q}}\right.\nonumber \\
 & \quad\phantom{\mathrel{=}\frac{\left(\frac{\lambda_{\nu;r;\epsilon}}{\omega_{\nu;\epsilon}}\right)^{2}}{1-g_{\nu;\epsilon}}\sum_{q_{1}=0}^{2n+1}\sum_{q_{2}=0}^{2n+1}\mathop{\times}}\left.\quad\mathop{+}\delta_{q_{1},q_{2}}+\Theta_{q_{1},q_{2}+1}e^{-i\Delta t\omega_{\nu;\epsilon}\sum_{q=q_{2}+1}^{q_{1}}\tilde{w}_{n;q}}+g_{\nu;\epsilon}\Theta_{q_{1},q_{2}+1}e^{i\Delta t\omega_{\nu;\epsilon}\sum_{q=q_{2}+1}^{q_{1}}\tilde{w}_{n;q}}\right)\label{eq:tilde gamma_=00007Bv; epsilon; n; 1, 1=00007D -- 2}\\
 & \quad=\frac{\left(\frac{\lambda_{\nu;r;\epsilon}}{\omega_{\nu;\epsilon}}\right)^{2}}{1-g_{\nu;\epsilon}}\sum_{q_{1}=0}^{2n+1}\sum_{q_{2}=0}^{2n+1}\mathcal{E}_{v;r;q_{1}}^{\left(\lambda\right)}\mathcal{E}_{v;r;q_{2}}^{\left(\lambda\right)}\sigma_{\nu;r;1;q_{1}}\sigma_{\nu;r;1;q_{2}}\nonumber \\
 & \quad\phantom{\mathrel{=}\frac{\left(\frac{\lambda_{\nu;r;\epsilon}}{\omega_{\nu;\epsilon}}\right)^{2}}{1-g_{\nu;\epsilon}}\sum_{q_{1}=0}^{2n+1}\sum_{q_{2}=0}^{2n+1}}\mathop{\times}\left(2\delta_{q_{1},q_{2}}\right.\nonumber \\
 & \quad\phantom{\mathrel{=}\frac{\left(\frac{\lambda_{\nu;r;\epsilon}}{\omega_{\nu;\epsilon}}\right)^{2}}{1-g_{\nu;\epsilon}}\sum_{q_{1}=0}^{2n+1}\sum_{q_{2}=0}^{2n+1}\mathop{\times}}\left.\quad\mathop{+}\Theta_{q_{1}-1,q_{2}}\left[g_{\nu;\epsilon}e^{i\Delta t\omega_{\nu;\epsilon}\sum_{q=q_{2}}^{q_{1}-1}\tilde{w}_{n;q}}+e^{-i\Delta t\omega_{\nu;\epsilon}\sum_{q=q_{2}}^{q_{1}-1}\tilde{w}_{n;q}}\right.\right.\nonumber \\
 & \quad\phantom{\mathrel{=}\frac{\left(\frac{\lambda_{\nu;r;\epsilon}}{\omega_{\nu;\epsilon}}\right)^{2}}{1-g_{\nu;\epsilon}}\sum_{q_{1}=0}^{2n+1}\sum_{q_{2}=0}^{2n+1}\mathop{\times}\quad\mathop{+}\Theta_{q_{1}-1,q_{2}}}\left.\left.\quad\mathop{+}e^{-i\Delta t\omega_{\nu;\epsilon}\sum_{q=q_{2}+1}^{q_{1}}\tilde{w}_{n;q}}+g_{\nu;\epsilon}e^{i\Delta t\omega_{\nu;\epsilon}\sum_{q=q_{2}+1}^{q_{1}}\tilde{w}_{n;q}}\right.\right]\nonumber \\
 & \quad\phantom{\mathrel{=}\frac{\left(\frac{\lambda_{\nu;r;\epsilon}}{\omega_{\nu;\epsilon}}\right)^{2}}{1-g_{\nu;\epsilon}}\sum_{q_{1}=0}^{2n+1}\sum_{q_{2}=0}^{2n+1}\mathop{\times}}\left.\quad\mathop{-}\Theta_{q_{1},q_{2}}\left[g_{\nu;\epsilon}e^{i\Delta t\omega_{\nu;\epsilon}\sum_{q=q_{2}}^{q_{1}}\tilde{w}_{n;q}}+e^{-i\Delta t\omega_{\nu;\epsilon}\sum_{q=q_{2}}^{q_{1}}\tilde{w}_{n;q}}\right]\right.\nonumber \\
 & \quad\phantom{\mathrel{=}\frac{\left(\frac{\lambda_{\nu;r;\epsilon}}{\omega_{\nu;\epsilon}}\right)^{2}}{1-g_{\nu;\epsilon}}\sum_{q_{1}=0}^{2n+1}\sum_{q_{2}=0}^{2n+1}\mathop{\times}}\left.\quad\mathop{-}\left\{ 1+g_{\nu;\epsilon}\right\} \delta_{q_{1}-1,q_{2}}\right.\nonumber \\
 & \quad\phantom{\mathrel{=}\frac{\left(\frac{\lambda_{\nu;r;\epsilon}}{\omega_{\nu;\epsilon}}\right)^{2}}{1-g_{\nu;\epsilon}}\sum_{q_{1}=0}^{2n+1}\sum_{q_{2}=0}^{2n+1}\mathop{\times}}\left.\quad\mathop{-}\Theta_{q_{1}-2,q_{2}}\left[e^{-i\Delta t\omega_{\nu;\epsilon}\sum_{q=q_{2}+1}^{q_{1}-1}\tilde{w}_{n;q}}+g_{\nu;\epsilon}e^{i\Delta t\omega_{\nu;\epsilon}\sum_{q=q_{2}+1}^{q_{1}-1}\tilde{w}_{n;q}}\right]\right)\nonumber \\
 & \quad=\frac{\left(\frac{\lambda_{\nu;r;\epsilon}}{\omega_{\nu;\epsilon}}\right)^{2}}{1-g_{\nu;\epsilon}}\sum_{q_{1}=0}^{2n+1}\sum_{q_{2}=0}^{2n+1}\mathcal{E}_{v;r;q_{1}}^{\left(\lambda\right)}\mathcal{E}_{v;r;q_{2}}^{\left(\lambda\right)}\sigma_{\nu;r;1;q_{1}}\sigma_{\nu;r;1;q_{2}}\nonumber \\
 & \quad\phantom{\mathrel{=}\frac{\left(\frac{\lambda_{\nu;r;\epsilon}}{\omega_{\nu;\epsilon}}\right)^{2}}{1-g_{\nu;\epsilon}}\sum_{q_{1}=0}^{2n+1}\sum_{q_{2}=0}^{2n+1}}\mathop{\times}\left(\delta_{q_{1},q_{2}}\left[2-g_{\nu;\epsilon}e^{i\Delta t\omega_{\nu;\epsilon}\tilde{w}_{n;q_{1}}}-e^{-i\Delta t\omega_{\nu;\epsilon}\tilde{w}_{n;q_{1}}}\right]\right.\nonumber \\
 & \quad\phantom{\mathrel{=}\frac{\left(\frac{\lambda_{\nu;r;\epsilon}}{\omega_{\nu;\epsilon}}\right)^{2}}{1-g_{\nu;\epsilon}}\sum_{q_{1}=0}^{2n+1}\sum_{q_{2}=0}^{2n+1}\mathop{\times}}\left.\quad\mathop{+}\Theta_{q_{1}-2,q_{2}}\left[g_{\nu;\epsilon}e^{i\Delta t\omega_{\nu;\epsilon}\sum_{q=q_{2}}^{q_{1}-1}\tilde{w}_{n;q}}-e^{-i\Delta t\omega_{\nu;\epsilon}\sum_{q=q_{2}+1}^{q_{1}-1}\tilde{w}_{n;q}}\right.\right.\nonumber 
\end{align}

\begin{align}
 & \quad\phantom{\mathrel{=}\frac{\left(\frac{\lambda_{\nu;r;\epsilon}}{\omega_{\nu;\epsilon}}\right)^{2}}{1-g_{\nu;\epsilon}}\sum_{q_{1}=0}^{2n+1}\sum_{q_{2}=0}^{2n+1}\mathop{\times}\quad\mathop{+}\Theta_{q_{1}-2,q_{2}}}\left.\left.\quad\mathord{-}g_{\nu;\epsilon}e^{i\Delta t\omega_{\nu;\epsilon}\sum_{q=q_{2}+1}^{q_{1}-1}\tilde{w}_{n;q}}-g_{\nu;\epsilon}e^{i\Delta t\omega_{\nu;\epsilon}\sum_{q=q_{2}}^{q_{1}}\tilde{w}_{n;q}}\right.\right.\nonumber \\
 & \quad\phantom{\mathrel{=}\frac{\left(\frac{\lambda_{\nu;r;\epsilon}}{\omega_{\nu;\epsilon}}\right)^{2}}{1-g_{\nu;\epsilon}}\sum_{q_{1}=0}^{2n+1}\sum_{q_{2}=0}^{2n+1}\mathop{\times}\quad\mathop{+}\Theta_{q_{1}-2,q_{2}}}\left.\left.\quad\mathop{-}e^{-i\Delta t\omega_{\nu;\epsilon}\sum_{q=q_{2}}^{q_{1}}\tilde{w}_{n;q}}-e^{-i\Delta t\omega_{\nu;\epsilon}\sum_{q=q_{2}}^{q_{1}-1}\tilde{w}_{n;q}}\right.\right.\nonumber \\
 & \quad\phantom{\mathrel{=}\frac{\left(\frac{\lambda_{\nu;r;\epsilon}}{\omega_{\nu;\epsilon}}\right)^{2}}{1-g_{\nu;\epsilon}}\sum_{q_{1}=0}^{2n+1}\sum_{q_{2}=0}^{2n+1}\mathop{\times}\quad\mathop{+}\Theta_{q_{1}-2,q_{2}}}\left.\left.\quad\mathop{+}e^{-i\Delta t\omega_{\nu;\epsilon}\sum_{q=q_{2}+1}^{q_{1}}\tilde{w}_{n;q}}+g_{\nu;\epsilon}e^{i\Delta t\omega_{\nu;\epsilon}\sum_{q=q_{2}+1}^{q_{1}}\tilde{w}_{n;q}}\right.\right]\nonumber \\
 & \quad\phantom{\mathrel{=}\frac{\left(\frac{\lambda_{\nu;r;\epsilon}}{\omega_{\nu;\epsilon}}\right)^{2}}{1-g_{\nu;\epsilon}}\sum_{q_{1}=0}^{2n+1}\sum_{q_{2}=0}^{2n+1}\mathop{\times}}\left.\quad\mathop{+}\delta_{q_{1}-1,q_{2}}\left[g_{\nu;\epsilon}e^{i\Delta t\omega_{\nu;\epsilon}\sum_{q=q_{2}}^{q_{1}-1}\tilde{w}_{n;q}}-\left\{ 1+g_{\nu;\epsilon}\right\} \right.\right.\nonumber \\
 & \quad\phantom{\mathrel{=}\frac{\left(\frac{\lambda_{\nu;r;\epsilon}}{\omega_{\nu;\epsilon}}\right)^{2}}{1-g_{\nu;\epsilon}}\sum_{q_{1}=0}^{2n+1}\sum_{q_{2}=0}^{2n+1}\mathop{\times}\quad\mathop{+}\delta_{q_{1}-1,q_{2}}}\left.\left.\quad\mathop{-}g_{\nu;\epsilon}e^{i\Delta t\omega_{\nu;\epsilon}\sum_{q=q_{2}}^{q_{1}}\tilde{w}_{n;q}}-e^{-i\Delta t\omega_{\nu;\epsilon}\sum_{q=q_{2}}^{q_{1}}\tilde{w}_{n;q}}\right.\right.\nonumber \\
 & \quad\phantom{\mathrel{=}\frac{\left(\frac{\lambda_{\nu;r;\epsilon}}{\omega_{\nu;\epsilon}}\right)^{2}}{1-g_{\nu;\epsilon}}\sum_{q_{1}=0}^{2n+1}\sum_{q_{2}=0}^{2n+1}\mathop{\times}\quad\mathop{+}\delta_{q_{1}-1,q_{2}}}\left.\left.\quad\mathop{+}e^{-i\Delta t\omega_{\nu;\epsilon}\sum_{q=q_{2}}^{q_{1}-1}\tilde{w}_{n;q}}+e^{-i\Delta t\omega_{\nu;\epsilon}\sum_{q=q_{2}+1}^{q_{1}}\tilde{w}_{n;q}}\right.\right.\nonumber \\
 & \quad\phantom{\mathrel{=}\frac{\left(\frac{\lambda_{\nu;r;\epsilon}}{\omega_{\nu;\epsilon}}\right)^{2}}{1-g_{\nu;\epsilon}}\sum_{q_{1}=0}^{2n+1}\sum_{q_{2}=0}^{2n+1}\mathop{\times}\quad\mathop{+}\delta_{q_{1}-1,q_{2}}}\left.\left.\left.\quad\mathop{+}g_{\nu;\epsilon}e^{i\Delta t\omega_{\nu;\epsilon}\sum_{q=q_{2}+1}^{q_{1}}\tilde{w}_{n;q}}\right.\right]\right)\nonumber \\
 & \quad=\frac{\left(\frac{\lambda_{\nu;r;\epsilon}}{\omega_{\nu;\epsilon}}\right)^{2}}{1-g_{\nu;\epsilon}}\sum_{q_{1}=0}^{2n+1}\sum_{q_{2}=0}^{2n+1}\mathcal{E}_{v;r;q_{1}}^{\left(\lambda\right)}\mathcal{E}_{v;r;q_{2}}^{\left(\lambda\right)}\sigma_{\nu;r;1;q_{1}}\sigma_{\nu;r;1;q_{2}}\nonumber \\
 & \quad\phantom{\mathrel{=}\frac{\left(\frac{\lambda_{\nu;r;\epsilon}}{\omega_{\nu;\epsilon}}\right)^{2}}{1-g_{\nu;\epsilon}}\sum_{q_{1}=0}^{2n+1}\sum_{q_{2}=0}^{2n+1}}\mathop{\times}\left(\delta_{q_{1},q_{2}}\left[1-g_{\nu;\epsilon}\right]+\delta_{q_{1},q_{2}}\left[1+g_{\nu;\epsilon}-g_{\nu;\epsilon}e^{i\Delta t\omega_{\nu;\epsilon}\tilde{w}_{n;q_{1}}}-e^{-i\Delta t\omega_{\nu;\epsilon}\tilde{w}_{n;q_{1}}}\right]\right.\nonumber \\
 & \quad\phantom{\mathrel{=}\frac{\left(\frac{\lambda_{\nu;r;\epsilon}}{\omega_{\nu;\epsilon}}\right)^{2}}{1-g_{\nu;\epsilon}}\sum_{q_{1}=0}^{2n+1}\sum_{q_{2}=0}^{2n+1}\mathop{\times}}\left.\quad\mathop{+}\Theta_{q_{1}-1,q_{2}}\left[g_{\nu;\epsilon}e^{i\Delta t\omega_{\nu;\epsilon}\sum_{q=q_{2}}^{q_{1}-1}\tilde{w}_{n;q}}-g_{\nu;\epsilon}e^{i\Delta t\omega_{\nu;\epsilon}\sum_{q=q_{2}+1}^{q_{1}-1}\tilde{w}_{n;q}}\right.\right.\nonumber \\
 & \quad\phantom{\mathrel{=}\frac{\left(\frac{\lambda_{\nu;r;\epsilon}}{\omega_{\nu;\epsilon}}\right)^{2}}{1-g_{\nu;\epsilon}}\sum_{q_{1}=0}^{2n+1}\sum_{q_{2}=0}^{2n+1}\mathop{\times}\quad\mathop{+}\Theta_{q_{1}-1,q_{2}}}\left.\left.\quad\mathop{-}g_{\nu;\epsilon}e^{i\Delta t\omega_{\nu;\epsilon}\sum_{q=q_{2}}^{q_{1}}\tilde{w}_{n;q}}+g_{\nu;\epsilon}e^{i\Delta t\omega_{\nu;\epsilon}\sum_{q=q_{2}+1}^{q_{1}}\tilde{w}_{n;q}}\right.\right.\nonumber \\
 & \quad\phantom{\mathrel{=}\frac{\left(\frac{\lambda_{\nu;r;\epsilon}}{\omega_{\nu;\epsilon}}\right)^{2}}{1-g_{\nu;\epsilon}}\sum_{q_{1}=0}^{2n+1}\sum_{q_{2}=0}^{2n+1}\mathop{\times}\quad\mathop{+}\Theta_{q_{1}-1,q_{2}}}\left.\left.\quad\mathop{+}e^{-i\Delta t\omega_{\nu;\epsilon}\sum_{q=q_{2}}^{q_{1}-1}\tilde{w}_{n;q}}-e^{-i\Delta t\omega_{\nu;\epsilon}\sum_{q=q_{2}+1}^{q_{1}-1}\tilde{w}_{n;q}}\right.\right.\nonumber \\
 & \quad\phantom{\mathrel{=}\frac{\left(\frac{\lambda_{\nu;r;\epsilon}}{\omega_{\nu;\epsilon}}\right)^{2}}{1-g_{\nu;\epsilon}}\sum_{q_{1}=0}^{2n+1}\sum_{q_{2}=0}^{2n+1}\mathop{\times}\quad\mathop{+}\Theta_{q_{1}-1,q_{2}}}\left.\left.\quad\mathop{-}e^{-i\Delta t\omega_{\nu;\epsilon}\sum_{q=q_{2}}^{q_{1}}\tilde{w}_{n;q}}+e^{-i\Delta t\omega_{\nu;\epsilon}\sum_{q=q_{2}+1}^{q_{1}}\tilde{w}_{n;q}}\right]\right),\hspace{2cm}\label{eq:tilde gamma_=00007Bv; epsilon; n; 1, 1=00007D -- 3}
\end{align}

\noindent which we can write more concisely as

\begin{align}
 & \tilde{\gamma}_{\nu;r;\epsilon;n;1,1}\left(\sigma_{\nu;r;1;q\in\left[0,2n+1\right]},\sigma_{\nu;r;1;q\in\left[0,2n+1\right]}\right)\nonumber \\
 & \quad=\left(\frac{\lambda_{\nu;r;\epsilon}}{\omega_{\nu;\epsilon}}\right)^{2}\sum_{q_{1}=0}^{2n+1}\sum_{q_{2}=0}^{q_{1}}\delta_{q_{1},q_{2}}\mathcal{E}_{v;r;q_{1}}^{\left(\lambda\right)}\mathcal{E}_{v;r;q_{2}}^{\left(\lambda\right)}\sigma_{\nu;r;1;q_{1}}\sigma_{\nu;r;1;q_{2}}+\nonumber \\
 & \quad\mathrel{\phantom{=}}\mathop{+}\sum_{q_{1}=0}^{2n+1}\sum_{q_{2}=0}^{q_{1}}\eta_{\nu;r;\epsilon;n;q_{1},q_{2}}\mathcal{E}_{v;r;q_{1}}^{\left(\lambda\right)}\mathcal{E}_{v;r;q_{2}}^{\left(\lambda\right)}\sigma_{\nu;r;1;q_{1}}\sigma_{\nu;r;1;q_{2}}\nonumber \\
 & \quad=\sum_{q=0}^{2n+1}\left\{ \mathcal{E}_{v;r;q}^{\left(\lambda\right)}\frac{\lambda_{\nu;r;\epsilon}}{\omega_{\nu;\epsilon}}\right\} ^{2}+\sum_{q_{1}=0}^{2n+1}\sum_{q_{2}=0}^{q_{1}}\mathcal{E}_{v;r;q_{1}}^{\left(\lambda\right)}\mathcal{E}_{v;r;q_{2}}^{\left(\lambda\right)}\tilde{\eta}_{\nu;r;\epsilon;n;q_{1},q_{2}}\sigma_{\nu;r;1;q_{1}}\sigma_{\nu;r;1;q_{2}},\label{eq:tilde gamma_=00007Bv; epsilon; n; 1, 1=00007D concise expression -- 1}
\end{align}

\noindent where

\begin{align}
\tilde{\eta}_{\nu;r;\epsilon;n;q_{1},q_{2}} & =-\frac{\left(\frac{\lambda_{\nu;r;\epsilon}}{\omega_{\nu;\epsilon}}\right)^{2}}{1-g_{\nu;\epsilon}}\left[-g_{\nu;\epsilon}e^{i\Delta t\omega_{\nu;\epsilon}\sum_{q=q_{2}}^{q_{1}-1}\tilde{w}_{n;q}}-e^{-i\Delta t\omega_{\nu;\epsilon}\sum_{q=q_{2}}^{q_{1}-1}\tilde{w}_{n;q}}\right.\nonumber \\
 & \phantom{=-\frac{\left(\frac{\lambda_{\nu;r;\epsilon}}{\omega_{\nu;\epsilon}}\right)^{2}}{1-g_{\nu;\epsilon}}}\left.\quad\mathord{+}g_{\nu;\epsilon}e^{i\Delta t\omega_{\nu;\epsilon}\sum_{q=q_{2}+1}^{q_{1}-1}\tilde{w}_{n;q}}+e^{-i\Delta t\omega_{\nu;\epsilon}\sum_{q=q_{2}+1}^{q_{1}-1}\tilde{w}_{n;q}}\right.\nonumber \\
 & \phantom{=-\frac{\left(\frac{\lambda_{\nu;r;\epsilon}}{\omega_{\nu;\epsilon}}\right)^{2}}{1-g_{\nu;\epsilon}}}\left.\quad\mathord{+}g_{\nu;\epsilon}e^{i\Delta t\omega_{\nu;\epsilon}\sum_{q=q_{2}}^{q_{1}}\tilde{w}_{n;q}}+e^{-i\Delta t\omega_{\nu;\epsilon}\sum_{q=q_{2}}^{q_{1}}\tilde{w}_{n;q}}\right.\nonumber \\
 & \phantom{=-\frac{\left(\frac{\lambda_{\nu;r;\epsilon}}{\omega_{\nu;\epsilon}}\right)^{2}}{1-g_{\nu;\epsilon}}}\left.\quad\mathord{-}g_{\nu;\epsilon}e^{i\Delta t\omega_{\nu;\epsilon}\sum_{q=q_{2}+1}^{q_{1}}\tilde{w}_{n;q}}-e^{-i\Delta t\omega_{\nu;\epsilon}\sum_{q=q_{2}+1}^{q_{1}}\tilde{w}_{n;q}}\right],\label{eq:tilde eta_=00007Bv; epsilon; n; q_1, q_2=00007D -- 1}
\end{align}

\noindent and as a reminder we use the convention

\begin{equation}
\sum_{q=a}^{b}F_{q}=\begin{cases}
0, & \text{if }a>b,\\
F_{a}+F_{a+1}+\ldots+F_{b} & \text{otherwise},
\end{cases}\label{eq:summing convention -- 2}
\end{equation}

\noindent with $F_{q}$ being an arbitrary sequence. Next, we calculate
the second term on the right-hand-side of Eq.~\eqref{eq:xi A psi -- 1}
{[}i.e. we calculate Eq.~\eqref{eq:tilde gamma_=00007Bv; epsilon; n; -1, -1=00007D -- 1}{]}.
Using Eqs.~\eqref{eq:xi_=00007Bv; r; epsilon; n; j=00003D0, 1; alpha=00003D+/-1=00007D -- 1},
\eqref{eq:C_=00007Bv; epsilon; n; 1, 1; -1, -1=00007D -- 3}, \eqref{eq:C_=00007Bv; epsilon; n; 0, 1; -1, -1=00007D -- 3},
\eqref{eq:C_=00007Bv; epsilon; n; 1, 0; -1, -1=00007D -- 3}, \eqref{eq:C_=00007Bv; epsilon; n; 0, 0; -1, -1=00007D -- 3},
\eqref{eq:tilde gamma_=00007Bv; epsilon; n; 1, 1=00007D -- 1}, \eqref{eq:tilde gamma_=00007Bv; epsilon; n; -1, -1=00007D -- 1}
and \eqref{eq:tilde gamma_=00007Bv; epsilon; n; 1, 1=00007D concise expression -- 1}
we get

\begin{align}
 & \tilde{\gamma}_{\nu;r;\epsilon;n;-1,-1}\left(\sigma_{\nu;r;-1;q\in\left[0,2n+1\right]},\sigma_{\nu;r;-1;q\in\left[0,2n+1\right]}\right)\nonumber \\
 & \quad=\left(\boldsymbol{\xi}_{\nu;r;\epsilon;n;0;-1}\right)^{\dagger}\mathbf{C}_{\nu;r;\epsilon;n;0,0;-1,-1}\boldsymbol{\xi}_{\nu;r;\epsilon;n;0;-1}+\left(\boldsymbol{\xi}_{\nu;r;\epsilon;n;0;-1}\right)^{\dagger}\mathbf{C}_{\nu;r;\epsilon;n;0,1;-1,-1}\boldsymbol{\xi}_{\nu;\epsilon;n;1;-1}\nonumber \\
 & \quad\mathrel{\phantom{=}}\mathop{+}\left(\boldsymbol{\xi}_{\nu;r;\epsilon;n;1;-1}\right)^{\dagger}\mathbf{C}_{\nu;r;\epsilon;n;1,0;-1,-1}\boldsymbol{\xi}_{\nu;r;\epsilon;n;0;-1}+\left(\boldsymbol{\xi}_{\nu;r;\epsilon;n;1;-1}\right)^{\dagger}\mathbf{C}_{\nu;r;\epsilon;n;1,1;-1,-1}\boldsymbol{\xi}_{\nu;\epsilon;n;1;-1}\nonumber \\
 & \quad=\left(\boldsymbol{\xi}_{\nu;r;\epsilon;n;0;-1}\right)^{\dagger}\mathbf{C}_{\nu;r;\epsilon;n;0,0;1,1}^{\dagger}\boldsymbol{\xi}_{\nu;r;\epsilon;n;0;-1}+\left(\boldsymbol{\xi}_{\nu;r;\epsilon;n;0;-1}\right)^{\dagger}\mathbf{C}_{\nu;r;\epsilon;n;1,0;1,1}^{\dagger}\boldsymbol{\xi}_{\nu;\epsilon;n;1;-1}\nonumber \\
 & \quad\mathrel{\phantom{=}}\mathop{+}\left(\boldsymbol{\xi}_{\nu;r;\epsilon;n;1;-1}\right)^{\dagger}\mathbf{C}_{\nu;r;\epsilon;n;0,1;1,1}^{\dagger}\boldsymbol{\xi}_{\nu;r;\epsilon;n;0;-1}+\left(\boldsymbol{\xi}_{\nu;r;\epsilon;n;1;-1}\right)^{\dagger}\mathbf{C}_{\nu;r;\epsilon;n;1,1;1,1}^{\dagger}\boldsymbol{\xi}_{\nu;\epsilon;n;1;-1}\nonumber \\
 & \quad=\left\{ \left(\boldsymbol{\xi}_{\nu;r;\epsilon;n;0;-1}\right)^{\dagger}\mathbf{C}_{\nu;r;\epsilon;n;0,0;1,1}\boldsymbol{\xi}_{\nu;r;\epsilon;n;0;-1}+\left(\boldsymbol{\xi}_{\nu;r;\epsilon;n;0;-1}\right)^{\dagger}\mathbf{C}_{\nu;r;\epsilon;n;0,1;1,1}\boldsymbol{\xi}_{\nu;\epsilon;n;1;-1}\right.\nonumber \\
 & \quad\mathrel{\phantom{=}}\left.\quad\mathop{+}\left(\boldsymbol{\xi}_{\nu;r;\epsilon;n;1;-1}\right)^{\dagger}\mathbf{C}_{\nu;r;\epsilon;n;1,0;1,1}\boldsymbol{\xi}_{\nu;r;\epsilon;n;0;-1}+\left(\boldsymbol{\xi}_{\nu;r;\epsilon;n;1;-1}\right)^{\dagger}\mathbf{C}_{\nu;r;\epsilon;n;1,1;1,1}\boldsymbol{\xi}_{\nu;\epsilon;n;1;-1}\right\} ^{*}\nonumber \\
 & \quad=\sum_{q=0}^{2n+1}\left\{ \mathcal{E}_{v;r;q}^{\left(\lambda\right)}\frac{\lambda_{\nu;r;\epsilon}}{\omega_{\nu;\epsilon}}\right\} ^{2}+\sum_{q_{1}=0}^{2n+1}\sum_{q_{2}=0}^{q_{1}}\mathcal{E}_{v;r;q_{1}}^{\left(\lambda\right)}\mathcal{E}_{v;r;q_{2}}^{\left(\lambda\right)}\tilde{\eta}_{\nu;r;\epsilon;n;q_{1},q_{2}}^{*}\sigma_{\nu;r;-1;q_{1}}\sigma_{\nu;r;-1;q_{2}}.\label{eq:tilde gamma_=00007Bv; epsilon; n; -1, -1=00007D concise expression -- 1}
\end{align}

\noindent Next, we calculate the third term on the right-hand-side
of Eq.~\eqref{eq:xi A psi -- 1} {[}i.e. we calculate the right-hand-side
of Eq.~\eqref{eq:tilde Gamma_=00007Bv; epsilon; n=00007D -- 1}{]}.
This is done in multiple steps, as there are many terms making up
the right-hand-side of Eq.~\eqref{eq:tilde Gamma_=00007Bv; epsilon; n=00007D -- 1}.
First, we calculate terms $1$ and $2$ in Eq.~\eqref{eq:tilde Gamma_=00007Bv; epsilon; n=00007D -- 1}.
Using Eqs.~\eqref{eq:xi_=00007Bv; r; epsilon; n; j=00003D0, 1; alpha=00003D+/-1=00007D -- 1},
\eqref{eq:C_=00007Bv; epsilon; n; 1, 1; -1, 1=00007D -- 3}, and \eqref{eq:C_=00007Bv; epsilon; n; 1, 1; 1, -1=00007D -- 3}
we obtain

\begin{align}
 & \left(\boldsymbol{\xi}_{\nu;r;\epsilon;n;1;-1}\right)^{\dagger}\mathbf{C}_{\nu;r;\epsilon;n;1,1;-1,1}\boldsymbol{\xi}_{\nu;r;\epsilon;n;1;1}+\left(\boldsymbol{\xi}_{\nu;r;\epsilon;n;1;1}\right)^{\dagger}\mathbf{C}_{\nu;r;\epsilon;n;1,1;1,-1}\boldsymbol{\xi}_{\nu;r;\epsilon;n;1;-1}\nonumber \\
 & \quad=\mathop{-}\frac{\left(\frac{\lambda_{\nu;r;\epsilon}}{\omega_{\nu;\epsilon}}\right)^{2}}{1-g_{\nu;\epsilon}}\sum_{q_{1}=0}^{2n+1}\sum_{q_{2}=0}^{2n+1}\mathcal{E}_{v;r;q_{1}}^{\left(\lambda\right)}\mathcal{E}_{v;r;q_{2}}^{\left(\lambda\right)}\sigma_{\nu;r;-1;q_{1}}\sigma_{\nu;r;1;q_{2}}\nonumber \\
 & \quad\phantom{\mathrel{=}\mathop{-}\frac{\left(\frac{\lambda_{\nu;r;\epsilon}}{\omega_{\nu;\epsilon}}\right)^{2}}{1-g_{\nu;\epsilon}}\sum_{q_{1}=0}^{2n+1}\sum_{q_{2}=0}^{2n+1}}\mathop{\times}\left(\delta_{q_{1},q_{2}}+\Theta_{q_{1},q_{2}+1}e^{-i\Delta t\omega_{\nu;\epsilon}\sum_{q=q_{2}+1}^{q_{1}}\tilde{w}_{n;q}}+\Theta_{q_{2},q_{1}+1}e^{i\Delta t\omega_{\nu;\epsilon}\sum_{q=q_{1}+1}^{q_{2}}\tilde{w}_{n;q}}\right)\nonumber \\
 & \quad\mathrel{\phantom{=}}\mathop{-}\frac{\left(\frac{\lambda_{\nu;r;\epsilon}}{\omega_{\nu;\epsilon}}\right)^{2}}{1-g_{\nu;\epsilon}}\sum_{q_{1}=0}^{2n+1}\sum_{q_{2}=0}^{2n+1}\mathcal{E}_{v;r;q_{1}}^{\left(\lambda\right)}\mathcal{E}_{v;r;q_{2}}^{\left(\lambda\right)}\sigma_{\nu;r;1;q_{1}}\sigma_{\nu;r;-1;q_{2}}\nonumber \\
 & \quad\phantom{\mathrel{=}\mathop{-}\frac{\left(\frac{\lambda_{\nu;r;\epsilon}}{\omega_{\nu;\epsilon}}\right)^{2}}{1-g_{\nu;\epsilon}}\sum_{q_{1}=0}^{2n+1}\sum_{q_{2}=0}^{2n+1}}\mathop{\times}\left(g_{\nu;\epsilon}\delta_{q_{1},q_{2}}+g_{\nu;\epsilon}\Theta_{q_{1},q_{2}+1}e^{-i\Delta t\omega_{\nu;\epsilon}\sum_{q=q_{2}+1}^{q_{1}}\tilde{w}_{n;q}}\right.\nonumber \\
 & \quad\phantom{\mathrel{=}\mathop{-}\frac{\left(\frac{\lambda_{\nu;r;\epsilon}}{\omega_{\nu;\epsilon}}\right)^{2}}{1-g_{\nu;\epsilon}}\sum_{q_{1}=0}^{2n+1}\sum_{q_{2}=0}^{2n+1}\mathop{\times}}\left.\quad\mathop{+}g_{\nu;\epsilon}\Theta_{q_{2},q_{1}+1}e^{i\Delta t\omega_{\nu;\epsilon}\sum_{q=q_{1}+1}^{q_{2}}\tilde{w}_{n;q}}\right)\nonumber \\
 & \quad=\mathop{-}\frac{\left(\frac{\lambda_{\nu;r;\epsilon}}{\omega_{\nu;\epsilon}}\right)^{2}}{1-g_{\nu;\epsilon}}\sum_{q_{1}=0}^{2n+1}\sum_{q_{2}=0}^{2n+1}\mathcal{E}_{v;r;q_{1}}^{\left(\lambda\right)}\mathcal{E}_{v;r;q_{2}}^{\left(\lambda\right)}\sigma_{\nu;r;-1;q_{1}}\sigma_{\nu;r;1;q_{2}}\nonumber \\
 & \quad\phantom{\mathrel{=}\mathop{-}\frac{\left(\frac{\lambda_{\nu;r;\epsilon}}{\omega_{\nu;\epsilon}}\right)^{2}}{1-g_{\nu;\epsilon}}\sum_{q_{1}=0}^{2n+1}\sum_{q_{2}=0}^{2n+1}}\mathop{\times}\left(\delta_{q_{1},q_{2}}+\Theta_{q_{1},q_{2}+1}e^{-i\Delta t\omega_{\nu;\epsilon}\sum_{q=q_{2}+1}^{q_{1}}\tilde{w}_{n;q}}+g_{\nu;\epsilon}\Theta_{q_{1},q_{2}+1}e^{i\Delta t\omega_{\nu;\epsilon}\sum_{q=q_{2}+1}^{q_{1}}\tilde{w}_{n;q}}\right)\nonumber \\
 & \quad\mathrel{\phantom{=}}\mathop{-}\frac{\left(\frac{\lambda_{\nu;r;\epsilon}}{\omega_{\nu;\epsilon}}\right)^{2}}{1-g_{\nu;\epsilon}}\sum_{q_{1}=0}^{2n+1}\sum_{q_{2}=0}^{2n+1}\mathcal{E}_{v;r;q_{1}}^{\left(\lambda\right)}\mathcal{E}_{v;r;q_{2}}^{\left(\lambda\right)}\sigma_{\nu;r;1;q_{1}}\sigma_{\nu;r;-1;q_{2}}\nonumber \\
 & \quad\phantom{\mathrel{=}\mathop{-}\frac{\left(\frac{\lambda_{\nu;r;\epsilon}}{\omega_{\nu;\epsilon}}\right)^{2}}{1-g_{\nu;\epsilon}}\sum_{q_{1}=0}^{2n+1}\sum_{q_{2}=0}^{2n+1}}\mathop{\times}\left(\delta_{q_{1},q_{2}}+\Theta_{q_{1},q_{2}+1}e^{i\Delta t\omega_{\nu;\epsilon}\sum_{q=q_{2}+1}^{q_{1}}\tilde{w}_{n;q}}+g_{\nu;\epsilon}\Theta_{q_{1},q_{2}+1}e^{-i\Delta t\omega_{\nu;\epsilon}\sum_{q=q_{2}+1}^{q_{1}}\tilde{w}_{n;q}}\right)\nonumber \\
 & \quad\mathrel{\phantom{=}}\mathop{+}\left(\frac{\lambda_{\nu;r;\epsilon}}{\omega_{\nu;\epsilon}}\right)^{2}\sum_{q_{1}=0}^{2n+1}\sum_{q_{2}=0}^{q_{1}}\delta_{q_{1},q_{2}}\mathcal{E}_{v;r;q_{1}}^{\left(\lambda\right)}\mathcal{E}_{v;r;q_{2}}^{\left(\lambda\right)}\sigma_{\nu;r;-1;q_{1}}\sigma_{\nu;r;1;q_{2}}.\label{eq:calculating tilde Gamma_=00007Bv; epsilon; n=00007D -- 1}
\end{align}

\noindent Next, we calculate terms $3$ and $4$ in Eq.~\eqref{eq:tilde Gamma_=00007Bv; epsilon; n=00007D -- 1}.
Using Eqs.~\eqref{eq:xi_=00007Bv; r; epsilon; n; j=00003D0, 1; alpha=00003D+/-1=00007D -- 1},
\eqref{eq:C_=00007Bv; epsilon; n; 0, 0; -1, 1=00007D -- 3}, and \eqref{eq:C_=00007Bv; epsilon; n; 0, 0; 1, -1=00007D -- 3}
we obtain

\begin{align}
 & \left(\boldsymbol{\xi}_{\nu;r;\epsilon;n;0;-1}\right)^{\dagger}\mathbf{C}_{\nu;r;\epsilon;n;0,0;-1,1}\boldsymbol{\xi}_{\nu;r;\epsilon;n;0;1}+\left(\boldsymbol{\xi}_{\nu;r;\epsilon;n;0;1}\right)^{\dagger}\mathbf{C}_{\nu;r;\epsilon;n;0,0;1,-1}\boldsymbol{\xi}_{\nu;r;\epsilon;n;0;-1}\nonumber \\
 & \quad=\mathop{-}\frac{\left(\frac{\lambda_{\nu;r;\epsilon}}{\omega_{\nu;\epsilon}}\right)^{2}}{1-g_{\nu;\epsilon}}\sum_{q_{1}=0}^{2n+1}\sum_{q_{2}=0}^{2n+1}\mathcal{E}_{v;r;q_{1}}^{\left(\lambda\right)}\mathcal{E}_{v;r;q_{2}}^{\left(\lambda\right)}\sigma_{\nu;r;-1;q_{1}}\sigma_{\nu;r;1;q_{2}}\nonumber \\
 & \quad\phantom{\mathrel{=}\mathop{-}\frac{\left(\frac{\lambda_{\nu;r;\epsilon}}{\omega_{\nu;\epsilon}}\right)^{2}}{1-g_{\nu;\epsilon}}\sum_{q_{1}=0}^{2n+1}\sum_{q_{2}=0}^{2n+1}}\mathop{\times}\left(\delta_{q_{1},q_{2}}+\Theta_{q_{1}-1,q_{2}}e^{-i\Delta t\omega_{\nu;\epsilon}\sum_{q=q_{2}}^{q_{1}-1}\tilde{w}_{n;q}}+\Theta_{q_{2}-1,q_{1}}e^{i\Delta t\omega_{\nu;\epsilon}\sum_{q=q_{1}}^{q_{2}-1}\tilde{w}_{n;q}}\right)\nonumber \\
 & \quad\mathrel{\phantom{=}}\mathop{-}\frac{\left(\frac{\lambda_{\nu;r;\epsilon}}{\omega_{\nu;\epsilon}}\right)^{2}}{1-g_{\nu;\epsilon}}\sum_{q_{1}=0}^{2n+1}\sum_{q_{2}=0}^{2n+1}\mathcal{E}_{v;r;q_{1}}^{\left(\lambda\right)}\mathcal{E}_{v;r;q_{2}}^{\left(\lambda\right)}\sigma_{\nu;r;1;q_{1}}\sigma_{\nu;r;-1;q_{2}}\nonumber \\
 & \quad\phantom{\mathrel{=}\mathop{-}\frac{\left(\frac{\lambda_{\nu;r;\epsilon}}{\omega_{\nu;\epsilon}}\right)^{2}}{1-g_{\nu;\epsilon}}\sum_{q_{1}=0}^{2n+1}\sum_{q_{2}=0}^{2n+1}}\mathop{\times}\left(g_{\nu;\epsilon}\delta_{q_{1},q_{2}}+g_{\nu;\epsilon}\Theta_{q_{1}-1,q_{2}}e^{-i\Delta t\omega_{\nu;\epsilon}\sum_{q=q_{2}}^{q_{1}-1}\tilde{w}_{n;q}}\right.\nonumber \\
 & \quad\phantom{\mathrel{=}\mathop{-}\frac{\left(\frac{\lambda_{\nu;r;\epsilon}}{\omega_{\nu;\epsilon}}\right)^{2}}{1-g_{\nu;\epsilon}}\sum_{q_{1}=0}^{2n+1}\sum_{q_{2}=0}^{2n+1}\mathop{\times}}\left.\quad\mathop{+}g_{\nu;\epsilon}\Theta_{q_{2}-1,q_{1}}e^{i\Delta t\omega_{\nu;\epsilon}\sum_{q=q_{1}}^{q_{2}-1}\tilde{w}_{n;q}}\right)\nonumber \\
 & \quad=\mathop{-}\frac{\left(\frac{\lambda_{\nu;r;\epsilon}}{\omega_{\nu;\epsilon}}\right)^{2}}{1-g_{\nu;\epsilon}}\sum_{q_{1}=0}^{2n+1}\sum_{q_{2}=0}^{2n+1}\mathcal{E}_{v;r;q_{1}}^{\left(\lambda\right)}\mathcal{E}_{v;r;q_{2}}^{\left(\lambda\right)}\sigma_{\nu;r;-1;q_{1}}\sigma_{\nu;r;1;q_{2}}\nonumber \\
 & \quad\phantom{\mathrel{=}\mathop{-}\frac{\left(\frac{\lambda_{\nu;r;\epsilon}}{\omega_{\nu;\epsilon}}\right)^{2}}{1-g_{\nu;\epsilon}}\sum_{q_{1}=0}^{2n+1}\sum_{q_{2}=0}^{2n+1}}\mathop{\times}\left(\delta_{q_{1},q_{2}}+\Theta_{q_{1}-1,q_{2}}e^{-i\Delta t\omega_{\nu;\epsilon}\sum_{q=q_{2}}^{q_{1}-1}\tilde{w}_{n;q}}+g_{\nu;\epsilon}\Theta_{q_{1}-1,q_{2}}e^{i\Delta t\omega_{\nu;\epsilon}\sum_{q=q_{2}}^{q_{1}-1}\tilde{w}_{n;q}}\right)\nonumber \\
 & \quad\mathrel{\phantom{=}}\mathop{-}\frac{\left(\frac{\lambda_{\nu;r;\epsilon}}{\omega_{\nu;\epsilon}}\right)^{2}}{1-g_{\nu;\epsilon}}\sum_{q_{1}=0}^{2n+1}\sum_{q_{2}=0}^{2n+1}\mathcal{E}_{v;r;q_{1}}^{\left(\lambda\right)}\mathcal{E}_{v;r;q_{2}}^{\left(\lambda\right)}\sigma_{\nu;r;1;q_{1}}\sigma_{\nu;r;-1;q_{2}}\nonumber \\
 & \quad\phantom{\mathrel{=}\mathop{-}\frac{\left(\frac{\lambda_{\nu;r;\epsilon}}{\omega_{\nu;\epsilon}}\right)^{2}}{1-g_{\nu;\epsilon}}\sum_{q_{1}=0}^{2n+1}\sum_{q_{2}=0}^{2n+1}}\mathop{\times}\left(\delta_{q_{1},q_{2}}+\Theta_{q_{1}-1,q_{2}}e^{i\Delta t\omega_{\nu;\epsilon}\sum_{q=q_{2}}^{q_{1}-1}\tilde{w}_{n;q}}+g_{\nu;\epsilon}\Theta_{q_{1}-1,q_{2}}e^{-i\Delta t\omega_{\nu;\epsilon}\sum_{q=q_{2}}^{q_{1}-1}\tilde{w}_{n;q}}\right)\nonumber \\
 & \quad\mathrel{\phantom{=}}\mathop{+}\left(\frac{\lambda_{\nu;r;\epsilon}}{\omega_{\nu;\epsilon}}\right)^{2}\sum_{q_{1}=0}^{2n+1}\sum_{q_{2}=0}^{q_{1}}\delta_{q_{1},q_{2}}\mathcal{E}_{v;r;q_{1}}^{\left(\lambda\right)}\mathcal{E}_{v;r;q_{2}}^{\left(\lambda\right)}\sigma_{\nu;r;1;q_{1}}\sigma_{\nu;r;-1;q_{2}}.\label{eq:calculating tilde Gamma_=00007Bv; epsilon; n=00007D -- 2}
\end{align}

\noindent Next, we calculate terms $5$ and $6$ in Eq.~\eqref{eq:tilde Gamma_=00007Bv; epsilon; n=00007D -- 1}.
Using Eqs.~\eqref{eq:xi_=00007Bv; r; epsilon; n; j=00003D0, 1; alpha=00003D+/-1=00007D -- 1},
\eqref{eq:C_=00007Bv; epsilon; n; 0, 1; -1, 1=00007D -- 3}, and \eqref{eq:C_=00007Bv; epsilon; n; 0, 1; 1, -1=00007D -- 3}
we obtain

\begin{align}
 & \left(\boldsymbol{\xi}_{\nu;r;\epsilon;n;0;-1}\right)^{\dagger}\mathbf{C}_{\nu;r;\epsilon;n;0,1;-1,1}\boldsymbol{\xi}_{\nu;r;\epsilon;n;1;1}+\left(\boldsymbol{\xi}_{\nu;r;\epsilon;n;0;1}\right)^{\dagger}\mathbf{C}_{\nu;r;\epsilon;n;0,1;1,-1}\boldsymbol{\xi}_{\nu;r;\epsilon;n;1;-1}\nonumber \\
 & \quad=\frac{\left(\frac{\lambda_{\nu;r;\epsilon}}{\omega_{\nu;\epsilon}}\right)^{2}}{1-g_{\nu;\epsilon}}\sum_{q_{1}=0}^{2n+1}\sum_{q_{2}=0}^{2n+1}\mathcal{E}_{v;r;q_{1}}^{\left(\lambda\right)}\mathcal{E}_{v;r;q_{2}}^{\left(\lambda\right)}\sigma_{\nu;r;-1;q_{1}}\sigma_{\nu;r;1;q_{2}}\nonumber \\
 & \quad\phantom{\mathrel{=}\frac{\left(\frac{\lambda_{\nu;r;\epsilon}}{\omega_{\nu;\epsilon}}\right)^{2}}{1-g_{\nu;\epsilon}}\sum_{q_{1}=0}^{2n+1}\sum_{q_{2}=0}^{2n+1}}\mathop{\times}\left(\delta_{q_{1}-1,q_{2}}+\Theta_{q_{1}-1,q_{2}+1}\prod_{q=q_{2}+1}^{q_{1}-1}\left\{ f_{\nu;\epsilon;n;1;q}\right\} +\Theta_{q_{2},q_{1}}\prod_{q=q_{1}}^{q_{2}}\left\{ f_{\nu;\epsilon;n;-1;q}\right\} \right)\nonumber \\
 & \quad\mathrel{\phantom{=}}\mathop{+}\frac{\left(\frac{\lambda_{\nu;r;\epsilon}}{\omega_{\nu;\epsilon}}\right)^{2}}{1-g_{\nu;\epsilon}}\sum_{q_{1}=0}^{2n+1}\sum_{q_{2}=0}^{2n+1}\mathcal{E}_{v;r;q_{1}}^{\left(\lambda\right)}\mathcal{E}_{v;r;q_{2}}^{\left(\lambda\right)}\sigma_{\nu;r;1;q_{1}}\sigma_{\nu;r;-1;q_{2}}\nonumber \\
 & \quad\phantom{\mathrel{=}\mathop{+}\frac{\left(\frac{\lambda_{\nu;r;\epsilon}}{\omega_{\nu;\epsilon}}\right)^{2}}{1-g_{\nu;\epsilon}}\sum_{q_{1}=0}^{2n+1}\sum_{q_{2}=0}^{2n+1}}\mathop{\times}\left(g_{\nu;\epsilon}\delta_{q_{1}-1,q_{2}}+g_{\nu;\epsilon}\Theta_{q_{1}-1,q_{2}+1}\prod_{q=q_{2}+1}^{q_{1}-1}\left\{ f_{\nu;\epsilon;n;1;q}\right\} +g_{\nu;\epsilon}\Theta_{q_{2},q_{1}}\prod_{q=q_{1}}^{q_{2}}\left\{ f_{\nu;\epsilon;n;-1;q}\right\} \right)\nonumber \\
 & \quad=\frac{\left(\frac{\lambda_{\nu;r;\epsilon}}{\omega_{\nu;\epsilon}}\right)^{2}}{1-g_{\nu;\epsilon}}\sum_{q_{1}=0}^{2n+1}\sum_{q_{2}=0}^{2n+1}\mathcal{E}_{v;r;q_{1}}^{\left(\lambda\right)}\mathcal{E}_{v;r;q_{2}}^{\left(\lambda\right)}\sigma_{\nu;r;-1;q_{1}}\sigma_{\nu;r;1;q_{2}}\nonumber \\
 & \quad\phantom{\mathrel{=}\frac{\left(\frac{\lambda_{\nu;r;\epsilon}}{\omega_{\nu;\epsilon}}\right)^{2}}{1-g_{\nu;\epsilon}}\sum_{q_{1}=0}^{2n+1}\sum_{q_{2}=0}^{2n+1}}\mathop{\times}\left(\delta_{q_{1}-1,q_{2}}+\Theta_{q_{1}-1,q_{2}+1}e^{-i\Delta t\omega_{\nu;\epsilon}\sum_{q=q_{2}+1}^{q_{1}-1}\tilde{w}_{n;q}}+g_{\nu;\epsilon}\Theta_{q_{1},q_{2}}e^{i\Delta t\omega_{\nu;\epsilon}\sum_{q=q_{2}}^{q_{1}}\tilde{w}_{n;q}}\right)\nonumber \\
 & \quad\mathrel{\phantom{=}}\mathop{+}\frac{\left(\frac{\lambda_{\nu;r;\epsilon}}{\omega_{\nu;\epsilon}}\right)^{2}}{1-g_{\nu;\epsilon}}\sum_{q_{1}=0}^{2n+1}\sum_{q_{2}=0}^{2n+1}\mathcal{E}_{v;r;q_{1}}^{\left(\lambda\right)}\mathcal{E}_{v;r;q_{2}}^{\left(\lambda\right)}\sigma_{\nu;r;1;q_{1}}\sigma_{\nu;r;-1;q_{2}}\nonumber \\
 & \quad\phantom{\mathrel{=}\mathop{+}\frac{\left(\frac{\lambda_{\nu;r;\epsilon}}{\omega_{\nu;\epsilon}}\right)^{2}}{1-g_{\nu;\epsilon}}\sum_{q_{1}=0}^{2n+1}\sum_{q_{2}=0}^{2n+1}}\mathop{\times}\left(g_{\nu;\epsilon}\delta_{q_{1}-1,q_{2}}+g_{\nu;\epsilon}\Theta_{q_{1}-1,q_{2}+1}e^{-i\Delta t\omega_{\nu;\epsilon}\sum_{q=q_{2}+1}^{q_{1}-1}\tilde{w}_{n;q}}+\Theta_{q_{1},q_{2}}e^{i\Delta t\omega_{\nu;\epsilon}\sum_{q=q_{2}}^{q_{1}}\tilde{w}_{n;q}}\right).\label{eq:calculating tilde Gamma_=00007Bv; epsilon; n=00007D -- 3}
\end{align}

\noindent Next, we calculate terms $7$ and $8$ in Eq.~\eqref{eq:tilde Gamma_=00007Bv; epsilon; n=00007D -- 1}.
Using Eqs.~\eqref{eq:xi_=00007Bv; r; epsilon; n; j=00003D0, 1; alpha=00003D+/-1=00007D -- 1},
\eqref{eq:C_=00007Bv; epsilon; n; 1, 0; -1, 1=00007D -- 3}, and \eqref{eq:C_=00007Bv; epsilon; n; 1, 0; 1, -1=00007D -- 3}
we obtain

\begin{align}
 & \left(\boldsymbol{\xi}_{\nu;r;\epsilon;n;1;-1}\right)^{\dagger}\mathbf{C}_{\nu;r;\epsilon;n;1,0;-1,1}\boldsymbol{\xi}_{\nu;r;\epsilon;n;0;1}+\left(\boldsymbol{\xi}_{\nu;r;\epsilon;n;1;1}\right)^{\dagger}\mathbf{C}_{\nu;r;\epsilon;n;1,0;1,-1}\boldsymbol{\xi}_{\nu;r;\epsilon;n;0;-1}\nonumber \\
 & \quad=\frac{\left(\frac{\lambda_{\nu;r;\epsilon}}{\omega_{\nu;\epsilon}}\right)^{2}}{1-g_{\nu;\epsilon}}\sum_{q_{1}=0}^{2n+1}\sum_{q_{2}=0}^{2n+1}\mathcal{E}_{v;r;q_{1}}^{\left(\lambda\right)}\mathcal{E}_{v;r;q_{2}}^{\left(\lambda\right)}\sigma_{\nu;r;-1;q_{1}}\sigma_{\nu;r;1;q_{2}}\nonumber \\
 & \quad\phantom{\mathrel{=}\frac{\left(\frac{\lambda_{\nu;r;\epsilon}}{\omega_{\nu;\epsilon}}\right)^{2}}{1-g_{\nu;\epsilon}}\sum_{q_{1}=0}^{2n+1}\sum_{q_{2}=0}^{2n+1}}\mathop{\times}\left(\delta_{q_{2}-1,q_{1}}+\Theta_{q_{1},q_{2}}e^{-i\Delta t\omega_{\nu;\epsilon}\sum_{q=q_{2}}^{q_{1}}\tilde{w}_{n;q}}+\Theta_{q_{2}-1,q_{1}+1}e^{i\Delta t\omega_{\nu;\epsilon}\sum_{q=q_{1}+1}^{q_{2}-1}\tilde{w}_{n;q}}\right)\nonumber \\
 & \quad\mathrel{\phantom{=}}\mathop{+}\frac{\left(\frac{\lambda_{\nu;r;\epsilon}}{\omega_{\nu;\epsilon}}\right)^{2}}{1-g_{\nu;\epsilon}}\sum_{q_{1}=0}^{2n+1}\sum_{q_{2}=0}^{2n+1}\mathcal{E}_{v;r;q_{1}}^{\left(\lambda\right)}\mathcal{E}_{v;r;q_{2}}^{\left(\lambda\right)}\sigma_{\nu;r;1;q_{1}}\sigma_{\nu;r;-1;q_{2}}\nonumber \\
 & \quad\phantom{\mathrel{=}\mathop{+}\frac{\left(\frac{\lambda_{\nu;r;\epsilon}}{\omega_{\nu;\epsilon}}\right)^{2}}{1-g_{\nu;\epsilon}}\sum_{q_{1}=0}^{2n+1}\sum_{q_{2}=0}^{2n+1}}\mathop{\times}\left(g_{\nu;\epsilon}\delta_{q_{2}-1,q_{1}}+g_{\nu;\epsilon}\Theta_{q_{1},q_{2}}e^{-i\Delta t\omega_{\nu;\epsilon}\sum_{q=q_{2}}^{q_{1}}\tilde{w}_{n;q}}+g_{\nu;\epsilon}\Theta_{q_{2}-1,q_{1}+1}e^{i\Delta t\omega_{\nu;\epsilon}\sum_{q=q_{1}+1}^{q_{2}-1}\tilde{w}_{n;q}}\right)\nonumber \\
 & \quad=\frac{\left(\frac{\lambda_{\nu;r;\epsilon}}{\omega_{\nu;\epsilon}}\right)^{2}}{1-g_{\nu;\epsilon}}\sum_{q_{1}=0}^{2n+1}\sum_{q_{2}=0}^{2n+1}\mathcal{E}_{v;r;q_{1}}^{\left(\lambda\right)}\mathcal{E}_{v;r;q_{2}}^{\left(\lambda\right)}\sigma_{\nu;r;-1;q_{1}}\sigma_{\nu;r;1;q_{2}}\nonumber \\
 & \quad\phantom{\mathrel{=}\frac{\left(\frac{\lambda_{\nu;r;\epsilon}}{\omega_{\nu;\epsilon}}\right)^{2}}{1-g_{\nu;\epsilon}}\sum_{q_{1}=0}^{2n+1}\sum_{q_{2}=0}^{2n+1}}\mathop{\times}\left(g_{\nu;\epsilon}\delta_{q_{1}-1,q_{2}}+\Theta_{q_{1},q_{2}}e^{-i\Delta t\omega_{\nu;\epsilon}\sum_{q=q_{2}}^{q_{1}}\tilde{w}_{n;q}}+g_{\nu;\epsilon}\Theta_{q_{1}-1,q_{2}+1}e^{i\Delta t\omega_{\nu;\epsilon}\sum_{q=q_{2}+1}^{q_{1}-1}\tilde{w}_{n;q}}\right)\nonumber \\
 & \quad\mathrel{\phantom{=}}\mathop{+}\frac{\left(\frac{\lambda_{\nu;r;\epsilon}}{\omega_{\nu;\epsilon}}\right)^{2}}{1-g_{\nu;\epsilon}}\sum_{q_{1}=0}^{2n+1}\sum_{q_{2}=0}^{2n+1}\mathcal{E}_{v;r;q_{1}}^{\left(\lambda\right)}\mathcal{E}_{v;r;q_{2}}^{\left(\lambda\right)}\sigma_{\nu;r;1;q_{1}}\sigma_{\nu;r;-1;q_{2}}\nonumber \\
 & \quad\phantom{\mathrel{=}\mathop{+}\frac{\left(\frac{\lambda_{\nu;r;\epsilon}}{\omega_{\nu;\epsilon}}\right)^{2}}{1-g_{\nu;\epsilon}}\sum_{q_{1}=0}^{2n+1}\sum_{q_{2}=0}^{2n+1}}\mathop{\times}\left(\delta_{q_{1}-1,q_{2}}+g_{\nu;\epsilon}\Theta_{q_{1},q_{2}}e^{-i\Delta t\omega_{\nu;\epsilon}\sum_{q=q_{2}}^{q_{1}}\tilde{w}_{n;q}}+\Theta_{q_{1}-1,q_{2}+1}e^{i\Delta t\omega_{\nu;\epsilon}\sum_{q=q_{2}+1}^{q_{1}-1}\tilde{w}_{n;q}}\right).\label{eq:calculating tilde Gamma_=00007Bv; epsilon; n=00007D -- 4}
\end{align}

\noindent From Eqs.~\eqref{eq:calculating tilde Gamma_=00007Bv; epsilon; n=00007D -- 1}-\eqref{eq:calculating tilde Gamma_=00007Bv; epsilon; n=00007D -- 4},
and \eqref{eq:tilde Gamma_=00007Bv; epsilon; n=00007D -- 1}, we observe
that

\begin{align}
\tilde{\Gamma}_{\nu;r;\epsilon;n}\left(\sigma_{\nu;r;1;q\in\left[0,2n+1\right]},\sigma_{\nu;r;-1;q\in\left[0,2n+1\right]}\right) & =\tilde{\gamma}_{\nu;r;\epsilon;n;-1,1}\left(\sigma_{\nu;r;-1;q\in\left[0,2n+1\right]},\sigma_{\nu;r;1;q\in\left[0,2n+1\right]}\right)\nonumber \\
 & \mathrel{\phantom{=}}\mathop{+}\tilde{\gamma}_{\nu;r;\epsilon;n;1,-1}\left(\sigma_{\nu;r;1;q\in\left[0,2n+1\right]},\sigma_{\nu;r;-1;q\in\left[0,2n+1\right]}\right),\label{eq:tilde Gamma_=00007Bv; epsilon; n=00007D -- 2}
\end{align}

\noindent where

\begin{align}
 & \tilde{\gamma}_{\nu;r;\epsilon;n;-1,1}\left(\sigma_{\nu;r;-1;q\in\left[0,2n+1\right]},\sigma_{\nu;r;1;q\in\left[0,2n+1\right]}\right)\nonumber \\
 & \quad=-\frac{\left(\frac{\lambda_{\nu;r;\epsilon}}{\omega_{\nu;\epsilon}}\right)^{2}}{1-g_{\nu;\epsilon}}\sum_{q_{1}=0}^{2n+1}\sum_{q_{2}=0}^{2n+1}\mathcal{E}_{v;r;q_{1}}^{\left(\lambda\right)}\mathcal{E}_{v;r;q_{2}}^{\left(\lambda\right)}\sigma_{\nu;r;-1;q_{1}}\sigma_{\nu;r;1;q_{2}}\nonumber \\
 & \quad\phantom{\mathrel{=}-\frac{\left(\frac{\lambda_{\nu;r;\epsilon}}{\omega_{\nu;\epsilon}}\right)^{2}}{1-g_{\nu;\epsilon}}\sum_{q_{1}=0}^{2n+1}\sum_{q_{2}=0}^{2n+1}}\mathop{\times}\left(\delta_{q_{1},q_{2}}+g_{\nu;\epsilon}\Theta_{q_{1}-1,q_{2}}e^{i\Delta t\omega_{\nu;\epsilon}\sum_{q=q_{2}}^{q_{1}-1}\tilde{w}_{n;q}}+\Theta_{q_{1}-1,q_{2}}e^{-i\Delta t\omega_{\nu;\epsilon}\sum_{q=q_{2}}^{q_{1}-1}\tilde{w}_{n;q}}\right.\nonumber \\
 & \quad\phantom{\mathrel{=}-\frac{\left(\frac{\lambda_{\nu;r;\epsilon}}{\omega_{\nu;\epsilon}}\right)^{2}}{1-g_{\nu;\epsilon}}\sum_{q_{1}=0}^{2n+1}\sum_{q_{2}=0}^{2n+1}\mathop{\times}}\left.\quad\mathop{-}\delta_{q_{1}-1,q_{2}}-\Theta_{q_{1}-1,q_{2}+1}e^{-i\Delta t\omega_{\nu;\epsilon}\sum_{q=q_{2}+1}^{q_{1}-1}\tilde{w}_{n;q}}-g_{\nu;\epsilon}\Theta_{q_{1},q_{2}}e^{i\Delta t\omega_{\nu;\epsilon}\sum_{q=q_{2}}^{q_{1}}\tilde{w}_{n;q}}\right.\nonumber \\
 & \quad\phantom{\mathrel{=}-\frac{\left(\frac{\lambda_{\nu;r;\epsilon}}{\omega_{\nu;\epsilon}}\right)^{2}}{1-g_{\nu;\epsilon}}\sum_{q_{1}=0}^{2n+1}\sum_{q_{2}=0}^{2n+1}\mathop{\times}}\left.\quad\mathop{-}g_{\nu;\epsilon}\delta_{q_{1}-1,q_{2}}-g_{\nu;\epsilon}\Theta_{q_{1}-1,q_{2}+1}e^{i\Delta t\omega_{\nu;\epsilon}\sum_{q=q_{2}+1}^{q_{1}-1}\tilde{w}_{n;q}}-\Theta_{q_{1},q_{2}}e^{-i\Delta t\omega_{\nu;\epsilon}\sum_{q=q_{2}}^{q_{1}}\tilde{w}_{n;q}}\right.\nonumber \\
 & \quad\phantom{\mathrel{=}-\frac{\left(\frac{\lambda_{\nu;r;\epsilon}}{\omega_{\nu;\epsilon}}\right)^{2}}{1-g_{\nu;\epsilon}}\sum_{q_{1}=0}^{2n+1}\sum_{q_{2}=0}^{2n+1}\mathop{\times}}\left.\quad\mathop{+}\delta_{q_{1},q_{2}}+\Theta_{q_{1},q_{2}+1}e^{-i\Delta t\omega_{\nu;\epsilon}\sum_{q=q_{2}+1}^{q_{1}}\tilde{w}_{n;q}}+g_{\nu;\epsilon}\Theta_{q_{1},q_{2}+1}e^{i\Delta t\omega_{\nu;\epsilon}\sum_{q=q_{2}+1}^{q_{1}}\tilde{w}_{n;q}}\right)\nonumber \\
 & \quad\mathrel{\phantom{=}}\mathop{+}\left(\frac{\lambda_{\nu;r;\epsilon}}{\omega_{\nu;\epsilon}}\right)^{2}\sum_{q_{1}=0}^{2n+1}\sum_{q_{2}=0}^{q_{1}}\delta_{q_{1},q_{2}}\mathcal{E}_{v;r;q_{1}}^{\left(\lambda\right)}\mathcal{E}_{v;r;q_{2}}^{\left(\lambda\right)}\sigma_{\nu;r;-1;q_{1}}\sigma_{\nu;r;1;q_{2}},\label{eq:tilde gamma_=00007Bv; epsilon; n; -1, 1=00007D -- 1}
\end{align}

\noindent and

\begin{align}
 & \tilde{\gamma}_{\nu;r;\epsilon;n;1,-1}\left(\sigma_{\nu;r;1;q\in\left[0,2n+1\right]},\sigma_{\nu;r;-1;q\in\left[0,2n+1\right]}\right)\nonumber \\
 & \quad=-\frac{\left(\frac{\lambda_{\nu;r;\epsilon}}{\omega_{\nu;\epsilon}}\right)^{2}}{1-g_{\nu;\epsilon}}\sum_{q_{1}=0}^{2n+1}\sum_{q_{2}=0}^{2n+1}\mathcal{E}_{v;r;q_{1}}^{\left(\lambda\right)}\mathcal{E}_{v;r;q_{2}}^{\left(\lambda\right)}\sigma_{\nu;r;1;q_{1}}\sigma_{\nu;r;-1;q_{2}}\nonumber \\
 & \quad\phantom{\mathrel{=}-\frac{\left(\frac{\lambda_{\nu;r;\epsilon}}{\omega_{\nu;\epsilon}}\right)^{2}}{1-g_{\nu;\epsilon}}\sum_{q_{1}=0}^{2n+1}\sum_{q_{2}=0}^{2n+1}}\mathop{\times}\left(\delta_{q_{1},q_{2}}+g_{\nu;\epsilon}\Theta_{q_{1}-1,q_{2}}e^{-i\Delta t\omega_{\nu;\epsilon}\sum_{q=q_{2}}^{q_{1}-1}\tilde{w}_{n;q}}+\Theta_{q_{1}-1,q_{2}}e^{i\Delta t\omega_{\nu;\epsilon}\sum_{q=q_{2}}^{q_{1}-1}\tilde{w}_{n;q}}\right.\nonumber \\
 & \quad\phantom{\mathrel{=}-\frac{\left(\frac{\lambda_{\nu;r;\epsilon}}{\omega_{\nu;\epsilon}}\right)^{2}}{1-g_{\nu;\epsilon}}\sum_{q_{1}=0}^{2n+1}\sum_{q_{2}=0}^{2n+1}\mathop{\times}}\left.\quad\mathop{-}\delta_{q_{1}-1,q_{2}}-\Theta_{q_{1}-1,q_{2}+1}e^{i\Delta t\omega_{\nu;\epsilon}\sum_{q=q_{2}+1}^{q_{1}-1}\tilde{w}_{n;q}}-g_{\nu;\epsilon}\Theta_{q_{1},q_{2}}e^{-i\Delta t\omega_{\nu;\epsilon}\sum_{q=q_{2}}^{q_{1}}\tilde{w}_{n;q}}\right.\nonumber \\
 & \quad\phantom{\mathrel{=}-\frac{\left(\frac{\lambda_{\nu;r;\epsilon}}{\omega_{\nu;\epsilon}}\right)^{2}}{1-g_{\nu;\epsilon}}\sum_{q_{1}=0}^{2n+1}\sum_{q_{2}=0}^{2n+1}\mathop{\times}}\left.\quad\mathop{-}g_{\nu;\epsilon}\delta_{q_{1}-1,q_{2}}-g_{\nu;\epsilon}\Theta_{q_{1}-1,q_{2}+1}e^{-i\Delta t\omega_{\nu;\epsilon}\sum_{q=q_{2}+1}^{q_{1}-1}\tilde{w}_{n;q}}-\Theta_{q_{1},q_{2}}e^{i\Delta t\omega_{\nu;\epsilon}\sum_{q=q_{2}}^{q_{1}}\tilde{w}_{n;q}}\right.\nonumber \\
 & \quad\phantom{\mathrel{=}-\frac{\left(\frac{\lambda_{\nu;r;\epsilon}}{\omega_{\nu;\epsilon}}\right)^{2}}{1-g_{\nu;\epsilon}}\sum_{q_{1}=0}^{2n+1}\sum_{q_{2}=0}^{2n+1}\mathop{\times}}\left.\quad\mathop{+}\delta_{q_{1},q_{2}}+\Theta_{q_{1},q_{2}+1}e^{i\Delta t\omega_{\nu;\epsilon}\sum_{q=q_{2}+1}^{q_{1}}\tilde{w}_{n;q}}+g_{\nu;\epsilon}\Theta_{q_{1},q_{2}+1}e^{-i\Delta t\omega_{\nu;\epsilon}\sum_{q=q_{2}+1}^{q_{1}}\tilde{w}_{n;q}}\right)\nonumber \\
 & \quad\mathrel{\phantom{=}}\mathop{+}\left(\frac{\lambda_{\nu;r;\epsilon}}{\omega_{\nu;\epsilon}}\right)^{2}\sum_{q_{1}=0}^{2n+1}\sum_{q_{2}=0}^{q_{1}}\delta_{q_{1},q_{2}}\mathcal{E}_{v;r;q_{1}}^{\left(\lambda\right)}\mathcal{E}_{v;r;q_{2}}^{\left(\lambda\right)}\sigma_{\nu;r;1;q_{1}}\sigma_{\nu;r;-1;q_{2}}.\label{eq:tilde gamma_=00007Bv; epsilon; n; 1, -1=00007D -- 1}
\end{align}

\noindent By comparing the expressions in Eqs.~\eqref{eq:tilde gamma_=00007Bv; epsilon; n; -1, 1=00007D -- 1}
and \eqref{eq:tilde gamma_=00007Bv; epsilon; n; 1, -1=00007D -- 1}
to that in Eq.~\eqref{eq:tilde gamma_=00007Bv; epsilon; n; 1, 1=00007D -- 2},
and using Eq.~\eqref{eq:tilde gamma_=00007Bv; epsilon; n; 1, 1=00007D concise expression -- 1},
one should be able to see that

\begin{equation}
\tilde{\gamma}_{\nu;r;\epsilon;n;-1,1}\left(\sigma_{\nu;r;-1;q\in\left[0,2n+1\right]},\sigma_{\nu;r;1;q\in\left[0,2n+1\right]}\right)=-\sum_{q_{1}=0}^{2n+1}\sum_{q_{2}=0}^{q_{1}}\mathcal{E}_{v;r;q_{1}}^{\left(\lambda\right)}\mathcal{E}_{v;r;q_{2}}^{\left(\lambda\right)}\tilde{\eta}_{\nu;r;\epsilon;n;q_{1},q_{2}}\sigma_{\nu;r;-1;q_{1}}\sigma_{\nu;r;1;q_{2}},\label{eq:tilde gamma_=00007Bv; epsilon; n; -1, 1=00007D concise expression -- 1}
\end{equation}

\noindent and

\begin{equation}
\tilde{\gamma}_{\nu;r;\epsilon;n;1,-1}\left(\sigma_{\nu;r;1;q\in\left[0,2n+1\right]},\sigma_{\nu;r;-1;q\in\left[0,2n+1\right]}\right)=-\sum_{q_{1}=0}^{2n+1}\sum_{q_{2}=0}^{q_{1}}\mathcal{E}_{v;r;q_{1}}^{\left(\lambda\right)}\mathcal{E}_{v;r;q_{2}}^{\left(\lambda\right)}\tilde{\eta}_{\nu;r;\epsilon;n;q_{1},q_{2}}^{*}\sigma_{\nu;r;1;q_{1}}\sigma_{\nu;r;-1;q_{2}}.\label{eq:tilde gamma_=00007Bv; epsilon; n; 1, -1=00007D concise expression -- 1}
\end{equation}

\noindent Next, using Eqs.~\eqref{eq:xi A psi -- 1}, \eqref{eq:tilde gamma_=00007Bv; epsilon; n; 1, 1=00007D concise expression -- 1}-\eqref{eq:tilde Gamma_=00007Bv; epsilon; n=00007D -- 2},
\eqref{eq:tilde gamma_=00007Bv; epsilon; n; -1, 1=00007D concise expression -- 1},
and \eqref{eq:tilde gamma_=00007Bv; epsilon; n; 1, -1=00007D concise expression -- 1}
we get

\begin{align}
 & \sum_{\epsilon}\left(\boldsymbol{\xi}_{\nu;r;\epsilon;n}\right)^{\dagger}\left(\mathbf{A}_{\nu;r;\epsilon;n}\right)^{-1}\mathbf{\boldsymbol{\psi}}_{\nu;r;\epsilon;n}\nonumber \\
 & \quad=-2\sum_{\epsilon}\sum_{q=0}^{2n+1}\left\{ \mathcal{E}_{v;r;q}^{\left(\lambda\right)}\frac{\lambda_{\nu;r;\epsilon}}{\omega_{\nu;\epsilon}}\right\} ^{2}-\sum_{q_{2}=0}^{2n+1}\sum_{q_{1}=0}^{q_{2}}\tilde{\gamma}_{\nu;r;n;q_{1},q_{2}}\left(\sigma_{\nu;r;1;q_{1}},\sigma_{\nu;r;-1;q_{1}},\sigma_{\nu;r;1;q_{2}},\sigma_{\nu;r;-1;q_{2}}\right),\label{eq:sum xi A psi -- 1}
\end{align}

\noindent where

\begin{align}
 & \tilde{\gamma}_{\nu;r;n;q_{1},q_{2}}\left(\sigma_{\nu;r;1;q_{1}},\sigma_{\nu;r;-1;q_{1}},\sigma_{\nu;r;1;q_{2}},\sigma_{\nu;r;-1;q_{2}}\right)\nonumber \\
 & \quad=\mathcal{E}_{\nu;r;q_{1}}^{\left(\lambda\right)}\mathcal{E}_{\nu;r;q_{2}}^{\left(\lambda\right)}\left(\sigma_{\nu;r;1;q_{2}}-\sigma_{\nu;r;-1;q_{2}}\right)\nonumber \\
 & \quad\mathrel{\phantom{=}}\mathop{\times}\left\{ \left(\sigma_{\nu;r;1;q_{1}}-\sigma_{\nu;r;-1;q_{1}}\right)\text{Re}\left[\tilde{\eta}_{\nu;r;n;q_{2},q_{1}}\right]+i\left(\sigma_{\nu;r;1;q_{1}}+\sigma_{\nu;r;-1;q_{1}}\right)\text{Im}\left[\tilde{\eta}_{\nu;r;n;q_{2},q_{1}}\right]\right\} ,\label{eq:gamma_=00007Bv; r; n; q_1, q_2=00007D -- 1}
\end{align}

\noindent and

\begin{equation}
\tilde{\eta}_{\nu;r;n;q_{1},q_{2}}=\sum_{\epsilon}\tilde{\eta}_{\nu;r;\epsilon;n;q_{1},q_{2}}.\label{eq:tilde eta_=00007Bv; r; n; q_1, q_2=00007D -- 1}
\end{equation}

Next, we derive explicit expressions for $\tilde{\eta}_{\nu;r;n;q_{1},q_{2}}$
for different values of $q_{1}$, and $q_{2}$. We do this by making
use of the spectral density of the noise coupled to the $\nu$-component
of the spin at site $r$ and temperature $T$, $A_{\nu;r;T}\left(\omega\right)$.
We define $A_{\nu;r;T}\left(\omega\right)$ in Sec.~\ref{sec:Spectral densities of noise}
and also derive the following expression for the function therein:

\begin{equation}
A_{\nu;r;T}\left(\omega\right)=\text{sign}\left(\omega\right)\frac{A_{\nu;r;T=0}\left(\left|\omega\right|\right)}{1-e^{-\beta\omega}},\label{eq:A_=00007Bv; r; T=00007D(omega) in path integral appendix -- 1}
\end{equation}

\noindent where

\begin{equation}
\text{sign}\left(\omega\right)=\begin{cases}
1 & \text{if }\omega>0,\\
-1 & \text{if }\omega<0,
\end{cases}\label{eq:sign function -- 1}
\end{equation}

\noindent and $A_{\nu;r;T=0}\left(\omega\right)$ is the zero-temperature
limit of $A_{\nu;r;T}\left(\omega\right)$, which we calculate in
Sec.~\ref{sec:Spectral densities of noise} to be:

\begin{equation}
A_{\nu;r;T=0}\left(\omega\right)=2\pi\sum_{\epsilon}\lambda_{\nu;r;\epsilon}^{2}\delta\left(\omega-\omega_{\nu;\epsilon}\right),\label{eq:A_=00007Bv; r; T=00003D0=00007D(omega) in path integral appendix -- 1}
\end{equation}

\noindent before taking the continuum limit. Note that 

\begin{equation}
\sum_{\epsilon}\lambda_{\nu;r;\epsilon}^{2}f\left(\omega_{\epsilon}\right)=\int_{0}^{\infty}\frac{d\omega}{2\pi}\,A_{\nu;r;T=0}\left(\omega\right)f\left(\omega\right).\label{eq:integral of A*f -- 1}
\end{equation}

\noindent where $f\left(\omega\right)$ is an arbitrary function.
Using Eq.~\eqref{eq:g_=00007Bv; epsilon=00007D -- 1}, \eqref{eq:tilde eta_=00007Bv; epsilon; n; q_1, q_2=00007D -- 1},
\eqref{eq:tilde eta_=00007Bv; r; n; q_1, q_2=00007D -- 1}, and \eqref{eq:integral of A*f -- 1}
we get

\begin{equation}
\tilde{\eta}_{\nu;n;k_{1},k_{2}}=\int_{0}^{\infty}\frac{d\omega}{2\pi}\,\frac{A_{\nu;T=0}\left(\omega\right)}{\omega^{2}\left\{ 1-e^{-\beta\omega}\right\} }\left\{ \tilde{f}_{\nu;n;k_{1},k_{2}}^{\left(\eta\right)}\left(\omega\right)+e^{-\beta\omega}\tilde{f}_{\nu;n;k_{1},k_{2}}^{\left(\eta\right)}\left(-\omega\right)\right\} ,\label{eq:tilde eta_=00007Bv; n; q_1, q_2=00007D -- 2}
\end{equation}

\noindent where

\begin{equation}
\tilde{f}_{\nu;n;q_{1},q_{2}}^{\left(\eta\right)}\left(\omega\right)=e^{-i\Delta t\omega\sum_{q=q_{2}}^{q_{1}-1}\tilde{w}_{n;q}}-e^{-i\Delta t\omega\sum_{q=k_{2}+1}^{q_{1}-1}\tilde{w}_{n;q}}-e^{-i\Delta t\omega\sum_{q=q_{2}}^{q_{1}}\tilde{w}_{n;q}}+e^{-i\Delta t\omega\sum_{q=q_{2}+1}^{q_{1}}\tilde{w}_{n;q}}.\label{eq:tilde f_=00007Bv; n; q_1, q_2=00007D^=00007B(eta)=00007D(omega) -- 1}
\end{equation}

\noindent Using Eq.~\eqref{eq:A_=00007Bv; r; T=00003D0=00007D(omega) in path integral appendix -- 1},
we can rewrite Eq.~\eqref{eq:tilde eta_=00007Bv; n; q_1, q_2=00007D -- 2}
as

\begin{align}
\tilde{\eta}_{\nu;r;n;q_{1},q_{2}} & =\int_{0}^{\infty}\frac{d\omega}{2\pi}\,\frac{A_{\nu;r;T=0}\left(\omega\right)}{\omega^{2}\left\{ 1-e^{-\beta\omega}\right\} }\tilde{f}_{\nu;n;q_{1},q_{2}}^{\left(\eta\right)}\left(\omega\right)+\int_{0}^{\infty}\frac{d\omega}{2\pi}\,\frac{A_{\nu;r;T=0}\left(\omega\right)}{\omega^{2}\left\{ 1-e^{-\beta\omega}\right\} }e^{-\beta\omega}\tilde{f}_{\nu;n;q_{1},q_{2}}^{\left(\eta\right)}\left(-\omega\right)\nonumber \\
 & =\int_{0}^{\infty}\frac{d\omega}{2\pi}\,\frac{A_{\nu;r;T=0}\left(\omega\right)}{\omega^{2}\left\{ 1-e^{-\beta\omega}\right\} }\tilde{f}_{\nu;n;q_{1},q_{2}}^{\left(\eta\right)}\left(\omega\right)+\int_{-\infty}^{0}\frac{d\omega}{2\pi}\,\frac{A_{\nu;r;T=0}\left(-\omega\right)}{\omega^{2}\left\{ 1-e^{\beta\omega}\right\} }e^{\beta\omega}\tilde{f}_{\nu;n;q_{1},q_{2}}^{\left(\eta\right)}\left(\omega\right)\nonumber \\
 & =\int_{0}^{\infty}\frac{d\omega}{2\pi}\,\frac{A_{\nu;r;T=0}\left(\omega\right)}{\omega^{2}\left\{ 1-e^{-\beta\omega}\right\} }\tilde{f}_{\nu;n;q_{1},q_{2}}^{\left(\eta\right)}\left(\omega\right)-\int_{-\infty}^{0}\frac{d\omega}{2\pi}\,\frac{A_{\nu;r;T=0}\left(-\omega\right)}{\omega^{2}\left\{ 1-e^{-\beta\omega}\right\} }\tilde{f}_{\nu;n;q_{1},q_{2}}^{\left(\eta\right)}\left(\omega\right)\nonumber \\
 & =\int_{0}^{\infty}\frac{d\omega}{2\pi}\,\frac{1}{\omega^{2}}\left[\text{sign}\left(\omega\right)\frac{A_{\nu;r;T=0}\left(\left|\omega\right|\right)}{1-e^{-\beta\omega}}\right]\tilde{f}_{\nu;n;q_{1},q_{2}}^{\left(\eta\right)}\left(\omega\right)\nonumber \\
 & \mathrel{\phantom{=}}\mathop{+}\int_{-\infty}^{0}\frac{d\omega}{2\pi}\,\frac{1}{\omega^{2}}\left[\text{sign}\left(\omega\right)\frac{A_{\nu;r;T=0}\left(\left|\omega\right|\right)}{1-e^{-\beta\omega}}\right]\tilde{f}_{\nu;n;q_{1},q_{2}}^{\left(\eta\right)}\left(\omega\right)\nonumber \\
 & =\int_{-\infty}^{\infty}\frac{d\omega}{2\pi}\,\frac{A_{\nu;r;T}\left(\omega\right)}{\omega^{2}}\tilde{f}_{\nu;n;q_{1},q_{2}}^{\left(\eta\right)}\left(\omega\right).\label{eq:tilde eta_=00007Bv; n; k_1, k_2=00007D -- 3}
\end{align}

\noindent Next, we simplify $\tilde{f}_{\nu;n;q_{1},q_{2}}^{\left(\eta\right)}\left(\omega\right)$.
Using Eq.~\eqref{eq:summing convention -- 1} and \eqref{eq:tilde f_=00007Bv; n; q_1, q_2=00007D^=00007B(eta)=00007D(omega) -- 1},
and noting that $q_{1}\ge q_{2}$ for all instances of $\tilde{f}_{\nu;n;q_{1},q_{2}}^{\left(\eta\right)}\left(\omega\right)$
in the path integral expression for $\hat{\rho}^{\left(A\right)}\left(t_{n}\right)$,
we can write

\begin{equation}
\tilde{f}_{\nu;n;q_{1},q_{2}}^{\left(\eta\right)}\left(\omega\right)=\begin{cases}
\tilde{f}_{\nu;n;q_{1}}^{\left(\eta,\Delta q=0\right)}\left(\omega\right), & \text{if }q_{1}-q_{2}=0,\\
\tilde{f}_{\nu;n;q_{1},q_{2}}^{\left(\eta,\Delta q=1\right)}\left(\omega\right), & \text{if }q_{1}-q_{2}=1,\\
\tilde{f}_{\nu;n;q_{1},q_{2}}^{\left(\eta,\Delta q\ge2\right)}\left(\omega\right), & \text{if }q_{1}-q_{2}\ge2,
\end{cases}\label{eq:tilde f_=00007Bv; n; q_1, q_2=00007D^=00007B(eta)=00007D(omega) -- 2}
\end{equation}

\noindent where

\begin{equation}
\tilde{f}_{\nu;n;q_{1}}^{\left(\eta,\Delta q=0\right)}\left(\omega\right)=1-e^{-i\Delta t\omega\tilde{w}_{n;q_{1}}},\label{eq:tilde f_=00007Bv; n; q_1=00007D^=00007B(eta, dq=00003D0)=00007D(omega) -- 1}
\end{equation}

\begin{equation}
\tilde{f}_{\nu;n;q_{1},q_{2}}^{\left(\eta,\Delta q=1\right)}\left(\omega\right)=e^{-i\Delta t\omega\tilde{w}_{n;q_{2}}}-1-e^{-i\Delta t\omega\left\{ \tilde{w}_{n;q_{1}}+\tilde{w}_{n;q_{2}}\right\} }+e^{-i\Delta t\omega\tilde{w}_{n;q_{1}}},\label{eq:tilde f_=00007Bv; n; q_1, q_2=00007D^=00007B(eta, dq=00003D1)=00007D(omega) -- 1}
\end{equation}

\noindent and

\begin{equation}
\tilde{f}_{\nu;n;q_{1},q_{2}}^{\left(\eta,\Delta q\ge2\right)}\left(\omega\right)=e^{-i\Delta t\omega\sum_{q=q_{2}}^{q_{1}-1}\tilde{w}_{n;q}}-e^{-i\Delta t\omega\sum_{q=q_{2}+1}^{q_{1}-1}\tilde{w}_{n;q}}-e^{-i\Delta t\omega\sum_{q=q_{2}}^{q_{1}}\tilde{w}_{n;q}}+e^{-i\Delta t\omega\sum_{q=q_{2}+1}^{q_{1}}\tilde{w}_{n;q}}.\label{eq:tilde f_=00007Bv; n; q_1, q_2=00007D^=00007B(eta, dq>=00003D2)=00007D(omega) -- 1}
\end{equation}

\noindent Noting that

\begin{equation}
4\sin\left(a\right)\sin\left(b\right)=4\left\{ \frac{e^{ia}-e^{-ia}}{2i}\right\} \left\{ \frac{e^{ib}-e^{-ib}}{2i}\right\} =e^{i\left\{ a-b\right\} }+e^{-i\left\{ a-b\right\} }-e^{i\left\{ a+b\right\} }-e^{-i\left\{ a+b\right\} },\label{eq:4sin(a)sin(b) expr -- 1}
\end{equation}

\noindent we can rewrite Eq.~\eqref{eq:tilde f_=00007Bv; n; q_1, q_2=00007D^=00007B(eta, dq>=00003D2)=00007D(omega) -- 1}
as follows:

\begin{align}
\tilde{f}_{\nu;n;q_{1},q_{2}}^{\left(\eta,\Delta q\ge2\right)}\left(\omega\right) & =e^{-i\Delta t\omega\sum_{q=q_{2}}^{q_{1}-1}\tilde{w}_{n;q}}-e^{-i\Delta t\omega\sum_{q=q_{2}+1}^{q_{1}-1}\tilde{w}_{n;q}}-e^{-i\Delta t\omega\sum_{q=q_{2}}^{q_{1}}\tilde{w}_{n;q}}+e^{-i\Delta t\omega\sum_{q=q_{2}+1}^{q_{1}}\tilde{w}_{n;q}}\nonumber \\
 & =e^{\frac{i}{2}\Delta t\omega\left\{ \tilde{w}_{n;q_{1}}-\tilde{w}_{n;q_{2}}\right\} }e^{-i\Delta t\omega\left\{ \frac{1}{2}\tilde{w}_{n;q_{1}}+\sum_{q=q_{2}+1}^{q_{1}-1}\tilde{w}_{n;q}+\frac{1}{2}\tilde{w}_{n;q_{2}}\right\} }\nonumber \\
 & \mathrel{\phantom{=}}\mathop{-}e^{\frac{i}{2}\Delta t\omega\left\{ \tilde{w}_{n;q_{1}}+\tilde{w}_{n;q_{2}}\right\} }e^{-i\Delta t\omega\left\{ \frac{1}{2}\tilde{w}_{n;q_{1}}+\sum_{q=q_{2}+1}^{q_{1}-1}\tilde{w}_{n;q}+\frac{1}{2}\tilde{w}_{n;q_{2}}\right\} }\nonumber \\
 & \mathrel{\phantom{=}}\mathop{-}e^{-\frac{i}{2}\Delta t\omega\left\{ \tilde{w}_{n;q_{1}}+\tilde{w}_{n;q_{2}}\right\} }e^{-i\Delta t\omega\left\{ \frac{1}{2}\tilde{w}_{n;q_{1}}+\sum_{q=q_{2}+1}^{q_{1}-1}\tilde{w}_{n;q}+\frac{1}{2}\tilde{w}_{n;q_{2}}\right\} }\nonumber \\
 & \mathrel{\phantom{=}}\mathop{+}e^{-\frac{i}{2}\Delta t\omega\left\{ \tilde{w}_{n;q_{1}}-\tilde{w}_{n;q_{2}}\right\} }e^{-i\Delta t\omega\left\{ \frac{1}{2}\tilde{w}_{n;q_{1}}+\sum_{q=q_{2}+1}^{q_{1}-1}\tilde{w}_{n;q}+\frac{1}{2}\tilde{w}_{n;q_{2}}\right\} }\nonumber \\
 & =e^{-i\Delta t\omega\left\{ \frac{1}{2}\tilde{w}_{n;q_{1}}+\sum_{q=q_{2}+1}^{q_{1}-1}\tilde{w}_{n;q}+\frac{1}{2}\tilde{w}_{n;q_{2}}\right\} }\nonumber \\
 & \mathrel{\phantom{=}}\mathop{\times}\left[e^{\frac{i}{2}\Delta t\omega\left\{ \tilde{w}_{n;q_{1}}-\tilde{w}_{n;q_{2}}\right\} }+e^{-\frac{i}{2}\Delta t\omega\left\{ \tilde{w}_{n;q_{1}}-\tilde{w}_{n;q_{2}}\right\} }\right.\nonumber \\
 & \phantom{\mathrel{=}\mathop{\times}}\left.\quad\mathop{-}e^{\frac{i}{2}\Delta t\omega\left\{ \tilde{w}_{n;q_{1}}+\tilde{w}_{n;q_{2}}\right\} }-e^{-\frac{i}{2}\Delta t\omega\left\{ \tilde{w}_{n;q_{1}}+\tilde{w}_{n;q_{2}}\right\} }\right]\nonumber \\
 & =4e^{-i\Delta t\omega\left\{ \frac{1}{2}\tilde{w}_{n;q_{1}}+\sum_{q=q_{2}+1}^{q_{1}-1}\tilde{w}_{n;q}+\frac{1}{2}\tilde{w}_{n;q_{2}}\right\} }\sin\left(\frac{\Delta t\omega\tilde{w}_{n;q_{1}}}{2}\right)\sin\left(\frac{\Delta t\omega\tilde{w}_{n;q_{2}}}{2}\right).\label{eq:tilde f_=00007Bv; n; q_1, q_2=00007D^=00007B(eta, dq>=00003D2)=00007D(omega) -- 2}
\end{align}

\noindent Moreover, using Eq.~\eqref{eq:summing convention -- 1},
we can rewrite similarly Eq.~\eqref{eq:tilde f_=00007Bv; n; q_1, q_2=00007D^=00007B(eta, dq=00003D1)=00007D(omega) -- 1}
as follows:

\begin{align}
\tilde{f}_{\nu;n;q_{1},q_{2}}^{\left(\eta,\Delta q=1\right)}\left(\omega\right) & =e^{-i\Delta t\omega\tilde{w}_{n;q_{2}}}-1-e^{-i\Delta t\omega\left\{ \tilde{w}_{n;q_{1}}+\tilde{w}_{n;q_{2}}\right\} }+e^{-i\Delta t\omega\tilde{w}_{n;q_{1}}}\nonumber \\
 & =e^{\frac{i}{2}\Delta t\omega\left\{ \tilde{w}_{n;q_{1}}-\tilde{w}_{n;q_{2}}\right\} }e^{-i\Delta t\omega\left\{ \frac{1}{2}\tilde{w}_{n;q_{1}}+\sum_{q=q_{2}+1}^{q_{1}-1}\tilde{w}_{n;q}+\frac{1}{2}\tilde{w}_{n;q_{2}}\right\} }\nonumber \\
 & \mathrel{\phantom{=}}\mathop{-}e^{\frac{i}{2}\Delta t\omega\left\{ \tilde{w}_{n;q_{1}}+\tilde{w}_{n;q_{2}}\right\} }e^{-i\Delta t\omega\left\{ \frac{1}{2}\tilde{w}_{n;q_{1}}+\sum_{q=q_{2}+1}^{q_{1}-1}\tilde{w}_{n;q}+\frac{1}{2}\tilde{w}_{n;q_{2}}\right\} }\nonumber \\
 & \mathrel{\phantom{=}}\mathop{-}e^{-\frac{i}{2}\Delta t\omega\left\{ \tilde{w}_{n;q_{1}}+\tilde{w}_{n;q_{2}}\right\} }e^{-i\Delta t\omega\left\{ \frac{1}{2}\tilde{w}_{n;q_{1}}+\sum_{q=q_{2}+1}^{q_{1}-1}\tilde{w}_{n;q}+\frac{1}{2}\tilde{w}_{n;q_{2}}\right\} }\nonumber \\
 & \mathrel{\phantom{=}}\mathop{+}e^{-\frac{i}{2}\Delta t\omega\left\{ \tilde{w}_{n;q_{1}}-\tilde{w}_{n;q_{2}}\right\} }e^{-i\Delta t\omega\left\{ \frac{1}{2}\tilde{w}_{n;q_{1}}+\sum_{q=q_{2}+1}^{q_{1}-1}\tilde{w}_{n;q}+\frac{1}{2}\tilde{w}_{n;q_{2}}\right\} }\nonumber \\
 & =e^{-i\Delta t\omega\left\{ \frac{1}{2}\tilde{w}_{n;q_{1}}+\sum_{q=q_{2}+1}^{q_{1}-1}\tilde{w}_{n;q}+\frac{1}{2}\tilde{w}_{n;q_{2}}\right\} }\nonumber \\
 & \mathrel{\phantom{=}}\mathop{\times}\left[e^{\frac{i}{2}\Delta t\omega\left\{ \tilde{w}_{n;q_{1}}-\tilde{w}_{n;q_{2}}\right\} }+e^{-\frac{i}{2}\Delta t\omega\left\{ \tilde{w}_{n;q_{1}}-\tilde{w}_{n;q_{2}}\right\} }\right.\nonumber \\
 & \phantom{\mathrel{=}\mathop{\times}}\left.\quad\mathop{-}e^{\frac{i}{2}\Delta t\omega\left\{ \tilde{w}_{n;q_{1}}+\tilde{w}_{n;q_{2}}\right\} }-e^{-\frac{i}{2}\Delta t\omega\left\{ \tilde{w}_{n;q_{1}}+\tilde{w}_{n;q_{2}}\right\} }\right]\nonumber \\
 & =4e^{-i\Delta t\omega\left\{ \frac{1}{2}\tilde{w}_{n;q_{1}}+\sum_{q=q_{2}+1}^{q_{1}-1}\tilde{w}_{n;q}+\frac{1}{2}\tilde{w}_{n;q_{2}}\right\} }\sin\left(\frac{\Delta t\omega\tilde{w}_{n;q_{1}}}{2}\right)\sin\left(\frac{\Delta t\omega\tilde{w}_{n;q_{2}}}{2}\right).\label{eq:tilde f_=00007Bv; n; q_1, q_2=00007D^=00007B(eta, dq=00003D1)=00007D(omega) -- 2}
\end{align}

\noindent Therefore, using Eqs.~\eqref{eq:tilde f_=00007Bv; n; q_1, q_2=00007D^=00007B(eta)=00007D(omega) -- 2},
\eqref{eq:tilde f_=00007Bv; n; q_1=00007D^=00007B(eta, dq=00003D0)=00007D(omega) -- 1},
\eqref{eq:tilde f_=00007Bv; n; q_1, q_2=00007D^=00007B(eta, dq>=00003D2)=00007D(omega) -- 2},
and \eqref{eq:tilde f_=00007Bv; n; q_1, q_2=00007D^=00007B(eta, dq=00003D1)=00007D(omega) -- 2}
we can express $\tilde{f}_{\nu;n;q_{1},q_{2}}^{\left(\eta\right)}\left(\omega\right)$
as

\begin{equation}
\tilde{f}_{\nu;n;q_{1},q_{2}}^{\left(\eta\right)}\left(\omega\right)=\begin{cases}
\tilde{f}_{\nu;n;q_{1}}^{\left(\eta,\Delta q=0\right)}\left(\omega\right), & \text{if }q_{1}-q_{2}=0,\\
\tilde{f}_{\nu;n;q_{1},q_{2}}^{\left(\eta,\Delta q\ge1\right)}\left(\omega\right), & \text{if }q_{1}-q_{2}\ge1,
\end{cases}\label{eq:tilde f_=00007Bv; n; q_1, q_2=00007D^=00007B(eta)=00007D(omega) -- 3}
\end{equation}

\noindent where 

\begin{equation}
\tilde{f}_{\nu;n;q_{1},q_{2}}^{\left(\eta,\Delta q\ge1\right)}\left(\omega\right)=4e^{-i\Delta t\omega\left\{ \frac{1}{2}\tilde{w}_{n;q_{1}}+\sum_{q=q_{2}+1}^{q_{1}-1}\tilde{w}_{n;q}+\frac{1}{2}\tilde{w}_{n;q_{2}}\right\} }\sin\left(\frac{\Delta t\omega\tilde{w}_{n;q_{1}}}{2}\right)\sin\left(\frac{\Delta t\omega\tilde{w}_{n;q_{2}}}{2}\right).\label{eq:tilde f_=00007Bv; n; q_1, q_2=00007D^=00007B(eta, dq=00003D>1)=00007D(omega) -- 1}
\end{equation}

Now we are ready to calculate the bosonic integral in Eq.~\eqref{eq:rho^(A) in path integral appendix -- 4}
in its entirety. Using Eqs.~\eqref{eq:Z_=00007Bu; v=00007D^(B) -- 2},
\eqref{eq:defining b measure -- 1}-\eqref{eq:defining F_=00007Bv; r; epsilon; n=00007D -- 1},
\eqref{eq:Gaussian integral identity -- 1}, \eqref{eq:det=00005BA_=00007Bv; epsilon; n=00007D=00005D -- 2},
and \eqref{eq:sum xi A psi -- 1} we get

\begin{align}
 & \int\mathcal{D}b\,F_{n\vphantom{\nu;r;\epsilon}}\left(\boldsymbol{\sigma}_{\nu\in\left\{ y,z\right\} ;\alpha=\pm1;q\in\left[0,2n+1\right]};b\right)\nonumber \\
 & \quad=e^{-2\sum_{\nu\in\left\{ y,z\right\} }\sum_{q=0}^{2n+1}\sum_{u,r,\epsilon}\left\{ \mathcal{E}_{v;r;q}^{\left(\lambda\right)}\frac{\lambda_{\nu;r;\epsilon}}{\omega_{\nu;\epsilon}}\right\} ^{2}}\prod_{u=-N}^{N}\left\{ \mathcal{Z}_{u;y}^{\left(B\right)}\mathcal{Z}_{u;z}^{\left(B\right)}\right\} \prod_{\nu\in\left\{ y,z\right\} }\left\{ \tilde{I}_{\nu;n}^{\left(\text{bath}\right)}\left(\boldsymbol{\sigma}_{\nu;\alpha=\pm1;q\in\left[0,2n+1\right]}\right)\right\} ,\label{eq:evaluating bosonic path integral -- 1}
\end{align}

\noindent where

\begin{equation}
\tilde{I}_{\nu;n}^{\left(\text{bath}\right)}\left(\boldsymbol{\sigma}_{\nu;\alpha=\pm1;q\in\left[0,2n+1\right]}\right)=\prod_{u=-N}^{N}\prod_{r=0}^{L-1}\left\{ \tilde{I}_{\nu;r+uL;n}^{\left(\text{bath}\right)}\left(\sigma_{\nu;r+uL;\alpha=\pm1;q\in\left[0,2n+1\right]}\right)\right\} ,\label{eq:I_=00007Bv; n=00007D^(bath) -- 1}
\end{equation}

\noindent with

\begin{equation}
\tilde{I}_{\nu;r;n}^{\left(\text{bath}\right)}\left(\sigma_{\nu;r;\alpha=\pm1;q\in\left[0,2n+1\right]}\right)=\prod_{q_{2}=0}^{2n+1}\prod_{q_{1}=0}^{q_{2}}\left\{ \tilde{I}_{\nu;r;n;q_{1},q_{2}}^{\left(\text{bath}\right)}\left(\sigma_{\nu;r;1;q_{1}},\sigma_{\nu;r;-1;q_{1}},\sigma_{\nu;r;1;q_{2}},\sigma_{\nu;r;-1;q_{2}}\right)\right\} ,\label{eq:I_=00007Bv; r; n=00007D^(bath) -- 1}
\end{equation}

\noindent and

\begin{equation}
\tilde{I}_{\nu;r;n;q_{1},q_{2}}^{\left(\text{bath}\right)}\left(\sigma_{\nu;r;1;q_{1}},\sigma_{\nu;r;-1;q_{1}},\sigma_{\nu;r;1;q_{2}},\sigma_{\nu;r;-1;q_{2}}\right)=e^{-\tilde{\gamma}_{\nu;r;n;q_{1},q_{2}}\left(\sigma_{\nu;r;1,q_{1}},\sigma_{\nu;r;-1,q_{1}},\sigma_{\nu;r;1,q_{2}},\sigma_{\nu;r;-1,q_{2}}\right)}.\label{eq:I_=00007Bv; r; n; q_1, q_2=00007D^(bath) -- 1}
\end{equation}

\noindent Using Eqs.~\eqref{eq:rho^(A) in path integral appendix -- 4}
and \eqref{eq:evaluating bosonic path integral -- 1}, we can rewrite
the path integral expression for $\hat{\rho}^{\left(A\right)}\left(t_{n}\right)$
as

\begin{align}
 & \rho^{\left(A\right)}\left(\boldsymbol{\sigma}_{z;1;2n+1},\boldsymbol{\sigma}_{z;-1;2n+1}\right)\nonumber \\
 & \quad\equiv\left\langle \boldsymbol{\sigma}_{z;1;2n+1}z\right|\hat{\rho}^{\left(A\right)}\left(t_{n}\right)\left|\boldsymbol{\sigma}_{z;-1;2n+1}z\right\rangle \nonumber \\
 & \quad=\prod_{\alpha=\pm1}\prod_{u=-N}^{N}\prod_{r=0}^{L-1}\left[\prod_{q=0}^{2n+1}\left\{ \sum_{\sigma_{y;r+uL;\alpha;q}=\pm1}\right\} \prod_{q=0}^{2n}\left\{ \sum_{\sigma_{z;r+uL;\alpha;q}=\pm1}\right\} \right]\nonumber \\
 & \quad\mathrel{\phantom{=}}\mathop{\times}\prod_{\alpha=\pm1}\prod_{u=-N}^{N}\prod_{r=0}^{L-1}\prod_{l=1}^{n}\left\{ \delta_{\sigma_{z;r+uL;\alpha;2l-1},\sigma_{z;r+uL;\alpha;2l}}\right\} e^{i\sum_{\alpha=\pm1}\tilde{\phi}_{\alpha;n}^{\left(\text{lcafc}\right)}\left(\boldsymbol{\sigma}_{z;\alpha;q\in\left[0,2n+1\right]}\right)}\nonumber \\
 & \quad\mathrel{\phantom{=}}\mathop{\times}\prod_{\alpha=\pm1}\left\{ \tilde{I}_{\alpha;n}^{\left(\text{tfc}\right)}\left(\boldsymbol{\sigma}_{y;\alpha;q\in\left[0,2n+1\right]}\right)\tilde{I}_{\alpha;n}^{\left(y\leftrightarrow z\right)}\left(\boldsymbol{\sigma}_{\nu\in\left\{ y,z\right\} ;\alpha;q\in\left[0,2n+1\right]}\right)\right\} \prod_{\nu\in\left\{ y,z\right\} }\left\{ \tilde{I}_{\nu;n}^{\left(\text{bath}\right)}\left(\boldsymbol{\sigma}_{\nu;\alpha=\pm1;q\in\left[0,2n+1\right]}\right)\right\} \nonumber \\
 & \quad\mathrel{\phantom{=}}\mathop{\times}\rho^{\left(i,A\right)}\left(\boldsymbol{\sigma}_{z;1;0},\boldsymbol{\sigma}_{z;-1;0}\right)+\mathcal{O}\left[\Delta t^{2}\right].\label{eq:rho^(A) in path integral appendix -- 5}
\end{align}

\noindent Next, we apply the $\delta_{\sigma_{z;r+uL;\alpha;2l-1},\sigma_{z;r+uL;\alpha;2l}}$
in Eq.~\eqref{eq:rho^(A) in path integral appendix -- 5} to reduce
the number of $\boldsymbol{\sigma}_{z;\alpha;q}$-degrees of freedom
by roughly half and make the following symbol substitutions:

\begin{align}
\sigma_{z;r;\alpha;q=2l} & \to\sigma_{z;r;\alpha;q=l},\quad\text{for }0\le l\le n,\label{eq:sigma-z symbol substitution -- 1}\\
\sigma_{z;r;\alpha;q=2n+1} & \to\sigma_{z;r;\alpha;q=n+1}.\label{eq:sigma-z symbol substitution -- 2}
\end{align}

\noindent Upon doing so, Eq.~\eqref{eq:rho^(A) in path integral appendix -- 5}
can be rewritten as

\begin{align}
 & \rho^{\left(A\right)}\left(\boldsymbol{\sigma}_{z;1;n+1},\boldsymbol{\sigma}_{z;-1;n+1}\right)\nonumber \\
 & \quad\equiv\left\langle \boldsymbol{\sigma}_{z;1;n+1}z\right|\hat{\rho}^{\left(A\right)}\left(t_{n}\right)\left|\boldsymbol{\sigma}_{z;-1;n+1}z\right\rangle \nonumber \\
 & \quad=\prod_{\alpha=\pm1}\prod_{u=-N}^{N}\prod_{r=0}^{L-1}\left[\prod_{q=0}^{\tilde{q}_{y;n}}\left\{ \sum_{\sigma_{y;r+uL;\alpha;q}=\pm1}\right\} \prod_{q=0}^{\tilde{q}_{z;n}-1}\left\{ \sum_{\sigma_{z;r+uL;\alpha;q}=\pm1}\right\} \right]e^{i\sum_{\alpha=\pm1}\phi_{\alpha;n}^{\left(\text{lcafc}\right)}\left(\boldsymbol{\sigma}_{z;\alpha;q\in\left[0,\tilde{q}_{z;n}\right]}\right)}\nonumber \\
 & \quad\mathrel{\phantom{=}}\mathop{\times}I_{n}^{\left(\text{yz-noise}\right)}\left(\boldsymbol{\sigma}_{\nu\in\left\{ y,z\right\} ;\alpha=\pm1;q\in\left[0,\tilde{q}_{\nu;n}\right]}\right)\rho^{\left(i,A\right)}\left(\boldsymbol{\sigma}_{z;1;0},\boldsymbol{\sigma}_{z;-1;0}\right)+\mathcal{O}\left[\Delta t^{2}\right],\label{eq:rho^(A) in path integral appendix -- 6}
\end{align}

\noindent where

\begin{equation}
\phi_{\alpha;n}^{\left(\text{lcafc}\right)}\left(\boldsymbol{\sigma}_{z;\alpha;q\in\left[0,\tilde{q}_{z;n}\right]}\right)=\sum_{q=0}^{\tilde{q}_{z;n}}\phi_{\alpha;n;k=q}^{\left(\text{lcafc}\right)}\left(\boldsymbol{\sigma}_{z;\alpha;q}\right),\label{eq:phi_=00007Balpha; n=00007D^(lcafc) in path integral appendix -- 1}
\end{equation}

\noindent with

\begin{equation}
\phi_{\alpha;n;k}^{\left(\text{lcafc}\right)}\left(\boldsymbol{\sigma}_{z;\alpha;q}\right)=-\frac{\alpha\Delta t}{2}\sum_{k^{\prime}=k-1}^{k}w_{n;k^{\prime}}H_{z}^{\left(A\right)}\left(t_{k^{\prime}};\boldsymbol{\sigma}_{z;\alpha;q}\right),\label{eq:phi_=00007Balpha; n; k=00007D^(lcafc) in path integral appendix -- 1}
\end{equation}

\begin{equation}
H_{z}^{\left(A\right)}\left(t_{k^{\prime}};\boldsymbol{\sigma}_{z;\alpha;q}\right)=\sum_{u=-N}^{N}H_{u;z}^{\left(A\right)}\left(t_{k^{\prime}};\boldsymbol{\sigma}_{z;\alpha;q}\right),\label{eq:H_z^(A)(t; sigma) in path integral appendix -- 1}
\end{equation}

\begin{equation}
H_{u;z}^{\left(A\right)}\left(t;\boldsymbol{\sigma}_{z;\alpha;q}\right)=\sum_{r=0}^{L-1}h_{z;r}\left(t\right)\sigma_{z;r+uL;\alpha;q}+\sum_{r=1}^{L-1}J_{z,z;r,r+1}\left(t\right)\sigma_{z;r+uL;\alpha;q}\sigma_{z;r+uL+1;\alpha;q},\label{eq:H_=00007Bu; z=00007D^(A)(t; sigma) in path integral appendix -- 1}
\end{equation}

\begin{equation}
w_{n;k}=\begin{cases}
0, & \text{if }k=-1,n+1,\\
\frac{1}{2}, & \text{if }k=0,n,\\
1, & \text{if }0<k<n;
\end{cases}\label{eq:k composite quadrature rule in path integral appendix -- 1}
\end{equation}

\noindent $I_{n}^{\left(\text{yz-noise}\right)}\left(\boldsymbol{\sigma}_{\nu\in\left\{ y,z\right\} ;\alpha=\pm1;q\in\left[0,\tilde{q}_{\nu;n}\right]}\right)$
is the total influence:

\begin{equation}
I_{n}^{\left(\text{yz-noise}\right)}\left(\boldsymbol{\sigma}_{\nu\in\left\{ y,z\right\} ;\alpha=\pm1;q\in\left[0,\tilde{q}_{\nu;n}\right]}\right)=\prod_{u=-N}^{N}\prod_{r=0}^{L-1}\left\{ I_{r+uL;n}^{\left(\text{yz-noise}\right)}\left(\sigma_{\nu\in\left\{ y,z\right\} ;r+uL;\alpha=\pm1;q\in\left[0,\tilde{q}_{\nu;n}\right]}\right)\right\} ,\label{eq:I_n for yz-noise in path integral appendix -- 1}
\end{equation}

\noindent with $I_{r;n}^{\left(\text{yz-noise}\right)}\left(\sigma_{\nu\in\left\{ y,z\right\} ;r;\alpha=\pm1;q\in\left[0,\tilde{q}_{\nu;n}\right]}\right)$
being the influence functional that couples the Ising spin variables\\
$\sigma_{\nu\in\left\{ y,z\right\} ;r;\alpha=\pm1;q\in\left[0,\tilde{q}_{\nu;n}\right]}$
with one another {[}including Ising spin variables associated with
different times{]}:

\begin{align}
I_{r;n}^{\left(\text{yz-noise}\right)}\left(\sigma_{\nu\in\left\{ y,z\right\} ;r;\alpha=\pm1;q\in\left[0,\tilde{q}_{\nu;n}\right]}\right) & =I_{n}^{\left(y\leftrightarrow z\right)}\left(\sigma_{\nu\in\left\{ y,z\right\} ;r;\alpha=\pm1;q\in\left[0,\tilde{q}_{\nu;n}\right]}\right)\mathop{\times}I_{y;r;n}^{\left(\text{tfc}\right)}\left(\sigma_{y;r;\alpha=\pm1;q\in\left[0,\tilde{q}_{y;n}\right]}\right)\nonumber \\
 & \mathrel{\phantom{=}}\mathop{\times}\prod_{\nu\in\left\{ y,z\right\} }\left\{ I_{\nu;r;n}^{\left(\text{bath}\right)}\left(\sigma_{\nu;r;\alpha=\pm1;q\in\left[0,\tilde{q}_{\nu;n}\right]}\right)\right\} ,\label{eq:I_=00007Br; n=00007D^(yz-noise) in path integral appendix -- 1}
\end{align}

\begin{align}
I_{n}^{\left(y\leftrightarrow z\right)}\left(\sigma_{\nu\in\left\{ y,z\right\} ;r;\alpha=\pm1;q\in\left[0,\tilde{q}_{\nu;n}\right]}\right) & =\prod_{l=0}^{n}\left\{ I^{\left(y\leftrightarrow z,1\right)}\left(\sigma_{y;r;1,2l+1},\sigma_{y;r;-1,2l+1},\sigma_{z;r;1,l+1},\sigma_{z;r;-1,l+1}\right)\right.\nonumber \\
 & \phantom{=\prod_{l=0}^{n}}\left.\quad\mathop{\times}I^{\left(y\leftrightarrow z,2\right)}\left(\sigma_{y;r;1,2l},\sigma_{y;r;-1,2l},\sigma_{z;r;1,l},\sigma_{z;r;-1,l}\right)\right\} ,\label{eq:I_n^(y<->z) in path integral appendix -- 1}
\end{align}

\begin{align}
 & I^{\left(y\leftrightarrow z,1\right)}\left(\sigma_{y;r;1,2l+1},\sigma_{y;r;-1,2l+1},\sigma_{z;r;1,l+1},\sigma_{z;r;-1,l+1}\right)\nonumber \\
 & \quad=I^{\left(y\to z\right)}\left(\sigma_{z;r;1,l+1},\sigma_{y;r;1,2l+1}\right)I^{\left(z\to y\right)}\left(\sigma_{y;r;-1,2l+1},\sigma_{z;r;-1,l+1}\right),\label{eq:I^(y<->z, 1) in path integral appendix -- 1}
\end{align}

\begin{align}
 & I^{\left(y\leftrightarrow z,2\right)}\left(\sigma_{y;r;1,2l},\sigma_{y;r;-1,2l},\sigma_{z;r;1,l},\sigma_{z;r;-1,l}\right)\nonumber \\
 & \quad=I^{\left(z\to y\right)}\left(\sigma_{y;r;1,2l},\sigma_{z;r;1,l}\right)I^{\left(y\to z\right)}\left(\sigma_{z;r;-1,l},\sigma_{y;r;-1,2l}\right),\label{eq:I^(y<->z, 2) in path integral appendix -- 1}
\end{align}

\begin{align}
I^{\left(z\to y\right)}\left(\sigma_{y;r;\alpha;q^{\vphantom{\prime}}},\sigma_{z;r;\alpha;q^{\prime}}\right) & =\left.\left\langle \sigma_{y;r;\alpha;q^{\vphantom{\prime}}}y\right|\sigma_{z;r;\alpha;q^{\prime}}z\right\rangle =\frac{1}{2\sqrt{2}}\left\{ 1-i\sigma_{y;r;\alpha;q^{\vphantom{\prime}}}+\sigma_{z;r;\alpha;q^{\prime}}+i\sigma_{y;r;\alpha;q^{\vphantom{\prime}}}\sigma_{z;r;\alpha;q^{\prime}}\right\} ,\label{eq:I^(z->y) in path integral appendix -- 1}\\
I^{\left(y\to z\right)}\left(\sigma_{z;r;\alpha;q^{\prime}},\sigma_{y;r;\alpha;q^{\vphantom{\prime}}}\right) & =\left.\left\langle \sigma_{z;r;\alpha;q^{\prime}}z\right|\sigma_{y;r;\alpha;q^{\vphantom{\prime}}}y\right\rangle =\left\{ I^{\left(z\to y\right)}\left(\sigma_{y;r;\alpha;q^{\vphantom{\prime}}},\sigma_{z;r;\alpha;q^{\prime}}\right)\right\} ^{*},\label{eq:I^(y->z) in path integral appendix -- 1}
\end{align}

\begin{equation}
I_{y;r;n}^{\left(\text{tfc}\right)}\left(\sigma_{y;r;\alpha=\pm1;q\in\left[0,\tilde{q}_{y;n}\right]}\right)=\prod_{l=0}^{n}\left\{ I_{y;r;n;k=l}^{\left(\text{tfc}\right)}\left(\sigma_{y;r;1;2l},\sigma_{y;r;-1;2l},\sigma_{y;r;1;2l+1},\sigma_{y;r;-1;2l+1}\right)\right\} ,\label{eq:I_=00007By; r; n=00007D^(tfc) in path integral appendix -- 1}
\end{equation}

\begin{equation}
I_{y;r;n;k}^{\left(\text{tfc}\right)}\left(\sigma_{y;r;1;q_{1}},\sigma_{y;r;-1;q_{1}},\sigma_{y;r;1;q_{2}},\sigma_{y;r;-1;q_{2}}\right)=I_{y;r;1;n;k}^{\left(\text{tfc}\right)}\left(\sigma_{y;r;1;q_{2}},\sigma_{y;r;1;q_{1}}\right)I_{y;r;-1;n;k}^{\left(\text{tfc}\right)}\left(\sigma_{y;r;-1;q_{1}},\sigma_{y;r;-1;q_{2}}\right),\label{eq:I_=00007Bv; r; n; k=00007D^(tfc) in path integral appendix -- 1}
\end{equation}

\begin{equation}
I_{y;r;\alpha;n;k}^{\left(\text{tfc}\right)}\left(\sigma_{y;r;\alpha;q_{1}},\sigma_{y;r;\alpha;q_{2}}\right)=\begin{cases}
\left\langle \sigma_{y;r;-1;q_{1}}y\right|e^{i\Delta tw_{n;k}h_{x;r}\left(t_{k}\right)\hat{\sigma}_{x;r}}\left|\sigma_{y;r;-1;q_{2}}y\right\rangle , & \text{if }\alpha=-1,\\
\left\langle \sigma_{y;r;1;q_{2}}y\right|e^{-i\Delta tw_{n;k}h_{x;r}\left(t_{k}\right)\hat{\sigma}_{x;r}}\left|\sigma_{y;r;1;q_{1}}y\right\rangle , & \text{if }\alpha=1,
\end{cases}\label{eq:I_=00007By; r; alpha; n; k=00007D^(tfc) in path integral appendix -- 1}
\end{equation}

\begin{equation}
I_{\nu;r;n}^{\left(\text{bath}\right)}\left(\sigma_{\nu;r;\alpha=\pm1;q\in\left[0,\tilde{q}_{\nu;n}\right]}\right)=\prod_{q_{2}=0}^{\tilde{q}_{\nu;n}}\prod_{q_{1}=0}^{q_{2}}\left\{ I_{\nu;r;n;q_{1},q_{2}}^{\left(\text{bath}\right)}\left(\sigma_{\nu;r;1;q_{1}},\sigma_{\nu;r;-1;q_{1}},\sigma_{\nu;r;1;q_{2}},\sigma_{\nu;r;-1;q_{2}}\right)\right\} ,\label{eq:I_=00007Bv; r; n=00007D^(bath) in path integral appendix -- 1}
\end{equation}

\begin{align}
 & I_{\nu;r;n;q_{1},q_{2}}^{\left(\text{bath}\right)}\left(\sigma_{\nu;r;1;q_{1}},\sigma_{\nu;r;-1;q_{1}},\sigma_{\nu;r;1;q_{2}},\sigma_{\nu;r;-1;q_{2}}\right)\nonumber \\
 & \quad=\prod_{\left(l_{1},l_{2}\right)\in\Upsilon_{\nu;n;q_{1},q_{2}}}\left\{ e^{-\gamma_{\nu;r;n;l_{1},l_{2}}\left(\sigma_{\nu;r;1;q_{1}},\sigma_{\nu;r;-1;q_{1}},\sigma_{\nu;r;1;q_{2}},\sigma_{\nu;r;-1;q_{2}}\right)}\right\} ,\label{eq:I_=00007Bv; r; n; q_1, q_2=00007D^(bath) in path integral appendix -- 1}
\end{align}

\begin{equation}
\Upsilon_{y;n;q_{1},q_{2}}=\left\{ \left(q_{1},q_{2}\right)\right\} ,\label{eq:Upsilon_=00007By; n; q_1; q_2=00007D in path integral appendix -- 1}
\end{equation}

\begin{align}
 & \Upsilon_{z;n;q_{1},q_{2}}\nonumber \\
 & \quad=\begin{cases}
\left\{ \left(0,0\right)\right\} , & \text{if }q_{1}=q_{2}=0,\\
\left\{ \left(0,2q_{2}-1\right),\left(0,2q_{2}\right)\right\} , & \text{if }q_{1}=0\text{ and }1\le q_{2}\le n,\\
\left\{ \left(0,2n+1\right)\right\} , & \text{if }q_{1}=0\text{ and }q_{2}=n+1,\\
\left\{ \left(2q_{1}-1,2q_{2}-1\right),\left(2q_{1},2q_{2}-1\right),\left(2q_{1}-1,2q_{2}\right),\left(2q_{1},2q_{2}\right)\right\} , & \text{if }1\le q_{1}<q_{2}\le n,\\
\left\{ \left(2q_{1}-1,2q_{2}-1\right),\left(2q_{1}-1,2q_{2}\right),\left(2q_{1},2q_{2}\right)\right\} , & \text{if }1\le q_{1}=q_{2}\le n,\\
\left\{ \left(2q_{1}-1,2n+1\right),\left(2q_{1},2n+1\right)\right\} , & \text{if }1\le q_{1}\le n\text{ and }q_{2}=n+1,\\
\left\{ \left(2n+1,2n+1\right)\right\} , & \text{if }q_{1}=q_{2}=n+1,
\end{cases}\label{eq:Upsilon_=00007Bz; n; q_1; q_2=00007D in path integral appendix -- 1}
\end{align}

\begin{align}
 & \gamma_{\nu;r;n;l_{1},l_{2}}\left(\sigma_{\nu;r;1;q_{1}},\sigma_{\nu;r;-1;q_{1}},\sigma_{\nu;r;1;q_{2}},\sigma_{\nu;r;-1;q_{2}}\right)\nonumber \\
 & \quad=\mathcal{E}_{\nu;r;l_{1}}^{\left(\lambda\right)}\mathcal{E}_{\nu;r;l_{2}}^{\left(\lambda\right)}\left(\sigma_{\nu;r;1;q_{2}}-\sigma_{\nu;r;-1;q_{2}}\right)\nonumber \\
 & \quad\mathrel{\phantom{=}}\mathop{\times}\left\{ \left(\sigma_{\nu;r;1;q_{1}}-\sigma_{\nu;r;-1;q_{1}}\right)\text{Re}\left[\eta_{\nu;r;n;l_{2},l_{1}}\right]+i\left(\sigma_{\nu;r;1;q_{1}}+\sigma_{\nu;r;-1;q_{1}}\right)\text{Im}\left[\eta_{\nu;r;n;l_{2},l_{1}}\right]\right\} ,\label{eq:gamma_=00007Bv; r; n; l_1; l_2=00007D in path integral appendix -- 1}
\end{align}

\begin{equation}
\eta_{\nu;r;n;l_{1},l_{2}}=\int_{-\infty}^{\infty}\frac{d\omega}{2\pi}\,A_{\nu;r;T}\left(\omega\right)\frac{f_{n;l_{1},l_{2}}^{\left(\eta\right)}\left(\omega\right)}{\omega^{2}},\label{eq:eta_=00007Bv; r; n; l_1, l_2=00007D in path integral appendix -- 1}
\end{equation}

\begin{equation}
f_{n;l_{1},l_{2}}^{\left(\eta\right)}\left(\omega\right)=\begin{cases}
f_{n;l_{1}}^{\left(\eta,\Delta l=0\right)}\left(\omega\right), & \text{if }l_{1}-l_{2}=0,\\
f_{n;l_{1},l_{2}}^{\left(\eta,\Delta l\ge1\right)}\left(\omega\right), & \text{if }l_{1}-l_{2}\ge1,
\end{cases}\label{eq:f_=00007Bn; l_1, l_2=00007D^=00007B(eta)=00007D(omega) in path integral appendix -- 1}
\end{equation}

\begin{equation}
f_{n;l_{1}}^{\left(\eta,\Delta l=0\right)}\left(\omega\right)=\text{Re}\left[1-e^{-i\Delta t\omega\tilde{w}_{n;l_{1}}}\right]=2\sin^{2}\left(\frac{\Delta t\omega\tilde{w}_{n;l_{1}}}{2}\right),\label{eq:f_=00007Bn; l_1, l_2=00007D^=00007B(eta, dl=00003D0)=00007D(omega) in path integral appendix -- 1}
\end{equation}

\begin{equation}
f_{n;l_{1},l_{2}}^{\left(\eta,\Delta l\ge1\right)}\left(\omega\right)=4e^{-i\Delta t\omega\left\{ \frac{1}{2}\tilde{w}_{n;l_{1}}+\sum_{l=l_{2}+1}^{l_{1}-1}\tilde{w}_{n;l}+\frac{1}{2}\tilde{w}_{n;l_{2}}\right\} }\sin\left(\frac{\Delta t\omega\tilde{w}_{n;l_{1}}}{2}\right)\sin\left(\frac{\Delta t\omega\tilde{w}_{n;l_{2}}}{2}\right),\label{eq:f_=00007Bn; l_1, l_2=00007D^=00007B(eta, dl=00003D>1)=00007D(omega) in path integral appendix -- 1}
\end{equation}

\begin{equation}
\tilde{w}_{n;l}=\begin{cases}
\frac{1}{4}, & \text{if }l=0,1,2n,2n+1,\\
\frac{1}{2}, & \text{if }1<l<2n.
\end{cases}\label{eq:l composite quadrature rule in path integral appendix -- 1}
\end{equation}

\noindent Note that one could set $f_{n;l_{1}}^{\left(\eta,\Delta l=0\right)}\left(\omega\right)$
to $1-e^{-i\Delta t\omega\tilde{w}_{n;l_{1}}}$ in Eq.~\eqref{eq:f_=00007Bn; l_1, l_2=00007D^=00007B(eta, dl=00003D0)=00007D(omega) in path integral appendix -- 1}
based on the discussions leading to Eq.~\eqref{eq:tilde f_=00007Bv; n; q_1=00007D^=00007B(eta, dq=00003D0)=00007D(omega) -- 1},
however only the real part is ever used as should be clear from Eq.~\eqref{eq:gamma_=00007Bv; r; n; l_1; l_2=00007D in path integral appendix -- 1}
and the fact that

\begin{equation}
\left(\sigma_{\nu;r;1;q_{1}}-\sigma_{\nu;r;-1;q_{1}}\right)\left(\sigma_{\nu;r;1;q_{1}}+\sigma_{\nu;r;-1;q_{1}}\right)=0.\label{eq:sigma product identity -- 1}
\end{equation}

All that remains is to show how the path integral expression for $\hat{\rho}^{\left(A\right)}\left(t_{n}\right)$
in Eq.~\eqref{eq:rho^(A) in path integral appendix -- 6} simplifies
when there is no coupling between the environment and the $y$-components
of the spins, i.e. when $A_{y;r;T}\left(\omega\right)=0$. Noting
that

\begin{equation}
I_{y;r;n}^{\left(\text{bath}\right)}\left(\sigma_{y;r;\alpha=\pm1;q\in\left[0,\tilde{q}_{y;n}\right]}\right)=1,\quad\text{if }A_{y;r;T}\left(\omega\right)=0,\label{eq:I_=00007By; r; n=00007D^(bath) when y-noise is zero -- 1}
\end{equation}

\noindent we can rewrite Eq.~\eqref{eq:rho^(A) in path integral appendix -- 6}
as

\begin{align}
\rho^{\left(A\right)}\left(\boldsymbol{\sigma}_{z;1;n+1},\boldsymbol{\sigma}_{z;-1;n+1}\right) & =\left\langle \boldsymbol{\sigma}_{z;1;n+1}z\right|\hat{\rho}^{\left(A\right)}\left(t_{n}\right)\left|\boldsymbol{\sigma}_{z;-1;n+1}z\right\rangle \nonumber \\
 & =\prod_{\alpha=\pm1}\prod_{u=-N}^{N}\prod_{r=0}^{L-1}\left[\prod_{q=0}^{\tilde{q}_{z;n}-1}\left\{ \sum_{\sigma_{z;r+uL;\alpha;q}=\pm1}\right\} \right]\nonumber \\
 & \mathrel{\phantom{=}}\mathop{\times}e^{i\sum_{\alpha=\pm1}\phi_{\alpha;n}^{\left(\text{lcafc}\right)}\left(\boldsymbol{\sigma}_{z;\alpha;q\in\left[0,\tilde{q}_{z;n}\right]}\right)}I_{n}^{\left(\text{z-noise}\right)}\left(\boldsymbol{\sigma}_{\nu\in\left\{ y,z\right\} ;\alpha=\pm1;q\in\left[0,\tilde{q}_{\nu;n}\right]}\right)\nonumber \\
 & \mathrel{\phantom{=}}\mathop{\times}\rho^{\left(i,A\right)}\left(\boldsymbol{\sigma}_{z;1;0},\boldsymbol{\sigma}_{z;-1;0}\right)+\mathcal{O}\left[\Delta t^{2}\right],\label{eq:rho^(A) in path integral appendix -- 7}
\end{align}

\noindent where

\begin{equation}
I_{r;n}^{\left(\text{z-noise}\right)}\left(\sigma_{z;r;\alpha=\pm1;q\in\left[0,\tilde{q}_{z;n}\right]}\right)=I_{z;r;n}^{\left(\text{tfc}\right)}\left(\sigma_{z;r;\alpha=\pm1;q\in\left[0,\tilde{q}_{z;n}\right]}\right)I_{z;r;n}^{\left(\text{bath}\right)}\left(\sigma_{z;r;\alpha=\pm1;q\in\left[0,\tilde{q}_{z;n}\right]}\right),\label{eq:I_=00007Br; n=00007D^(z-noise) in path integral appendix -- 1}
\end{equation}

\noindent with

\begin{equation}
I_{z;r;n}^{\left(\text{tfc}\right)}\left(\sigma_{y;r;\alpha=\pm1;q\in\left[0,\tilde{q}_{y;n}\right]}\right)=\prod_{l=0}^{n}\left\{ I_{z;r;n;k=l}^{\left(\text{tfc}\right)}\left(\sigma_{z;r;1;l},\sigma_{z;r;-1;l},\sigma_{z;r;1;l+1},\sigma_{z;r;-1;l+1}\right)\right\} ,\label{eq:I_=00007Bz; r; n=00007D^(tfc) in path integral appendix -- 1}
\end{equation}

\noindent and

\begin{align}
I_{z;r;n;k}^{\left(\text{tfc}\right)}\left(\sigma_{z;r;1;l},\sigma_{z;r;-1;l},\sigma_{z;r;1;l+1},\sigma_{z;r;-1;l+1}\right) & =\prod_{\alpha=\pm1}\prod_{u=-N}^{N}\prod_{r=0}^{L-1}\prod_{q=0}^{\tilde{q}_{y;n}}\left\{ \sum_{\sigma_{y;r+uL;\alpha;q}=\pm1}\right\} \nonumber \\
 & \mathrel{\phantom{=}}\mathop{\times}I^{\left(y\leftrightarrow z,2\right)}\left(\sigma_{y;r;1,2l},\sigma_{y;r;-1,2l},\sigma_{z;r;1,l},\sigma_{z;r;-1,l}\right)\nonumber \\
 & \mathrel{\phantom{=}}\mathop{\times}I_{y;r;n;k=l}^{\left(\text{tfc}\right)}\left(\sigma_{y;r;1;2l},\sigma_{y;r;-1;2l},\sigma_{y;r;1;2l+1},\sigma_{y;r;-1;2l+1}\right)\nonumber \\
 & \mathrel{\phantom{=}}\mathop{\times}I^{\left(y\leftrightarrow z,1\right)}\left(\sigma_{y;r;1,2l+1},\sigma_{y;r;-1,2l+1},\sigma_{z;r;1,l+1},\sigma_{z;r;-1,l+1}\right).\label{eq:I_=00007Bz; r; n; k=00007D^(tfc) in path integral appendix -- 1}
\end{align}

\noindent From Eqs.~\eqref{eq:I^(y<->z, 1) in path integral appendix -- 1}-\eqref{eq:I_=00007By; r; alpha; n; k=00007D^(tfc) in path integral appendix -- 1}
we see that the $I^{\left(y\leftrightarrow z,2\right)}$ and $I^{\left(y\leftrightarrow z,1\right)}$
in Eq.~\eqref{eq:I_=00007Bz; r; n; k=00007D^(tfc) in path integral appendix -- 1}
transform $I_{y;r;n;k=l}^{\left(\text{tfc}\right)}$ from the $\sigma_{y;r;\alpha;q}$-basis
to the $\sigma_{z;r;\alpha;q}$-basis. Therefore we can rewrite Eq.~\eqref{eq:I_=00007Bz; r; n; k=00007D^(tfc) in path integral appendix -- 1}
as

\begin{align}
 & I_{z;r;n;k}^{\left(\text{tfc}\right)}\left(\sigma_{z;r;1;q_{1}},\sigma_{z;r;-1;q_{1}},\sigma_{z;r;1;q_{2}},\sigma_{z;r;-1;q_{2}}\right)\nonumber \\
 & \quad=I_{z;r;1;n;k}^{\left(\text{tfc}\right)}\left(\sigma_{z;r;1;q_{2}},\sigma_{z;r;1;q_{1}}\right)I_{z;r;-1;n;k}^{\left(\text{tfc}\right)}\left(\sigma_{z;r;-1;q_{1}},\sigma_{z;r;-1;q_{2}}\right),\label{eq:I_=00007Bz; r; n; k=00007D^(tfc) in path integral appendix -- 2}
\end{align}

\noindent where

\begin{equation}
I_{z;r;\alpha;n;k}^{\left(\text{tfc}\right)}\left(\sigma_{z;r;\alpha;q_{1}},\sigma_{y;r;\alpha;q_{2}}\right)=\begin{cases}
\left\langle \sigma_{z;r;-1;q_{1}}z\right|e^{i\Delta tw_{n;k}h_{x;r}\left(t_{k}\right)\hat{\sigma}_{x;r}}\left|\sigma_{z;r;-1;q_{2}}z\right\rangle , & \text{if }\alpha=-1,\\
\left\langle \sigma_{z;r;1;q_{2}}z\right|e^{-i\Delta tw_{n;k}h_{x;r}\left(t_{k}\right)\hat{\sigma}_{x;r}}\left|\sigma_{z;r;1;q_{1}}z\right\rangle , & \text{if }\alpha=1.
\end{cases}\label{eq:I_=00007Bz; r; alpha; n; k=00007D^(tfc) in path integral appendix -- 1}
\end{equation}

\noindent Using Eqs.~\eqref{eq:I_=00007Bz; r; n; k=00007D^(tfc) in path integral appendix -- 1}
and \eqref{eq:I_=00007Bz; r; alpha; n; k=00007D^(tfc) in path integral appendix -- 1},
and the exponential identity \citep{Gottfried1}:

\begin{equation}
e^{-\frac{i}{1}\theta\hat{\sigma}_{x;r}}=\cos\left(\frac{\theta}{2}\right)\hat{1}_{r}-i\hat{\sigma}_{x;r}\sin\left(\frac{\theta}{2}\right),\label{eq:pauli matrix identity -- 1}
\end{equation}

\noindent we can write

\begin{align}
I_{\nu;r;\alpha;n;k}^{\left(\text{tfc}\right)}\left(\sigma_{y;r;\alpha;q_{1}},\sigma_{y;r;\alpha;q_{2}}\right) & =\frac{1}{4}\left(\sigma_{\nu;r;\alpha;q_{1}}+\sigma_{\nu;r;\alpha;q_{2}}\right)^{2}\cos\left(\frac{\theta_{r;n;k}}{2}\right)\nonumber \\
 & \phantom{=}\mathop{+}i^{1+c_{\nu}}\alpha\left\{ \frac{1}{2}\left(\sigma_{\nu;r;\alpha;q_{1}}-\sigma_{\nu;r;\alpha;q_{2}}\right)\right\} ^{c_{\nu}}\sin\left(\frac{\theta_{r;n;k}}{2}\right),\label{eq:I_=00007Bv; r; alpha; n; k=00007D^(tfc) in path integral appendix -- 1}
\end{align}

\begin{equation}
c_{\nu}=\begin{cases}
1, & \text{if }\nu=y,\\
2, & \text{if }\nu=z,
\end{cases}\label{eq:c_v in path integral appendix -- 1}
\end{equation}

\begin{equation}
\theta_{r;n;k}=2\Delta tw_{n;k}h_{x;r}\left(t_{k}\right).\label{eq:theta_=00007Br; n; k=00007D in path integral appendix -- 1}
\end{equation}

In summary, the path integral expression for $\hat{\rho}^{\left(A\right)}\left(t_{n}\right)$
in Eq.~\eqref{eq:rho^(A) in path integral appendix -- 7} can be
written concisely as:

\begin{align}
\rho^{\left(A\right)}\left(\boldsymbol{\sigma}_{z;1;n+1},\boldsymbol{\sigma}_{z;-1;n+1}\right) & =\left\langle \boldsymbol{\sigma}_{z;1;n+1}z\right|\hat{\rho}^{\left(A\right)}\left(t_{n}\right)\left|\boldsymbol{\sigma}_{z;-1;n+1}z\right\rangle \nonumber \\
 & =\prod_{\alpha=\pm1}\prod_{u=-N}^{N}\prod_{r=0}^{L-1}\left[\prod_{q=0}^{q_{y;n}}\left\{ \sum_{\sigma_{y;r+uL;\alpha;q}=\pm1}\right\} \prod_{q=0}^{q_{z;n}-1}\left\{ \sum_{\sigma_{z;r+uL;\alpha;q}=\pm1}\right\} \right]\nonumber \\
 & \mathrel{\phantom{=}}\mathop{\times}e^{i\sum_{\alpha=\pm1}\phi_{\alpha;n}^{\left(\text{lcafc}\right)}\left(\boldsymbol{\sigma}_{z;\alpha;q\in\left[0,q_{z;n}\right]}\right)}I_{n}\left(\boldsymbol{\sigma}_{\nu\in\left\{ y,z\right\} ;\alpha=\pm1;q\in\left[0,q_{\nu;n}\right]}\right)\nonumber \\
 & \mathrel{\phantom{=}}\mathop{\times}\rho^{\left(i,A\right)}\left(\boldsymbol{\sigma}_{z;1;0},\boldsymbol{\sigma}_{z;-1;0}\right)+\mathcal{O}\left[\Delta t^{2}\right],\label{eq:rho^(A) in path integral appendix -- 8}
\end{align}

\noindent where

\begin{equation}
q_{y;n}=\begin{cases}
-1, & \text{if }A_{y;r;T}\left(\omega\right)=0,\\
2n+1, & \text{otherwise},
\end{cases}\label{eq:q_=00007By; n=00007D in path integral appendix -- 1}
\end{equation}

\begin{equation}
q_{z;n}=n+1,\label{eq:q_=00007Bz; n=00007D in path integral appendix -- 1}
\end{equation}

\noindent and

\begin{equation}
I_{r;n}\left(\sigma_{\nu\in\left\{ y,z\right\} ;r;\alpha=\pm1;q\in\left[0,q_{\nu;n}\right]}\right)=\begin{cases}
I_{r;n}^{\left(\text{z-noise}\right)}\left(\sigma_{z;r;\alpha=\pm1;q\in\left[0,q_{z;n}\right]}\right), & \text{if }A_{y;r;T}\left(\omega\right)=0,\\
I_{r;n}^{\left(\text{yz-noise}\right)}\left(\sigma_{\nu\in\left\{ y,z\right\} ;r;\alpha=\pm1;q\in\left[0,q_{\nu;n}\right]}\right), & \text{otherwise},
\end{cases}\label{eq:I_=00007Br; n=00007D in path integral appendix -- 1}
\end{equation}

\noindent which is the result presented in Sec.~\ref{subsec:Reduced state operator of the device -- 1}.

\newpage{}

\section{Evaluating numerically the spectral densities of noise\label{sec:Evaluating numerically the spectral densities of noise}}

As discussed in Sec.~\ref{sec:Spectral densities of noise}, the
spectral density of the noise coupled to the $\nu$-component of the
spin at site $r$ and temperature $T$, $A_{\nu;r;T}\left(\omega\right)$,
will sometimes naturally breakdown into multiple components:

\begin{equation}
A_{\nu;r;T}\left(\omega\right)=\sum_{\varsigma}A_{\nu;r;T;\varsigma}\left(\omega\right),\label{eq:A_=00007Bv; r; T=00007D(omega) breakdown into subcomponent -- 1}
\end{equation}

\noindent where 

\begin{equation}
A_{\nu;r;T;\varsigma}\left(\omega\right)=\text{sign}\left(\omega\right)\frac{A_{\nu;r;T=0;\varsigma}\left(\left|\omega\right|\right)}{1-e^{-\beta\omega}},\label{eq:A_=00007Bv; r; T; varsigma=00007D(omega) -- 1}
\end{equation}

\noindent where $\text{sign}\left(\omega\right)$ is the sign function
{[}defined in Eq.~\eqref{eq:sign function in main manuscript -- 1}{]},
$\beta=1/\left(k_{B}T\right)$, and where $A_{\nu;r;T=0;\varsigma}\left(\omega\right)$
is the $\varsigma^{\text{th}}$ component of $A_{\nu;r;T=0}\left(\omega\right)$.
In this appendix, we discuss how to evaluate numerically $A_{\nu;r;T;\varsigma}\left(\omega\right)$.
For convenience, we introduce the following shorthand notation:

\begin{equation}
f\left(a^{\pm}\right)\equiv\lim_{\omega\to a^{\pm}}f\left(\omega\right),\label{eq:limit shorthand -- 1}
\end{equation}

\noindent where $f\left(\omega\right)$ is an arbitrary function,
and $a$ is a real number. Since $A_{\nu;r;T;\varsigma}\left(\omega\right)$
has a removable singularity at $\omega=0$, we calculate $A_{\nu;r;T;\varsigma}\left(0^{\pm}\right)$
using L'Hôpital's rule:

\begin{equation}
A_{\nu;r;T;\varsigma}\left(0^{\pm}\right)=\pm\frac{\left.\frac{d}{d\omega}\left\{ A_{\nu;r;T=0;\varsigma}\left(\pm\omega\right)\right\} \right|_{\omega\to0^{+}}}{\left.\frac{d}{d\omega}\left\{ 1-e^{-\beta\omega}\right\} \right|_{\omega\to0^{\pm}}}=\frac{\left.\frac{d}{d\omega}\left\{ A_{\nu;r;T=0}\left(\omega\right)\right\} \right|_{\omega\to0^{+}}}{\beta}.\label{eq:A_=00007Bv; r; T; varsigma=00007D(0^=00007B+/-=00007D) -- 1}
\end{equation}

\noindent For finite $\omega$ in the neighbourhood of $\omega=0$,
we perform a Taylor expansion of the denominator of Eq.~\eqref{eq:A_=00007Bv; r; T; varsigma=00007D(omega) -- 1}
to fifth order in $\beta\omega$:

\begin{align}
A_{\nu;r;T;\varsigma}\left(\omega\neq0\right) & =\text{sign}\left(\omega\right)\frac{A_{\nu;T=0;\varsigma}\left(\left|\omega\right|\right)}{1-e^{-\beta\omega}}\nonumber \\
 & \approx\text{sign}\left(\omega\right)\frac{A_{\nu;r;T=0;\varsigma}\left(\left|\omega\right|\right)}{-\left\{ -\beta\omega+\frac{1}{2}\left(\beta\omega\right)^{2}-\frac{1}{6}\left(\beta\omega\right)^{3}+\frac{1}{24}\left(\beta\omega\right)^{4}-\frac{1}{120}\left(\beta\omega\right)^{5}\right\} }\nonumber \\
 & =\text{sign}\left(\omega\right)\frac{A_{\nu;r;T=0;\varsigma}\left(\omega\right)}{\beta\omega\left\{ 1-\frac{1}{2}\beta\omega+\frac{1}{6}\left(\beta\omega\right)^{2}-\frac{1}{24}\left(\beta\omega\right)^{3}+\frac{1}{120}\left(\beta\omega\right)^{4}\right\} }\nonumber \\
 & \approx\text{sign}\left(\omega\right)\frac{A_{\nu;r;T=0;\varsigma}\left(\left|\omega\right|\right)}{\left(\beta\omega\right)}\left\{ 1+\left[\frac{1}{2}\left(\beta\omega\right)-\frac{1}{6}\left(\beta\omega\right)^{2}+\frac{1}{24}\left(\beta\omega\right)^{3}-\frac{1}{120}\left(\beta\omega\right)^{4}\right]\right.\nonumber \\
 & \phantom{\mathrel{\approx}\text{sign}\left(\omega\right)\frac{A_{\nu;r;T=0;\varsigma}\left(\left|\omega\right|\right)}{\left(\beta\omega\right)}}\left.\quad\mathop{+}\left[\frac{1}{2}\left(\beta\omega\right)-\frac{1}{6}\left(\beta\omega\right)^{2}+\frac{1}{24}\left(\beta\omega\right)^{3}-\frac{1}{120}\left(\beta\omega\right)^{4}\right]^{2}\right.\nonumber \\
 & \phantom{\mathrel{\approx}\text{sign}\left(\omega\right)\frac{A_{\nu;r;T=0;\varsigma}\left(\left|\omega\right|\right)}{\left(\beta\omega\right)}}\left.\quad\mathop{+}\left[\frac{1}{2}\left(\beta\omega\right)-\frac{1}{6}\left(\beta\omega\right)^{2}+\frac{1}{24}\left(\beta\omega\right)^{3}-\frac{1}{120}\left(\beta\omega\right)^{4}\right]^{3}\right.\nonumber \\
 & \phantom{\mathrel{\approx}\text{sign}\left(\omega\right)\frac{A_{\nu;r;T=0;\varsigma}\left(\left|\omega\right|\right)}{\left(\beta\omega\right)}}\left.\quad\mathop{+}\left[\frac{1}{2}\left(\beta\omega\right)-\frac{1}{6}\left(\beta\omega\right)^{2}+\frac{1}{24}\left(\beta\omega\right)^{3}-\frac{1}{120}\left(\beta\omega\right)^{4}\right]^{4}\right\} \nonumber \\
 & \approx\text{sign}\left(\omega\right)\frac{A_{\nu;r;T=0;\varsigma}\left(\left|\omega\right|\right)}{\left(\beta\omega\right)}\left\{ 1+\left[\frac{1}{2}\left(\beta\omega\right)-\frac{1}{6}\left(\beta\omega\right)^{2}+\frac{1}{24}\left(\beta\omega\right)^{3}-\frac{1}{120}\left(\beta\omega\right)^{4}\right]\right.\nonumber \\
 & \phantom{\mathrel{\approx}\text{sign}\left(\omega\right)\frac{A_{\nu;r;T=0;\varsigma}\left(\left|\omega\right|\right)}{\left(\beta\omega\right)}}\left.\quad\mathop{+}\left[\frac{1}{4}\left(\beta\omega\right)^{2}-\frac{1}{6}\left(\beta\omega\right)^{3}+\frac{1}{36}\left(\beta\omega\right)^{4}+\frac{1}{24}\left(\beta\omega\right)^{4}\right]\right.\nonumber \\
 & \phantom{\mathrel{\approx}\text{sign}\left(\omega\right)\frac{A_{\nu;r;T=0;\varsigma}\left(\left|\omega\right|\right)}{\left(\beta\omega\right)}}\left.\quad\mathop{+}\left[\frac{1}{8}\left(\beta\omega\right)^{3}-\frac{1}{12}\left(\beta\omega\right)^{4}-\frac{1}{24}\left(\beta\omega\right)^{4}\right]+\left[\frac{1}{16}\left(\beta\omega\right)^{4}\right]\right\} \nonumber \\
 & =\text{sign}\left(\omega\right)\frac{A_{\nu;r;T=0;\varsigma}\left(\left|\omega\right|\right)}{\left(\beta\omega\right)}\left\{ 1+\left[\frac{1}{2}\left(\beta\omega\right)\right]+\left[\frac{1}{4}\left(\beta\omega\right)^{2}-\frac{1}{6}\left(\beta\omega\right)^{2}\right]\right.\nonumber \\
 & \phantom{\mathrel{\approx}\text{sign}\left(\omega\right)\frac{A_{\nu;r;T=0;\varsigma}\left(\left|\omega\right|\right)}{\left(\beta\omega\right)}}\left.\quad\mathop{+}\left[\frac{1}{24}\left(\beta\omega\right)^{3}+\frac{1}{8}\left(\beta\omega\right)^{3}-\frac{1}{6}\left(\beta\omega\right)^{3}\right]\right.\nonumber \\
 & \phantom{\mathrel{\approx}\text{sign}\left(\omega\right)\frac{A_{\nu;r;T=0;\varsigma}\left(\left|\omega\right|\right)}{\left(\beta\omega\right)}}\left.\quad\mathop{+}\left[-\frac{1}{120}\left(\beta\omega\right)^{4}+\frac{1}{36}\left(\beta\omega\right)^{4}+\frac{1}{24}\left(\beta\omega\right)^{4}\right.\right.\nonumber \\
 & \phantom{\mathrel{\approx}\text{sign}\left(\omega\right)\frac{A_{\nu;r;T=0;\varsigma}\left(\left|\omega\right|\right)}{\left(\beta\omega\right)}}\left.\phantom{\quad\mathop{+}}\left.\quad\mathop{-}\frac{1}{12}\left(\beta\omega\right)^{4}-\frac{1}{24}\left(\beta\omega\right)^{4}+\frac{1}{16}\left(\beta\omega\right)^{4}\right]\right\} \nonumber \\
 & =\text{sign}\left(\omega\right)\frac{A_{\nu;r;T=0;\varsigma}\left(\left|\omega\right|\right)}{\left(\beta\omega\right)}\left\{ 1+\frac{1}{2}\left(\beta\omega\right)+\frac{1}{12}\left(\beta\omega\right)^{2}-\frac{1}{720}\left(\beta\omega\right)^{4}\right\} .\label{eq:A_=00007Bv; r; T; varsigma=00007D(omega) for small omega-- 1}
\end{align}

\noindent In practice, we approximate $A_{\nu;r;T;\varsigma}\left(\omega\right)$
by Eq.~\eqref{eq:A_=00007Bv; r; T; varsigma=00007D(omega) for small omega-- 1}
for $0<\left|\omega\right|<10^{-3}/\beta$ to avoid overflow problems.
For $\omega\le-10^{-3}/\beta$, we rewrite $A_{\nu;r;T;\varsigma}\left(\omega\right)$
as

\begin{equation}
A_{\nu;r;T;\varsigma}\left(\omega\le-10^{-3}/\beta\right)=-\frac{e^{\beta\omega}A_{\nu;r;T=0;\varsigma}\left(-\omega\right)}{e^{\beta\omega}-1},\label{eq:A_=00007Bv; r; T; varsigma=00007D(omega<=00003D-0.001/beta) for small omega-- 1}
\end{equation}

\noindent to also avoid overflow problems. For $\omega\ge10^{-3}/\beta$,
we evaluate $A_{\nu;r;T;\varsigma}\left(\omega\right)$ using Eq.~\eqref{eq:A_=00007Bv; r; T; varsigma=00007D(omega) -- 1}:

\begin{equation}
A_{\nu;r;T;\varsigma}\left(\omega\ge10^{-3}/\beta\right)=\frac{A_{\nu;r;T=0;\varsigma}\left(\omega\right)}{1-e^{-\beta\omega}}.\label{eq:A_=00007Bv; r; T; varsigma=00007D(omega>=00003D0.001/beta) for small omega-- 1}
\end{equation}

\noindent It is also useful to specify hard infrared and ultraviolet
cutoff frequencies, $\omega_{\nu;\varsigma}^{\left(\text{IR}\right)}$
and $\omega_{\nu;\varsigma}^{\left(\text{UV}\right)}$, for the $A_{\nu;r;T=0;\varsigma}\left(\omega\right)$
such that

\begin{equation}
A_{\nu;r;T=0;\varsigma}\left(\left|\omega\right|<\omega_{\nu;\varsigma}^{\left(\text{IR}\right)}\right)=0,\label{eq:introducing IR cutoff -- 1}
\end{equation}

\begin{equation}
A_{\nu;r;T=0;\varsigma}\left(\left|\omega\right|>\omega_{\nu;\varsigma}^{\left(\text{UV}\right)}\right)=0,\label{eq:introducing UV cutoff -- 1}
\end{equation}

\noindent where

\begin{equation}
0\le\omega_{\nu;\varsigma}^{\left(\text{IR}\right)}\le\omega_{\nu;\varsigma}^{\left(\text{UV}\right)}.\label{eq:cutoff freq inequality -- 1}
\end{equation}

\noindent We can summarize the results of Eqs.~\eqref{eq:A_=00007Bv; r; T; varsigma=00007D(0^=00007B+/-=00007D) -- 1}-\eqref{eq:cutoff freq inequality -- 1}
by writing

\begin{equation}
A_{\nu;r;T;\varsigma}\left(\omega\right)=\begin{cases}
\frac{\left.\frac{d}{d\omega}\left\{ A_{\nu;r;T=0;\varsigma}\left(\omega\right)\right\} \right|_{\omega\to0^{+}}}{\beta}, & \text{if }\omega=0^{\pm},\\
\text{sign}\left(\omega\right)\frac{A_{\nu;r;T=0;\varsigma}\left(\left|\omega\right|\right)}{\left(\beta\omega\right)}\left\{ 1+\frac{1}{2}\left(\beta\omega\right)+\frac{1}{12}\left(\beta\omega\right)^{2}-\frac{1}{720}\left(\beta\omega\right)^{4}\right\} , & \text{if }0<\left|\omega\right|<10^{-3}/\beta,\\
-\frac{e^{\beta\omega}A_{\nu;r;T=0;\varsigma}\left(-\omega\right)}{e^{\beta\omega}-1}, & \text{if }\omega\le-10^{-3}/\beta,\\
\frac{A_{\nu;r;T=0;\varsigma}\left(\omega\right)}{1-e^{-\beta\omega}}, & \text{if }\omega\ge10^{-3}/\beta,
\end{cases}\label{eq:A_=00007Bv; r; T; varsigma=00007D(omega) in numerical evaluation summary -- 1}
\end{equation}

\noindent where

\begin{equation}
A_{\nu;r;T=0;\varsigma}\left(\left|\omega\right|<\omega_{\nu;\varsigma}^{\left(\text{IR}\right)}\right)=0,\label{eq:introducing IR cutoff -- 2}
\end{equation}

\begin{equation}
A_{\nu;r;T=0;\varsigma}\left(\left|\omega\right|>\omega_{\nu;\varsigma}^{\left(\text{UV}\right)}\right)=0,\label{eq:introducing UV cutoff -- 2}
\end{equation}

\noindent with

\begin{equation}
0\le\omega_{\nu;\varsigma}^{\left(\text{IR}\right)}\le\omega_{\nu;\varsigma}^{\left(\text{UV}\right)}.\label{eq:cutoff freq inequality -- 2}
\end{equation}

\newpage{}

\section{How to evaluate numerically $\eta_{\nu;r;n;l_{1},l_{2}}$\label{sec:How to evaluate eta -- 1}}

Considerable care must be taken when evaluating $\eta_{\nu;r;n;l_{1},l_{2}}$,
which can be expressed as {[}see Eq.~\eqref{eq:eta_=00007Bv; r; n; l_1, l_2=00007D in main manuscript -- 1}{]}:

\begin{equation}
\eta_{\nu;r;n;l_{1},l_{2}}=\int_{-\infty}^{\infty}\frac{d\omega}{2\pi}\,A_{\nu;r;T}\left(\omega\right)\frac{f_{n;l_{1},l_{2}}^{\left(\eta\right)}\left(\omega\right)}{\omega^{2}},\label{eq:eta in numerical evaluation appendix -- 1}
\end{equation}

\noindent where $A_{\nu;r;T}\left(\omega\right)$ is the spectral
density of the noise coupled to the $\nu$-component of the spin at
site $r$ and temperature $T$, and

\begin{equation}
f_{n;l_{1},l_{2}}^{\left(\eta\right)}\left(\omega\right)=\begin{cases}
f_{n;l_{1}}^{\left(\eta,\Delta l=0\right)}\left(\omega\right), & \text{if }l_{1}-l_{2}=0,\\
f_{n;l_{1},l_{2}}^{\left(\eta,\Delta l\ge1\right)}\left(\omega\right), & \text{if }l_{1}-l_{2}\ge1,
\end{cases}\label{eq:f_=00007Bn; l_1, l_2=00007D^=00007B(eta)=00007D(omega) -- 3}
\end{equation}

\noindent with

\begin{equation}
f_{\nu;n;l_{1}}^{\left(\eta,\Delta l=0\right)}\left(\omega\right)=2\sin^{2}\left(\frac{\Delta t\omega\tilde{w}_{n;l_{1}}}{2}\right),\label{eq:f_=00007Bn; l_1, l_2=00007D^=00007B(eta, dl=00003D0)=00007D(omega) -- 3}
\end{equation}

\begin{equation}
f_{\nu;n;l_{1},l_{2}}^{\left(\eta,\Delta l\ge1\right)}\left(\omega\right)=4e^{-i\Delta t\omega\left\{ \frac{1}{2}\tilde{w}_{n;l_{1}}+\sum_{l=l_{2}+1}^{l_{1}-1}\tilde{w}_{n;l}+\frac{1}{2}\tilde{w}_{n;l_{2}}\right\} }\sin\left(\frac{\Delta t\omega\tilde{w}_{n;l_{1}}}{2}\right)\sin\left(\frac{\Delta t\omega\tilde{w}_{n;l_{2}}}{2}\right),\label{eq:f_=00007Bv; n; k_1, k_2=00007D^=00007B(eta, dl=00003D>1)=00007D(omega) -- 3}
\end{equation}

\begin{equation}
\tilde{w}_{n;l}=\begin{cases}
\frac{1}{4}, & \text{if }l=0,1,2n,2n+1,\\
\frac{1}{2}, & \text{if }1<l<2n,
\end{cases}\label{eq:l composite quadrature rule -- 1}
\end{equation}

\begin{equation}
\sum_{l=a}^{b}F_{l}=\begin{cases}
0, & \text{if }a>b,\\
F_{a}+F_{a+1}+\ldots+F_{b} & \text{otherwise},
\end{cases}\label{eq:summing convention -- 1}
\end{equation}

\noindent and $F_{l}$ being an arbitrary sequence. In $\texttt{spinbosonchain}$,
the $\eta_{\nu;r;n;l_{1},l_{2}}$ are represented by \\
$\texttt{spinbosonchain.\_influence.eta.Eta}$ objects. In this appendix,
we discuss in detail how to calculate $\eta_{\nu;r;n;l_{1},l_{2}}$.

\subsection{Caching $\eta_{\nu;r;n;l_{1},l_{2}}$\label{subsec:Caching eta -- 1}}

Given how computationally expensive it is to evaluate the $\eta_{\nu;r;n;l_{1},l_{2}}$,
it is beneficial to calculate only the $\eta_{\nu;r;n;l_{1},l_{2}}$
that are required for a full simulation, cache them, and then reuse
them wherever possible. As discussed briefly in Sec.~\ref{subsec:Memory of the system -- 1},
there exists an integer $K_{\tau}>0$ such that the $\eta_{\nu;r;n;l_{1},l_{2}}$
for $2n+1\ge l_{1}\ge l_{2}+2K_{\tau}$ contribute negligibly to the
dynamics. This is related to the ``memory'' of the system, $\tau$,
which we informally defined in Sec.~\ref{subsec:Memory of the system -- 1}.
$\eta_{\nu;r;n;l_{1},l_{2}}$ can be viewed as a kind of two-time
{[}or two-point{]} function where $l_{1}$ and $l_{2}$ are associated
with the times

\begin{equation}
t_{1}=\Delta t\sum_{k=0}^{k_{1}}\tilde{w}_{n;l},\label{eq:l_1 to t_1 -- 1}
\end{equation}

\begin{equation}
t_{2}=\Delta t\sum_{k=0}^{k_{2}}\tilde{w}_{n;l},\label{eq:l_2 to t_2 -- 1}
\end{equation}

\noindent respectively, with $\tilde{w}_{n;k}$ being defined in Eq.~\eqref{eq:l composite quadrature rule in main manuscript -- 1},
and $\Delta t$ being the time step size introduced in the QUAPI formalism.
The time difference $t_{1}-t_{2}$ is therefore

\begin{equation}
t_{1}-t_{2}=\Delta t\sum_{l=l_{2}+1}^{l_{1}}\tilde{w}_{n;l}.\label{eq:t_1 - t_2 -- 1}
\end{equation}

\noindent If we let

\begin{equation}
K_{\tau}=\max\left(0,\left\lceil \frac{\tau-\frac{7}{4}\Delta t}{\Delta t}\right\rceil \right)+3,\label{eq:K_=00007Btau=00007D -- 1}
\end{equation}

\noindent then for all $2n+1\ge l_{1}\ge l_{2}+2K_{\tau}$, $\eta_{\nu;n;l_{1},l_{2}}$
will contribute negligibly to the dynamics of the system since $t_{1}-t_{2}\ge\tau$.
The proof is relatively straightforward:

\begin{align}
t_{1}-t_{2} & =\Delta t\sum_{l=l_{2}+1}^{l_{1}}\tilde{w}_{n;l}\nonumber \\
 & \ge\Delta t\sum_{l=l_{2}+1}^{l_{2}+2K_{\tau}}\tilde{w}_{n;l}\nonumber \\
 & =\Delta t\left\{ \tilde{w}_{n;l_{2}+2K_{\tau}}+\tilde{w}_{n;l_{2}+2K_{\tau}-1}+\sum_{l=l_{2}+2}^{l_{2}+2K_{\tau}-2}\tilde{w}_{n;l}+\tilde{w}_{n;l_{2}+1}\right\} \nonumber \\
 & =\Delta t\left\{ \tilde{w}_{n;l_{2}+2K_{\tau}}+\tilde{w}_{n;l_{2}+2K_{\tau}-1}+\tilde{w}_{n;l_{2}+1}\vphantom{\frac{1}{2}\left(\left[l_{2}+2K_{\tau}-2\right]\right)}\right.\nonumber \\
 & \phantom{=\Delta t}\left.\quad\mathord{+}\frac{1}{2}\left(\left[l_{2}+2K_{\tau}-2\right]-\left[l_{2}+2\right]+1\right)\right\} \nonumber \\
 & =\Delta t\left\{ \tilde{w}_{n;l_{2}+2K_{\tau}}+\tilde{w}_{n;l_{2}+2K_{\tau}-1}+\tilde{w}_{n;l_{2}+1}+\frac{1}{2}\left(2K_{\tau}-3\right)\right\} \nonumber \\
 & \ge\Delta t\left\{ \tilde{w}_{n;2n+1}+\tilde{w}_{n;2n}+\tilde{w}_{n;1}+\frac{1}{2}\left(2K_{\tau}-3\right)\right\} \nonumber \\
 & =\Delta t\left\{ \frac{3}{4}+\frac{1}{2}\left(2K_{\tau}-3\right)\right\} \hphantom{\Delta t\left\{ \frac{3}{4}+\left(\left\lceil \frac{\tau-\frac{7}{4}\Delta t}{\Delta t}\right\rceil +\frac{3}{2}\right)\right\} }\hspace{2.8cm}\nonumber \\
 & =\Delta t\left\{ \frac{3}{4}+\frac{1}{2}\left(2K_{\tau}-6\right)+\frac{3}{2}\right\} \nonumber \\
 & =\Delta t\left\{ \frac{7}{4}+\left(K_{\tau}-3\right)\right\} \nonumber \\
 & =\Delta t\left\{ \frac{7}{4}+\max\left(0,\left\lceil \frac{\tau-\frac{7}{4}\Delta t}{\Delta t}\right\rceil \right)\right\} \nonumber \\
 & =\Delta t\left\{ \max\left(\frac{7}{4},\left\lceil \frac{\tau-\frac{7}{4}\Delta t}{\Delta t}\right\rceil +\frac{7}{4}\right)\right\} \nonumber \\
 & \ge\tau.\label{eq:proof of lower bound on K_tau -- 1}
\end{align}

\noindent Thus far we have shown that at most we need only calculate
$\eta_{\nu;r;n;l_{1},l_{2}}$ for $2n+1\ge l_{1}\ge l_{2}+2K_{\tau}$.
That being said, as we will demonstrate below, the $f_{n;l_{1},l_{2}}^{\left(\eta\right)}\left(\omega\right)$
{[}Eq.~\eqref{eq:f_=00007Bn; l_1, l_2=00007D^=00007B(eta)=00007D(omega) in main manuscript -- 1}{]}
that appear in $\eta_{\nu;r;n;l_{1},l_{2}}$ {[}Eq.~\eqref{eq:eta_=00007Bv; r; n; l_1, l_2=00007D in main manuscript -- 1}{]}
possess certain symmetries that enable us to calculate an even smaller
set of $\eta_{\nu;r;n;l_{1},l_{2}}$ while still capturing essentially
exact dynamics.

Let us simplify $f_{n;l_{1},l_{2}}^{\left(\eta\right)}\left(\omega\right)$
for a variety of different values of $l_{1}$ and $l_{2}$. If $l_{1}=l_{2}$,
we have

\begin{equation}
f_{n;l_{1},l_{1}}^{\left(\eta\right)}\left(\omega\right)=2\sin^{2}\left(\frac{\Delta t\omega}{8}\right),\quad\text{if }l_{1}\in\left\{ 0,1,2n,2n+1\right\} ,\label{eq:f, if l_1 in =00007B0, 1, 2n, 2n+1=00007D -- 1}
\end{equation}

\begin{equation}
f_{n;l_{1},l_{1}}^{\left(\eta\right)}\left(\omega\right)=2\sin^{2}\left(\frac{\Delta t\omega}{4}\right),\quad\text{if }2\le l_{1}\le2n-1,\label{eq:f, if 2<=00003Dl_1<=00003D2n-1 -- 1}
\end{equation}

\noindent If $l_{1}=l_{2}+1=1$,  or $l_{1}=l_{2}+1=2n+1$, we have

\begin{equation}
f_{n;1,0}^{\left(\eta\right)}\left(\omega\right)=f_{n;2n+1,2n}^{\left(\eta\right)}\left(\omega\right)=4e^{-\frac{i}{4}\Delta t\omega}\sin^{2}\left(\frac{\Delta t\omega}{8}\right),\label{eq:f, if l_1=00003Dl_2+1=00003D1 or 2n+1 -- 1}
\end{equation}

\noindent If $2n-1\ge l_{1}\ge2$ and $l_{2}=0$, we have

\begin{align}
f_{n;l_{1},0}^{\left(\eta\right)}\left(\omega\right) & =4e^{-i\Delta t\omega\left\{ \frac{1}{4}+\sum_{l=1}^{l_{1}-1}\tilde{w}_{n;l}+\frac{1}{8}\right\} }\sin\left(\frac{\Delta t\omega}{4}\right)\sin\left(\frac{\Delta t\omega}{8}\right)\nonumber \\
 & =4e^{-i\Delta t\omega\left\{ \frac{1}{4}+\sum_{l=2}^{l_{1}-1}\frac{1}{2}+\tilde{w}_{n;1}+\frac{1}{8}\right\} }\sin\left(\frac{\Delta t\omega}{4}\right)\sin\left(\frac{\Delta t\omega}{8}\right)\nonumber \\
 & =4e^{-i\Delta t\omega\left\{ \frac{1}{4}+\frac{1}{2}\left(l_{1}-2\right)+\frac{1}{4}+\frac{1}{8}\right\} }\sin\left(\frac{\Delta t\omega}{4}\right)\sin\left(\frac{\Delta t\omega}{8}\right)\nonumber \\
 & =4e^{\frac{3i}{8}\Delta t\omega}e^{-\frac{i}{2}\Delta t\omega l_{1}}\sin\left(\frac{\Delta t\omega}{4}\right)\sin\left(\frac{\Delta t\omega}{8}\right),\quad\text{if }2n-1\ge l_{1}\ge2.\label{eq:f, if 2n-1>=00003Dl_1>=00003D2 and l_2=00003D0 -- 1}
\end{align}

\noindent If $l_{1}=2n$ and $l_{2}=0$, we have

\begin{align}
f_{n;2n,0}^{\left(\eta\right)}\left(\omega\right) & =4e^{-i\Delta t\omega\left\{ \frac{1}{8}+\sum_{l=1}^{2n-1}\tilde{w}_{n;l}+\frac{1}{8}\right\} }\sin^{2}\left(\frac{\Delta t\omega}{8}\right)\nonumber \\
 & =4e^{-i\Delta t\omega\left\{ \frac{1}{8}+\sum_{l=2}^{2n-1}\frac{1}{2}+\tilde{w}_{n;1}+\frac{1}{8}\right\} }\sin^{2}\left(\frac{\Delta t\omega}{8}\right)\nonumber \\
 & =4e^{-i\Delta t\omega\left\{ \frac{1}{8}+\frac{1}{2}\left(2n-2\right)+\frac{1}{4}+\frac{1}{8}\right\} }\sin^{2}\left(\frac{\Delta t\omega}{8}\right)\nonumber \\
 & =4e^{\frac{i}{2}\Delta t\omega}e^{-\frac{i}{2}\Delta t\omega\left\{ 2n\right\} }\sin^{2}\left(\frac{\Delta t\omega}{8}\right).\label{eq:f, if l_1=00003D2n and l_2=00003D0 -- 1}
\end{align}

\noindent If $l_{1}=2n+1$ and $l_{2}=0$, we have

\begin{align}
f_{n;2n+1,0}^{\left(\eta\right)}\left(\omega\right) & =4e^{-i\Delta t\omega\left\{ \frac{1}{8}+\sum_{l=1}^{2n}\tilde{w}_{n;l}+\frac{1}{8}\right\} }\sin^{2}\left(\frac{\Delta t\omega}{8}\right)\nonumber \\
 & =4e^{-i\Delta t\omega\left\{ \frac{1}{8}+\tilde{w}_{n;2n}+\sum_{l=2}^{2n-1}\frac{1}{2}+\tilde{w}_{n;1}+\frac{1}{8}\right\} }\sin^{2}\left(\frac{\Delta t\omega}{8}\right)\nonumber \\
 & =4e^{-i\Delta t\omega\left\{ \frac{1}{8}+\frac{1}{4}+\frac{1}{2}\left(2n-2\right)+\frac{1}{4}+\frac{1}{8}\right\} }\sin^{2}\left(\frac{\Delta t\omega}{8}\right)\nonumber \\
 & =4e^{-i\Delta t\omega\left\{ \frac{1}{8}+\frac{1}{4}-\frac{3}{2}+\frac{1}{4}+\frac{1}{8}\right\} }e^{-\frac{i}{2}\Delta t\omega\left\{ 2n+1\right\} }\sin^{2}\left(\frac{\Delta t\omega}{8}\right)\nonumber \\
 & =4e^{\frac{3i}{4}\Delta t\omega}e^{-\frac{i}{2}\Delta t\omega\left\{ 2n+1\right\} }\sin^{2}\left(\frac{\Delta t\omega}{8}\right).\label{eq:f, if l_1=00003D2n+1 and l_2=00003D0 -- 1}
\end{align}

\noindent If $2n-1\ge l_{1}\ge2$ and $l_{2}=1$, we have

\begin{align}
f_{n;l_{1},1}^{\left(\eta\right)}\left(\omega\right) & =4e^{-i\Delta t\omega\left\{ \frac{1}{4}+\sum_{l=2}^{l_{1}-1}\frac{1}{2}+\frac{1}{8}\right\} }\sin\left(\frac{\Delta t\omega}{4}\right)\sin\left(\frac{\Delta t\omega}{8}\right)\nonumber \\
 & =4e^{-i\Delta t\omega\left\{ \frac{3}{8}+\frac{1}{2}\left(l_{1}-2\right)\right\} }\sin\left(\frac{\Delta t\omega}{4}\right)\sin\left(\frac{\Delta t\omega}{8}\right)\nonumber \\
 & =4e^{\frac{i}{8}\Delta t\omega}e^{-\frac{i}{2}\Delta t\omega\left\{ l_{1}-1\right\} }\sin\left(\frac{\Delta t\omega}{4}\right)\sin\left(\frac{\Delta t\omega}{8}\right),\quad\text{if }2n-1\ge l_{1}\ge2.\label{eq:f, if 2n-1>=00003Dl_1>=00003D2 and l_2=00003D1 -- 1}
\end{align}

\noindent If $l_{1}=2n$ and $l_{2}=1$, we have

\begin{align}
f_{n;2n,1}^{\left(\eta\right)}\left(\omega\right) & =4e^{-i\Delta t\omega\left\{ \frac{1}{8}+\sum_{l=2}^{2n-1}\frac{1}{2}+\frac{1}{8}\right\} }\sin^{2}\left(\frac{\Delta t\omega}{8}\right)\nonumber \\
 & =4e^{-i\Delta t\omega\left\{ \frac{1}{4}+\frac{1}{2}\left(2n-2\right)\right\} }\sin^{2}\left(\frac{\Delta t\omega}{8}\right)\nonumber \\
 & =4e^{\frac{i}{4}\Delta t\omega}e^{-\frac{i}{2}\Delta t\omega\left\{ 2n-1\right\} }\sin^{2}\left(\frac{\Delta t\omega}{8}\right)\nonumber \\
 & =4e^{\frac{i}{4}\Delta t\omega}e^{-\frac{i}{2}\Delta t\omega\left\{ 2n-1\right\} }\sin^{2}\left(\frac{\Delta t\omega}{8}\right).\label{eq:f, if l_1=00003D2n and l_2=00003D1 -- 1}
\end{align}

\noindent If $l_{1}=2n+1$ and $l_{2}=1$, we have

\begin{align}
f_{n;2n+1,1}^{\left(\eta\right)}\left(\omega\right) & =4e^{-i\Delta t\omega\left\{ \frac{1}{8}+\sum_{l=2}^{2n}\tilde{w}_{n;l}+\frac{1}{8}\right\} }\sin^{2}\left(\frac{\Delta t\omega}{8}\right)\nonumber \\
 & =4e^{-i\Delta t\omega\left\{ \frac{1}{8}+\tilde{w}_{n;2n}+\sum_{l=2}^{2n-1}\frac{1}{2}+\frac{1}{8}\right\} }\sin^{2}\left(\frac{\Delta t\omega}{8}\right)\nonumber \\
 & =4e^{-i\Delta t\omega\left\{ \frac{1}{8}+\frac{1}{4}+\frac{1}{2}\left(2n-2\right)+\frac{1}{8}\right\} }\sin^{2}\left(\frac{\Delta t\omega}{8}\right)\nonumber \\
 & =4e^{\frac{i}{2}\Delta t\omega}e^{-\frac{i}{2}\Delta t\omega\left\{ 2n\right\} }\sin^{2}\left(\frac{\Delta t\omega}{8}\right).\label{eq:f, if l_1=00003D2n+1 and l_2=00003D1 -- 1}
\end{align}

\noindent If $2n-1\ge l_{1}>l_{2}\ge2$, we have

\begin{align}
f_{n;l_{1},l_{2}}^{\left(\eta\right)}\left(\omega\right) & =4e^{-i\Delta t\omega\left\{ \frac{1}{4}+\sum_{l=l_{2}+1}^{l_{1}-1}\frac{1}{2}+\frac{1}{4}\right\} }\sin^{2}\left(\frac{\Delta t\omega}{4}\right)\nonumber \\
 & =4e^{-i\Delta t\omega\left\{ \frac{1}{2}+\frac{1}{2}\left(l_{1}-l_{2}-1\right)\right\} }\sin^{2}\left(\frac{\Delta t\omega}{4}\right)\nonumber \\
 & =4e^{-\frac{i}{2}\Delta t\omega\left\{ l_{1}-l_{2}\right\} }\sin^{2}\left(\frac{\Delta t\omega}{4}\right),\quad\text{if }2n-1\ge l_{1}>l_{2}\ge2.\label{eq:f, if 2n-1>=00003Dl_1>l_2>=00003D2 -- 1}
\end{align}

\noindent If $l_{1}=2n$, and $2n-1\ge l_{2}\ge2$, we have

\begin{align}
f_{n;2n,l_{2}}^{\left(\eta\right)}\left(\omega\right) & =4e^{-i\Delta t\omega\left\{ \frac{1}{8}+\sum_{l=l_{2}+1}^{2n-1}\frac{1}{2}+\frac{1}{4}\right\} }\sin\left(\frac{\Delta t\omega}{4}\right)\sin\left(\frac{\Delta t\omega}{8}\right)\nonumber \\
 & =4e^{-i\Delta t\omega\left\{ \frac{3}{8}+\frac{1}{2}\left(2n-l_{2}-1\right)\right\} }\sin\left(\frac{\Delta t\omega}{4}\right)\sin\left(\frac{\Delta t\omega}{8}\right)\nonumber \\
 & =4e^{\frac{i}{8}\Delta t\omega}e^{-\frac{i}{2}\Delta t\omega\left\{ 2n-l_{2}\right\} }\sin\left(\frac{\Delta t\omega}{4}\right)\sin\left(\frac{\Delta t\omega}{8}\right),\quad\text{if }2n-1\ge l_{2}\ge2.\label{eq:f, if l_1=00003D2n and 2n-1>=00003Dl_2>=00003D2 -- 1}
\end{align}

\noindent If $l_{1}=2n+1$, and $2n-1\ge l_{2}\ge2$, we have

\begin{align}
f_{n;2n+1,l_{2}}^{\left(\eta\right)}\left(\omega\right) & =4e^{-i\Delta t\omega\left\{ \frac{1}{8}+\sum_{l=l_{2}+1}^{2n}\tilde{w}_{n;l}+\frac{1}{4}\right\} }\sin\left(\frac{\Delta t\omega}{4}\right)\sin\left(\frac{\Delta t\omega}{8}\right)\nonumber \\
 & =4e^{-i\Delta t\omega\left\{ \frac{1}{8}+\tilde{w}_{n;2n}+\sum_{l=l_{2}+1}^{2n-1}\frac{1}{2}+\frac{1}{4}\right\} }\sin\left(\frac{\Delta t\omega}{4}\right)\sin\left(\frac{\Delta t\omega}{8}\right)\nonumber \\
 & =4e^{-i\Delta t\omega\left\{ \frac{1}{8}+\frac{1}{4}+\sum_{l=l_{2}+1}^{2n-1}\frac{1}{2}+\frac{1}{4}\right\} }\sin\left(\frac{\Delta t\omega}{4}\right)\sin\left(\frac{\Delta t\omega}{8}\right)\nonumber \\
 & =4e^{-i\Delta t\omega\left\{ \frac{5}{8}+\frac{1}{2}\left(2n-l_{2}-1\right)\right\} }\sin\left(\frac{\Delta t\omega}{4}\right)\sin\left(\frac{\Delta t\omega}{8}\right)\nonumber \\
 & =4e^{\frac{3i}{8}\Delta t\omega}e^{-\frac{i}{2}\Delta t\omega\left\{ 2n+1-l_{2}\right\} }\sin\left(\frac{\Delta t\omega}{4}\right)\sin\left(\frac{\Delta t\omega}{8}\right),\quad\text{if }2n-1\ge l_{2}\ge2.\label{eq:f, if l_1=00003D2n+1 and 2n-1>=00003Dl_2>=00003D2 -- 1}
\end{align}

\noindent From Eqs.~\eqref{eq:f, if l_1 in =00007B0, 1, 2n, 2n+1=00007D -- 1}-\eqref{eq:f, if l_1=00003D2n+1 and 2n-1>=00003Dl_2>=00003D2 -- 1},
we see that the following statements are true {[}some of which are
trivial{]}:

\begin{equation}
f_{n;l_{1},0}^{\left(\eta\right)}\left(\omega\right)=f_{K_{\tau};2K_{\tau}+1,2K_{\tau}+1-l_{1}}^{\left(\eta\right)}\left(\omega\right),\quad\text{if }\min\left(2K_{\tau}-1,2n-1\right)\ge l_{1}\ge0,\label{eq:f summary -- 1}
\end{equation}

\begin{equation}
f_{n;2n,0}^{\left(\eta\right)}\left(\omega\right)=f_{n;2n,0}^{\left(\eta\right)}\left(\omega\right),\quad\text{if }K_{\tau}-1\ge n\ge1,\label{eq:f summary -- 2}
\end{equation}

\begin{equation}
f_{n;2n+1,0}^{\left(\eta\right)}\left(\omega\right)=f_{n;2n+1,0}^{\left(\eta\right)}\left(\omega\right),\quad\text{if }K_{\tau}-1\ge n\ge1,\label{eq:f summary -- 3}
\end{equation}

\begin{equation}
f_{n;l_{1},1}^{\left(\eta\right)}\left(\omega\right)=f_{K_{\tau}+1;2K_{\tau}+2,2K_{\tau}+3-l_{1}}^{\left(\eta\right)}\left(\omega\right),\quad\text{if }\min\left(2K_{\tau},2n-1\right)\ge l_{1}\ge1,\label{eq:f summary -- 4}
\end{equation}

\begin{equation}
f_{n;2n,1}^{\left(\eta\right)}\left(\omega\right)=f_{n;2n,1}^{\left(\eta\right)}\left(\omega\right),\quad\text{if }K_{\tau}\ge n\ge1,\label{eq:f summary -- 5}
\end{equation}

\begin{equation}
f_{n;2n+1,1}^{\left(\eta\right)}\left(\omega\right)=f_{n;2n,0}^{\left(\eta\right)}\left(\omega\right),\quad\text{if }K_{\tau}-1\ge n\ge1,\label{eq:f summary -- 6}
\end{equation}

\begin{equation}
f_{n;l_{1},l_{2}}^{\left(\eta\right)}\left(\omega\right)=f_{K_{\tau}+2;2K_{\tau}+3,2K_{\tau}+3-l_{1}+l_{2}}^{\left(\eta\right)}\left(\omega\right),\quad\text{if }2n-1\ge l_{1}\ge l_{2}\ge2n-2K_{\tau},\label{eq:f summary -- 7}
\end{equation}

\begin{equation}
f_{n;2n,l_{2}}^{\left(\eta\right)}\left(\omega\right)=f_{K_{\tau}+1;2K_{\tau}+2,2K_{\tau}+2-2n+l_{2}}^{\left(\eta\right)}\left(\omega\right),\quad\text{if }n\ge K_{\tau}+1,\text{ and }2n\ge l_{2}\ge2n+1-2K_{\tau},\label{eq:f summary -- 8}
\end{equation}

\noindent and

\begin{equation}
f_{n;2n+1,l_{2}}^{\left(\eta\right)}\left(\omega\right)=f_{K_{\tau};2K_{\tau}+1,2K_{\tau}-2n+l_{2}}^{\left(\eta\right)}\left(\omega\right),\quad\text{if }n\ge K_{\tau},\text{ and }2n+1\ge l_{2}\ge2n+2-2K_{\tau}.\label{eq:f summary -- 9}
\end{equation}

Based on the above results {[}Eqs.~\eqref{eq:f summary -- 1}-\eqref{eq:f summary -- 9}{]},
we define the following cache arrays which we use to calculate the
$\eta_{\nu;r;n;l_{1},l_{2}}$ that are required to capture essentially
exact dynamics:

\begin{equation}
\eta_{\nu;r;K_{\tau};a}^{\left(\text{cache-1}\right)}=\eta_{\nu;r;a+1;2a+2,0},\quad0\le a\le K_{\tau}-2,\label{eq:eta cache -- 1}
\end{equation}

\begin{equation}
\eta_{\nu;r;K_{\tau};a}^{\left(\text{cache-2}\right)}=\eta_{\nu;r;a+1;2a+3,0},\quad0\le a\le K_{\tau}-2,\label{eq:eta cache -- 2}
\end{equation}

\begin{equation}
\eta_{\nu;r;K_{\tau};a}^{\left(\text{cache-3}\right)}=\eta_{\nu;r;a+1;2a+2,1},\quad0\le a\le K_{\tau}-1,\label{eq:eta cache -- 3}
\end{equation}

\begin{equation}
\eta_{\nu;r;K_{\tau};a}^{\left(\text{cache-4}\right)}=\eta_{\nu;r;K_{\tau}+2;2K_{\tau}+3,a+4},\quad0\le a\le2K_{\tau}-1,\label{eq:eta cache -- 4}
\end{equation}

\begin{equation}
\eta_{\nu;r;K_{\tau};a}^{\left(\text{cache-5}\right)}=\eta_{\nu;r;K_{\tau}+1;2K_{\tau}+2,a+3},\quad0\le a\le2K_{\tau}-1,\label{eq:eta cache -- 5}
\end{equation}

\noindent and

\begin{equation}
\eta_{\nu;r;K_{\tau};a}^{\left(\text{cache-6}\right)}=\eta_{\nu;r;K_{\tau};2K_{\tau}+1,a+2},\quad0\le a\le2K_{\tau}-1.\label{eq:eta cache -- 6}
\end{equation}

\noindent The required $\eta_{\nu;r;n;k_{1},k_{2}}$ are calculated
using the above cache arrays as follows:

\begin{equation}
\eta_{\nu;r;n;l_{1},0}=\eta_{\nu;r;K_{\tau};2K_{\tau}-1-l_{1}}^{\left(\text{cache-6}\right)},\quad\text{if }\min\left(2K_{\tau}-1,2n-1\right)\ge l_{1}\ge0,\label{eq:eta summary -- 1}
\end{equation}

\begin{equation}
\eta_{\nu;r;n;2n,0}=\eta_{\nu;r;K_{\tau};n-1}^{\left(\text{cache-1}\right)},\quad\text{if }K_{\tau}-1\ge n\ge1,\label{eq:eta summary -- 2}
\end{equation}

\begin{equation}
\eta_{\nu;r;n;2n+1,0}=\eta_{\nu;r;K_{\tau};n-1}^{\left(\text{cache-2}\right)},\quad\text{if }K_{\tau}-1\ge n\ge1,\label{eq:eta summary -- 3}
\end{equation}

\begin{equation}
\eta_{\nu;r;n;l_{1},1}=\eta_{\nu;r;K_{\tau};2K_{\tau}-l_{1}}^{\left(\text{cache-5}\right)},\quad\text{if }\min\left(2K_{\tau},2n-1\right)\ge l_{1}\ge1,\label{eq:eta summary -- 4}
\end{equation}

\begin{equation}
\eta_{\nu;r;n;2n,1}=\eta_{\nu;r;K_{\tau};n-1}^{\left(\text{cache-3}\right)},\quad\text{if }K_{\tau}\ge n\ge1,\label{eq:eta summary -- 5}
\end{equation}

\begin{equation}
\eta_{\nu;r;n;2n+1,1}=\eta_{\nu;r;K_{\tau};n-1}^{\left(\text{cache-1}\right)},\quad\text{if }K_{\tau}-1\ge n\ge1,\label{eq:eta summary -- 6}
\end{equation}

\begin{equation}
\eta_{\nu;r;n;l_{1},l_{2}}=\eta_{\nu;r;K_{\tau};2K_{\tau}-1-l_{1}+l_{2}}^{\left(\text{cache-4}\right)},\quad\text{if }2n-1\ge l_{1}\ge2,\text{ and }2K_{\tau}-1\ge l_{1}-l_{2},\label{eq:eta summary -- 7}
\end{equation}

\begin{equation}
\eta_{\nu;r;n;2n,l_{2}}=\eta_{\nu;r;K_{\tau};2K_{\tau}-2n-1+l_{2}}^{\left(\text{cache-5}\right)},\quad\text{if }2n\ge l_{2}\ge\max\left(2,2n+1-2K_{\tau}\right),\label{eq:eta summary -- 8}
\end{equation}

\noindent and

\begin{equation}
\eta_{\nu;r;n;2n+1,l_{2}}=\eta_{\nu;r;K_{\tau};2K_{\tau}-2n-2+l_{2}}^{\left(\text{cache-6}\right)},\quad\text{if }2n+1\ge l_{2}\ge\max\left(2,2n+2-2K_{\tau}\right).\label{eq:eta summary -- 9}
\end{equation}

\subsection{How to evaluate frequency integrals numerically\label{subsec:How to evaluate numerically frequency integrals -- 1}}

\noindent The evaluation of $\eta_{\nu;r;n;l_{1},l_{2}}$ {[}see Eq.~\eqref{eq:eta in numerical evaluation appendix -- 1}{]}
involves computing a frequency integral with an integrand that contains
$f_{n;l_{1},l_{2}}^{\left(\eta\right)}\left(\omega\right)$ which
is highly oscillatory, and as such special care must be taken. Moreover,
the integrand contains a removable singularity at $\omega=0$ and
thus requires further special care. In this section, we discuss how
this should be done.

We begin by using

\begin{equation}
4\sin\left(a\right)\sin\left(b\right)=4\left\{ \frac{e^{ia}-e^{-ia}}{2i}\right\} \left\{ \frac{e^{ib}-e^{-ib}}{2i}\right\} e^{i\left\{ a-b\right\} }+e^{-i\left\{ a-b\right\} }-e^{i\left\{ a+b\right\} }-e^{-i\left\{ a+b\right\} },\label{eq:trig identity -- 1}
\end{equation}

\noindent to rewrite $f_{n;l_{1},l_{2}}^{\left(\eta\right)}\left(\omega\right)$
{[}Eq.~\eqref{eq:f_=00007Bn; l_1, l_2=00007D^=00007B(eta)=00007D(omega) in main manuscript -- 1}{]}
as follows:

\begin{equation}
f_{n;l_{1},l_{2}}^{\left(\eta\right)}\left(\omega\right)=\begin{cases}
f_{n;l_{1}}^{\left(\eta,\Delta l=0\right)}\left(\omega\right), & \text{if }l_{1}-l_{2}=0,\\
f_{n;l_{1},l_{2}}^{\left(\eta,\Delta k\ge1\right)}\left(\omega\right), & \text{if }l_{1}-l_{2}\ge1,
\end{cases}\label{eq:rewritting f for numerical integration -- 1}
\end{equation}

\begin{align}
f_{n;l_{1}}^{\left(\eta,\Delta l=0\right)}\left(\omega\right) & =2\sin^{2}\left(\frac{\Delta t\omega\tilde{w}_{n;l_{1}}}{2}\right)\nonumber \\
 & =\frac{1}{2}\left[1+1-e^{i\Delta t\tilde{w}_{n;l_{1}}\omega}-e^{-i\Delta t\tilde{w}_{n;l_{1}}\omega}\right]\nonumber \\
 & =1-\cos\left(\Delta t\tilde{w}_{n;l_{1}}\omega\right),\label{eq:rewritting f^=00007B(dl=00003D0)=00007D for numerical integration -- 1}
\end{align}

\noindent and

\begin{align}
f_{n;l_{1},l_{2}}^{\left(\eta,\Delta l\ge1\right)}\left(\omega\right) & =4e^{-i\Delta t\omega\left\{ \frac{1}{2}\tilde{w}_{n;l_{1}}+\sum_{l=l_{2}+1}^{l_{1}-1}\tilde{w}_{n;l}+\frac{1}{2}\tilde{w}_{n;l_{2}}\right\} }\sin\left(\frac{\Delta t\omega\tilde{w}_{n;l_{1}}}{2}\right)\sin\left(\frac{\Delta t\omega\tilde{w}_{n;l_{2}}}{2}\right)\nonumber \\
 & =e^{-i\Delta t\omega\left\{ \frac{1}{2}\tilde{w}_{n;l_{1}}+\sum_{l=l_{2}+1}^{l_{1}-1}\tilde{w}_{n;l}+\frac{1}{2}\tilde{w}_{n;l_{2}}\right\} }\nonumber \\
 & \mathrel{\phantom{=}}\mathop{\times}\left[e^{\frac{i}{2}\Delta t\left\{ \tilde{w}_{n;l_{1}}-\tilde{w}_{n;l_{2}}\right\} \omega}+e^{-\frac{i}{2}\Delta t\left\{ \tilde{w}_{n;l_{1}}-\tilde{w}_{n;l_{2}}\right\} \omega}\right.\nonumber \\
 & \phantom{\mathrel{=}\mathop{\times}}\left.\quad\mathop{-}e^{\frac{i}{2}\Delta t\left\{ \tilde{w}_{n;l_{1}}+\tilde{w}_{n;l_{2}}\right\} \omega}-e^{-\frac{i}{2}\Delta t\left\{ \tilde{w}_{n;l_{1}}+\tilde{w}_{n;l_{2}}\right\} \omega}\right]\nonumber \\
 & =\left[e^{-i\left\{ \Delta t\sum_{l=l_{2}}^{l_{1}-1}\tilde{w}_{n;l}\right\} \omega}+e^{-i\left\{ \Delta t\sum_{l=l_{2}+1}^{l_{1}}\tilde{w}_{n;l}\right\} \omega}\right.\nonumber \\
 & \mathrel{\phantom{=}}\left.\quad\mathop{-}e^{-i\left\{ \Delta t\sum_{l=l_{2}+1}^{l_{1}-1}\tilde{w}_{n;l}\right\} \omega}-e^{-i\left\{ \Delta t\sum_{l=l_{2}}^{l_{1}}\tilde{w}_{n;l}\right\} \omega}\right]\nonumber \\
 & =e^{-iW_{n;l_{1},l_{2}}^{\left(0\right)}\omega}+e^{-iW_{n;l_{1},l_{2}}^{\left(1\right)}\omega}-e^{-iW_{n;l_{1},l_{2}}^{\left(2\right)}\omega}-e^{-iW_{n;l_{1},l_{2}}^{\left(3\right)}\omega},\label{eq:rewritting f^=00007B(dl>=00003D1)=00007D for numerical integration -- 1}
\end{align}

\noindent with

\begin{align}
W_{n;l_{1},l_{2}}^{\left(0\right)} & =\Delta t\sum_{l=l_{2}}^{l_{1}-1}\tilde{w}_{n;l},\label{eq:W_=00007Bn; l_1, l_2=00007D^=00007B(0)=00007D -- 1}\\
W_{n;l_{1},l_{2}}^{\left(1\right)} & =\Delta t\sum_{l=l_{2}+1}^{l_{1}}\tilde{w}_{n;l},\label{eq:W_=00007Bn; l_1, l_2=00007D^=00007B(1)=00007D -- 1}\\
W_{n;l_{1},l_{2}}^{\left(2\right)} & =\Delta t\sum_{l=l_{2}+1}^{l_{1}-1}\tilde{w}_{n;l},\label{eq:W_=00007Bn; l_1, l_2=00007D^=00007B(2)=00007D -- 1}\\
W_{n;l_{1},l_{2}}^{\left(3\right)} & =\Delta t\sum_{l=l_{2}}^{l_{1}}\tilde{w}_{n;l}.\label{eq:W_=00007Bn; l_1, l_2=00007D^=00007B(3)=00007D -- 1}
\end{align}

\noindent Next, we decompose $f_{n;l_{1}}^{\left(\eta,\Delta l=0\right)}\left(\omega\right)$
and $f_{n;l_{1},l_{2}}^{\left(\eta,\Delta l\ge1\right)}\left(\omega\right)$
into their real and imaginary parts:

\begin{equation}
f_{n;l_{1}}^{\left(\eta,\Delta l=0\right)}\left(\omega\right)=\text{Re}\left[f_{n;l_{1}}^{\left(\eta,\Delta l=0\right)}\left(\omega\right)\right]+i\text{Im}\left[f_{n;l_{1}}^{\left(\eta,\Delta l=0\right)}\left(\omega\right)\right],\label{eq:decomposing f^=00007B(dl=00003D0)=00007D -- 1}
\end{equation}

\noindent with

\begin{equation}
\text{Re}\left[f_{n;l_{1}}^{\left(\eta,\Delta l=0\right)}\left(\omega\right)\right]=1-\cos\left(\left\{ \Delta t\tilde{w}_{n;l_{1}}\right\} \omega\right),\label{eq:Real part of f^=00007B(dl=00003D0)=00007D -- 1}
\end{equation}

\noindent and

\begin{equation}
\text{Im}\left[f_{n;l_{1}}^{\left(\eta,\Delta l=0\right)}\left(\omega\right)\right]=0;\label{eq:Imaginary part of f^=00007B(dl=00003D0)=00007D -- 1}
\end{equation}

\noindent and

\begin{equation}
f_{n;l_{1},l_{2}}^{\left(\eta,\Delta l\ge1\right)}\left(\omega\right)=\text{Re}\left[f_{n;l_{1},l_{2}}^{\left(\eta,\Delta l\ge1\right)}\left(\omega\right)\right]+i\text{Im}\left[f_{n;l_{1},l_{2}}^{\left(\eta,\Delta l\ge1\right)}\left(\omega\right)\right],\label{eq:decomposing f^=00007B(dl>=00003D1)=00007D -- 1}
\end{equation}

\noindent with

\begin{equation}
\text{Re}\left[f_{n;l_{1},l_{2}}^{\left(\eta,\Delta l\ge1\right)}\left(\omega\right)\right]=\cos\left(W_{n;l_{1},l_{2}}^{\left(0\right)}\omega\right)+\cos\left(W_{n;l_{1},l_{2}}^{\left(1\right)}\omega\right)-\cos\left(W_{n;l_{1},l_{2}}^{\left(2\right)}\omega\right)-\cos\left(W_{n;l_{1},l_{2}}^{\left(3\right)}\omega\right),\label{eq:Real part of f^=00007B(dl>=00003D1)=00007D -- 1}
\end{equation}

\noindent and

\begin{equation}
\text{Im}\left[f_{n;l_{1},l_{2}}^{\left(\eta,\Delta l\ge1\right)}\left(\omega\right)\right]=-\sin\left(W_{n;l_{1},l_{2}}^{\left(0\right)}\omega\right)-\sin\left(W_{n;l_{1},l_{2}}^{\left(1\right)}\omega\right)+\sin\left(W_{n;l_{1},l_{2}}^{\left(2\right)}\omega\right)+\sin\left(W_{n;l_{1},l_{2}}^{\left(3\right)}\omega\right).\label{eq:Imaginary part of f^=00007B(dl>=00003D1)=00007D -- 1}
\end{equation}

\noindent Note that we can rewrite Eq.~\eqref{eq:Real part of f^=00007B(dl=00003D0)=00007D -- 1}
as

\begin{equation}
\text{Re}\left[f_{n;l_{1}}^{\left(\eta,\Delta l=0\right)}\left(\omega\right)\right]=\cos\left(W_{n;l_{1},l_{2}}^{\left(0\right)}\omega\right)+\cos\left(W_{n;l_{1},l_{2}}^{\left(1\right)}\omega\right)-\cos\left(W_{n;l_{1},l_{2}}^{\left(2\right)}\omega\right)-\cos\left(W_{n;l_{1},l_{2}}^{\left(3\right)}\omega\right),\label{eq:Real part of f^=00007B(dl=00003D0)=00007D -- 2}
\end{equation}

\noindent using the summing convention of Eq.~\eqref{eq:summing convention -- 1}
such that the right-hand-side is effectively identical in form to
that of Eq.~\eqref{eq:Real part of f^=00007B(dl>=00003D1)=00007D -- 1}.

As mentioned at the beginning of this section, the integrand of the
frequency integral that is computed in order to evaluate $\eta_{\nu;r;n;l_{1},l_{2}}$
has a removable singularity at $\omega=0$ {[}see Eqs.~\eqref{eq:eta_=00007Bv; r; n; l_1, l_2=00007D in main manuscript -- 1}-\eqref{eq:sign function in main manuscript -- 1}{]}.
As a result, special care needs to be taken in evaluating the integrand
in the neighbourhood of $\omega=0$. First, we Taylor expand $\frac{1}{\omega^{2}}\text{Re}\left[f_{n;l_{1},l_{2}}^{\left(\eta\right)}\left(\omega\right)\right]$
{[}using Eqs.~\eqref{eq:Real part of f^=00007B(dl>=00003D1)=00007D -- 1}
and \eqref{eq:Real part of f^=00007B(dl=00003D0)=00007D -- 2} in
the process{]}:

\begin{equation}
\frac{1}{\omega^{2}}\text{Re}\left[f_{n;l_{1},l_{2}}^{\left(\eta\right)}\left(\omega\right)\right]=\sum_{m=1}^{\infty}\frac{\left(-1\right)^{m}\omega^{2m-2}}{\left(2m\right)!}\left[\left\{ W_{n;l_{1},l_{2}}^{\left(0\right)}\right\} ^{2m}+\left\{ W_{n;l_{1},l_{2}}^{\left(1\right)}\right\} ^{2m}-\left\{ W_{\nu n;l_{1},l_{2}}^{\left(2\right)}\right\} ^{2m}-\left\{ W_{n;l_{1},l_{2}}^{\left(3\right)}\right\} ^{2m}\right].\label{eq:Taylor expand Re=00005Bf=00005D / w^2 -- 1}
\end{equation}

\noindent Next, we Taylor expand $\frac{1}{\omega^{2}}\text{Im}\left[f_{n;l_{1},l_{2}}^{\left(\eta,\Delta l\ge1\right)}\left(\omega\right)\right]$
{[}using Eq.~\eqref{eq:Imaginary part of f^=00007B(dl>=00003D1)=00007D -- 1}
in the process{]}:

\begin{align}
\frac{1}{\omega^{2}}\text{Im}\left[f_{n;l_{1},l_{2}}^{\left(\eta,\Delta l\ge1\right)}\left(\omega\right)\right] & =-\sum_{m=1}^{\infty}\frac{\left(-1\right)^{m}\left(\Delta t\right)^{2m+1}\left(\omega\right)^{2m-1}}{\left(2m+1\right)!}\left[\left\{ W_{n;l_{1},l_{2}}^{\left(0\right)}\right\} ^{2m+1}+\left\{ W_{n;l_{1},l_{2}}^{\left(1\right)}\right\} ^{2m+1}\right.\nonumber \\
 & \phantom{\mathrel{=}-\sum_{m=1}^{\infty}\frac{\left(-1\right)^{m}\left(\Delta t\right)^{2m+1}\left(\omega\right)^{2m-1}}{\left(2m+1\right)!}}\left.\quad\mathop{-}\left\{ W_{n;l_{1},l_{2}}^{\left(2\right)}\right\} ^{2m+1}-\left\{ W_{n;l_{1},l_{2}}^{\left(3\right)}\right\} ^{2m+1}\right].\label{eq:Taylor expand Im=00005Bf=00005D / w^2 -- 1}
\end{align}

\noindent In practice, we perform both Taylor expansions up to $m=3$
in the neighbourhood $\left|\omega\right|<10^{-3}/W_{n;l_{1},l_{2}}^{\left(\max\right)}$
where

\begin{equation}
W_{n;l_{1},l_{2}}^{\left(\max\right)}=\max\left(\left\{ W_{n;l_{1},l_{2}}^{\left(j\right)}\right\} _{u=0}^{3}\right),\label{eq:W_=00007Bn; l_1, l_2=00007D^=00007B(max)=00007D -- 1}
\end{equation}

\noindent otherwise we use Eqs\@.~\eqref{eq:Imaginary part of f^=00007B(dl=00003D0)=00007D -- 1}
and \eqref{eq:Real part of f^=00007B(dl>=00003D1)=00007D -- 1}-\eqref{eq:Real part of f^=00007B(dl=00003D0)=00007D -- 2}. 

From Eq.~\eqref{eq:eta in numerical evaluation appendix -- 2}, we
see that one must evaluate $A_{\nu;r;T}\left(\omega\right)$ \textendash{}
which is the spectral density of the noise coupled to the $\nu$-component
of the spin at site $r$ and temperature $T$ \textendash{} in order
to evaluate $\eta_{\nu;r;n;l_{1},l_{2}}$. As discussed in Sec.~\ref{sec:Spectral densities of noise},
$A_{\nu;r;T}\left(\omega\right)$ will comprise of multiple components:

\begin{equation}
A_{\nu;r;T}\left(\omega\right)=\sum_{\varsigma}A_{\nu;r;T;\varsigma}\left(\omega\right),\label{eq:A_=00007Bv; T=00007D(omega) breakdown into subcomponent -- 1}
\end{equation}

\noindent When $A_{\nu;r;T}\left(\omega\right)$ contains multiple
components as described above, we can express $\eta_{\nu;r;n;l_{1},l_{2}}$
as

\begin{equation}
\eta_{\nu;r;n;l_{1},l_{2}}=\sum_{\varsigma}\eta_{\nu;r;n;l_{1},l_{2};\varsigma},\label{eq:eta in numerical evaluation appendix -- 2}
\end{equation}

\noindent where

\begin{equation}
\eta_{\nu;r;n;l_{1},l_{2};\varsigma}=\int_{-\infty}^{\infty}\frac{d\omega}{2\pi}\,A_{\nu;r;T;\varsigma}\left(\omega\right)\frac{f_{n;l_{1},l_{2}}^{\left(\eta\right)}\left(\omega\right)}{\omega^{2}}.\label{eq:eta_=00007Bv; n; k_1, k_2; varsigma=00007D -- 1}
\end{equation}

\noindent In practice, it is best to evaluate each $\eta_{\nu;r;n;l_{1},l_{2};\varsigma}$
according to Eq.~\eqref{eq:eta_=00007Bv; n; k_1, k_2; varsigma=00007D -- 1}
and then add them all together according to Eq.~\eqref{eq:eta in numerical evaluation appendix -- 2}
to get $\eta_{\nu;r;n;l_{1},l_{2}}$.

Next, using Eqs.~\eqref{eq:decomposing f^=00007B(dl=00003D0)=00007D -- 1}-\eqref{eq:W_=00007Bn; l_1, l_2=00007D^=00007B(max)=00007D -- 1},
we partition the integral in Eq.~\eqref{eq:eta_=00007Bv; n; k_1, k_2; varsigma=00007D -- 1}
as:

\begin{equation}
\eta_{\nu;r;n;l_{1},l_{2};\varsigma}\approx\sum_{a=0}^{8}\left\{ \mathcal{I}_{\nu;r;n;l_{1},l_{2};\varsigma;a}^{\left(\eta,1\right)}+i\mathcal{I}_{\nu;r;n;l_{1},l_{2};\varsigma;a}^{\left(\eta,2\right)}\right\} ,\label{eq:partitioning eta integral -- 1}
\end{equation}

\noindent where

\begin{equation}
\mathcal{I}_{\nu;r;n;l_{1},l_{2};\varsigma;a\in\left\{ 2,3,5,6\right\} }^{\left(\eta,1\right)}=\int_{\omega_{\nu;r;n;l_{1};l_{2};\varsigma;a}}^{\omega_{\nu;r;n;l_{1};l_{2};\varsigma;a+1}}\frac{d\omega}{2\pi}\,A_{\nu;r;T;\varsigma}\left(\omega\right)F_{n;l_{1},l_{2}}^{\left(\eta,1\right)}\left(\omega\right),\label{eq:real part of eta integral partitions a=00003D2,3,5,6 -- 1}
\end{equation}

\begin{equation}
\mathcal{I}_{\nu;r;n;l_{1},l_{2};\varsigma;a\in\left\{ 2,3,5,6\right\} }^{\left(\eta,2\right)}=\int_{\omega_{\nu;r;n;l_{1};l_{2};\varsigma;a}}^{\omega_{\nu;r;n;l_{1};l_{2};\varsigma;a+1}}\frac{d\omega}{2\pi}\,A_{\nu;r;T;\varsigma}\left(\omega\right)F_{n;l_{1},l_{2}}^{\left(\eta,2\right)}\left(\omega\right),\label{eq:imaginary part of eta integral partitions a=00003D2,3,5,6 -- 1}
\end{equation}

\begin{equation}
\mathcal{I}_{\nu;r;n;l_{1},l_{2};\varsigma;a\in\left\{ 0,1,4,7,8\right\} }^{\left(\eta,1\right)}=\sum_{j=0}^{3}\mathcal{I}_{\nu;r;n;l_{1},l_{2};\varsigma;a;j}^{\left(\eta,1\right)},\label{eq:real part of eta integral partitions a=00003D0,1,4,7,8 -- 1}
\end{equation}

\begin{equation}
\mathcal{I}_{\nu;r;n;l_{1},l_{2};\varsigma;a\in\left\{ 0,1,4,7,8\right\} }^{\left(\eta,2\right)}=\sum_{j=0}^{3}\mathcal{I}_{\nu;r;n;l_{1},l_{2};\varsigma;a;j}^{\left(\eta,2\right)},\label{eq:imaginary part of eta integral partitions a=00003D0,1,4,7,8 -- 1}
\end{equation}

\noindent with

\begin{equation}
\omega_{\nu;r;n;l_{1};l_{2};\varsigma;a=0}=-\omega_{\nu;\varsigma}^{\left(\text{UV}\right)},\label{eq:eta integration pt a=00003D0 -- 1}
\end{equation}

\begin{equation}
\omega_{\nu;r;n;l_{1};l_{2};\varsigma;a=2}=\min\left(\max\left(-\omega_{\nu;\varsigma}^{\left(\text{UV}\right)},\omega^{\left(A\right)}\right),-\omega_{\nu;\varsigma}^{\left(\text{IR}\right)}\right),\label{eq:eta integration pt a=00003D2 -- 1}
\end{equation}

\begin{equation}
\omega_{\nu;r;n;l_{1};l_{2};\varsigma;a=4}=-\omega_{\nu;\varsigma}^{\left(\text{IR}\right)},\label{eq:eta integration pt a=00003D4 -- 1}
\end{equation}

\begin{equation}
\omega_{\nu;r;n;l_{1};l_{2};\varsigma;a=1}=\begin{cases}
\omega^{\left(B\right)}, & \text{if }\omega_{\nu;r;n;l_{1};l_{2};\varsigma;a=0}<\omega^{\left(B\right)}<\omega_{\nu;r;n;l_{1};l_{2};\varsigma;a=2},\\
\omega_{\nu;r;n;l_{1};l_{2};\varsigma;a=0}, & \text{otherwise},
\end{cases},\label{eq:eta integration pt a=00003D1 -- 1}
\end{equation}

\begin{equation}
\omega_{\nu;r;n;l_{1};l_{2};\varsigma;a=3}=\begin{cases}
\omega^{\left(B\right)}, & \text{if }\omega_{\nu;r;n;l_{1};l_{2};\varsigma;a=2}<\omega^{\left(B\right)}<\omega_{\nu;r;n;l_{1};l_{2};\varsigma;a=4},\\
\omega_{\nu;r;n;l_{1};l_{2};\varsigma;a=4}, & \text{otherwise},
\end{cases},\label{eq:eta integration pt a=00003D3 -- 1}
\end{equation}

\begin{equation}
\omega_{\nu;r;n;l_{1};l_{2};\varsigma;a}=-\omega_{\nu;r;n;l_{1};l_{2};\varsigma;9-a},\quad\text{if }a\ge5.\label{eq:eta integration pt a>=00003D5 -- 1}
\end{equation}

\begin{equation}
\omega^{\left(A\right)}=\begin{cases}
-\frac{\pi}{W_{n;l_{1},l_{2}}^{\left(\max\right)}}, & \text{if }W_{n;l_{1},l_{2}}^{\left(\max\right)}\neq0,\\
-\infty, & \text{otherwise},
\end{cases}\label{eq:omega^(A) -- 1}
\end{equation}

\begin{equation}
\omega^{\left(B\right)}=-\frac{25}{\beta},\label{omega^(B) -- 1}
\end{equation}

\noindent $\beta=1/\left(k_{B}T\right)$ being the inverse temperature,
$\omega_{\nu;\varsigma}^{\left(\text{IR}\right)}$ and $\omega_{\nu;\varsigma}^{\left(\text{UV}\right)}$
being the hard infrared and ultraviolet cutoff frequencies of $A_{\nu;r;T}\left(\omega\right)$
{[}see Eqs.~\eqref{eq:introducing IR cutoff -- 2}-\eqref{eq:cutoff freq inequality -- 2}{]},

\begin{align}
\left.F_{n;l_{1},l_{2}}^{\left(\eta,1\right)}\left(\omega\right)\right|_{\left|\omega\right|<10^{-3}/W_{n;l_{1},l_{2}}^{\left(\max\right)}} & =\sum_{m=1}^{3}\frac{\left(-1\right)^{m}\omega^{2m-2}}{\left(2m\right)!}\left[\left\{ W_{n;l_{1},l_{2}}^{\left(0\right)}\right\} ^{2m}+\left\{ W_{n;l_{1},l_{2}}^{\left(1\right)}\right\} ^{2m}\right.\nonumber \\
 & \phantom{\mathrel{=}\sum_{m=1}^{\infty}\frac{\left(-1\right)^{m}\omega^{2m-2}}{\left(2m\right)!}}\left.\quad\mathop{-}\left\{ W_{n;l_{1},l_{2}}^{\left(2\right)}\right\} ^{2m}-\left\{ W_{n;l_{1},l_{2}}^{\left(3\right)}\right\} ^{2m}\right],\label{eq:Re=00005BF=00005D in numerical evaluation summary -- 1}
\end{align}

\begin{align}
\left.F_{n;l_{1},l_{2}}^{\left(\eta,1\right)}\left(\omega\right)\right|_{\left|\omega\right|\ge10^{-3}/W_{n;l_{1},l_{2}}^{\left(\max\right)}} & =\frac{1}{\omega^{2}}\left[\cos\left(W_{n;l_{1},l_{2}}^{\left(0\right)}\omega\right)+\cos\left(W_{n;l_{1},l_{2}}^{\left(1\right)}\omega\right)\right.\nonumber \\
 & \phantom{\mathrel{=}\frac{1}{\omega^{2}}}\left.\quad\mathop{-}\cos\left(W_{n;l_{1},l_{2}}^{\left(2\right)}\omega\right)-\cos\left(W_{n;l_{1},l_{2}}^{\left(3\right)}\omega\right)\right],\label{eq:Re=00005BF=00005D in numerical evaluation summary -- 2}
\end{align}

\begin{align}
\left.F_{n;l_{1},l_{2}}^{\left(\eta,2\right)}\left(\omega\right)\right|_{\left|\omega\right|<10^{-3}/W_{n;l_{1},l_{2}}^{\left(\max\right)}} & =-\left(1-\delta_{l_{1},l_{2}}\right)\sum_{m=1}^{\infty}\frac{\left(-1\right)^{m}\left(\omega\right)^{2m-1}}{\left(2m+1\right)!}\left[\left\{ W_{n;l_{1},l_{2}}^{\left(0\right)}\right\} ^{2m+1}+\left\{ W_{n;l_{1},l_{2}}^{\left(1\right)}\right\} ^{2m+1}\right.\nonumber \\
 & \phantom{\mathrel{=}-\left(1-\delta_{l_{1},l_{2}}\right)\sum_{m=1}^{\infty}\frac{\left(-1\right)^{m}\left(\omega\right)^{2m-1}}{\left(2m+1\right)!}}\left.\quad\mathop{-}\left\{ W_{n;l_{1},l_{2}}^{\left(2\right)}\right\} -\left\{ W_{n;l_{1},l_{2}}^{\left(3\right)}\right\} ^{2m+1}\right],\label{eq:Im=00005BF=00005D in numerical evaluation summary -- 1}
\end{align}

\begin{align}
\left.F_{n;l_{1},l_{2}}^{\left(\eta,2\right)}\left(\omega\right)\right|_{\left|\omega\right|\ge10^{-3}/W_{n;l_{1},l_{2}}^{\left(\max\right)}} & =\frac{\left(1-\delta_{l_{1},l_{2}}\right)}{\omega^{2}}\left[-\sin\left(W_{n;l_{1},l_{2}}^{\left(0\right)}\omega\right)-\sin\left(W_{n;l_{1},l_{2}}^{\left(1\right)}\omega\right)\right.\nonumber \\
 & \phantom{\mathrel{=}\frac{\left(1-\delta_{l_{1},l_{2}}\right)}{\omega^{2}}}\left.\quad\mathop{+}\sin\left(W_{n;l_{1},l_{2}}^{\left(2\right)}\omega\right)+\sin\left(W_{n;l_{1},l_{2}}^{\left(3\right)}\omega\right)\right],\label{eq:Im=00005BF=00005D in numerical evaluation summary -- 2}
\end{align}

\begin{equation}
\mathcal{I}_{\nu;r;n;l_{1},l_{2};\varsigma;a\in\left\{ 0,1,4,7,8\right\} ;j}^{\left(\eta,1\right)}=\left(-1\right)^{\left\lfloor j/2\right\rfloor }\int_{\omega_{\nu;r;n;l_{1};l_{2};\varsigma;a}}^{\omega_{\nu;r;n;l_{1};l_{2};\varsigma;a+1}}\frac{d\omega}{2\pi}\,\frac{A_{\nu;r;T;\varsigma}\left(\omega\right)}{\omega^{2}}\cos\left(W_{n;l_{1},l_{2}}^{\left(j\right)}\omega\right),\label{eq:real part of eta integral partitions a=00003D0,1,4,7,8 -- 2}
\end{equation}

\begin{equation}
\mathcal{I}_{\nu;r;n;l_{1},l_{2};\varsigma;a\in\left\{ 0,1,4,7,8\right\} ;j}^{\left(\eta,2\right)}=-\left(-1\right)^{\left\lfloor j/2\right\rfloor }\left(1-\delta_{l_{1},l_{2}}\right)\int_{\omega_{\nu;r;n;l_{1};l_{2};\varsigma;a}}^{\omega_{\nu;r;n;l_{1};l_{2};\varsigma;a+1}}\frac{d\omega}{2\pi}\,\frac{A_{\nu;r;T;\varsigma}\left(\omega\right)}{\omega^{2}}\sin\left(W_{n;l_{1},l_{2}}^{\left(j\right)}\omega\right),\label{eq:imaginary part of eta integral partitions a=00003D0,1,4,7,8 -- 2}
\end{equation}

\noindent In practice, it is best to evaluate each $\mathcal{I}_{\nu;r;n;l_{1},l_{2};\varsigma;a\in\left\{ 2,3,5,6\right\} }^{\left(\eta,1\right)}$,
$\mathcal{I}_{\nu;r;n;l_{1},l_{2};\varsigma;a\in\left\{ 2,3,5,6\right\} }^{\left(\eta,2\right)}$,
\\
$\mathcal{I}_{\nu;r;n;l_{1},l_{2};\varsigma;a\in\left\{ 0,1,4,7,8\right\} ;j}^{\left(\eta,1\right)}$,
and $\mathcal{I}_{\nu;r;n;l_{1},l_{2};\varsigma;a\in\left\{ 0,1,4,7,8\right\} ;j}^{\left(\eta,2\right)}$
according to Eqs.~\eqref{eq:real part of eta integral partitions a=00003D2,3,5,6 -- 1}-\eqref{eq:imaginary part of eta integral partitions a=00003D0,1,4,7,8 -- 2},
and then add them all together according to Eqs.~\eqref{eq:eta in numerical evaluation appendix -- 2}
and \eqref{eq:partitioning eta integral -- 1} to get $\eta_{\nu;r;n;l_{1},l_{2}}$.

In $\texttt{spinbosonchain}$, the integrals in Eqs.~\eqref{eq:real part of eta integral partitions a=00003D2,3,5,6 -- 1}
and \eqref{eq:imaginary part of eta integral partitions a=00003D2,3,5,6 -- 1}
are evaluated using $\texttt{Python}$'s \\
$\texttt{scipy.integrate.quad}$'s default adaptive integrator since
the corresponding integrands in these cases are not highly oscillatory,
considering the limits of integrations. An upper bound of $2000$
subintervals is used in the integrator algorithm, which should yield
a negligible integration error compared to the other sources of error
in our QUAPI approach {[}e.g. the Trotterization error{]}. When $W_{n;l_{1},l_{2}}^{\left(j\right)}=0$,
the integral in Eq.~\eqref{eq:imaginary part of eta integral partitions a=00003D0,1,4,7,8 -- 2}
evaluates to zero hence no explicit numerical integration is required,
whereas the integral in Eq.~\eqref{eq:real part of eta integral partitions a=00003D0,1,4,7,8 -- 2}
can be evaluated using $\texttt{scipy.integrate.quad}$'s default
adaptive integrator with an upper bound of $2000$ subintervals used
in the integrator algorithm, since the integrand is not highly oscillatory
\textendash{} if at all. When $W_{n;l_{1},l_{2}}^{\left(j\right)}>0$,
the integrands in Eqs.~\eqref{eq:real part of eta integral partitions a=00003D0,1,4,7,8 -- 2}
and \eqref{eq:imaginary part of eta integral partitions a=00003D0,1,4,7,8 -- 2}
are highly oscillatory. Because of this, the integral in Eq.~\eqref{eq:real part of eta integral partitions a=00003D0,1,4,7,8 -- 2}
is evaluated using $\texttt{scipy.integrate.quad}$ with the keyword
arguments $\texttt{weight=`cos'}$ and $\texttt{wvar=}W_{n;l_{1},l_{2}}^{\left(j\right)}$,
and in Eq.~\eqref{eq:imaginary part of eta integral partitions a=00003D0,1,4,7,8 -- 2}
the integral is evaluated with the keyword arguments $\texttt{weight=`sin'}$
and $\texttt{wvar=}W_{n;l_{1},l_{2}}^{\left(j\right)}$. In both cases,
the upper bound of subintervals used are proportional to the number
of completed oscillations across the range of integration. In $\texttt{spinbosonchain}$
we set the upper bound of subintervals to:

\begin{equation}
\texttt{limit}=2000\left\lfloor \frac{\omega_{\nu;r;n;l_{1};l_{2};\varsigma;a+1}-\omega_{\nu;r;n;l_{1};l_{2};\varsigma;a}}{\omega_{\nu;r;n;l_{1};l_{2};\varsigma;a}}+1\right\rfloor ,\label{eq:upper bound of integration subintervals -- 1}
\end{equation}

\noindent where $\texttt{limit}$ refers to the keyword argument in
$\texttt{scipy.integrate.quad}$ that specifies the upper bound of
subintervals used.

\newpage{}

\section{MPS compression techniques used in $\texttt{spinbosonchain}$\label{sec:MPS compression techniques used in sbc}}

As discussed in Sec.~\ref{subsec:Tensor network representation of influence functional -- 1},
$\texttt{spinbosonchain}$ represents the local path influence functionals
$\mathring{I}_{r;n}\left(\cdots\right)$ {[}Eq.~\eqref{eq:ring I_=00007Br; n=00007D -- 1}{]}
by a set of MPS's. Similarly, the quantities $\mathring{\rho}_{m}^{\left(\vdash\right)}\left(\mathbf{b}_{m}\right)$,
$\mathring{\rho}_{n;m}\left(\mathbf{b}_{m}\right)$ {[}introduced
in Sec.~\ref{subsec:Rearranging terms in the path integral expression of rho^(A) further -- 1}{]}
and the system's reduced density matrix $\mathring{\rho}_{n}^{\left(A\right)}\left(\mathbf{j}_{\left(n+1\right)\Delta m}\right)$
are also represented by MPS's. Each of the aforementioned MPS's are
updated by applying to it a MPO, which increases subsequently the
bond dimensions, thus increasing memory requirements and the simulation
runtime. To address this, $\texttt{spinbosonchain}$ offers a variety
of MPS compression methods that can be used to improve the computationally
efficiency of the simulations.

For finite MPS's, users can apply either the ``zip-up'' method \citep{Stoudenmire1}
or the ``direct'' method \citep{Stoudenmire1}, the latter consists
of performing a singular value decomposition (SVD) or QR sweep in
one direct without truncating the Schmidt spectra, followed by a SVD
sweep in the opposite direction with truncation. By truncating the
Schmidt spectra, one truncates the bond dimensions. Of course, such
an operation introduces a truncation error, however this can be controlled
to some degree. For a detailed discussion on SVD truncation sweeps,
see e.g. Ref.~\citep{Schollwock1}. For detailed discussions on the
``zip-up'' method, see e.g. Refs.~\citep{Paeckel1,Stoudenmire1}.

One can improve the compression of a finite MPS by performing a subsequent
set of variational compression sweeps. This may be worthwhile for
a MPS representing a local path influence functional, however it is
likely to be too costly for a MPS that spans space (e.g. one representing
the system state) rather than time (e.g. a local path influence functional).
It is recommended to consider only this feature for MPS's spanning
time. Moreover, this variational compression feature is not available
for infinite MPS's, which are MPS's with infinitely many unit cells.

For infinite MPS's with single site unit cells, $\texttt{spinbosonchain}$
employs the compression method described in Ref.~\citep{Orus2}.
For infinite MPS's with multi-site unit cells, $\texttt{spinbosonchain}$
employs the compression method described in Ref.~\citep{Orus2,McCulloch1}.

In $\texttt{spinbosonchain}$, users specify MPS compression related
parameters using the \\
$\texttt{spinbosonchain.compress.Params}$ class. Basic SVD and QR
related operations are found in the modules \\
$\texttt{spinbosonchain.\_svd}$ and $\texttt{spinbosonchain.\_qr}$
respectively. All compression methods in $\texttt{spinbosonchain}$
are implemented in the $\texttt{spinbosonchain.\_mpomps}$ module.
This module also includes implementations for MPO applications to
MPS's.

\newpage{}

\bibliographystyle{apsrev4-1}
\bibliography{manuscript-revtex}

\begin{thebibliography}{21}%
\makeatletter
\providecommand \@ifxundefined [1]{%
 \@ifx{#1\undefined}
}%
\providecommand \@ifnum [1]{%
 \ifnum #1\expandafter \@firstoftwo
 \else \expandafter \@secondoftwo
 \fi
}%
\providecommand \@ifx [1]{%
 \ifx #1\expandafter \@firstoftwo
 \else \expandafter \@secondoftwo
 \fi
}%
\providecommand \natexlab [1]{#1}%
\providecommand \enquote  [1]{``#1''}%
\providecommand \bibnamefont  [1]{#1}%
\providecommand \bibfnamefont [1]{#1}%
\providecommand \citenamefont [1]{#1}%
\providecommand \href@noop [0]{\@secondoftwo}%
\providecommand \href [0]{\begingroup \@sanitize@url \@href}%
\providecommand \@href[1]{\@@startlink{#1}\@@href}%
\providecommand \@@href[1]{\endgroup#1\@@endlink}%
\providecommand \@sanitize@url [0]{\catcode `\\12\catcode `\$12\catcode
  `\&12\catcode `\#12\catcode `\^12\catcode `\_12\catcode `\%12\relax}%
\providecommand \@@startlink[1]{}%
\providecommand \@@endlink[0]{}%
\providecommand \url  [0]{\begingroup\@sanitize@url \@url }%
\providecommand \@url [1]{\endgroup\@href {#1}{\urlprefix }}%
\providecommand \urlprefix  [0]{URL }%
\providecommand \Eprint [0]{\href }%
\providecommand \doibase [0]{http://dx.doi.org/}%
\providecommand \selectlanguage [0]{\@gobble}%
\providecommand \bibinfo  [0]{\@secondoftwo}%
\providecommand \bibfield  [0]{\@secondoftwo}%
\providecommand \translation [1]{[#1]}%
\providecommand \BibitemOpen [0]{}%
\providecommand \bibitemStop [0]{}%
\providecommand \bibitemNoStop [0]{.\EOS\space}%
\providecommand \EOS [0]{\spacefactor3000\relax}%
\providecommand \BibitemShut  [1]{\csname bibitem#1\endcsname}%
\let\auto@bib@innerbib\@empty
%</preamble>
\bibitem [{\citenamefont {{D-Wave Systems}}(2021)}]{sbc1}%
  \BibitemOpen
  \bibfield  {author} {\bibinfo {author} {\bibnamefont {{D-Wave Systems}}},\
  }\href@noop {} {\enquote {\bibinfo {title} {spinbosonchain: an open source
  library for simulating the dynamics of a generalized spin-boson chain
  model},}\ } (\bibinfo {year} {2021}),\ \bibinfo {note} {code available from
  \url{https://github.com/dwavesystems/spin-boson-chain}}\BibitemShut {NoStop}%
\bibitem [{\citenamefont {{Strathearn}}\ \emph {et~al.}(2018)\citenamefont
  {{Strathearn}}, \citenamefont {{Kirton}}, \citenamefont {{Kilda}},
  \citenamefont {{Keeling}},\ and\ \citenamefont {{Lovett}}}]{Strathearn1}%
  \BibitemOpen
  \bibfield  {author} {\bibinfo {author} {\bibfnamefont {A.}~\bibnamefont
  {{Strathearn}}}, \bibinfo {author} {\bibfnamefont {P.}~\bibnamefont
  {{Kirton}}}, \bibinfo {author} {\bibfnamefont {D.}~\bibnamefont {{Kilda}}},
  \bibinfo {author} {\bibfnamefont {J.}~\bibnamefont {{Keeling}}}, \ and\
  \bibinfo {author} {\bibfnamefont {B.~W.}\ \bibnamefont {{Lovett}}},\ }\href
  {\doibase 10.1038/s41467-018-05617-3} {\bibfield  {journal} {\bibinfo
  {journal} {Nature Communications}\ }\textbf {\bibinfo {volume} {9}},\
  \bibinfo {eid} {3322} (\bibinfo {year} {2018})},\ \Eprint
  {http://arxiv.org/abs/1711.09641} {arXiv:1711.09641 [quant-ph]} \BibitemShut
  {NoStop}%
\bibitem [{\citenamefont {Suzuki}\ \emph {et~al.}(2019)\citenamefont {Suzuki},
  \citenamefont {Oshiyama},\ and\ \citenamefont {Shibata}}]{Suzuki1}%
  \BibitemOpen
  \bibfield  {author} {\bibinfo {author} {\bibfnamefont {S.}~\bibnamefont
  {Suzuki}}, \bibinfo {author} {\bibfnamefont {H.}~\bibnamefont {Oshiyama}}, \
  and\ \bibinfo {author} {\bibfnamefont {N.}~\bibnamefont {Shibata}},\ }\href
  {\doibase 10.7566/JPSJ.88.061003} {\bibfield  {journal} {\bibinfo  {journal}
  {Journal of the Physical Society of Japan}\ }\textbf {\bibinfo {volume}
  {88}},\ \bibinfo {pages} {061003} (\bibinfo {year} {2019})},\ \Eprint
  {http://arxiv.org/abs/https://doi.org/10.7566/JPSJ.88.061003}
  {https://doi.org/10.7566/JPSJ.88.061003} \BibitemShut {NoStop}%
\bibitem [{\citenamefont {Oshiyama}\ \emph {et~al.}(2020)\citenamefont
  {Oshiyama}, \citenamefont {Shibata},\ and\ \citenamefont
  {Suzuki}}]{Oshiyama1}%
  \BibitemOpen
  \bibfield  {author} {\bibinfo {author} {\bibfnamefont {H.}~\bibnamefont
  {Oshiyama}}, \bibinfo {author} {\bibfnamefont {N.}~\bibnamefont {Shibata}}, \
  and\ \bibinfo {author} {\bibfnamefont {S.}~\bibnamefont {Suzuki}},\ }\href
  {\doibase 10.7566/JPSJ.89.104002} {\bibfield  {journal} {\bibinfo  {journal}
  {Journal of the Physical Society of Japan}\ }\textbf {\bibinfo {volume}
  {89}},\ \bibinfo {pages} {104002} (\bibinfo {year} {2020})},\ \Eprint
  {http://arxiv.org/abs/https://doi.org/10.7566/JPSJ.89.104002}
  {https://doi.org/10.7566/JPSJ.89.104002} \BibitemShut {NoStop}%
\bibitem [{Note1()}]{Note1}%
  \BibitemOpen
  \bibinfo {note} {The computational resources required to achieve a particular
  target error will depend on the system being studied. In some scenarios, e.g.
  systems dominated by strong low-frequency noise, the target error may scale
  poorly with computer memory and/or wall time, especially if one is interested
  in long-time dynamics.}\BibitemShut {Stop}%
\bibitem [{Ten(2019)}]{TensorNetworks1}%
  \BibitemOpen
  \href@noop {} {\enquote {\bibinfo {title} {Introducing tensornetwork, an open
  source library for efficient tensor calculations},}\ }\bibinfo {howpublished}
  {\url{https://ai.googleblog.com/2019/06/introducing-tensornetwork-open-source.html}}
  (\bibinfo {year} {2019})\BibitemShut {NoStop}%
\bibitem [{\citenamefont {Lehoucq}\ and\ \citenamefont
  {Sorensen}()}]{Lehoucq1}%
  \BibitemOpen
  \bibfield  {author} {\bibinfo {author} {\bibfnamefont {R.}~\bibnamefont
  {Lehoucq}}\ and\ \bibinfo {author} {\bibfnamefont {D.}~\bibnamefont
  {Sorensen}},\ }\href@noop {} {\enquote {\bibinfo {title} {Implicitly
  restarted arnoldi method},}\ }\bibinfo {note} {Available online at
  \url{http://www.netlib.org/utk/people/JackDongarra/etemplates/node220.html}}\BibitemShut
  {NoStop}%
\bibitem [{\citenamefont {{Stefanucci}}\ and\ \citenamefont {{van
  Leeuwen}}(2013)}]{Stefanucci1}%
  \BibitemOpen
  \bibfield  {author} {\bibinfo {author} {\bibfnamefont {G.}~\bibnamefont
  {{Stefanucci}}}\ and\ \bibinfo {author} {\bibfnamefont {R.}~\bibnamefont
  {{van Leeuwen}}},\ }\href {\doibase 10.1017/CBO9781139023979} {\emph
  {\bibinfo {title} {{Nonequilibrium Many-Body Theory of Quantum Systems}}}}\
  (\bibinfo  {publisher} {Cambridge University Press},\ \bibinfo {address} {New
  York, NY},\ \bibinfo {year} {2013})\BibitemShut {NoStop}%
\bibitem [{\citenamefont {Makri}\ and\ \citenamefont
  {Makarov}(1995{\natexlab{a}})}]{Makri1}%
  \BibitemOpen
  \bibfield  {author} {\bibinfo {author} {\bibfnamefont {N.}~\bibnamefont
  {Makri}}\ and\ \bibinfo {author} {\bibfnamefont {D.~E.}\ \bibnamefont
  {Makarov}},\ }\href {\doibase 10.1063/1.469508} {\bibfield  {journal}
  {\bibinfo  {journal} {J. Chem. Phys.}\ }\textbf {\bibinfo {volume} {102}},\
  \bibinfo {pages} {4600} (\bibinfo {year} {1995}{\natexlab{a}})},\ \Eprint
  {http://arxiv.org/abs/https://doi.org/10.1063/1.469508}
  {https://doi.org/10.1063/1.469508} \BibitemShut {NoStop}%
\bibitem [{\citenamefont {Makri}\ and\ \citenamefont
  {Makarov}(1995{\natexlab{b}})}]{Makri2}%
  \BibitemOpen
  \bibfield  {author} {\bibinfo {author} {\bibfnamefont {N.}~\bibnamefont
  {Makri}}\ and\ \bibinfo {author} {\bibfnamefont {D.~E.}\ \bibnamefont
  {Makarov}},\ }\href {\doibase 10.1063/1.469509} {\bibfield  {journal}
  {\bibinfo  {journal} {J. Chem. Phys.}\ }\textbf {\bibinfo {volume} {102}},\
  \bibinfo {pages} {4611} (\bibinfo {year} {1995}{\natexlab{b}})},\ \Eprint
  {http://arxiv.org/abs/https://doi.org/10.1063/1.469509}
  {https://doi.org/10.1063/1.469509} \BibitemShut {NoStop}%
\bibitem [{\citenamefont {Schollw{\"o}ck}(2011)}]{Schollwock1}%
  \BibitemOpen
  \bibfield  {author} {\bibinfo {author} {\bibfnamefont {U.}~\bibnamefont
  {Schollw{\"o}ck}},\ }\href {\doibase
  https://doi.org/10.1016/j.aop.2010.09.012} {\bibfield  {journal} {\bibinfo
  {journal} {Annals of Physics}\ }\textbf {\bibinfo {volume} {326}},\ \bibinfo
  {pages} {96 } (\bibinfo {year} {2011})},\ \bibinfo {note} {january 2011
  Special Issue}\BibitemShut {NoStop}%
\bibitem [{\citenamefont {Or\'us}\ and\ \citenamefont {Vidal}(2008)}]{Orus2}%
  \BibitemOpen
  \bibfield  {author} {\bibinfo {author} {\bibfnamefont {R.}~\bibnamefont
  {Or\'us}}\ and\ \bibinfo {author} {\bibfnamefont {G.}~\bibnamefont {Vidal}},\
  }\href {\doibase 10.1103/PhysRevB.78.155117} {\bibfield  {journal} {\bibinfo
  {journal} {Phys. Rev. B}\ }\textbf {\bibinfo {volume} {78}},\ \bibinfo
  {pages} {155117} (\bibinfo {year} {2008})}\BibitemShut {NoStop}%
\bibitem [{\citenamefont {McCulloch}(2008)}]{McCulloch1}%
  \BibitemOpen
  \bibfield  {author} {\bibinfo {author} {\bibfnamefont {I.~P.}\ \bibnamefont
  {McCulloch}},\ }\href@noop {} {\  (\bibinfo {year} {2008})},\ \Eprint
  {http://arxiv.org/abs/0804.2509} {arXiv:0804.2509 [cond-mat.str-el]}
  \BibitemShut {NoStop}%
\bibitem [{\citenamefont {{Chen}}\ and\ \citenamefont {{Wu}}(2003)}]{Chen1}%
  \BibitemOpen
  \bibfield  {author} {\bibinfo {author} {\bibfnamefont {K.}~\bibnamefont
  {{Chen}}}\ and\ \bibinfo {author} {\bibfnamefont {L.-A.}\ \bibnamefont
  {{Wu}}},\ }\href@noop {} {\bibfield  {journal} {\bibinfo  {journal} {Quant.
  Info. Comput.}\ }\textbf {\bibinfo {volume} {3}},\ \bibinfo {pages} {193}
  (\bibinfo {year} {2003})}\BibitemShut {NoStop}%
\bibitem [{\citenamefont {Rudolph}(2003)}]{Rudolph1}%
  \BibitemOpen
  \bibfield  {author} {\bibinfo {author} {\bibfnamefont {O.}~\bibnamefont
  {Rudolph}},\ }\href {\doibase 10.1103/PhysRevA.67.032312} {\bibfield
  {journal} {\bibinfo  {journal} {Phys. Rev. A}\ }\textbf {\bibinfo {volume}
  {67}},\ \bibinfo {pages} {032312} (\bibinfo {year} {2003})}\BibitemShut
  {NoStop}%
\bibitem [{\citenamefont {{Negele}}\ and\ \citenamefont
  {{Orland}}(1998)}]{Negele1}%
  \BibitemOpen
  \bibfield  {author} {\bibinfo {author} {\bibfnamefont {J.~W.}\ \bibnamefont
  {{Negele}}}\ and\ \bibinfo {author} {\bibfnamefont {H.}~\bibnamefont
  {{Orland}}},\ }\href@noop {} {\emph {\bibinfo {title} {{Quantum Many Particle
  Systems}}}}\ (\bibinfo  {publisher} {Addison-Wesley},\ \bibinfo {address}
  {Reading, MA},\ \bibinfo {year} {1998})\BibitemShut {NoStop}%
\bibitem [{\citenamefont {Ballentine}(2014)}]{Ballentine}%
  \BibitemOpen
  \bibfield  {author} {\bibinfo {author} {\bibfnamefont {L.~E.}\ \bibnamefont
  {Ballentine}},\ }\href@noop {} {\emph {\bibinfo {title} {Quantum Mechanics: A
  Modern Development, Second Edition}}}\ (\bibinfo  {publisher} {World
  Scientific Publishing Company},\ \bibinfo {year} {2014})\BibitemShut
  {NoStop}%
\bibitem [{\citenamefont {Altland}\ and\ \citenamefont
  {Simons}(2010)}]{Altland1}%
  \BibitemOpen
  \bibfield  {author} {\bibinfo {author} {\bibfnamefont {A.}~\bibnamefont
  {Altland}}\ and\ \bibinfo {author} {\bibfnamefont {B.}~\bibnamefont
  {Simons}},\ }\href {https://books.google.ca/books?id=GpF0Pgo8CqAC} {\emph
  {\bibinfo {title} {Condensed Matter Field Theory}}},\ Cambridge books online\
  (\bibinfo  {publisher} {Cambridge University Press},\ \bibinfo {year}
  {2010})\BibitemShut {NoStop}%
\bibitem [{\citenamefont {Gottfried}\ and\ \citenamefont
  {Yan}(2003)}]{Gottfried1}%
  \BibitemOpen
  \bibfield  {author} {\bibinfo {author} {\bibfnamefont {K.}~\bibnamefont
  {Gottfried}}\ and\ \bibinfo {author} {\bibfnamefont {T.~M.}\ \bibnamefont
  {Yan}},\ }\href@noop {} {\emph {\bibinfo {title} {Quantum Mechanics:
  Fundamentals}}}\ (\bibinfo  {publisher} {Springer New York},\ \bibinfo {year}
  {2003})\BibitemShut {NoStop}%
\bibitem [{\citenamefont {Stoudenmire}\ and\ \citenamefont
  {White}(2010)}]{Stoudenmire1}%
  \BibitemOpen
  \bibfield  {author} {\bibinfo {author} {\bibfnamefont {E.~M.}\ \bibnamefont
  {Stoudenmire}}\ and\ \bibinfo {author} {\bibfnamefont {S.~R.}\ \bibnamefont
  {White}},\ }\href {\doibase 10.1088/1367-2630/12/5/055026} {\bibfield
  {journal} {\bibinfo  {journal} {New J. Phys.}\ }\textbf {\bibinfo {volume}
  {12}},\ \bibinfo {pages} {055026} (\bibinfo {year} {2010})}\BibitemShut
  {NoStop}%
\bibitem [{\citenamefont {{Paeckel}}\ \emph {et~al.}(2019)\citenamefont
  {{Paeckel}}, \citenamefont {{K{\"o}hler}}, \citenamefont {{Swoboda}},
  \citenamefont {{Manmana}}, \citenamefont {{Schollw{\"o}ck}},\ and\
  \citenamefont {{Hubig}}}]{Paeckel1}%
  \BibitemOpen
  \bibfield  {author} {\bibinfo {author} {\bibfnamefont {S.}~\bibnamefont
  {{Paeckel}}}, \bibinfo {author} {\bibfnamefont {T.}~\bibnamefont
  {{K{\"o}hler}}}, \bibinfo {author} {\bibfnamefont {A.}~\bibnamefont
  {{Swoboda}}}, \bibinfo {author} {\bibfnamefont {S.~R.}\ \bibnamefont
  {{Manmana}}}, \bibinfo {author} {\bibfnamefont {U.}~\bibnamefont
  {{Schollw{\"o}ck}}}, \ and\ \bibinfo {author} {\bibfnamefont
  {C.}~\bibnamefont {{Hubig}}},\ }\href {\doibase 10.1016/j.aop.2019.167998}
  {\bibfield  {journal} {\bibinfo  {journal} {Annals of Physics}\ }\textbf
  {\bibinfo {volume} {411}},\ \bibinfo {eid} {167998} (\bibinfo {year}
  {2019})},\ \Eprint {http://arxiv.org/abs/1901.05824} {arXiv:1901.05824
  [cond-mat.str-el]} \BibitemShut {NoStop}%
\end{thebibliography}%

\end{document}